\documentclass[longauth]{aa}

\usepackage{natbib}
\bibpunct{(}{)}{;}{a}{}{,} % to follow the A&A style

\usepackage[varg]{txfonts}

\usepackage{bm}

\usepackage{enumerate}
\usepackage{xcolor} % use with \textcolor
\usepackage[T1]{fontenc}
\usepackage{ae,aecompl}
\usepackage{graphicx}   % Including figure files
\usepackage{amsmath}    % Advanced maths commands
\usepackage{amssymb}    % Extra maths symbols

\usepackage{enumitem}

\usepackage{array}
\newcolumntype{P}[1]{>{\raggedright\arraybackslash}p{#1}}

\newcommand{\cig}{\texttt{CIGALE}}
\newcommand{\magp}{\texttt{MAGPHYS}}
\newcommand{\gra}{\texttt{GRASIL}}
\newcommand{\hers}{{\it Herschel}}
\newcommand{\spit}{{\it Spitzer}}

\newcommand{\logoh}{12$+$log(O/H)}
\newcommand{\micron}{$\mu$m}

\newcommand{\zsun}{$Z_\odot$}
\newcommand{\msun}{$M_\odot$}
\newcommand{\lsun}{$L_\odot$}
\newcommand{\ergs}{erg\,s$^{-1}$}
\newcommand{\msunyr}{$M_\odot$\,yr$^{-1}$}

\newcommand{\hi}{H{\sc i}}

\newcommand{\mstar}{M$_{\rm star}$}
\newcommand{\mdust}{M$_{\rm dust}$}
\newcommand{\chisq}{$\chi^2$}
\newcommand{\redchisq}{$\chi^2_\nu$}
\newcommand{\rms}{{\it rms}} 
\newcommand{\rmsw}{{\it rms}$_w$} 

\newcommand{\av}{$A_{\rm V}$}
\newcommand{\afuv}{$A_{\rm FUV}$}
\newcommand{\tauinf}{$\tau_{\rm inf}$}
\newcommand{\tgal}{$t_{\rm gal}$}
\newcommand{\tesc}{$t_{\rm esc}$}
\newcommand{\rmc}{$R_{\rm gmc}$}
\newcommand{\rc}{$R_{\rm gal}$}
\newcommand{\fmol}{$f_{\rm mol}$}
\newcommand{\ebv}{E(B$-$V)}
\newcommand{\qpah}{$q_{\rm PAH}$}
\newcommand{\umin}{$U_{\rm min}$}
\newcommand{\umax}{$U_{\rm max}$}
\newcommand{\uion}{$U_{\rm ion}$}
\newcommand{\ha}{H$\alpha$}

\newcommand{\nii}{[N{\sc ii}]}

\newcommand{\ltir}{L$_{\rm TIR}$}
\newcommand{\lmipsone}{L$_{24}$}
\newcommand{\lks}{L$_{\rm Ks}$}
\newcommand{\liracone}{L$_{3.6}$}
\newcommand{\lwiseone}{L$_{\mathrm W1}$}
\newcommand{\upsiirac}{$\Upsilon^{[3.6]}$}
\newcommand{\upsiwise}{$\Upsilon^{{\mathrm W1}}$}

\newcommand{\lfuv}{L$_{\rm FUV}$}

\begin{document}

%\title[SEDs of nearby galaxies]{Fitting spectral energy distributions from 0.1\,\micron\ to 1\,mm of nearby star-forming galaxies}
\title{Comprehensive comparison of models for spectral energy distributions from 0.1\,\micron\ to 1\,mm of nearby star-forming galaxies }

\author{L.~K. Hunt \inst{\ref{inst:hunt}}
\and 
I.~De Looze \inst{\ref{inst:delooze_a},\ref{inst:delooze_b}}
\and
M.~Boquien \inst{\ref{inst:boquien}}
\and
R.~Nikutta \inst{\ref{inst:nikutta_a},\ref{inst:nikutta_b}}
\and
A.~Rossi \inst{\ref{inst:rossi}}
\and
S.~Bianchi \inst{\ref{inst:hunt}}
\and
D.~A. Dale \inst{\ref{inst:dale}}
\and
G.~L. Granato \inst{\ref{inst:granato}}
\and
R.~C. Kennicutt \inst{\ref{inst:rob_a},\ref{inst:rob_b},\ref{inst:rob_c}}
\and
L.~Silva \inst{\ref{inst:granato}}
\and
L.~Ciesla \inst{\ref{inst:ciesla_a},\ref{inst:ciesla_b}}
\and
M.~Rela{\~n}o \inst{\ref{inst:relano_a},\ref{inst:relano_b}}
\and
S.~Viaene \inst{\ref{inst:delooze_b},\ref{inst:viaene_b}}
\and
B.~Brandl \inst{\ref{inst:brandl}}
\and
D.~Calzetti \inst{\ref{inst:calzetti}}
\and
K.~V. Croxall \inst{\ref{inst:croxall_a},\ref{inst:croxall_b}}
\and
B.~T. Draine \inst{\ref{inst:draine}}
\and
M.~Galametz \inst{\ref{inst:galametz_a},\ref{inst:ciesla_a},\ref{inst:ciesla_b}}
\and
K.~D. Gordon \inst{\ref{inst:gordon_a},\ref{inst:delooze_b}}
\and
B.~A. Groves \inst{\ref{inst:groves}}
\and
G.~Helou \inst{\ref{inst:helou}}
\and
R.~Herrera-Camus \inst{\ref{inst:herreracamus}}
\and
J.~L. Hinz \inst{\ref{inst:rob_b}}
\and
J.~Koda \inst{\ref{inst:koda}}
\and
S.~Salim \inst{\ref{inst:salim}}
\and
K.~M. Sandstrom \inst{\ref{inst:sandstrom}}
\and
J.~D. Smith \inst{\ref{inst:smith}}
\and
C.~D. Wilson \inst{\ref{inst:wilson}}
\and
S.~Zibetti \inst{\ref{inst:hunt}}
%\and
%members of the broader KINGFISH collaboration \inst{\ref{inst:world}}
}

\offprints{L. K. Hunt}
\institute{INAF/Osservatorio Astrofisico di Arcetri, Largo Enrico Fermi 5, 50125 Firenze, Italy \label{inst:hunt}
\email{hunt@arcetri.astro.it}
\and 
Department of Physics and Astronomy, University College London, Gower Street, London WC1E 6BT, UK \label{inst:delooze_a}
\and
Sterrenkundig Observatorium, Universiteit Gent, Krijgslaan 281 S9, B-9000 Gent, Belgium \label{inst:delooze_b}
\and
%Unidad de Astronom{\'i}a, Fac. Cs. B{\'a}sicas, Universidad de Antofagasta, Avda. U. de Antofagasta 02800, Antofagasta, Chile 
Centro de Astronomía (CITEVA), Universidad de Antofagasta, Avenida Angamos 601, Antofagasta, Chile \label{inst:boquien}
\and
National Optical Astronomy Observatory, 950 N. Cherry Ave., Tucson, AZ 85719, USA \label{inst:nikutta_a}
\and
Instituto de Astrof{\'i}sica, Facultad de F{\'i}sica, Pontificia, 
Universidad Cat{\'o}lica de Chile, 306, Santiago 22, Chile \label{inst:nikutta_b}
\and
INAF/Osservatorio di Astrofisica e Scienza dello Spazio di Bologna, Via Gobetti 93/3, 40129 Bologna, Italy
\label{inst:rossi}
\and
Department of Physics and Astronomy, University of Wyoming, Laramie, WY 82071, USA \label{inst:dale}
\and
INAF/Osservatorio Astronomico di Trieste, Via G. B. Tiepolo 11, 34143 Trieste, Italy \label{inst:granato}
\and
Institute of Astronomy, University of Cambridge, Madingley Road, Cambridge CB3 0HA, UK \label{inst:rob_a}
\and
Steward Observatory, University of Arizona, 933 North Cherry Avenue, Tucson, AZ 85721, USA \label{inst:rob_b}
\and
Department of Physics \& Astronomy, Texas A\&M University, College Station, TX  777843 \label{inst:rob_c}
\and
IRFU, CEA, Universit\'e Paris-Saclay, F-91191 Gif-sur-Yvette, France \label{inst:ciesla_a}
\and
Universit\'e Paris Diderot, AIM, Sorbonne Paris Cit\'e, CEA, CNRS, F-91191 Gif-sur-Yvette, France \label{inst:ciesla_b}
\and
Dept. F{\'i}sica Te{\'o}rica y del Cosmos, Universidad de Granada, Granada, Spain \label{inst:relano_a}
\and
Instituto Universitario Carlos I de F{\'i}sica Te{\'o}rica y Computacional, Universidad de Granada, %18071, 
Granada, Spain \label{inst:relano_b} 
\and
Centre for Astrophysics Research, University of Hertfordshire, College Lane, Hatfield AL10 9AB, UK \label{inst:viaene_b}
\and
Leiden Observatory, Leiden University, P.O. Box 9513, 2300 RA Leiden, The Netherlands \label{inst:brandl}
\and
Department of Astronomy, University of Massachusetts, Amherst, MA 01003, USA \label{inst:calzetti}
\and
Department of Astronomy, The Ohio State University, 4051 McPherson Laboratory, 140 West 18th Avenue, Columbus, OH 43210, USA \label{inst:croxall_a}
\and
Illumination Works LLC, 5650 Blazer Parkway, Suite 152, Dublin OH 43017, USA \label{inst:croxall_b}
\and
Princeton University Observatory, Peyton Hall, Princeton, NJ 08544-1001, USA \label{inst:draine}
\and
European Southern Observatory, Karl-Schwarzschild-Strasse 2, D-85748, Garching, Germany \label{inst:galametz_a}
\and
%AIM, CEA, CNRS, Universit\'e Paris-Saclay, Universit\'e Paris Diderot, Sorbonne Paris Cit\'e, F-91191 Gif-sur-Yvette, France \label{inst:galametz_b}
%\and
Space Telescope Science Institute, 3700 San Martin Dr., Baltimore, MD 21218, USA \label{inst:gordon_a}
% \and
% Sterrenkundig Observatorium, Universiteit Gent, Gent, Belgium \label{inst:gordon_b}
\and
Research School of Astronomy and Astrophysics, The Australian National University, Canberra, ACT 2611, Australia\label{inst:groves}
\and
IPAC, California Institute of Technology, 1200 E. California Blvd, Pasadena, CA 91125, USA\label{inst:helou}
\and
Max-Planck-Institut f\"ur Extraterrestrische Physik, Giessenbachstr., D-85748 Garching, Germany \label{inst:herreracamus}
\and
Department of Physics and Astronomy, Stony Brook University, Stony Brook, NY 11794-3800, USA \label{inst:koda}
\and
Department of Astronomy, Indiana University, Bloomington, IN 47404, USA \label{inst:salim}
\and
Center for Astrophysics and Space Sciences, Department of Physics, University of California, San Diego, 9500 Gilman Drive, La Jolla, CA 92093, USA \label{inst:sandstrom}
\and
Department of Physics \& Astronomy, University of Toledo, 2801 W. Bancroft Street, Toledo, OH 43606, USA \label{inst:smith}
\and
Department of Physics and Astronomy, McMaster University, 1280 Main St. W., Hamilton, Ontario L8S 4M1, Canada \label{inst:wilson}
%\and
%Institutes throughout the world \label{inst:world}
}

\date{draft version \today}

 \titlerunning{SED fitting of nearby galaxies}
   \authorrunning{Hunt et al.}

\abstract{We have fit the 
far-ultraviolet (FUV) to sub-millimeter (850\,\micron) spectral energy distributions (SEDs) of the
61 galaxies from the ``Key Insights on Nearby Galaxies: A Far-Infrared Survey with \hers\,"
(KINGFISH).
The fitting has been performed using three models: 
the Code for Investigating GALaxy Evolution (CIGALE),
the GRAphite-SILicate approach (GRASIL),
and the Multiwavelength Analysis of Galaxy PHYSical properties (MAGPHYS).
We have analyzed the results of the three codes in terms of the SED shapes,
and by comparing the derived quantities 
with simple ``recipes'' %``reference'' values 
for stellar mass (\mstar), star-formation rate (SFR),
dust mass (\mdust), and monochromatic luminosities.
Although the algorithms rely on different assumptions for star-formation history,
dust attenuation and dust reprocessing, they all well approximate the observed SEDs
and are in generally good agreement for the associated quantities.
However, the three codes show very different behavior in the mid-infrared regime: 
% referee
in the 5--10\,\micron\ region dominated by PAH emission, and also
between 25 and 70\,\micron\ where there are no observational constraints for the KINGFISH sample.
% referee
We find that different algorithms give discordant SFR estimates for galaxies with low specific SFR,
and that the standard recipes for calculating FUV absorption overestimate the %FUV
extinction compared to the SED-fitting results. 
Results also suggest that assuming a ``standard'' constant stellar mass-to-light ratio overestimates
\mstar\ relative to the SED fitting, and 
we provide new SED-based formulations %for the determination of IRX from monochromatic FIR luminosities, and 
for estimating \mstar\ from %IRAC 3.6\,\micron\ luminosities.
WISE W1 (3.4\,\micron) luminosities and colors.
From a principal component analysis of \mstar, SFR, \mdust, and O/H, we reproduce previous
scaling relations among \mstar, SFR, and O/H, and find that \mdust\ can be predicted to within $\sim 0.3$\,dex
using only \mstar\ and SFR.
}

\maketitle

%---------------------------------------------------------------

\section{Introduction}
\label{sec:intro}

As galaxies form and evolve, their spectral energy distributions (SEDs) are characterized by different shapes.
Dust grains reprocess stellar radiation to a degree which depends on many factors, but mainly on the galaxy's evolutionary state 
and its star-formation history (SFH).
Stars form in dense, cool giant molecular clouds and complexes (GMCs), and heat the surrounding dust; % to relatively high temperatures;
as the stars age, the dust cools, and the stars emerge from their natal clouds.
As evolution proceeds, the dust in the diffuse interstellar medium (ISM) is heated by the more quiescent interstellar
radiation field (ISRF) 
of an older stellar population. %more evolved stars. 
Thus dust emission is a fundamental probe of the SFH of a galaxy, and the current phase of its evolution.
A direct comparison of luminosity emitted by dust compared to that by stars shows that, overall, roughly half of
the stellar light is reprocessed by dust over cosmic time \citep{hauser01,dole06,franceschini08}.

Over the last two decades, the increasing availability of data from ultraviolet (UV) to far-infrared (FIR) wavelengths 
has led to the development of several physically-motivated models for fitting galaxy SEDs.
Among these are the Code for Investigating GALaxy Evolution \citep[CIGALE,][]{noll09},
the ``GRAphite-SILicate'' approach \citep[GRASIL,][]{silva98}, and
the Multi-wavelength Analysis of Galaxy PHYSical Properties \citep[MAGPHYS,][]{dacunha08}. 
These algorithms rely on somewhat different assumptions for inferring SFH, extinction curves,
dust reprocessing, and dust emission, and are all widely used for deriving fundamental quantities such as stellar mass \mstar,
star-formation rate SFR and total IR (TIR) luminosity \ltir\ from galaxy SEDs 
% CIGALE
%\citep{burgarella11}
%\citep{giovannoli11}
%\citep{buat12}
% GRASIL
%\citep{iglesias07}
%\citep{michalowski08}
%\citep{lofaro13}
% MAGPHYS
%\citep{smith12}
%\citep{berta13}
%\citep{pereira15}
\citep[e.g.,][]{iglesias07,michalowski08,burgarella11,giovannoli11,buat12,smith12,berta13,lofaro13,pereira15}.
While comparisons with simulations show that the codes are generally able to reproduce observed SEDs
\citep[e.g.,][]{hayward15}, 
little systematic comparison has been done of the codes themselves
\citep[although see][]{pappalardo16}. 
In this paper, we perform such a comparison using updated photometry \citep{dale17} from the UV to 
sub-millimeter (submm) of a sample
of galaxies from the Key Insights on Nearby Galaxies: A FIR Survey with \hers\
\citep[KINGFISH,][]{kennicutt11}.

The KINGFISH sample of 61 galaxies is ideal for comparing SED fitting algorithms, as 
there is a wealth of photometric and spectroscopic data over a wide range of wavelengths
\citep[see][]{dale17}.
KINGFISH galaxies are selected to be nearby ($\la$30\,Mpc) and
to span the wide range of morphology, stellar mass, dust opacity, and SFR
observed in the Local Universe.
57 of the 61 galaxies also are part of the 
SIRTF Infrared Nearby Galaxy Survey \citep[SINGS,][]{kennicutt03}.
Although the KINGFISH sample is somewhat biased toward star-forming galaxies,
several host low-luminosity active galactic nuclei 
(e.g., NGC\,3627, NGC\,4594, NGC\,4569, NGC\,4579, NGC\,4736, NGC\,4826), and
% $\sim$16\% are early types, ellipticals or lenticulars.
ten galaxies are early types, ellipticals or lenticulars.
As we shall see in more detail in the following sections,
KINGFISH stellar masses span $\sim$\,5 orders of magnitude from $\sim 10^6$
to $10^{11}$\,\msun, and most %(with the exception of NGC\,1097 and NGC\,2146)
are along the ``main-sequence'' relation of SFR and \mstar\ 
\citep[SFMS,][]{brinchmann04,salim07}.

The rest of the paper is structured as follows:
the three SED-modeling codes are described in Sect. \ref{sec:codes}.
In Sect. \ref{sec:comparison}, we analyze differences in the SEDs from the three algorithms and
compare the %inferred 
fitted galaxy parameters to independently-derived quantities.
The ramifications of the different
assumptions made in the models are discussed in Sect. \ref{sec:assumptions}.
Sect. \ref{sec:scaling} presents general scaling relations, 
together with refined ``recipes'' for calculating stellar mass,
and a principal component analysis (PCA) to ameliorate the effect of
mutual correlations among the parameters.
%Results from the KINGFISH PCA are discussed in terms of a local benchmark applicable
%in the context of high-redshift galaxy populations in Sect. \ref{sec:highz}. 
We summarize our conclusions in Sect. \ref{sec:conclusions}.

\section{The SED-fitting codes}
\label{sec:codes}

All the codes rely on a given SFH, with stellar emission defined by
an Initial Mass Function (IMF) \citep[here][]{chabrier03}, applied
to Single-age Stellar Populations \citep[SSPs, here][]{bruzual03}.
However, the assumed SFHs are code dependent as are the assumptions for 
calculating dust extinction and dust emission, and the relative ratio of
stars to dust.

The codes share the aim of solving the Bayesian parameter inference problem:
\begin{equation}
\label{eq:bayes}
%P(\boldmath\theta|\boldmath D) \propto P(\boldmath\theta) P(\boldmath D|\boldmath\theta)
P({\bm\theta}|\bm D) \propto P(\bm\theta) P(\bm D|\bm\theta)
\end{equation}
seeking to derive the full \emph{posterior} probability distribution
$P(\bm\theta|\bm D)$ of galaxy physical parameter vector $\bm\theta$ given the data vector $\bm D$ (the
observed SED). This posterior is proportional to the product of the
\emph{prior} $P(\bm\theta)$ on all model parameters (the probability
of a model being drawn before seeing the data), and the
\emph{likelihood} $P(\bm D|\bm\theta)$ that the data are compatible
with a model generated by the parameters\footnote{Although the parameter inference
problem is the same for all the codes, the model physical parameter vector $\bm\theta$
is different for each model. }. If the data carry Gaussian
uncertainties, the likelihood is proportional to $\exp(-\chi^2/2)$
\citep[see e.g.,][]{Trotta2008,Nikutta2012phd}.

In the following, we describe each model in some detail,
and give a summary of the different assumptions in Table \ref{tab:models}. 
Conceptual differences and possible ramifications for the various approaches
will be discussed in Sect. \ref{sec:assumptions}.
For all three codes, the uncertainties of the inferred parameters correspond to %$\pm 1\sigma$, namely
the 16\% and the 84\% percentiles ($\pm 1\sigma$ confidence intervals)
of their marginalized posterior Probability Distribution Functions (PDFs).

\subsection{CIGALE}
\label{sec:cigale}

The \cig \footnote{\url{http://cigale.lam.fr}} \citep[Code Investigating GALaxy Evolution;][]{noll09,ciesla16,boquien18} 
code is built around two central concepts to model galaxies and estimate their physical properties:
%\begin{enumerate}
\begin{enumerate}[leftmargin=*,itemsep=0pt,itemindent=0pt]
\item 
\cig\ assumes that the energy that is absorbed by the dust from the UV to the near-infrared (NIR) is re-emitted self-consistently in the mid- (MIR) and far-infrared (FIR).
%The computation of the spectra from the ultraviolet to the radio is based on an energy-balance principle: the energy that is absorbed by the dust from the ultraviolet to the near-infrared is re-emitted self-consistently in the mid- and far-infrared. This consistent approach helps break some of the degeneracies, such as the age--attenuation degeneracy, that can plague the modeling of galaxies in the absence of constraints on dust emission.
\item %The physical properties and the associated uncertainties are estimated in a Bayesian-like way from the probability distribution function (PDF). 
%The latter is computed from the likelihood over a systematic grid that can comprise upward $10^8$ models.
The physical properties and the associated uncertainties are estimated in a Bayesian-like way over a systematic grid.
%  that comprises $\ga 10^8$ models (here $7.1\times10^8$).
% added 22/1/2018
\end{enumerate}  

In practice the models are built combining several components: an SFH that can be analytic or arbitrary, single-age stellar populations, templates of ionized gas including lines and continuum (free-free, free-bound, and 2-photon processes), a flexible dust attenuation curve, dust emission templates,
synchrotron emission, and finally the effect of the intervening intergalactic medium. 
Each component is computed by an independent module; 
% in the form of an independent module; 
different modules are available. For instance, stellar populations can be modeled alternatively with the \cite{bruzual03} or the \cite{maraston05} models. 
For this run, we have used the following modules and sets of parameters:
%\begin{itemize}
\begin{itemize}[leftmargin=*,itemsep=0pt,itemindent=0pt]
\item The star-formation history is modeled following a so-called ``delayed'' parametrization \citep[e.g.,][]{ciesla16}: 
%(SFR$(t)\propto t\times\exp(-t/\tau)$). 
\begin{equation}
{\rm SFR}(t) \propto\ \begin{cases}
t\ \exp(-t/\tau) & \text{when $t\leq t_{trunc}$} \\
r_{\rm SFR} \  {\rm SFR}(t\,=\,t_{\rm trunc}) & \text{when $t> t_{trunc}$} \quad .
\end{cases}
\label{eqn:sfr_cigale}
\end{equation}
%We assumed 
%{\bf \cig\ assumes} that star formation started 8, 10, or 12\,Gyr ago, with a timescale of 0.5, 1, 3, 5, or 10\,Gyr. To take into account a recent quenching or burst of star formation, we also included a step-like increase/decrease ($\times$0.00, 0.25, 0.50, 0.75, 1.0, 2.5, 5.0, 10) starting 10, 100, or 450\,Myr ago;
The second case\footnote{$r_{\rm SFR}$ [=\,SFR($t>t_{\rm trunc}$)/SFR($t_{\rm trunc}$)] is a constant that quantifies 
%the dip of the SFR 
a steady SFR at and after time $t_{\rm trunc}$, 
that could be higher or lower than the SFR at $t\,=\,t_{\rm trunc}$.}, with $r_{\rm SFR}$, 
%takes into account quenching 
considers reduced SFR for $t>t_{\rm trunc}$ (e.g., quenching), 
or an increase of star 
formation occurring at time $t_{\rm trunc}$. % \citep[see also][]{ciesla16}.
\item The stellar emission is computed adopting the \citet{bruzual03} SSPs with a metallicity Z\,=\,0.02 and a \citet{chabrier03} IMF;
\item With the stellar spectrum computed, the nebular emission is included based on the production rate of Lyman continuum photons. 
\cig\ employs templates computed using CLOUDY models, with the same metallicity as the stellar population. 
%We assume 
We fixed the \cig\ 
ionisation parameter log \uion\,=\,$-2$, and assumed that 100\% of the 
Lyman continuum photons ionise the gas, that is, the escape fraction is zero %identically zero}
and Lyman continuum photons do not contribute directly to dust heating;
\item To account for the absorption of stellar and nebular radiation by interstellar dust, 
\cig\ adopts a modified starburst attenuation law \citep[e.g.,][]{calzetti00}
that considers differential reddening of stellar populations of different ages:
the baseline law is multiplied by a power law in wavelength $\lambda^\delta$, 
with the slope $\delta$ ranging from $-0.5$ and 0.0 with steps of 0.1. 
The normalisation \ebv\ for stars younger than 10\,Myr ranges from 0.01 mag to 0.60 mag. 
%To account for the difference in attenuation for stars of different ages, \cig\ includes a differential reddening factor 
%for stars older than 10\,Myr (that we defined to be 0.25, 0.50, or 0.75, 
%with 0.44 for a bona-fide starburst attenuation curve). 
To account for
the difference in attenuation for stars of different ages 
\citep[e.g.,][]{charlot00},
%(e.g., Charlot & Fall 2000), 
\cig\ includes an attenuation reduction factor
for stars older than 10\,Myr; here we set it to 0.25, 0.50, or 0.75.
Finally, \cig\ adds a variable bump in the attenuation curve at 0.2175\,\micron\ with a strength of 0.0 (no bump), 1.5, or 3.0 (Milky-Way-like);
\item With the total luminosity absorbed by the dust, 
the energy is re-emitted self-consistently adopting the \citet{draineli07} and \citet{draine14} dust models,
assuming that the dust emission is optically thin.
\cig\ considers possible variations of the polycyclic aromatic hydrocarbon (PAH) abundance (\qpah=0.47, 2.50, 4.58, or 6.62\%), 
of the minimum radiation field intensity (\umin=0.10, 0.25, 0.50, 1.0, 2.5, 5.0, 10, or 25), and 
the fraction of the dust mass $\gamma$ %illuminated by a variable radiation field intensity
heated by a power-law distribution of ISRF intensities ($U^{-\alpha}$)
with log\,$\gamma$ ranging from $-3.0$ to $-0.3$ in 10 steps. 
The maximum starlight intensity \umax\ is fixed to $10^7$, and
$\alpha$, the power-law index is fixed to 2.0.
\end{itemize}

With 11 variables sampled as described, the total grid consists of 80,870,400 model templates. 
%Each model is fitted to the observation by computing the $\chi^2$ on all valid bands, with 
%upper limits (ULs) being discarded for consistency with the other codes. 
Each model is fitted to the observations by computing the $\chi^2$ on all valid bands;
data points with only upper limits were discarded for consistency with the
other codes that cannot accommodate them.
Data are fitted in $f_\nu$ (linear) space. 
Finally, the output parameters are obtained by computing the likelihood of the models,
and the likelihood-weighted means and standard deviations to estimate the physical properties and the associated uncertainties.
%\begin{enumerate}
%\begin{enumerate}[leftmargin=*,itemsep=0pt,itemindent=0pt]
% 11/1/2018 now not needed
%\item selecting models at least 0.1\% as likely as the best fitting model to reproduce the observations;
%{\bf xxx Mederic, can you please clarify the meaning of this?}
%\item computing the likelihood of these models;
%\item computing the likelihood-weighted means and standard deviations to estimate the physical properties and the associated uncertainties.
%\end{enumerate}

%%%%%%%%
\begin{table*} 
\caption{Summary of model assumptions for SED fitting}
%\resizebox{\textwidth}{!}{
\resizebox{0.95\textwidth}{!}{
%\begin{tabular}{P{0.2\textwidth}P{0.31\textwidth}P{0.31\textwidth}P{0.31\textwidth}}
\begin{tabular}{P{0.17\textwidth}P{0.32\textwidth}P{0.32\textwidth}P{0.32\textwidth}}
\hline 
\multicolumn{1}{c}{Property} &
\multicolumn{1}{c}{\cig} &
\multicolumn{1}{c}{\gra$^{\mathrm a}$} &
\multicolumn{1}{c}{\magp} \\
\hline 
\\
SFH & SFR(\tgal) delayed$+$truncation [defined by Eqn. (\ref{eqn:sfr_cigale})] with 
\tgal\,=\,(8, 10, 12)\,Gyr; \linebreak
%$\tau$\,=\,(0.5, 1, 3, 5, 10)\,Gyr; \linebreak
%$r_{\rm SFR}$\,=\,(0.0, 0.25, 0.5, 0.75, 1, 2.5, 5, 10);
%${\rm age}_{\rm trunc}$\,=\,(10, 100, 450)\,Myr$^{\mathrm b}$.  
$\tau$\,=\,(0.5, 1, 2, 4, 8)\,Gyr; \linebreak
$r_{\rm SFR}$\,=\,(0.01, 0.05, 0.1, 0.5, 1, 5, 10);
${\rm age}_{\rm trunc}$\,=\,(10, 100, 1000)\,Myr$^{\mathrm b}$.  
& 
SFR$(t_{\rm gal})\,=\,\nu\,M_{\rm gas}(t_{\rm gal})^k$ 
with primordial gas infall described as
$\dot{M}_{\rm gas}\propto \exp(-t/\tau_{\rm inf})$;
$k\,=\,1$;
(NSS)$^{\mathrm c}$ $\nu$\,=\,(0.3, 0.5, 0.8, 2.3, 8.0, 23.0)\,Gyr$^{-1}$;
(NSS)$^{\mathrm c}$ $\tau_{\rm inf}$\,=\,(0.01, 0.1, 0.5, 1, 2, 5, 10)\,Gyr;
(NSS)$^{\mathrm c}$ $t_{\rm gal}$\,=\,(0.01, 0.02, 0.05, 0.1, 0.2, 0.5, 1, 2, 4, 7, 10, 13)\,Gyr.
& 
SFR$(t_{\rm gal})\,=\,\exp(-\gamma\,t_{\rm gal})$ 
with random bursts potentially occurring at all times
with amplitude $A\,=\,M_{\rm burst}/M_{\rm const}$, the 
ratio of the stellar masses in the burst and exponentially declining component; 
$t_{\rm gal} \in [0.1,13.5]$\,Gyr;
burst duration $\in [3,30]$\,Myr.  \\
\\
\hline 
\\
Geometry & None & 
Two geometries: (NSS) spheroid with King profiles for stars and dust, and
(NSD) disk radial$+$vertical exponential profiles for stars and dust;
GMCs are randomly embedded within each of these structural components; \linebreak
stellar radial scalelength (NSS)$^{\mathrm c}$ \rc\,=\,(0.04, 0.14, 0.52, 1.9, 7.2, 26.6)\,kpc; \linebreak
(NSD) inclination angle $i$ such that $\cos(i)\,=\,(1, 0.8, 0.6, 0.4, 0.2, 0)$.
& None \\
\\
\hline 
\\
Stellar populations & 
\citet{bruzual03} SSPs with \citet{chabrier03} IMF, and solar metallicity ($Z$\,=\,\zsun).
& 
\citet{bruzual03} SSPs with \citet{chabrier03} IMF, and metallicities ranging from $Z\,=\,0.01$\,\zsun\ to $Z\,=\,2.5$\,\zsun.
& 
\citet{bruzual03} SSPs with \citet{chabrier03} IMF, and metallicities ranging from $Z\,=\,0.02$\,\zsun\ to $Z\,=\,2$\,\zsun.\\
\\
\hline 
\\
Ionized gas? & Yes$^{\mathrm d}$ & No & No \\
\\
\hline 
\\
Dust attenuation & 
Modified starburst attenuation law with power-law slope \linebreak
$\delta$\,=\,($-0.5, -0.4, -0.3, -0.2, -0.1, 0.0$);
normalization \ebv\ for stars younger than 10\,Myr $\in$[0.01,0.60] mag; 
differential \ebv\ factor \ebv$_{\rm old}$/\ebv$_{\rm young}$\,=\,(0.25, 0.50, 0.75); 
variable 0.2175\,\micron\ bump with strength of 0.0 (no bump), 1.5, 3.0 (Milky-Way-like).
& Attenuation law as a consequence of 
geometry, grain opacities from \citet{laor93} mediated over grain size distributions
from 0.001\,\micron\ to 10\,\micron, and radiative transfer of the
GMC and diffuse dust components. 
Free parameters are:
\rmc\,=\,(6.1, 14.5, 22.2, 52.2)\,pc; \linebreak
\fmol\,=\,(0.1, 0.3, 0.5, 0.9); \linebreak
\tesc\,=\,(0.001, 0.005, 0.015, 0.045, 0.105)\,Gyr.
& 
Two-component (BC, ambient ISM) dust attenuation \citep{charlot00} as in Eqn. (\ref{eqn:tau_magphys}) with 
$\mu \in$[0,1], 
drawn from the probability density function p($\mu)\,=\,1-\tanh(8\mu\,-\,6)$; \linebreak
$\hat{\tau}_{V}$ parametrized according to the probability density function p($\hat{\tau}_{V})\,=\,1-\tanh(1.5\hat{\tau}_{V}\,-\,6.7)$. 
Optical depth $\hat{\tau}_{V}$ is time-dependent as in Eqn. (\ref{eqn:tau_time}).
\\
\\
\hline 
\\
Dust emission & 
Overall dust luminosity defined by energy-balance considerations with
SED shape governed by the dust models of \citet{draine07,draine14}.
With the exception of one ($\alpha\,\equiv\,$2.0),
parameters of these models are left to vary: \linebreak
\qpah\,=\,(0.47, 2.50, 4.58, 6.62)\%;  \linebreak
\umin\,=\,(0.10, 0.25, 0.50, 1.0, 2.5, 5.0, 10, 25);  \linebreak
log$\gamma$\,=\,[$-3.0$,$-0.3$] in 10 steps. 
Dust emission is assumed to be optically thin;
the DL07 models used in \cig\ have 
$\kappa_{\rm abs}\,=\,0.38$\,cm$^2$\,g$^{-1}$ at 850\,\micron. 
& 
Overall dust luminosity and SED shape governed by geometry,
grain opacities from \citet{laor93} mediated over grain size distributions
from 0.001\,\micron\ to 10\,\micron, and radiative transfer of the
GMC and diffuse dust components. 
The dust column is assumed to be proportional to the metallicity of the given SFH,
and the consistent relation between extinction and emission ensures
energy conservation.
The same variable parameters for dust extinction govern dust emission
through radiative transfer.
% see mkdustopacities_grasil.txt
%Dust emissivity $\kappa_{\rm abs}\,=\,6.4$\,cm$^2$\,g$^{-1}$ at 250\,\micron.
Dust opacity $\kappa_{\rm abs}\,=\,0.56$\,cm$^2$\,g$^{-1}$ at 850\,\micron\
\citep{laor93}.
& 
Overall dust luminosity defined by energy-balance considerations with
SED shape governed by
four species of dust emitters in two environments (BC, ambient ISM),
with both having PAH$+$hot$+$warm grains
($\xi^{\text{BC}}_{\text{PAH}}$, $\xi^{\text{BC}}_{\text{MIR}}$, $\xi^{\text{BC}}_{\text{W}}$,
$\xi^{\text{ISM}}_{\text{PAH}}$,
$\xi^{\text{ISM}}_{\text{MIR}}$,
$\xi^{\text{ISM}}_{\text{W}}$),
but an additional cold-dust component for the ambient ISM 
($\xi^{\text{ISM}}_{\text{C}}$).
In addition to ensuring unity
($\xi^{\text{BC}}_{\text{PAH}} + \xi^{\text{BC}}_{\text{MIR}} + \xi^{\text{BC}}_{\text{W}}\,=\,1$,
$\xi^{\text{ISM}}_{\text{PAH}} + \xi^{\text{ISM}}_{\text{MIR}} + \xi^{\text{ISM}}_{\text{W}} + \xi^{\text{ISM}}_{\text{C}}\,=\,1$)
fixed parameters are:
$\xi^{\text{ISM}}_{\text{PAH}}\,=\,0.550(1-\xi^{\text{ISM}}_{\text{C}})$; 
$\xi^{\text{ISM}}_{\text{MIR}}\,=\,0.275(1-\xi^{\text{ISM}}_{\text{C}})$; and 
$\xi^{\text{ISM}}_{\text{W}}\,=\,0.175(1-\xi^{\text{ISM}}_{\text{C}})$. \linebreak
Parameters left to vary are:
$\xi^{\text{BC}}_{\text{W}} \in$[0,1];
$\xi^{\text{BC}}_{\text{MIR}} \in[0,1-\xi^{\text{BC}}_{\text{W}}]$;
$\xi^{\text{ISM}}_{\text{C}} \in$[0,1];
$T^{\text{BC}}_{\text{W}} \in$[30,70]\,K;
$T^{\text{ISM}}_{\text{W}} \in$[30,70]\,K;
$T^{\text{ISM}}_{\text{C}} \in$[10,30]\,K.
Dust emission is assumed to be optically thin;
dust opacity $\kappa_{\rm abs}\,=\,0.77$\,cm$^2$\,g$^{-1}$ at 850\,\micron\ \citep{dunne00}. 
%dust emissivity $\kappa_{\rm abs}\,=\,8.9$\,cm$^2$\,g$^{-1}$ at 250\,\micron\ \citep{dunne00}. 
\\
\\
\hline 
\\
Free parameters & 11 with \linebreak
SFH (\tgal, $\tau$, $r_{\rm SFR}$, age$_{\rm trunc}$); \linebreak 
dust attenuation ($\delta$, normalization \ebv, differential \ebv, variable 0.2175\,\micron\ bump strength); \linebreak
dust emission (\qpah, \umin, $\gamma$).
& 
7 for NSS templates with \linebreak
SFH (\tgal, \tauinf, $\nu$); geometry (\rc); \linebreak
dust attenuation (\rmc, \fmol, \tesc); \linebreak
dust emission (same as for dust attenuation). \linebreak 
8 for NSD templates with the addition of galaxy inclination (viewing angle).
& 
12 with \linebreak 
SFH ($\gamma$, \tgal, $A$, $Z_{\rm star}$); \linebreak
dust attenuation ($\mu$, and $\hat{\tau}_V$); \linebreak
dust emission
($\xi^{\text{BC}}_{\text{W}}$, $\xi^{\text{BC}}_{\text{MIR}}$, $T^{\text{BC}}_{\text{W}}$, 
$\xi^{\text{ISM}}_{\text{C}}$, $T^{\text{ISM}}_{\text{C}}$, $T^{\text{ISM}}_{\text{W}}$). 
\\
\\
\hline 
\label{tab:models} 
\end{tabular} 
}
%\vspace{-\baselineskip}
\vspace{-1.5\baselineskip}
{\footnotesize
\begin{flushleft}
$^{\mathrm a}$~The parameter ranges for \gra\ are sampling points only; ultimately
the best-fit parameters are interpolated so that essentially the entire range of
parameters is covered making the symbol $\in$ more appropriate (see text for more details). \\
$^{\mathrm b}$~Age of the truncation at $t\,=\,t_{\rm trunc}$.\\
$^{\mathrm c}$~For the NSD library: $\nu$\,=\,(0.3, 0.9, 3, 9); 
$\tau_{\rm inf}$\,=\,(0.1, 0.5, 1, 5, 10)\,Gyr;
$t_{\rm gal}$\,=\,(0.5, 1.25, 2, 5, 13)\,Gyr.
\rc\,=\,(0.1, 0.3, 1, 3, 10)\,kpc;
\rmc\,=\,(6.1, 13.1, 19.3, 28.1)\,pc;
\fmol\,=\,(0.1, 0.2, 0.5, 0.9);
\tesc\,=\,(0.001, 0.002, 0.005, 0.02, 0.05, 0.1)\,Gyr.\\
$^{\mathrm d}$~See Sect. \ref{sec:cigale} for details.
\end{flushleft}
}
\end{table*}

\subsection{GRASIL}
\label{sec:grasil}

The \gra \footnote{\url{http://adlibitum.oats.inaf.it/silva/grasil/grasil.html}}
chemo-spectrophotometric self-consistent models 
\citep{silva98} rely on a chemical evolution code that follows the SFR, the gas fraction, and the metallicity, 
comprising the basic ingredients for a stellar population synthesis. 
The stellar populations are simulated through a grid of integrated spectra of SSPs of different ages and metallicities.
%Padua stellar models from
%\citet{bressan98,bressan02}, with the addition of the effects of dusty circumstellar shells around asymptotic giant branch (AGB) stars. 
The newest version of the code adopted here relies on a \citet{chabrier03} IMF, and is
based on the \citet{bruzual03} populations.

The chemical evolution process is modeled through a separate code
\citep[CHE\_EVO,][]{silva99} that considers the
infall of primordial (metal-free) gas with an exponential folding timescale (\tauinf) in order to 
simulate the cold-accretion phase of galaxy formation
($\dot{M}_{\rm gas}\propto \exp(-t/\tau_{\rm inf})$).
The SFR scheme is a Schmidt/Kennicutt-type law \citep[e.g.,][]{schmidt59,kennicutt98}, 
with SFR\,=\,$\nu\,\times\,M_{\rm gas}^k$, where $M_{\rm gas}$ is the available gas mass, and $\nu$ the SF efficiency.
%For starbursts, one or more bursts of SF are added to the generally smoothly varying SFH. 
Thus the model describes a SFH according to the variations of the input parameters
\tauinf\ and $\nu$: %(starbursts are not currently included).
the current version of the code includes 49 SFHs
for the spheroids (see below), and 20 SFHs for the disks.
The smaller range for \tauinf\ and $\nu$ (see Table \ref{tab:models}) is sufficient for the disks
\citep[e.g.,][]{calura09}.

The effects of dust on SEDs depend on the relative spatial distribution of stars and dust. 
Hence \gra\ relies on three components: star-forming GMCs, stars that have already emerged from these clouds, 
and diffuse gas$+$dust (e.g., cirrus-like). 
Disk galaxies are described through a double exponential (radial, vertical), 
assuming that the dust is distributed radially like the stars, but has a smaller vertical scale height
\citep[specifically 0.3 times the stellar vertical scale, see][]{bianchi07}. 
The vertical stellar scaleheight is taken to be 0.1 of the stellar radial scale length.
Spheroidal systems %(early-type galaxies) 
are quantified by \citet{king62} profiles for both the 
stars and the dust. 
For both geometries, GMCs are embedded within these structural components.
Once the geometry is given, radiative transfer is performed %for the three dust components separately 
through the GMCs and the diffuse medium 
assuming the \citet{laor93} opacities
for grain sizes from 0.001\,\micron\ to 10\,\micron,
mediated over the grain size distribution given by \citet{silva98}.
The relative contribution of dust and gas, namely the dust-to-gas ratio, is taken to be proportional to the metallicity of the given SFH.
More details are found in \citet{silva98}.

%For this work, we have computed with 
%\gra\ $\sim 3\times10^6$ spheroidal SED templates
%(New Star-forming Spheroids, NSS), and $1.2\times10^6$ disk templates (New Star-forming Disks, NSD). 
%The best fit %(in log space) 
%is identified through a Bayesian approach, relying on a Monte-Carlo Markov chain algorithm, implemented in Python. 
%The exact form of the best-fit model is obtained through interpolation of the specific SED templates in a 7-dimensional parameter space (for spheroids) and an 8-dimensional one for disks (since inclination or viewing angle needs to be included). 
%The 7 free parameters are:
% Robert 17/5/2018
For this work, we have computed with GRASIL $\sim3\times10^6$
spheroidal SED templates (New Star-forming Spheroids, NSS), and
$1.2\times10^6$ disk templates (New Star-forming Disks, NSD),
corresponding to the full Cartesian product of all values sampled per
model parameter. The common seven free parameters for both NSS and NSD are:

% from Robert 17/5/2018 why wavelength?
%\vspace{-\baselineskip}
\begin{itemize}[leftmargin=*,itemsep=0pt,itemindent=0pt]
\item
for the SFH: 
% referee
the exponential folding timescale (\tauinf), 
the SF efficiency ($\nu$), galaxy age (\tgal); 
%\item
%radius of the molecular clouds, \rmc, that since cloud mass is fixed, defines optical depth of the GMCs %(assumed to vary with time)
%\item
%molecular gas mass fraction, \fmol; 
%\item
%escape time for the stars to emerge from GMCs, \tesc; 
%\item
%radial scale lengths, \rc (vertical dimension scales with this);
  \item radius \rmc\ of the molecular clouds that, since cloud
    mass is fixed, defines optical depth of the GMCs;
  \item molecular gas mass fraction (\fmol);
  \item escape time (\tesc) for the stars to emerge from GMCs;
  \item radial scale lengths (\rc) (vertical dimension scales
    with this).
%  \item wavelength $\lambda$
\end{itemize}
\vspace{-0.5\baselineskip}
An additional free parameter is needed for the NSD library: galaxy inclination or viewing angle $i$. 

% Robert, 17/5/2018
%We reshape the generated SED sets into 8- (NSS) and 9-dimensional
%We reshape the generated SED sets 
Operationally, the SED library is reshaped 
into seven- (NSS) and eight-dimensional (NSD) hypercubes. 
The wavelength axis of the SEDs is considered as an additional dimension in the cube, and 
the cube axes represent the model parameters. 
% Both hypercubes are rectilinear and fully populated, i.e., all axes are orthogonal 
% in the sense that the parameters are (ideally) independent, and there are no gaps at any grid vertex. 
The vertices of the hypercubes correspond to unique combinations of model parameters; % (the sampling along each grid axis need not be uniform). 
%In our case this sampling is
here the sampling is
either linear or logarithmic (per-axis), ensuring that %we cover realistically the often many orders of magnitude spanned by parameters.
the parameter space is sufficiently covered by the sampling.

The hypercubes enable multidimensional interpolation of SEDs at
\emph{any} continuous vector of model parameter values
%\hbox{$\bm\theta=\{\theta_j\},\ j \in (\tau_{\rm inf}, \nu, t_{\rm
%    gal}, R_{\rm gmc}, f_{\rm mol}, t_{\rm esc}, R_{\rm gal}, (i))$}
\hbox{$\bm\theta=\{\theta_j\},\ j \in (\tau_{\rm inf}, \nu, t_{\rm
    gal}, R_{\rm gmc}, f_{\rm mol}, t_{\rm esc}, R_{\rm gal}, (i))$}
within the envelope spanned by the parameter axes, that is not just at
the discrete grid vertices. An important assumption in this scheme is
that every parameter axis is sampled \emph{finely enough} so as not to 
miss important features in the output SED.

To fit an observed SED we run a Markov Chain Monte-Carlo code
originally developed in \citet{Nikutta2012phd}. It invokes a
Metropolis-Hastings sampler \citep{Metropolis+1953,Hastings_1970}
which at every step samples from log-uniform priors $P(\bm\theta)$ on
all free model parameters (except $\lambda$, for which we use the
observed set of wavelengths). The model SED is interpolated from the
hypercube on the fly using the sampled $\bm\theta$ as input, and
compared to the observed SED, logging the likelihood. A long chain of
samples is recorded in the run, which by construction of the MCMC
algorithm converges toward the posterior distribution
$P(\bm\theta|\bm D)$.

The histograms of the chains are the marginalized one-dimensional posterior
distributions. Their analysis can include, e.g., determining the
maximum-a-posteriori (MAP) vector $\bm\theta_{\mathrm MAP}$, computing the mode of
the distribution (location of the distribution peak), or the median (mean)
$\pm$ confidence ranges around it. %For the purpose of this work we content ourselves with 
Here for the SED best fit, we use a model generated by the vector %the MAP value, i.e. the vector
$\bm\theta_{\mathrm MAP}$ of free parameters values that together maximize
the likelihood.
%Note that this need not be the mode of every
%marginalized posterior distribution. 
Derived parameters (stellar mass, \mstar, dust mass, \mdust, SFR, and metallicity) are median
values of their posterior PDFs.
These posteriors are not modeled directly, but rather computed from the full sample of
SEDs produced in the MCMC run.
%obtained from the best-fit SFH at the median of the age distribution \tgal; 
Uncertainties are then inferred by
computing the $\pm 1\sigma$ confidence ranges 
% of the derived parameters allowed by the marginalized posteriors of the model parameters $\theta$.
around the median values.
$A_{\rm V}$ and $A_{\rm FUV}$ are the ratios, at the
respective wavelengths, of attenuated to unattenuated light.
We run MCMC twice for every
galaxy, once with the NSS and NSD model hypercubes. The best-fit model
is then chosen between the NSS and NSD libraries according to the
lowest \emph{rms} residual.

\subsection{MAGPHYS}
\label{sec:magphys}

\magp \footnote{\url{http://www.iap.fr/magphys/}} is an %Bayesian 
analysis tool to fit multiwavelength SEDs of galaxies
\citep{dacunha08}. 
Based on a Bayesian approach, the median PDFs
of a set of physical parameters characterising the stars and dust
in a galaxy are derived. The emission of stars is modeled using 
\citet{bruzual03} SSP models, assuming a \citet{chabrier03} IMF. 
An analytic prescription of the SFH is coupled with randomly superimposed bursts
to approximate realistic SFHs.
More specifically, the exponentially declining SFH model is parametrized as 
%$\psi$(t)~$\propto$~$\exp$(-$\gamma$t), characterised by an 
SFR$({\mathrm t})~\propto \exp(-\gamma {\mathrm t})$, characterized by an 
age %$t_{\text{g}}$ 
\tgal\
of the galaxy and star formation time-scale $\gamma^{-1}$. %The parameter %$t_{\text{g}}$ 
Throughout the galaxy's lifetime, random bursts are set to occur with equal probability at all times,
with an amplitude defined by the stellar mass ratios in the burst and the exponentially declining component. 
The SFR is assumed to be constant throughout the burst with a duration of the bursts ranging between %3$\times$10$^7$ and 3$\times$10$^8$\,yr. 
30\,Myr and 300\,Myr.
The stellar metallicity $Z_{\rm star}$ is varied uniformly between 0.02 and 2\,\zsun.
The probability of random bursts is set so that half of the SFH templates in the stellar library have experienced a burst during the last 2\,Gyr.
  
Dust attenuation is modeled using the two-phase model of \citet{charlot00}, which
accounts for the increased level of attenuation of young stars ($< 10$\,Myr) that %($< 10^7$ yr) that
were born in dense molecular clouds. 
Thus, young stars experience obscuration
from dust in their birth clouds and the ambient ISM while stars older than 10\,Myr %$10^7$ yr 
are attenuated only by the ambient ISM. 
Consequently, the attenuation of starlight is time dependent:
\begin{equation}
\hat{\tau}_\lambda \,=\, \begin{cases}
\tau_\lambda^{\rm BC} +  \tau_\lambda^{\rm ISM} & \text{for $t^\prime \leq t_0$ } \\
\tau_\lambda^{\rm ISM} & \text{for $t^\prime > t_0$ } \\
\end{cases}
\label{eqn:tau_time}
\end{equation}
\noindent
where $\hat{\tau}_\lambda$ is the ``effective'' absorption optical depth of the stars at
time $t^\prime$, and $t_0$ is defined to be $10^7$\,yr.
The wavelength dependence of dust
attenuation is modeled based on the following relations: 
\begin{eqnarray}
\tau_\lambda^{\rm BC}  & = & (1-\mu)\ \hat{\tau}_V\ (\lambda/5500A)^{-1.3} \nonumber \\
\tau_\lambda^{\rm ISM} & = & \mu\ \hat{\tau}_V\ (\lambda/5500A)^{-0.7} 
\label{eqn:tau_magphys}
\end{eqnarray}
where $\mu$ is the fraction of the %total 
$V$-band optical depth contributed by the diffuse
ISM (and thus $f_\mu\,\equiv\,1-\mu$ is the fraction of obscuration in birth clouds, BC). 

The total infrared luminosity is a combination of the infrared emission from birth clouds and the ambient ISM:
\begin{eqnarray}
L^{\text{tot}}_{\lambda,\text{d}}~=~L^{\text{BC}}_{\lambda,\text{d}}~+~L^{\text{ISM}}_{\lambda,\text{d}}.
\end{eqnarray}
The dust emission in birth clouds and the ambient ISM 
is modeled using a combination of dust-emission mechanisms:
PAHs and hot$+$warm dust grains in birth clouds and similar dust species in the ambient ISM, but with an additional cold-dust component.
As described in \citet{dacunha08}, 
the hot grains consist of single-temperature modified black bodies (MBBs) with fixed temperatures;
the warm and cold dust temperatures are allowed to vary,
with cold dust temperatures between 10\,K and 30\,K (for ambient ISM only) and warm dust temperatures between 30\,K and 70\,K,
using the extended dust libraries from \citet{viaene14}. 
The opacity curves are assumed to be power laws, and different emissivity indices are assigned to the different dust components. 
All emission is assumed to be optically thin.
%See \citet{dacunha08} for more details.

The prior for the
parameter, $f_{\mu}$, which sets the relative contribution of birth clouds and the ambient ISM, 
is assumed to be uniformly distributed between 0 and 1. 
A similar uniform distribution between 0 and 1 is assumed for the fractional contribution of 
warm dust emission to BC IR luminosity,
$\xi^{\text{BC}}_{\text{W}}$, in birth clouds. 
For the ambient ISM, the fractional contribution of cold dust emission to the ISM IR luminosity, $\xi^{\text{ISM}}_{\text{C}}$, 
is assumed to be uniformly distributed between 0.5 and 1. 
The fractional contributions to the IR emission of the ambient ISM of PAHs ($\xi^{\text{ISM}}_{\text{PAH}}$), 
the hot MIR continuum ($\xi^{\text{ISM}}_{\text{MIR}}$),
and warm grains ($\xi^{\text{ISM}}_{\text{W}}$) are fixed 
to average ratios with  $\xi^{\text{ISM}}_{\text{C}}$ for the Milky Way
\citep[for more details, see Table \ref{tab:models} and][]{dacunha08}.
The dust temperatures for warm and cold dust grains are assumed to be uniformly distributed within their temperature ranges. 
To summarize, \magp\ has 6 free parameters 
($\xi^{\text{BC}}_{\text{W}}$, $\xi^{\text{BC}}_{\text{MIR}}$, $T^{\text{BC}}_{\text{W}}$, 
$\xi^{\text{ISM}}_{\text{C}}$, $T^{\text{ISM}}_{\text{C}}$, $T^{\text{ISM}}_{\text{W}}$) 
to model the infrared SED emission,
and 6 free parameters
($\gamma$, \tgal, $A$, $Z_{\rm star}$, $\mu$, and $\hat{\tau}_V$)
to model the  stellar emission and dust attenuation.

%%%%%%%%
% see KINGFISH_WavelengthSummary.srt
\begin{table} 
%\begin{center} 
\caption{Wavelength coverage for KINGFISH SEDs}
\label{tab:wavelengths}
\begin{tabular}{lcc}
\hline 
\multicolumn{1}{c}{Filter} &
\multicolumn{1}{c}{Wavelength} &
\multicolumn{1}{c}{Number} \\
 & \multicolumn{1}{c}{(\micron)} &
\multicolumn{1}{c}{galaxies$^{\mathrm a}$} \\
\hline 
GALEX\_FUV     & 0.152     & 57 \\
GALEX\_NUV     & 0.227     & 58 \\
SDSS\_u        & 0.354     & 40 \\
B\_Bessel      & 0.440     & 61 \\
SDSS\_g        & 0.477     & 40 \\
V\_Bessel      & 0.550     & 59 \\
SDSS\_r        & 0.623     & 40 \\
R\_Cousins     & 0.640     & 56 \\
SDSS\_i        & 0.762     & 40 \\
I\_Cousins     & 0.790     & 56 \\
SDSS\_z        & 0.913     & 40 \\
2MASS\_J       & 1.235     & 61 \\
2MASS\_H       & 1.662     & 61 \\
2MASS\_Ks      & 2.159     & 61 \\
WISE\_W1       & 3.353     & 59 \\
IRAC\_CH1      & 3.550     & 61 \\
IRAC\_CH2      & 4.490     & 61 \\
WISE\_W2       & 4.603     & 57 \\
IRAC\_CH3      & 5.730     & 61 \\
IRAC\_CH4      & 7.870     & 61 \\
WISE\_W3      & 11.561     & 61 \\
WISE\_W4$^{\mathrm b}$      & 22.088     & 54 \\
% detections 
% MIPS\_24      & 23.70     & 61 \\
% PACS\_70      & 71.11     & 58 \\
% MIPS\_70      & 71.42     & 58 \\
% PACS\_100    & 101.20     & 58 \\
% MIPS\_160    & 155.90     & 58 \\
% PACS\_160    & 162.70     & 58 \\
% SPIRE\_250   & 249.40     & 57 \\
% SPIRE\_350   & 349.90     & 57 \\
% SPIRE\_500   & 503.70     & 56 \\
% SCUBA\_850   & 850.0     & 21 \\
% detections or ULs
MIPS\_24      & 23.70     & 61 \\
PACS\_70      & 71.11     & 61 \\
MIPS\_70      & 71.42     & 61 \\
PACS\_100    & 101.20     & 61 \\
MIPS\_160    & 155.90     & 61 \\
PACS\_160    & 162.70     & 61 \\
SPIRE\_250   & 249.40     & 61 \\
SPIRE\_350   & 349.90     & 61 \\
SPIRE\_500   & 503.70     & 61 \\
SCUBA\_850   & 850.0     & 21 \\
% Radio\_20cm & 200000.0     & 50 \\
\hline 
\end{tabular}
\begin{flushleft}
$^{\mathrm a}$~These numbers give the sum of the detections and upper limits. \\
$^{\mathrm b}$~Shifting this effective wavelength to the modified value given by \citet{brown14b}
would bring the WISE W4 and MIPS\,24\,\micron\ fluxes into closer agreement.
\end{flushleft}
\end{table}

By varying the star formation history, stellar metallicity and dust
attenuation, a library of 50,000 stellar population models are generated. An
additional set of 50,000 dust SED templates is generated with a range of dust
temperatures and varying relative abundances for the various dust components.
To link the stellar radiation that was absorbed by dust to the thermal dust
emission, the code assumes a dust energy balance, namely the amount of stellar
energy that is absorbed by dust is re-emitted in the infrared (with a 15\%
margin to allow for model uncertainties arising from geometry effects, etc.).

To derive the best fitting parameters in the model, the observed luminosities are compared to the luminosities of each model j and the goodness of each model fit is characterized by:
\begin{equation}
\chi^{2}_{j} = \sum_{i} \left(\frac{L_{\nu}^{i}-w_{j} L_{\nu,j}^{i}}{\sigma_{i}} \right)^{2}
\end{equation}
with the observed and model luminosities, $L_{\nu}^{i}$ and $L_{\nu,j}^{i}$, and observational uncertainties, $\sigma_{i}$ in the i$^{th}$ waveband, and a model scaling factor, $w_{j}$, to minimise $\chi^{2}_{j}$ for each model j. All models are convolved with the appropriate response curves prior to comparison with the observed fluxes for each filter. 
Under the assumption of Gaussian uncertainties (see above),
the PDF for every parameter is derived by weighting a specific parameter value with the probability $\exp$(-$\chi^{2}_{j}$/2) of every model $j$;
the %best-fit values 
output model parameters correspond to the median of the PDF.

%%%%%%%
\begin{figure*}
\vspace{\baselineskip}
\vspace{\baselineskip}
\includegraphics[width=0.95\textwidth]{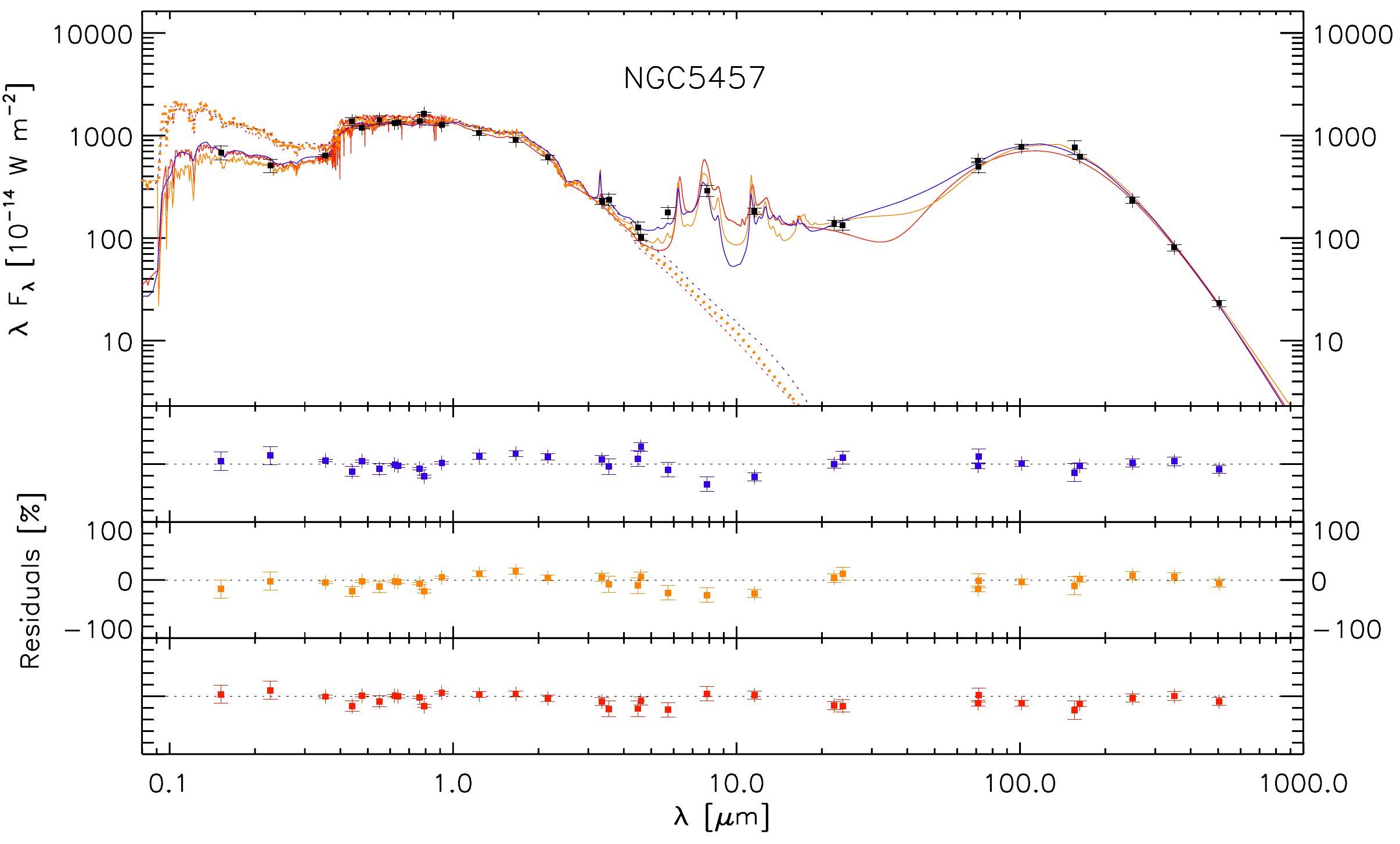}
\caption{Panchromatic SED for NGC\,5457 (M\,101) based on the photometry measurements from \citet{dale17}
overlaid with the best-fitting SED model inferred from the SED fitting tools 
%MagPhys 
\magp\ (red curve), %CIGALE 
% referee
\cig\ (dark-orange curve) and %GRASIL 
% \cig\ (green curve) and %GRASIL 
\gra\ (blue curve). The dashed curves represent the (unattenuated) intrinsic model emission for each SED fitting method (using the same color coding). 
The bottom part of each panel shows the residuals for each of these models compared to the observed fluxes in each waveband.
%Gray arrows points show the upper limits when available. 
}
\label{fig:n5457_sed}
\end{figure*}

% DOES solve the hyperlink problem
%\vspace{-1.5\baselineskip}
\section{Comparison of SED models}
\label{sec:comparison}

The SED algorithms have been applied to the KINGFISH sample of 61 galaxies 
which have a wide range of multiwavelength photometric observational constraints.
The fits have been performed independently, by different individuals,
in order to avoid potential biases in the outcome.
In the ideal case, the multiwavelength SED emission of a single galaxy is constrained by %twenty-nine 
32 photometric data points % (not considering the 20cm radio point)
across the UV-to-submm wavelength range. 
We refer to \citet{dale17} for a detailed description of the data reduction and aperture photometry techniques used in each of those bands. 
Before fitting, the data
from \citet{dale17} has been corrected for foreground
Galactic extinction according to \av\ measurements by
\citet{schlafly11} and the extinction curve of \citet{draine03}.

Not all KINGFISH galaxies have complete observational coverage, and some observations have resulted in non-detections 
(upper limits are not accounted for in the SED modeling), which results in an inhomogeneous data coverage for the entire sample of KINGFISH galaxies. 
While this inhomogeneity of photometric data points might bias the quantities derived from the SED modeling 
\citep[e.g.,][]{ciesla14}, the main interest of this paper is to compare the SED model output from the three different codes 
(which have been constrained by the same set of data).
Any inhomogeneity in the photometric constraints for different galaxies will not affect the main goal of this work. 
Table \ref{tab:wavelengths} gives an overview of the filters and central wavelength of the wavebands 
used to constrain the SED models, and the number of galaxies 
for which measurements (detections or upper limits) are available \citep[see also][]{dale17}.
%that have been observed and detected in each of these wavebands.
In all galaxies, the number of data points significantly exceeds the number of free parameters
in the models.

Some of the filters also cover the same wavelength range (e.g., SDSS\,$ugriz$ and $BVRI$, MIPS and PACS) but show offsets in their absolute photometry. 
To avoid preferentially biasing any individual source of photometry, we have opted to use all 
available photometric constraints available for every single galaxy. 
%The relative comparison of the SED output parameters should however not be affected by a 
%possible incompatibility of flux measurements at similar wavelengths. 
As long as there is no preferred spectral region in the models, 
the relative comparison of the SED output quantities should not be strongly affected by 
inconsistencies of flux measurements at similar wavelengths. 

In the remainder of this section, we compare the SED results from the three different codes.
First, we assess
the capability of the model to reproduce the {\it shape} of the data SED
(Sect. \ref{sec:comparison:shape});
then, we
confront fitted results against an independently-derived set of ``reference'' or recipe 
quantities (Sect. \ref{sec:comparison:reference}).

% DOES solve the hyperlink problem
%\vspace{-1.8\baselineskip}
\subsection{Comparison of SED shapes}
\label{sec:comparison:shape}

\begin{figure*}
\vspace{\baselineskip}
\includegraphics[width=0.95\textwidth]{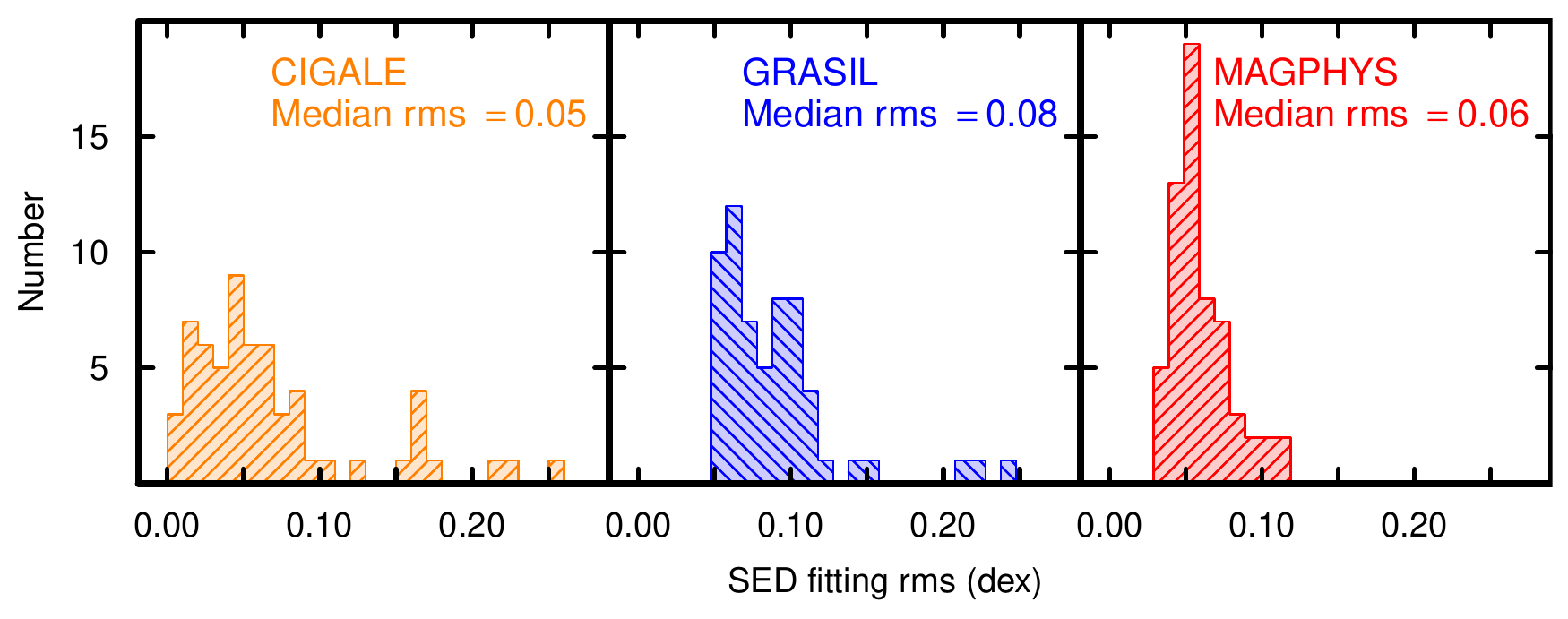}
\caption{Distribution of the root-mean-square residuals for the three
SED-fitting algorithms, \cig, \gra, \magp.
%The best-fit Gaussian is shown as a solid line, with the width as given
%in each panel by $\sigma$.
The \rms\ is calculated as the square root of the mean of the sum of squares;
the values are comparable to the typical uncertainty in the fluxes themselves.
}
\label{fig:rms}
\end{figure*}

The best-fit\footnote{Unlike the derived quantities from the inferred SED (that are
mean of the PDF for \cig, and median of the PDF for \gra\ and \magp),
these are the maximum-likelihood solutions.} 
SEDs for all three models are overlaid on the observed SED 
for a representative KINGFISH galaxy (NGC\,5457 = M\,101) in Fig. \ref{fig:n5457_sed}, and for the remaining galaxies 
in Fig. \ref{fig:all_seds} in Appendix \ref{app:seds}.
We have assessed the quality of the three SED-fitting algorithms using three
criteria: reduced \chisq, \redchisq;
the root-mean-square residuals in logarithmic space [i.e., log(flux)], \rms; and
the weighted root-mean-square residuals \rmsw.
Figure \ref{fig:rms} shows the comparison of the \rms\ values;
the \rms\ is calculated as the square root of the mean of the sum of squares.
All algorithms provide quite good approximations of the observed SED across all wavelengths, 
typically with \rms\ \la 0.08\,dex; %or better of the data. 
such values are typical of the uncertainties in the fluxes themselves \citep[see][]{dale17}.
Interestingly, the outliers with large \rms\ for \cig\ and \gra\ are not the same galaxies;
\cig\ has more problems with dwarf galaxies
(e.g., DDO\,053, IC\,2574, NGC\,5408) while \gra\ struggles with early types
(e.g., NGC\,584, NGC\,1316, NGC\,4594).

%%%%%%%

In the optical and FIR-to-submm wavelength domains, the three SED models show similar SED shapes. 
Despite the different SFH prescriptions for the three codes, the SED models are all able to reproduce the stellar emission of 
intermediate-aged and old stars in the KINGFISH galaxies. 
Also the emission of the colder dust components in the models seems to agree well with a similar slope for the Rayleigh-Jeans 
tail of the SED in all three models. 
This is not surprising because
the average FIR-submm dust emissivity indices of the three models are very close:
$\beta$\,=\,2.08 \citep{bianchi13} for the % $\beta$\,=\,1.8 for the 
\citet[][hereafter DL07]{draineli07} dust model used in \cig; 
$\beta$\,=\,2.02 (see Sect. \ref{sec:comparison:mdust}) for the \citet{laor93} dust used in \gra; and
$\beta$\,=\,2 assumed for the cold dust component in \magp. 
The KINGFISH galaxies without observational constraints at FIR wavelengths (e.g., DDO\,154 and DDO\,165) 
show strong variations in their fitted dust SEDs, indicating that dust energy balance models 
cannot constrain the dust component in galaxies based only on UV and optical information on the dust extinction.

In the UV and FIR wavelength domains, the three models are also generally in good agreement, %for most galaxies. 
although for some galaxies \gra\ overestimates the observed UV emission.
However, the predictions of the three models sometimes differ significantly at NIR and MIR wavelengths. 
Since \magp\ applies a fixed PAH emission template for their models (see Table \ref{tab:models}), 
% referee
the relative changes in PAH abundance are not always reflected in the best-fit model (e.g., NGC\,3190). 
Several galaxies show little or no PAH emission in their \spit/IRS spectra \citep[e.g., Ho\,II, NGC\,1266, NGC\,1377, NGC\,2841,
NGC\,4594:][]{roussel06,smith07}, but have significant PAH emission modeled by \cig\ and \magp.
A detailed study of the PAH emission in NGC\,1377 has shown that it is suppressed by 
dust that is optically thick at $\sim$10\,\micron\
\citep{roussel06}.
On the other hand, the strong PAH features in some galaxy spectra observed with IRS 
\citep[e.g., NGC\,3190, NGC\,3521, NGC\,4569:][]{smith07} are not reproduced by \gra.
% referee
PAH emission in galaxies seems to be a source of significant disagreement among the models,
and the models in many cases are unable to adequately approximate the detailed shape 
of the emission features.

At MIR wavelengths, there are also some continuum variations among the three different models. 
\citet{ciesla14} have already shown that MIR photometry is required to constrain the emission of warmer dust in SED models. 
But even with the MIPS\,24\,$\mu$m data point, the shape of the three SED models between 24\,$\mu$m and 70\,$\mu$m can be very different. 
The \magp\ models tend to have a bump in their SED shape in between 24\,$\mu$m and 70\,$\mu$m for some galaxies (e.g., NGC\,3773, NGC\,4236), 
possibly due to the addition of a warm component unconstrained by data.
\gra\ shows a more constant SED slope at those wavelengths, while there is typically a small dip in emission in the \cig\ models;
this dip can cause difficulties in fitting the mid- and far-IR emission of some galaxies (e.g., NGC\,5408). 

\begin{figure}
\vspace{\baselineskip}
\includegraphics[width=0.48\textwidth]{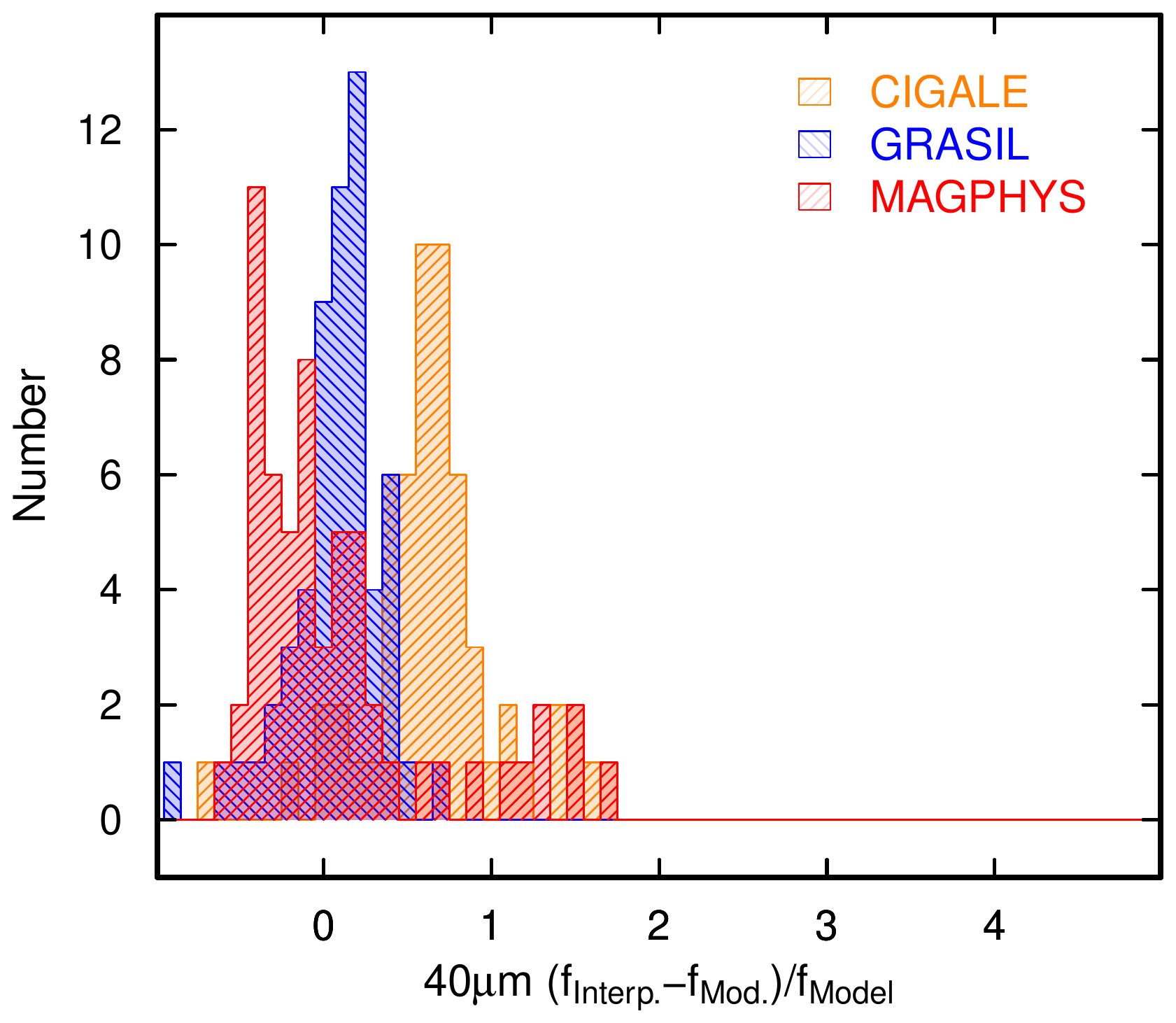}
\caption{Distributions of the residuals, $(f_{\rm Interpolated}-f_{\rm Model})/f_{\rm Model}$, at $\sim$40\,\micron\ for the three
SED-fitting algorithms, \cig, \gra, \magp.
As noted in the legend, \cig\ residuals are shown in dark orange, \gra\ in blue, and \magp\ in red.
}
\label{fig:histo40}
\end{figure}

The changing behavior of the models in the MIR regime is illustrated in Fig. \ref{fig:histo40} where we show
the 40\,\micron\ residuals 
($(f_{\rm Interpolated}-f_{\rm Model})/f_{\rm Model}$)
calculated by linearly interpolating the observed flux between 24\,\micron\ and 70\,\micron.
\magp\ tends to overestimate the interpolated 40\,\micron\ emission (by median $\sim$\,13\%, but with large spread),
while \cig\ underestimates the emission ($\sim$\,71\%);
\gra\ also underestimates but by less ($\la$\,11\%). 
Although it is tempting to assume that small differences mean an accurate model,
the true shape of the SED in this wavelength region is highly uncertain.
% referee
Our linear interpolation is only providing a ``fiducial'' against which to compare the
models;
here our aim is to compare the different models with a common
reference, rather than infer ``truth''.

%%%%%%%

\citet{wu10} show that over a wide range of IR luminosities, ratios of the 70\,\micron\ and
24\,\micron\ fluxes can vary by a factor of 5, and the variations seem to depend on the flux
ratios at shorter wavelengths. 
Indeed, for \magp, the flux ratio between MIPS 24\,\micron\ (or WISE 22\,\micron) and the 12\,\micron\ WISE
band seems to play a role;
if high, then the model wants to include more warm dust resulting in a 40\,\micron\ ``bump".
For \cig, the power-law index $\alpha$ governing the variation of the ISRF $U$, is
important; increasing $\alpha$ to 
2.5 results in an increase in 40\,\micron\ flux of 30\%,
not enough to compensate completely, but bringing the models closer to the observations. 
Finally, \gra\ seems to do a decent job of reproducing the observations, except in
one highly discrepant galaxy;
for NGC\,4594 (the Sombrero), \gra's estimate of the 24\,\micron\ flux ($f_\nu(24)$) is lower than the observations by a factor of 4.
\gra\ underestimates the diffuse dust component that is responsible for the mid-IR emission, possibly
as a consequence of the geometry of this galaxy; the best \gra\ fit is for 
a spheroid, but the dust in the Sombrero is found in a conspicuous dust lane.

\subsection{Reference quantities for comparison}
\label{sec:comparison:reference}

To compare the three different SED models, 
we have devised a set of six quantities representative of a galaxy's general properties that will be used to quantify any model deviations. 
As fundamental quantities descriptive of galaxies, we have opted to compare stellar mass (\mstar), star formation rate (SFR), 
dust mass (\mdust), total-infrared luminosity (\ltir), 
intrinsic (dust-corrected) far-ultraviolet (FUV) luminosity (\lfuv) %(\lfuv(unatt)) 
and FUV attenuation (\afuv). 
\mstar, SFR, and \mdust\ are quantities directly output by \cig\ and \magp; 
for \magp\ they are derived as the median values of the PDFs
based on Bayesian statistics for the derived model quantities, while
for \cig, they are the means (see Sect. \ref{sec:codes}).
For \gra, the marginalized posteriors of the model parameters SFH and age of the galaxy \tgal\ are output
(together with the other fitted parameters, see Table \ref{tab:models});
\mstar, SFR, and \mdust\ are not directly fit, but rather obtained from the medians of the PDFs 
allowed by the %(renormalized) quantities of the SED allowed by the PDFs 
PDFs and confidence levels of SFH and \tgal. 
For \cig, \ltir, \lfuv, and \afuv\ are computed in the same way as the other quantities,
from the likelihood-weighted means.
For \gra\ and \magp,
the luminosities and \afuv\ are derived from a convolution (with the appropriate response curves) and integration of the 
best-fit (maximum likelihood) model SED. 
Attenuation is inferred
at a given wavelength through the ratio of intrinsic to observed (attenuated) emission. 

For an independent measure of these six ``fundamental" galaxy quantities, 
we have computed estimates using recipes based on one or two photometric bands
(with the exception of the updated DL07 models for \mdust\ also included in the analysis). 
The methodology for these derivations is described in detail in Appendix \ref{app:ref}
and summarized in Table \ref{tab:app:methods}; the resulting values are reported in Table \ref{tab:app:parameters}.
We emphasize that the quantities calculated by these recipes are almost certainly
not the truth, but rather ``poor-person'' estimates, necessary when multiwavelength coverage is missing. 
The problem is that truth is unknown and here elusive. 
It could be reflected by one or more of the SED algorithms that 
certainly do better than the simple recipes based on a restricted photometric regime;
it is almost certainly not reflected by these recipes since the whole idea of fitting SEDs is
to improve our understanding of these parameters and their interrelation.
The following sections attempt to maintain this philosophy.

%%%%%%%%
\begin{table*} 
\caption{Correlations of SED-derived vs. independently-derived recipe quantities: ~$y\,=\,a\ + b\, x$}
\resizebox{\textwidth}{!}{
\begin{tabular}{lllcllr}
\hline 
\multicolumn{1}{c}{Quantity} &
\multicolumn{1}{c}{{\it x} method} &
\multicolumn{1}{c}{{\it y} method} &
\multicolumn{1}{c}{Number} &
\multicolumn{1}{c}{$a$} &
\multicolumn{1}{c}{$b$} &
\multicolumn{1}{c}{RMS} \\
&&&
\multicolumn{1}{c}{galaxies} &
&&
\multicolumn{1}{c}{residual } \\
\hline 
% 26/7/2018
% 8/9/2018
\hline
Log[\mstar/$10^9$\,\msun] & var M/L$_{3.6}$ (Wen+ 2013) & \cig  & 61 & -0.106 $\pm$\   0.02 &  0.979 $\pm$\   0.02 &  0.121 \\
  &   & \gra  & 61 & -0.325 $\pm$\   0.03 &  1.026 $\pm$\   0.02 &  0.181 \\
  &   & \magp & 61 & -0.232 $\pm$\   0.03 &  0.999 $\pm$\   0.02 &  0.192 \\
\hline
Log[\mstar/$10^9$\,\msun] & fix M/L$_{3.6}$ & \cig  & 61 & -0.305 $\pm$\   0.02 &  1.046 $\pm$\   0.01 &  0.120 \\
  &   & \gra  & 61 & -0.534 $\pm$\   0.04 &  1.102 $\pm$\   0.03 &  0.152 \\
  &   & \magp & 61 & -0.427 $\pm$\   0.03 &  1.071 $\pm$\   0.02 &  0.216 \\
\hline
%\cig  &  Log[\mstar\,(const.)/$10^9$\,\msun)] & Log[\mstar\,(SED)/$10^9$\,\msun] & 61 & -0.305 $\pm$\   0.02 &  1.046 $\pm$\   0.01 &  0.120 \\
%\gra  &    &   & 61 & -0.534 $\pm$\   0.04 &  1.102 $\pm$\   0.03 &  0.152 \\
%\magp &    &   & 61 & -0.427 $\pm$\   0.03 &  1.071 $\pm$\   0.02 &  0.216 \\
\hline
%\cig  &  Log[SFR(FUV+TIR)/\msunyr ] & Log[SFR\,(SED)/\msunyr] & 61 & -0.034 $\pm$\   0.03 &  0.981 $\pm$\   0.04 &  0.204 \\
%\gra  &    &   & 61 &  0.117 $\pm$\   0.01 &  0.971 $\pm$\   0.01 &  0.065 \\
%\magp &    &   & 61 & -0.277 $\pm$\   0.03 &  0.935 $\pm$\   0.03 &  0.206 \\
%\hline
%\cig  &  Log[SFR(H$\alpha$+24)/\msunyr ] & Log[SFR\,(SED)/\msunyr] & 60 & -0.040 $\pm$\   0.04 &  0.957 $\pm$\   0.04 &  0.251 \\
%\gra  &    &   & 60 &  0.125 $\pm$\   0.03 &  0.941 $\pm$\   0.03 &  0.186 \\
%\magp &    &   & 60 & -0.274 $\pm$\   0.04 &  0.955 $\pm$\   0.04 &  0.256 \\
Log[SFR/\msunyr] & FUV$+$TIR/\msunyr ] & \cig  & 61 & -0.034 $\pm$\   0.03 &  0.981 $\pm$\   0.04 &  0.204 \\
   &   & \gra  & 61 &  0.117 $\pm$\   0.01 &  0.971 $\pm$\   0.01 &  0.065 \\
   &   & \magp & 61 & -0.277 $\pm$\   0.03 &  0.935 $\pm$\   0.03 &  0.206 \\
\hline
Log[SFR/\msunyr] & H$\alpha$+24 & \cig  & 60 & -0.040 $\pm$\   0.04 &  0.957 $\pm$\   0.04 &  0.251 \\
   &   & \gra  & 60 &  0.125 $\pm$\   0.03 &  0.941 $\pm$\   0.03 &  0.186 \\
   &   & \magp & 60 & -0.274 $\pm$\   0.04 &  0.955 $\pm$\   0.04 &  0.256 \\
\hline
%\cig  &  Log[\mdust\,(MBB)/$10^7$\,\msun] & Log[\mdust\,(SED)/$10^7$\,\msun] & 58 &  0.018 $\pm$\   0.01 &  1.029 $\pm$\   0.01 &  0.103 \\
%\gra  &    &   & 58 &  0.315 $\pm$\   0.02 &  0.991 $\pm$\   0.02 &  0.149 \\
%\magp &    &   & 58 & -0.399 $\pm$\   0.01 &  0.974 $\pm$\   0.01 &  0.056 \\
%\hline
%\cig  &  Log[\mdust\,(DL07)/$10^7$\,\msun] & Log[\mdust\,(SED)/$10^7$\,\msun] & 54 &  0.123 $\pm$\   0.02 &  0.952 $\pm$\   0.02 &  0.173 \\
%\gra  &    &   & 54 &  0.416 $\pm$\   0.03 &  0.917 $\pm$\   0.02 &  0.161 \\
%\magp &    &   & 54 & -0.292 $\pm$\   0.02 &  0.901 $\pm$\   0.02 &  0.120 \\
\hline
Log[\mdust/$10^7$\,\msun] & MBB fit & \cig  & 58 &  0.018 $\pm$\   0.01 &  1.029 $\pm$\   0.01 &  0.103 \\
  &   & \gra  & 58 &  0.315 $\pm$\   0.02 &  0.991 $\pm$\   0.02 &  0.149 \\
  &   & \magp & 58 & -0.399 $\pm$\   0.01 &  0.974 $\pm$\   0.01 &  0.056 \\
\hline
Log[\mdust/$10^7$\,\msun] & DL07 fit & \cig  & 54 &  0.123 $\pm$\   0.02 &  0.952 $\pm$\   0.02 &  0.173 \\
  &   & \gra  & 54 &  0.416 $\pm$\   0.03 &  0.917 $\pm$\   0.02 &  0.161 \\
  &   & \magp & 54 & -0.292 $\pm$\   0.02 &  0.901 $\pm$\   0.02 &  0.120 \\
%\cig  &  Log[\ltir\,(DL07)/$10^9$\,\lsun] & Log[\ltir\,(SED)/$10^9$\,\lsun] & 58 & -0.087 $\pm$\   0.01 &  1.001 $\pm$\   0.01 &  0.046 \\
%\gra  &    &   & 58 & -0.072 $\pm$\   0.01 &  0.981 $\pm$\   0.01 &  0.053 \\
%\magp &    &   & 58 & -0.063 $\pm$\   0.01 &  0.972 $\pm$\   0.01 &  0.059 \\
%\hline
%\cig  &  Log[\ltir\,(G13)/$10^9$\,\lsun] & Log[\ltir\,(SED)/$10^9$\,\lsun] & 58 & -0.046 $\pm$\   0.01 &  1.014 $\pm$\   0.01 &  0.051 \\
%\gra  &    &   & 58 & -0.040 $\pm$\   0.01 &  0.995 $\pm$\   0.01 &  0.032 \\
%\magp &    &   & 58 & -0.022 $\pm$\   0.01 &  0.983 $\pm$\   0.01 &  0.042 \\
\hline
Log[\ltir/$10^7$\,\lsun] & DL07 formulation & \cig  & 58 & -0.087 $\pm$\   0.01 &  1.001 $\pm$\   0.01 &  0.046 \\
  &   & \gra  & 58 & -0.072 $\pm$\   0.01 &  0.981 $\pm$\   0.01 &  0.053 \\
  &   & \magp & 58 & -0.063 $\pm$\   0.01 &  0.972 $\pm$\   0.01 &  0.059 \\
\hline
Log[\ltir/$10^7$\,\lsun] & GL13 formulation & \cig  & 58 & -0.046 $\pm$\   0.01 &  1.014 $\pm$\   0.01 &  0.051 \\
  &   & \gra  & 58 & -0.040 $\pm$\   0.01 &  0.995 $\pm$\   0.01 &  0.032 \\
  &   & \magp & 58 & -0.022 $\pm$\   0.01 &  0.983 $\pm$\   0.01 &  0.042 \\
\hline
%\cig  &  Log[\lfuv/$10^9$\,\lsun] & Log[\lfuv\,(SED)/$10^9$\,\lsun] & 54 &  0.001 $\pm$\   0.02 &  1.041 $\pm$\   0.02 &  0.115 \\
%\gra  &    &   & 54 &  0.098 $\pm$\   0.01 &  0.981 $\pm$\   0.01 &  0.060 \\
%\magp &    &   & 54 & -0.048 $\pm$\   0.02 &  0.962 $\pm$\   0.03 &  0.127 \\
Log[\lfuv/$10^9$\,\lsun] & IRX correction Murphy+ (2011) & \cig  & 54 &  0.001 $\pm$\   0.02 &  1.041 $\pm$\   0.02 &  0.115 \\
  &   & \gra  & 54 &  0.098 $\pm$\   0.01 &  0.981 $\pm$\   0.01 &  0.060 \\
  &   & \magp & 54 & -0.048 $\pm$\   0.02 &  0.962 $\pm$\   0.03 &  0.127 \\
\hline
%\cig  &  A$_{\rm FUV}$ (Murphy+ 2011) & A$_{\rm FUV}$ (SED) & 54 &  0.181 $\pm$\   0.05 &  0.933 $\pm$\   0.02 &  0.218 \\
%\gra  &    &   & 54 &  0.136 $\pm$\   0.04 &  0.852 $\pm$\   0.02 &  0.150 \\
%\magp &    &   & 54 &  0.211 $\pm$\   0.06 &  0.807 $\pm$\   0.03 &  0.210 \\
A$_{\rm FUV}$ & Murphy+ (2011) & \cig  & 54 &  0.181 $\pm$\   0.05 &  0.933 $\pm$\   0.02 &  0.218 \\
  &   & \gra  & 54 &  0.136 $\pm$\   0.04 &  0.852 $\pm$\   0.02 &  0.150 \\
  &   & \magp & 54 &  0.211 $\pm$\   0.06 &  0.807 $\pm$\   0.03 &  0.210 \\
\hline
\label{tab:corr} 
\end{tabular} 
}
%\begin{flushleft}
%$^{\mathrm a}$ ~$y\,=\,a\ + b\, x$ \\
%\end{flushleft}
%\vspace{-\baselineskip}
\end{table*}

\subsection{SED model derived quantities compared}
\label{sec:comparison:parameters}

Here we perform linear regressions on the results from the three SED codes
with respect to the derived recipe quantities.
In principle, such an analysis will give insight as to the relative performance
of the codes, but more importantly will enable an independent assessment of the accuracy of
the reference quantities that are in truth simplified recipes that cannot be
as accurate as a complete multiwavelength SED fitting.
We calculated the regressions using a ``robust" estimator \citep[see][]{li06,fox08}, as implemented
in the {\it R} statistical package\footnote{R is a free software environment for statistical computing and graphics (\url{https://www.r-project.org/}).}.
In Figures \ref{fig:mstar}-\ref{fig:fuv}, the best-fit correlations are indicated with solid lines. 
Table \ref{tab:corr} gives the results of the correlation analysis for the
comparison of the results of the SED modeling and the reference recipe values;
the normalizations for \mstar, \mdust, \ltir, and \lfuv\ for both axes are non-zero 
because otherwise the non-unit slopes would exaggerate the deviations for small $x$ values. 
A discussion of results and disagreements is given in Sect. \ref{sec:assumptions}.

\subsubsection{Comparison of stellar masses}
\label{sec:comparison:mstar}

The comparison of the SED results with the stellar masses determined from two IRAC-based independent methods
(see Appendix \ref{app:mstar}) is illustrated in Fig. \ref{fig:mstar}.
There is %an excellent correlation 
good agreement between the values of \mstar\ inferred from SED fitting and \mstar\ from
%both methods based on 3.6\,\micron\ luminosities; the \rms\ residuals are between 0.12 and 0.19\,dex for
both methods based on 3.6\,\micron\ luminosities; the \rms\ deviations are between 0.12 and 0.19\,dex for
the \citet{wen13} method with a luminosity-dependent $\Upsilon_*$ (mass-to-light ratio, M/L),
and between 0.12 to 0.22\,dex for constant $\Upsilon_*$.
However, both the \citet{wen13} luminosity-dependent $\Upsilon_*$ and the constant $\Upsilon_*$ \citep{mcgaugh14} 
formulations overestimate \mstar\ relative to the SED fitting algorithms:
the discrepancy is $\sim$0.1--0.3\,dex for the former, and 
$\sim$0.3--0.5\,dex for the latter. %relative to constant $\Upsilon_*$.
For constant $\Upsilon_*$, the deviation seems to depend on \mstar, since the power-law slopes are generally
significantly greater than unity.
% These deviations are larger than the \rms\ residuals (see Table \ref{tab:corr}), and may be telling us something
These discrepancies are larger than the \rms\ deviations (see Table \ref{tab:corr}), and may be telling us something
about the limitations of the assumption of constant $\Upsilon_*$ even at 3.6\,\micron\
\citep[e.g.,][]{eskew12,norris14}.
A new formulation based on our SED-fitting results for converting %3.6\,\micron\ 
3.4$-$3.6\,\micron\ luminosities into stellar masses
is discussed in Sect. \ref{sec:stellarmassrecipe}.

\begin{figure*}
\includegraphics[width=0.95\textwidth]{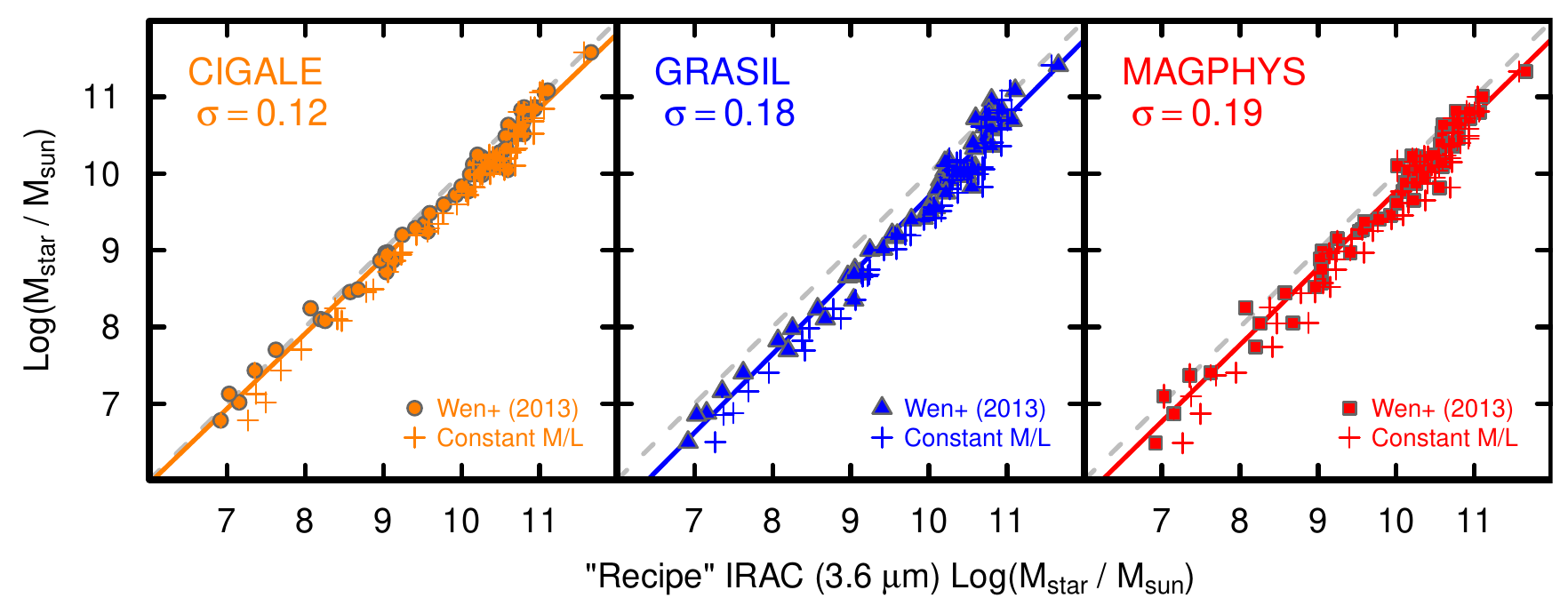}
\vspace{-0.6\baselineskip}
\caption{SED-derived \mstar\ plotted vs. independently determined \mstar\ from
the recipe IRAC 3.6\,\micron\ luminosities
(see Sect. \ref{sec:comparison:mstar} for details).
The \citet{wen13} \mstar\ values are shown by filled (dark-orange) circles (\cig),
filled (blue) triangles (\gra), and filled (red) squares (\magp),
and the constant M/L ones by $+$;
the $\sigma$ values shown in the upper left corner of each panel correspond
to the mean deviations from the fit of \mstar\ with the \citet{wen13} method (see Table \ref{tab:corr}).
Similarly, SED-fitting uncertainties are shown as vertical lines only for the \citet{wen13} $x$ values,
and are usually smaller than the symbol size.
The robust correlation relative to the \citet{wen13} values is shown as a solid line, and the identity relation by a (gray) dashed one. 
}
\label{fig:mstar}
%\end{figure*}
%\begin{figure*}
%\includegraphics[width=\textwidth]{KINGFISH_SFR_cigale_magphys_grasil-crop.pdf}
%\includegraphics[width=\textwidth]{BC03KINGFISH_SFR_cigale_magphys_grasil_NEW-crop.pdf}
%\includegraphics[width=\textwidth]{BC03KINGFISH_SFRSymbols_cigale_magphys_grasil_NEW-crop.pdf}
%\includegraphics[width=\textwidth]{BC03KINGFISH_SFRBoth_Symbols_cigale_magphys_grasil-crop.pdf}
%\includegraphics[width=\textwidth]{BC03KINGFISH_SFRBothNEWLKH_Symbols_cigale_magphys_grasil-crop.pdf}
\includegraphics[width=0.95\textwidth]{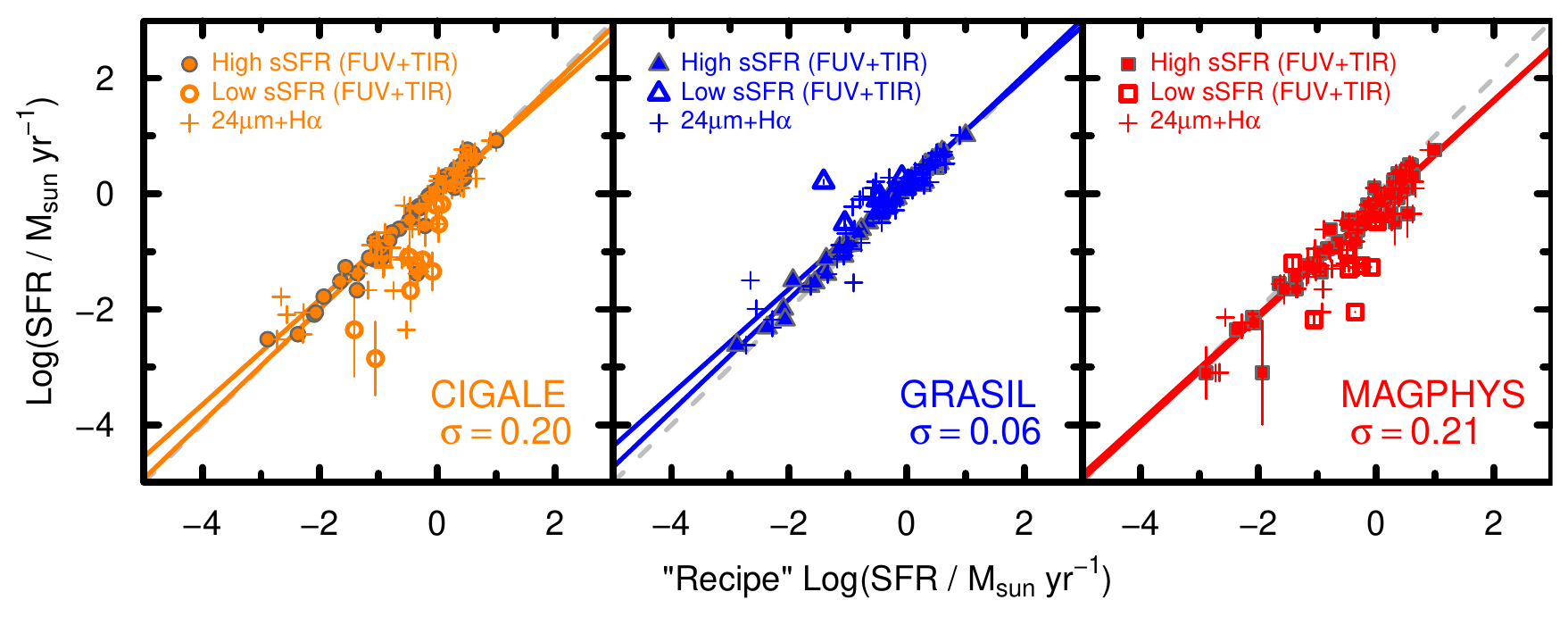}
\vspace{-0.6\baselineskip}
\caption{SED-derived SFR plotted vs. independently determined recipe SFR 
(see Sect. \ref{sec:comparison:sfr} for details).
Two different SFR tracers are shown: FUV$+$TIR and \ha$+$24\micron\ luminosity;
see Appendix \ref{app:ref} for details.
Symbols (dark-orange circles for \cig,
blue triangles for \gra, and red squares for \magp) are calculated with SFR(FUV$+$TIR);
plus signs show the recipe SFR(\ha$+$24\micron) luminosity.
Filled symbols correspond to ``high" specific SFR (Log(sSFR/yr$^{-1}$) $> -10.6$), 
and open ones to ``low" specific SFR (Log(sSFR/yr$^{-1}$) $\leq -10.6$, calculated with SFR(FUV$+$TIR). 
The robust correlations are shown as solid lines, and the identity relation by a (gray) dashed one;
%The lines are as in Fig. \ref{fig:mstar};
in each panel, the steeper power-law slope corresponds to the fit to SFR(FUV$+$TIR) and the shallower one to SFR(\ha$+$24\micron)
(see Table \ref{tab:corr} for details). 
The \rms\ deviations for the fit of SED-derived quantities vs. the reference ones [for SFR(FUV$+$TIR)] are shown by the $\sigma$
value in the lower right corner of each panel;
similarly, SED-fitting uncertainties are shown as vertical lines only for SFR(FUV$+$TIR) $x$ values.
\rms\ deviations for SFR(\ha$+$24\micron) are
0.25\,dex, 0.18\,dex, and 0.26\,dex for \cig, \gra, and \magp, respectively (see Table \ref{tab:corr}).
}
\label{fig:sfr}
%\end{figure*}
%\begin{figure*}
%\includegraphics[width=\textwidth]{KINGFISH_Mdust_cigale_magphys_grasil-crop.pdf}
% \includegraphics[width=\textwidth]{NEWKINGFISH_NewGRASILMdust_cigale_magphys_grasil-crop.pdf}
%\includegraphics[width=\textwidth]{BC03KINGFISH_Mdust_cigale_magphys_grasil_AGAIN-crop.pdf}
\includegraphics[width=0.95\textwidth]{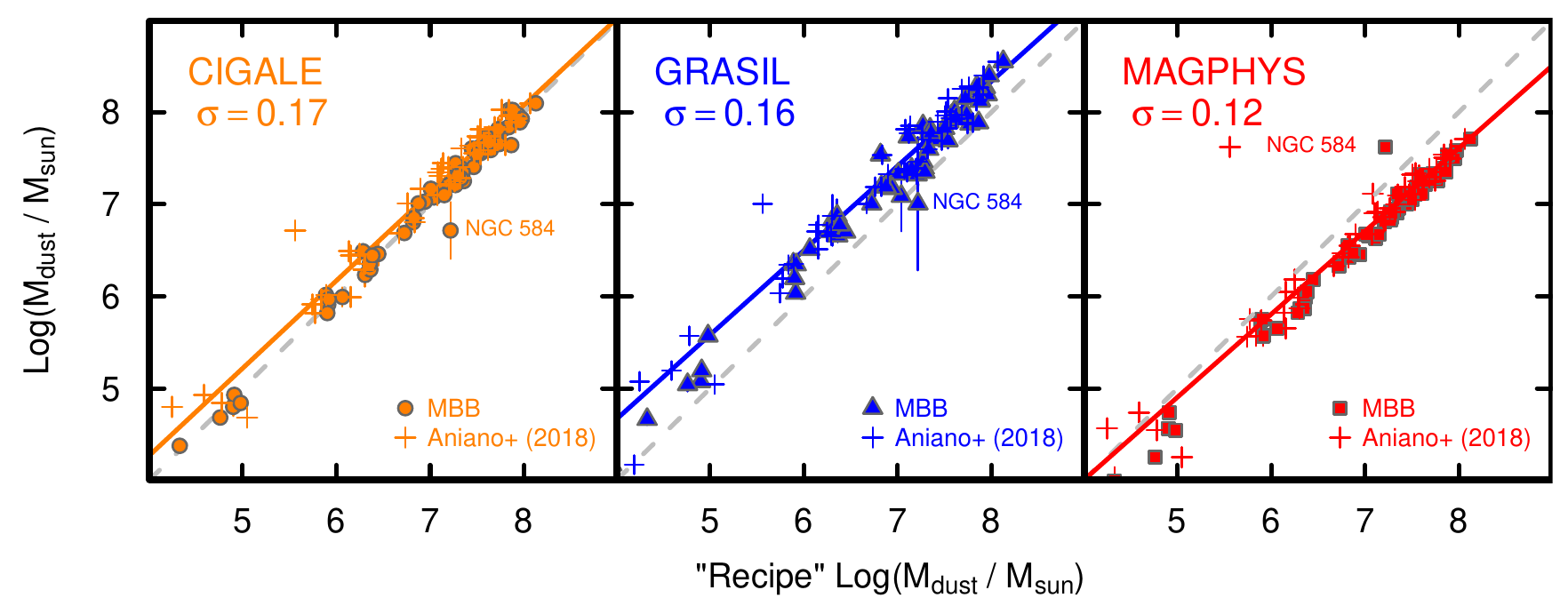}
\vspace{-0.6\baselineskip}
\caption{SED-derived \mdust\ plotted vs. independently determined \mdust\ 
(see Sect. \ref{sec:comparison:mdust} for details).
%The robust correlation is shown as a solid line, and the identity relation by a (gray) dashed one. 
%The lines are as in Fig. \ref{fig:mstar}.
The identity relation by a (gray) dashed lines, and  
the robust correlation relative to the DL07 values is shown as a solid line; 
the mean deviations for the fit of SED-derived quantities vs. those 
%by \citet{bianchi13} 
%from an MBB fit using the same methods as in \citet{bianchi13} 
using DL07 are shown by the $\sigma$ value in the upper left corner of each panel.
SED-fitting uncertainties are shown as vertical lines only for MBB $x$ values.
}
\label{fig:mdust}
\end{figure*}

\subsubsection{Comparison of star-formation rates}
\label{sec:comparison:sfr}

SED fitting typically gives more than one value of SFR;
here we have compared the SED SFR averaged over the last 100\,Myr %$10^8$\,yr
with our two choices of reference hybrid SFRs 
estimated from FUV$+$TIR luminosities and \ha$+$24\,\micron\ luminosities 
% \lmipsone\ corresponds to L24
(see Appendix \ref{app:sfr}).
The (robust) regression parameters are reported in Table \ref{tab:corr} as before,
and the comparisons of SED-inferred SFRs with reference ones are shown in Fig. \ref{fig:sfr}.

For \cig\ and \magp, the
agreement with SED fitting and independently-derived SFRs is slightly worse than with \mstar;
mean deviations are $\sim$0.2\,dex, and there are several galaxies 
for which reference values are much higher than the SED-derived values.
%with much lower predicted SFRs than estimated by the reference values. 
On the other hand, \gra\ SFRs are relatively close to the reference values, with
the exception of NGC\,1404, an early-type galaxy for which 
the recipe SFR is roughly 10 times lower than the \gra\ prediction; %predicts a SFR roughly 10 times higher;
there are no FIR detections for this galaxy so SFR is not as well constrained as with IR data.

Because SFRs are a sensitive function of SFH, it is possible that some of the SED fitting algorithms are unable to identify
the most suitable SFH because of degeneracies;
similarly good SED fits may be obtained with a variety of different SFHs.
Virtually all of the deviant galaxies for \cig\ and \magp\ are early types and/or lenticulars with low levels of specific SFR
(sSFR\,=\,SFR/\mstar) where the FUV may be indicating older stellar
populations rather than young stars \citep[e.g.,][]{rich05}. 
However,  SFR(\ha$+$\lmipsone) also shows a discrepancy relative to the fitting algorithms,
although the  scatter is slightly larger than for
%both SFR tracers present discrepancies relative to the fitting algorithms,
%although for all models SED-derived SFRs compared with SFR(\ha$+$\lmipsone) show a slightly larger scatter than 
SFR(FUV$+$TIR) (see Table \ref{tab:corr}). 
This discrepancy could also be due to the SFR we chose for comparison, namely the 100\,Myr average;
because of timescales, this estimate is expected to be more consistent with FUV$+$TIR than with \ha$+$\lmipsone.

As noticed by \citet{schiminovich07}, it is very difficult to probe star formation
at levels below sSFR$\la 10^{-12}$\,yr$^{-1}$, % suggested by JD since SFR/Mstar has units of yr^-1
and there are four KINGFISH galaxies with sSFRs at roughly this level (NGC\,1404, NGC\,584, NGC\,1316, NGC\,4594).
%In any case, the power-law slopes for both tracers are always significantly below unity, and the most significant sub-unity slope
%is also associated with the most pronounced (negative) deviation (see Table \ref{tab:corr}). %, although still $\la$0.2\,dex.
The problem of tracing SFR in low-sSFR (mainly early-type) galaxies
will be discussed further in Sect. \ref{sec:sfragain}.

\subsubsection{Comparison of dust masses}
\label{sec:comparison:mdust}

As shown in Fig. \ref{fig:mdust},
the single-temperature modified blackbody (MBB) dust masses 
estimated from the updated \hers\ photometry using the methods of \citet[][see Appendix \ref{app:mdust}]{bianchi13}
are able to reproduce fairly well the SED models. 
The scatter is quite low with mean \rms\ deviations between 0.06 and 0.15\,dex (see Table \ref{tab:corr}), 
although dominated by early-type NGC\,584, which is a problematic galaxy
for all the codes (see also Fig. \ref{fig:all_seds}).
However, the MBB offsets are sometimes significant; 
MBB estimates for \mdust\ are generally higher than \magp\ estimates, and
%\magp\ tends to generally underestimate \mdust\ relative to the MBB fits, but 
more so at high dust masses. 
Conversely, MBB estimates are below those of \gra,
%while \gra\ overestimates \mdust, 
and more so at low masses. 
%(see significant sub-unity slopes for both codes in Table \ref{tab:corr}).
The MBB estimates (with the DL07 opacities) show the best
agreement with the \cig\ models \citep[based on the updated version of the DL07 models,][]{draine14}. 
The power-law slope of the comparison is consistent with unity ($1.029\,\pm\,0.01$),
and the mean offset is virtually zero (see Table \ref{tab:corr}), consistent with
the \rms\ deviations for \cig\ of $\sim$0.10\,dex. % so this offset is % well within the uncertainty of the comparison.  perfectly consistent.

\begin{figure*}
\includegraphics[width=0.95\textwidth]{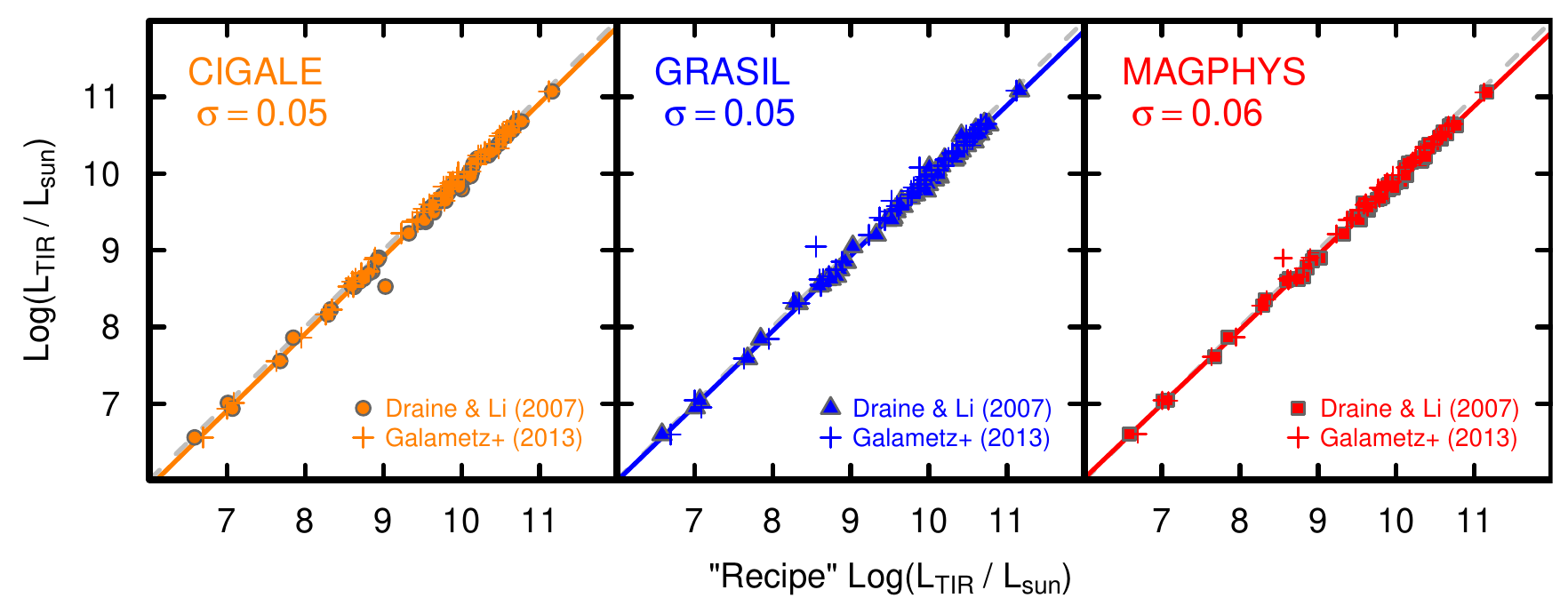}
\vspace{-0.5\baselineskip}
\caption{SED-derived \ltir\ plotted vs. independently determined \ltir\ from
\spit\ and \hers\ photometric data
(see Sect. \ref{sec:comparison:lums} for details).
In each panel,
the DL07 \ltir\ values are shown by filled circles (\cig),
filled triangles (\gra), and filled squares (\magp),
and those from \citet{galametz13} by $+$.
The $\sigma$ values shown in the upper left corner of each panel correspond
to the mean deviations of the \ltir\ fit with the DL07 values (see Table \ref{tab:corr}).
%The robust correlation is shown as a solid line, and the identity relation by a (gray) dashed one. 
The lines are as in Fig. \ref{fig:mstar}, and 
SED-fitting uncertainties are shown as vertical lines only for the DL07 $x$ values
(but they are typically smaller than the symbol size).
}
\label{fig:tir}
%\end{figure*}
%\begin{figure*}
%\includegraphics[width=\textwidth]{BC03KINGFISH_LFUV_Murphy_cigale_magphys_grasil_AGAIN-crop.pdf}
\includegraphics[width=0.95\textwidth]{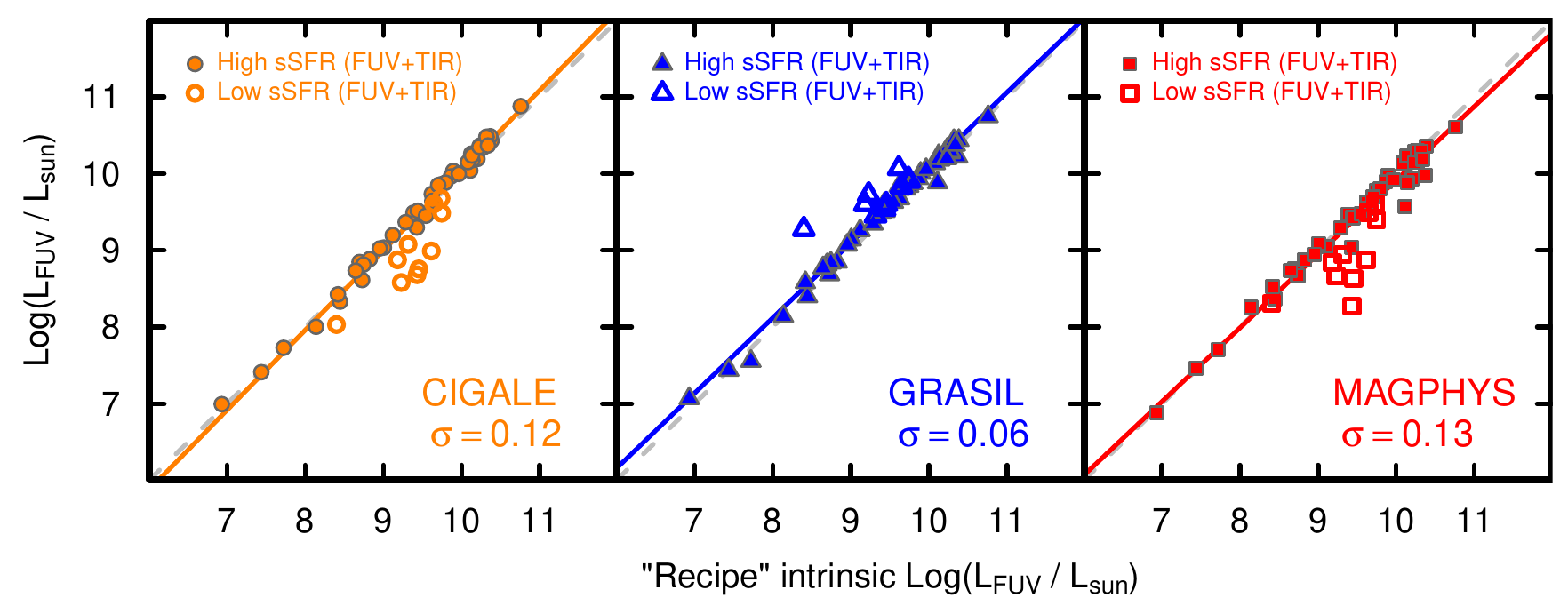}
\vspace{-0.5\baselineskip}
\caption{SED-derived \lfuv\ plotted vs. independently determined \lfuv\ with extinction corrections derived
from \spit\ and \hers\ photometric data
(see Sect. \ref{sec:comparison:lums} for details).
The $\sigma$ values shown in the lower right corner of each panel correspond
to the mean deviations of the \lfuv\ fit (see Table \ref{tab:corr}).
%The robust correlation is shown as a solid line, and the identity relation by a (gray) dashed one. 
The lines are as in Fig. \ref{fig:mstar}.
As in Fig. \ref{fig:sfr}, filled symbols correspond to high specific SFR (Log(sSFR/yr$^{-1}$) $> -10.6$), 
and open ones to low specific SFR (Log(sSFR/yr$^{-1}$) $\leq -10.6$),
as noted in the legend in the upper left corners. 
This sSFR limit corresponds roughly to the lowest quantile in the KINGFISH
galaxies, and also to the inflection in the SFMS by \citet{salim07}.
}
\label{fig:fuv}
%\end{figure*}
%\begin{figure*}
%\includegraphics[width=\textwidth]{KINGFISH_AFUV_G13_cigale_magphys_grasil-crop.pdf}
%\includegraphics[width=\textwidth]{BC03KINGFISH_AFUV_G13Murphy_SSFR_cigale_magphys_grasil-crop.pdf}
\includegraphics[width=0.95\textwidth]{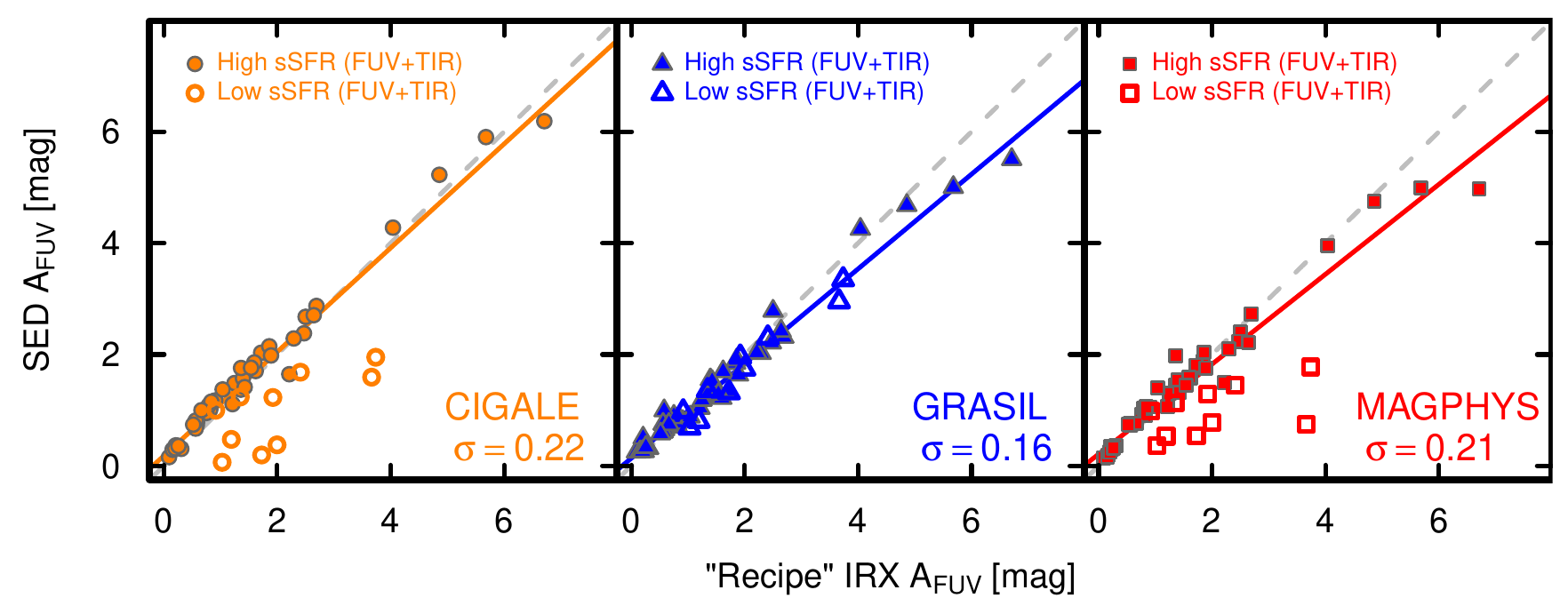}
\vspace{-0.5\baselineskip}
\caption{SED-derived \afuv\ plotted vs. \afuv\ derived according to 
\citet[][see Sect. \ref{sec:comparison:lums} for details]{murphy11}.
% The robust correlation is shown as a solid line, and the identity relation by a (gray) dashed one. 
%The mean residuals for the fit of SED-derived \afuv\ vs. \afuv\ derived as in \citet{hao11} 
The lines are as in Fig. \ref{fig:mstar}.
The mean deviations for the fit of SED-derived \afuv\ vs. \afuv\ derived as in \citet{murphy11} 
are shown by the $\sigma$
value in the lower right corner of each panel.
As in Fig. \ref{fig:sfr}, filled symbols correspond to high specific SFR (Log(sSFR/yr$^{-1}$) $> -10.6$), 
and open ones to low specific SFR (Log(sSFR/yr$^{-1}$) $\leq -10.6$),
as noted in the legend in the upper left corners. 
}
\label{fig:afuv}
\end{figure*}

The \rms\ deviations of the DL07 \citep{aniano18} models compared to the SED fitting \mdust\ are
higher than for the MBB fits, ranging from 0.12 to 0.17\,dex (shown in Fig. \ref{fig:mdust}).
The offset in the comparison of the DL07 models used here for \cig\ with the DL07 \mdust\ values given by
\citet{aniano18} is consistent with the renormalization applied by those authors.
This renormalization, based on the results by \citet{planck16}, lowers the DL07 dust mass by
a small amount that depends on \umin, the minimum ISRF heating the dust.
On average, this correction amounts to $\sim$12\% (see Table \ref{tab:corr}).
The \mdust\ estimates of the different codes show similar behavior relative to the DL07 values by \citet{aniano18}
as for the MBB values calculated here:
namely \cig\ shows the best agreement,
while the DL07 values are low compared to \gra\ and high compared to \magp.

We investigated whether discrepancies in \mdust\ between the \gra\ and \magp\ SED models and 
the reference dust estimates %MBB estimates
could be attributed to differences in the adopted dust opacity.
The \magp\ models \citep{dacunha08} assume a fiducial dust opacity at 850\,\micron\
of   $\kappa_{\rm abs}\,=\,0.77$\,cm$^2$\,g$^{-1}$ \citep{dunne00}, 
and a spectral index $\beta$ of 1.5 or 2.0 %depending on the temperature of the dust.
for the warm and cold component, respectively.
At 850\,\micron, the DL07 models have $\kappa_{\rm abs}\,=\,0.38$\,cm$^2$\,g$^{-1}$, roughly a factor
of two lower than the value used by \magp\ \citep[and $\sim$1.5 times lower than the value at 
850\,\micron\ of the opacities of][]{laor93}.
Thus, the observed underestimate of $\la$ 2 would be consistent with
the different assumed dust opacities of \magp\ relative to the DL07 values used by \citet{bianchi13}.
However, the observed significant sub-unity slope (see Table \ref{tab:corr}),
and the use of a flatter $\beta$ according to the temperature of the dust,
contribute to the discrepancy which increases with increasing \mdust.
% {\bf lkh tried to investigate this further with single-T MBBs, but would need the temperatures and relative
% luminosities of the \magp\ dust components in order to proceed;
% also need to confirm the dust emissivities which lkh took from the da Cunha 
% definition paper of 2008.}

\gra\ dust is based on the \citet{laor93} opacity curves, which for the combined grain
populations gives
$\kappa_{\rm abs}\,=\,6.4$\,cm$^2$\,g$^{-1}$ at 250\,\micron, roughly 60\% higher than the
value of $\kappa_{\rm abs}\,=\,4.0$\,cm$^2$\,g$^{-1}$ used here \citep[see][]{bianchi13}, based on
the dust models by DL07.
This would imply that the \gra\ \mdust\ values should be {\it underestimated} by a factor of
$\sim$1.6 ($\sim$0.2\,dex) relative to the MBB values,
but, instead, they tend to be overestimated.
Another difference between the DL07 models and the \citet{laor93} dust used by \gra\ is the 
mean emissivity power-law index, $\beta$.
If the DL07 opacities are fitted with a wavelength-dependent power law 
between 70 and 700\,\micron, the power-law index $\beta$\,=\,2.08 \citep{bianchi13};
for the \citet{laor93} grains the fitted index over the same wavelength range is slightly smaller, $\beta$\,=\,2.02.
Although seemingly a minor difference, 
% shallower $\beta$ values tend to increase \mdust,
% Ilse complains rightly, so need to reword
because most of the dust mass resides in the cooler dust that emits at longer wavelengths,
and because the absolute emissivity at the fiducial wavelength is fixed,
shallower $\beta$ values cause an increase 
%in the fitted flux ratios and thus, incrementally, the dust mass.
in the submm emission and thus, incrementally, lower estimated dust mass when matching
to observed fluxes.
Between 100 and 500\,\micron, this tiny difference in $\beta$ causes an increase in fitted flux
at longer wavelengths, and thus of \mdust, of $\sim$10\%; % together with the differences in dust opacities 
this could partially compensate the differences in adopted dust opacities.

%However, the differences in opacities for \gra\ go in the opposite direction from the observed trend,
%as we would expect dust masses to be underestimated rather than overestimated relative to the MBB fits 
% would expect dust masses to be smaller rather than larger relative to 
% the \gra\ dust mass estimates go in the opposite direction from the observed trend,
However, the MBB fits (and the DL07 values from \citealt{aniano18}) are lower than the \gra\ dust-mass estimates,
contrary to what would be expected from the differences in opacities. 
It is interesting that the only one of the three SED models that includes realistic geometries of dust and stars
%{\bf
%and computes a range of grain temperatures given the impinging ISRF, 
%}
generally gives dust masses that are higher than the single-temperature MBB fits.
It is possible that the true dust mass needed to shape the SED with the combined effects of dust extinction and emission
is larger than what would be inferred from the simple MBB assumption
\citep[e.g.,][]{dale02,galliano11,magdis12,santini14}.

\subsubsection{Comparison of luminosities and FUV attenuation}
\label{sec:comparison:lums}

Figure \ref{fig:tir} compares \ltir\ derived from SED fitting with the two photometric formulations described
in Appendix \ref{app:lums}: DL07 and \citet[][hereafter G13]{galametz13}. 
\ltir\ is the most robust parameter compared with SED fitting, with \rms\ deviations relative to the analytical expressions
between 0.03 and 0.06\,dex.
Nevertheless, both formulations slightly overestimate \ltir\ relative to
the SED models. 
Taking DL07 which relies only on \spit\ photometry, the discrepancy is % ranges from 0.09\,dex (\cig) to
$\sim$0.06--0.09\,dex; % (\gra, \magp); 
the agreement is better for the G13 formulation which incorporates \hers\
photometry (0.05\,dex for \cig; 0.04\,dex for \gra; 0.02\,dex for \magp).
The power-law slopes relative to both estimates of \ltir\ are unity to within the
uncertainties for all the SED models, with the possible exception of \magp\ (relative to DL07).
% The \cig\ power-law slope relative to DL07 is unity, 
% probably a consequence of the incorporation of the same models
% for the infrared-portion of the SED; however it is also within unity (1.014$\pm$0.01) for the G13 expression.
% On the other hand,
% \gra\ shows a unit slope for G13, but a slightly sub-unit slope (0.981$\pm$0.01) for DL07;
% the \magp\ slopes are also slightly sub-unity.
Overall, the ultimate agreement with the SED-derived values 
is within 3--5\% for G13 and within $\sim 6-9$\% for DL07.

%The discrepant point for both \gra\ and \magp\ models at \ltir$\sim 10^9$\,\lsun\ is Fornax\,A (NGC\,1316),
%an early-type radio galaxy at the center of the Fornax cluster;
%NGC\,1316 is also a merger remnant and has a significant dust component acquired externally, in addition
%to a luminosity contribution from its Active Galactic Nucleus
%\citep[AGN,][]{lanz10}.

The intrinsic FUV luminosities \lfuv\ from SED fitting and from the corrected observed luminosity are compared in Fig. \ref{fig:fuv}.
As described in Appendix \ref{app:lums}, 
we have derived the reference \lfuv\ by correcting
observed FUV fluxes for attenuation using \afuv\ calculated according to \citet{murphy11}\footnote{Here we use \ltir\ from the formulation of G13.}.
Instead of FUV colors \citep[e.g.,][]{boquien12}, this correction relies on
IRX, $\log_{10}$ of the ratio of \ltir\ and \lfuv\ \citep[e.g.,][]{buat05}.
As in previous figures,
the open symbols in Fig. \ref{fig:fuv} correspond to low sSFR\,=\,SFR/\mstar\ (Log(sSFR/yr$^{-1}$) $\leq -10.6$),
roughly the inflection or turnover point in the SFMS by \citet{salim07}, and also
approximately to the lowest quantile of the KINGFISH sample.
\lfuv\ estimated by all the SED algorithms is very close to the photometric estimate 
using the \citet{murphy11} recipe for the extinction correction, with
mean deviations between $\sim$0.08--0.13\,dex.
Results are unchanged if we incorporate, instead, the recipe by \citet{hao11}.

Interestingly, for the problematic early-type galaxy, NGC\,584 
(the discrepant open triangle in the middle panel of Fig. \ref{fig:fuv}),
the recipe \lfuv\ is much lower than the \gra\ estimate,
while recipe \lfuv\ for galaxies with low sSFR
%two early-type galaxies, NGC\,1404 (see also Fig. \ref{fig:sfr})
tends to exceed the \cig\ and \magp\ values
(see also Sect. \ref{sec:comparison:sfr}).
As discussed above, these discrepancies are almost certainly due to different approaches in associating a 
specific SFH with a given SED, and we will elaborate on this further in Sect. \ref{sec:assumptions}.

The SED models derive extinction at a given wavelength through the ratio of intrinsic to observed (attenuated) emission,
while the reference \afuv\ is derived through IRX rather than UV colors (see Appendix \ref{app:lums}).
Figure \ref{fig:afuv} shows the comparison of the SED-derived \afuv\ and \afuv\ calculated %from IRX according to \citet{hao11}. 
according to \citet{murphy11}.
In all cases, there is a discrepancy between photometric and SED fitting results,
with photometric \afuv\ exceeding the SED \afuv\ values with fairly large scatter, $\sim$0.2\,dex.
The discrepancy increases with increasing attenuation (see significant sub-unity slopes in
Table \ref{tab:corr}), and can be $\ga$ 1\,mag at high \afuv. %$\sim$4\,mag.
However, at low \afuv\ (and low \lfuv, see above), the disagreements for \cig\ and \magp\ %all SED-fitting algorithms 
are apparently associated with low sSFR (shown by open symbols in Fig. \ref{fig:afuv}).
This association probably results from two potential problems with the usual (photometric) estimates of \afuv:
dust heating from longer-lived low-mass stars %evolved stars 
may contribute to IR emission and thus spuriously increase IRX 
%\textcolor{green}{[Maybe mention Boquien+16 as it shows a nice variation with the relative radiation field from old stellar populations]}, 
causing \afuv\ to be overestimated
\citep[e.g.,][]{boquien16}.
Conversely, FUV emission from post-Asymptotic Giant Branch (pAGB) stars may contribute to FUV luminosity
and cause \afuv\ to be underestimated.
%{\bf need to discuss additional possible reasons why and how they apply to the different algorithms...} 
%\textcolor{green}{[Maybe it would be worth noting that AGN can pose a problem as they have not been included in either model (cigale handles AGN but I did not activate this feature for this paper)]}
For \gra, these factors may also be problematic, but the disagreements are not
so clearly associated with galaxies having low sSFR;
we will discuss this point further below.

\begin{figure*}
\includegraphics[width=\textwidth]{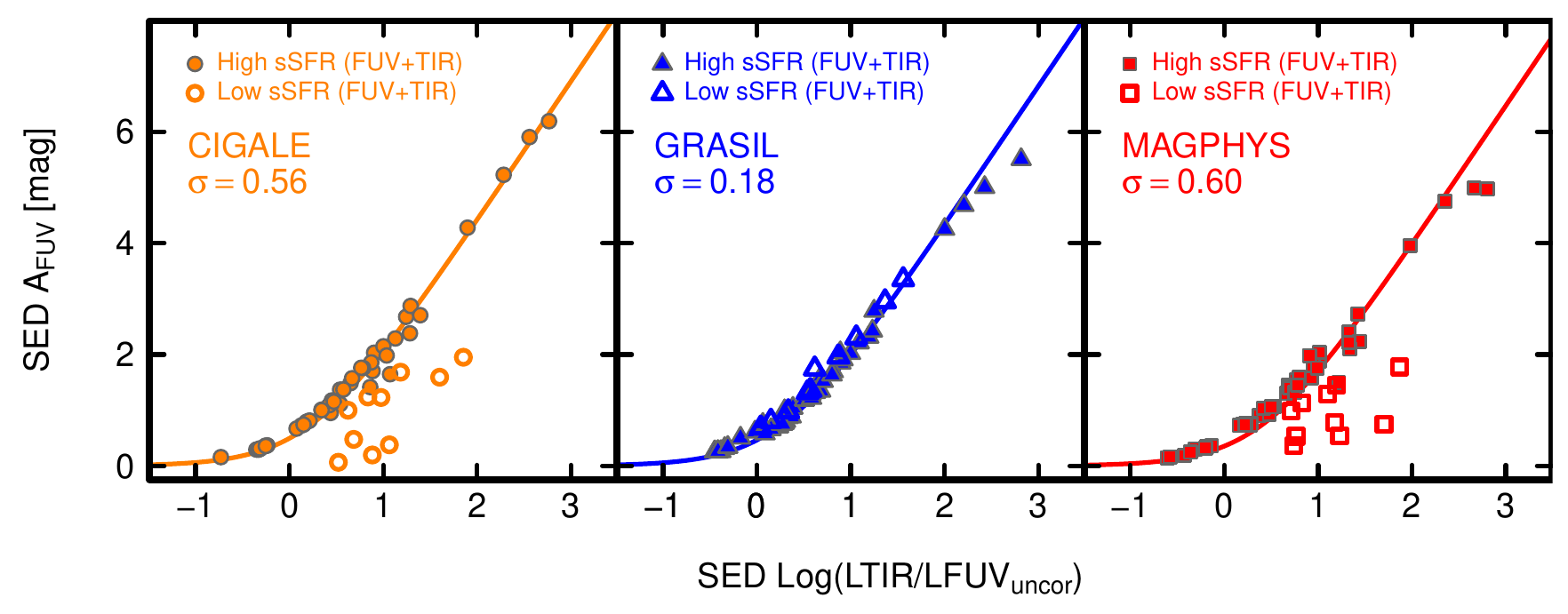}
\vspace{-\baselineskip}
\caption{SED-derived \afuv\ plotted against SED-derived IRX [$\log_{10}$(\ltir/\lfuv)].
The solid curve shows the fit obtained by adopting the formulation in
Eqn. (\ref{eqn:afuv}); as described in the text, the best-fitting $a_{\rm FUV}$ values
are estimated using only the galaxies with (Log(sSFR/yr$^{-1}$) $> -10.6$).
The mean deviations comparing the SED-derived \afuv\ and the fitted ones from SED-derived IRX 
(now including all galaxies) are shown by the $\sigma$
value in each panel. %the upper right corner of each panel.
As in previous figures, filled symbols correspond to high specific SFR (Log(sSFR/yr$^{-1}$) $> -10.6$), 
and open ones to low specific SFR (Log(sSFR/yr$^{-1}$) $\leq -10.6$),
as noted in the legend in the upper left corners. 
}
\label{fig:afuvsed}
\end{figure*}

The shallower slope of \afuv\ relative to the reference values common to all three
%codes may provide valuable input to the \citet{hao11} formulation:
codes may provide valuable input to the \citet{murphy11} or (equivalent) \citet[][]{hao11} formulations:
\begin{equation}
A_{\rm FUV}\,=\,2.5\ \log_{10}\,(1 + a_{\rm FUV} \times 10^{\rm IRX})
\label{eqn:afuv}
\end{equation}
\noindent
where $a_{\rm FUV}$ is a scale parameter, 
related to the fraction of the bolometric luminosity emitted in the FUV, $\eta_{\rm FUV}$
%\begin{equation}
($L_{\rm FUV}({\rm cor})\,=\,\eta_{\rm FUV}\,L_{\rm bol}$). % \quad 
%\end{equation}
%roughly equal to the product of the
%inverse bolometric correction $\eta_{\rm FUV}$ 
%and the attenuation ratio 
%$(1-e^{-\tau_{\rm FUV}})/(1-e^{-\bar{\tau}})$,
%where $\bar{\tau}$ is the effective opacity for the starlight heating the dust.
The FUV optical depth $\tau_{\rm FUV}$ (FUV attenuation in magnitudes \afuv\,=\,1.086\,$\tau_{\rm FUV}$)
is defined by:
\begin{equation}
L_{\rm FUV}({\rm obs})\,=\,L_{\rm FUV}({\rm cor})\,e^{-\tau_{\rm FUV}}\quad 
\end{equation}
and the effective opacity of the dust-heating starlight $\bar{\tau}$:
\begin{equation}
L_{\rm TIR}\,=\,L_{\rm bol}\,(1 - e^{-\bar{\tau}})\quad   .
\end{equation}
where $L_{\rm bol}$ is the bolometric luminosity.
As shown by \citet{hao11},
\begin{equation}
\tau_{\rm FUV}\,=\,\ln \left[ 1 + \eta_{\rm FUV}\, \frac{L_{\rm TIR}}{L_{\rm FUV}({\rm obs})}\, \frac{1-e^{-\tau_{\rm FUV}}}
{1-e^{-\bar{\tau}}} \right]
\end{equation}
implying that
$a_{\rm FUV}\,=\,\eta_{\rm FUV}\,(1-e^{-\tau_{\rm FUV}})/(1-e^{-\bar{\tau}})$
since IRX\,=\,$\log_{10}(L_{\rm TIR}/L_{\rm FUV}({\rm obs}))$.

%\citet{hao11} find $a_{\rm FUV}\,=\,0.46\pm0.12$ for their sample (which is very similar
\citet{murphy11} find $a_{\rm FUV}\,=\,0.43$ for the KINGFISH sample
studied here, while
\citet{hao11} find $a_{\rm FUV}\,=\,0.46$ for a similar sample.
We have estimated new values of $a_{\rm FUV}$ for each of the SED algorithms by
fitting Eqn. (\ref{eqn:afuv}) to the comparison of SED-derived \afuv\ and the IRX values of the best-fit SED (using the
\ltir\ shown in the ordinate of Fig. \ref{fig:tir} combined with ``observed'', extinguished, values of \lfuv,
i.e., not the corrected values shown in the ordinate of Fig. \ref{fig:fuv}). 
The fits have been performed using only galaxies with ``high sSFR''
[Log(sSFR/yr$^{-1}$) $> -10.6$].

Figure \ref{fig:afuvsed} shows the results of this exercise;
the \rms\ deviations of the fits given in %the upper right-hand corner of 
each panel correspond to all the galaxies, including all
values of sSFR.
We find that \cig\ 
% 14/12/2017 checked
% 2/2/2018 checked again
($a_{\rm FUV}\,=\,0.59\pm 0.02$)
and \gra\ 
% 2/2/2018 checked again
($a_{\rm FUV}\,=\,0.52\pm 0.02$)
% 0.51441    0.02241
prefer higher values of $a_{\rm FUV}$, 
while the best fit for \magp\ gives a lower value
% 2/2/2018 checked again
($a_{\rm FUV}\,=\,0.40\pm 0.02$).
That $a_{\rm FUV}$ is generally larger than the recipe-derived value ($0.43-0.46$)
is possibly counter-intuitive,
given the sub-unity slope comparing SED- and recipe \afuv\
seen in Fig. \ref{fig:afuv}.
However, the SED-derived IRX tend to be 0.1--0.2\,dex smaller than the
photometric values of IRX, and, except for \gra, are related with a super-unity power-law
index; thus in some sense the effects compensate one another and result in a slightly
larger $a_{\rm FUV}$.

Although the extinction curve assumptions in \cig\ and \magp\ differ substantially from
the geometry of stars and dust contemplated by \gra,
the SED shapes of all three algorithms are well approximated by Eqn. (\ref{eqn:afuv}).
Except for \gra, the scatter is large for galaxies with low sSFR, but 
the general agreement is encouraging, both because SEDs generated from diverse
complex algorithms are consistent, and also because the simplistic photometric approach is
a realistic approximation of galaxy SEDs, at least for the KINGFISH galaxies.
%Our estimate of $a_{\rm FUV}$ from the SED codes is $\sim 0.2-0.3$.
%This smaller value, related to the sub-unity slopes shown in Fig. \ref{fig:afuv},
%varies from code to code with \gra\ giving the smallest value
%$0.19\,\pm\,0.02$,
%and \cig\ the largest $0.32\,\pm\,0.03$ (\magp: $0.24\,\pm\,0.04$).
%Including only the galaxies with sSFR above a certain threshold (filled symbols in Fig. \ref{fig:afuv}),
%we refit the SED-derived \afuv\ values using the \citet{hao11} as before but with the new fitted $a_{\rm FUV}$ coefficients.
%The scatter relative to Fig. \ref{fig:afuv} is much reduced, ranging
%from $\sigma\,=\,0.16$ for \gra\ to $\sigma\,=\,0.29$ for \magp\  (\cig: $\sigma\,=\,0.20$).

\section{Impact of different model assumptions}
\label{sec:assumptions}

In the previous section, we have compared results for fundamental quantities derived
from SED fitting to those obtained from simpler methods (recipes).
Sometimes the agreement both among the models and with the %reference 
recipe quantities is
excellent (e.g., \ltir); in other cases, there are slight (\mstar) or severe (\afuv)
discrepancies of the recipe parameter relative to all the models. 
Finally, there are some cases where a particular model %does better than the
is in closer agreement than others relative to the reference parameter. 
%(e.g., SFR for \gra; \mdust\ for \cig).
Some of this behavior may arise from the
assumptions behind the photometric methods used to derive the reference values,
which may or may not be also incorporated in the models
(e.g., optically thin dust, FUV light from active star-forming stellar populations, etc.).

%\textcolor{blue}{
It is also true that SED fitting inherently suffers from degeneracies; 
%\textcolor{green}{
%[They suffer from degeneracies but they are taken into account through the PDF. 
%Empirical recipes ``hide'' this by being much less flexible. I think it should be made clear that it is not 
%because \cig/\magp/\gra\ give different results compared to empirical estimators that they are wrong.]
%}
a similar SED may emerge from radically different SFHs, while small variations
of other parameters may cause very different SEDs \citep[e.g., the dust optical depth, see][]{takagi03}.
In part, the inclusion of the IR regime helps to break the age-attenuation degeneracy
inherent in optical SED fitting \citep[e.g.,][]{lotz00,lofaro13},
distinguishing between dusty star-forming galaxies and evolved stellar populations
\citep[e.g.,][]{pozzetti00}.
Nevertheless, it is important to examine how the different SED-fitting algorithms
treat possible degeneracies in order to achieve the best-fit SED.
%}

As pointed out by \citet{michalowski14}, the scatter of SED-derived values
is probably an inherent limit for the accuracy of SED models because of
the necessary simplifications 
(e.g., galaxy geometry and the form of the dust-attenuation wavelength dependence).
On the other hand, the quality of SED fitting is strongly affected by the set and 
quality of data at our disposal. 
Ultimately, the quantity and quality of the data are the defining factors in the 
reliability of SED-fitting models.
Given the broad wavelength coverage and good quality of the KINGFISH photometry
\citep{dale17}, we can assess differences in the results of SED fitting better
than previously possible.
Below we discuss some of the assumptions intrinsic to each of
the SED models, and how these could impact the derived results.

\subsection{Star-formation history, stellar mass, and SFR}
\label{sec:assumptions:sfh}

Perhaps the most critical parameter in the SED fitting is the assumed SFH.
All SED-fitting algorithms rely on a grid of SFHs,
but which differ in their formulation (see Table \ref{tab:models}).
The version of \cig\ used here defines a ``delayed'' SFH at early times, with a step-like
change of the SFR added at more recent times;
\magp\ adopts an exponentially declining SFR 
with random bursts of SF activity superimposed uniformly over the lifetime of the galaxy.
\gra\ approaches the problem from a different point of view,
namely to model the timescale of gas inflow and leave as an additional free parameter the
efficiency of the conversion of gas into stars;
the age of the galaxy results from the best-fitting SED.
These differences in SFH among the models propagate to differences % in their ability to accommodate some of the 
between the photometric recipes and model-derived quantities.
 
It has been argued that a necessary ingredient for deriving accurate stellar masses at
high redshift is
a bi-modal SFH, that is one with more than one episode of star formation \citep{michalowski14}.
On the other hand, \citet{lofaro13} find that most of the IR-luminous galaxies at $z\sim 1-2$
modeled with \gra\ %(a modified version of the code we use here) 
do not require a two-component SFH.
\citet{conroy10} analyzed the impact of SFH on the (UV-NIR) SED of simulated galaxies;
the simulated galaxies were characterized by basically one star-formation episode each, but 
at different ages to distinguish passive from star-forming galaxies.
\cig\ and \magp\ both have bi-modal SFHs, with one or more recent bursts of star formation
superimposed on an older episode; however 
the \gra\ libraries we use here have only
a single episode whose timescale and efficiency are fitted parameters.

Despite the different approaches to SFH,
the three codes generally give similar stellar masses and even SFRs.
Given that the three codes result in similar \mstar\ values, 
%the reference \mstar\ \citep{wen13} is fairly well reproduced by all three, 
it is likely that the
three different recipes for SFH are equally effective for the nearby KINGFISH galaxies.
Stellar populations for all codes are modeled with SSPs from \citet{bruzual03}, 
and use the \citet{chabrier03} IMF.
As discussed in Sect. \ref{sec:comparison:mstar},
the best agreement with SED-derived \mstar\ and the reference \mstar\ values
is for the \citet{wen13} luminosity-dependent  $\Upsilon_*$ formulation, rather than a constant $\Upsilon_*$ \citep[e.g.,][]{mcgaugh14}.
The latter gives \mstar\ that, on average, is 0.3--0.5\,dex larger than 
derived from SED fitting, while the estimates using the luminosity-dependent $\Upsilon_*$ \citep{wen13} tend to
be $\sim$0.1--0.3\,dex too large.

% DOES solve the hyperlink problem
%\vspace{\baselineskip}

Part of the discrepancy of the recipe \mstar\ may be
from the contribution of warm dust to the 3.6\,\micron\ continuum \citep[][]{meidt12,meidt14},
which we did not correct for here (although we do correct for nebular contamination, see Appendix \ref{app:mstar});
nevertheless, globally, the warm-dust component is expected to be rather small,
\citep[$\sim$3--10\%,][]{meidt12} so probably cannot explain the systematic difference.
In addition, our assumption that IRAC 3.6\,\micron\ and WISE W1 fluxes are the same
may also be incorrect in some cases; however, 
judging from our own photometry, they cannot be more than a few percent discrepant. 
Stellar masses derived from SED fitting are almost certainly superior,
when there is sufficient data coverage (here also IR). 
Moreover, the relatively good agreement among the codes suggests
that stellar masses can be consistently determined
even under the rather different assumptions inherent to each of the models.
Different formulations of SFH, SSPs, and extinction (see below) do not greatly affect
the determinations of stellar mass, at least when IR data are included.

An important difference in the \cig\ modeling is that 
SFHs are included with a strong diminution 
of star formation in the recent past to allow for quenching
\citep[see also][]{ciesla16}. 
Thus it is possible to model passive galaxies now forming few to no stars at all. 
However, the characterization of a very low level of star formation is particularly difficult. 
A SFR of $10^{-7}$ \msunyr\ will give an SED very similar to that obtained with an 
SFR of $10^{-3}$ \msunyr\ as in either case, older stellar populations will contribute a 
large fraction of total dust heating.
Indeed, for galaxies with sSFR$\la 3 \times 10^{-11}$\,yr$^{-1}$ [Log(sSFR/\,yr$^{-1}$)\,=\,$-10.6$],
both \cig\ and \magp\ show differences in the estimates of SFR, \lfuv, and 
\afuv\ compared to empirical recipes and to \gra. 
The differences in inferred SFR are evident even when the recipe SFR tracer relies on \ha$+$24\micron,
rather than FUV$+$TIR.

\begin{figure}[!h]
\vbox{
\includegraphics[width=0.48\textwidth]{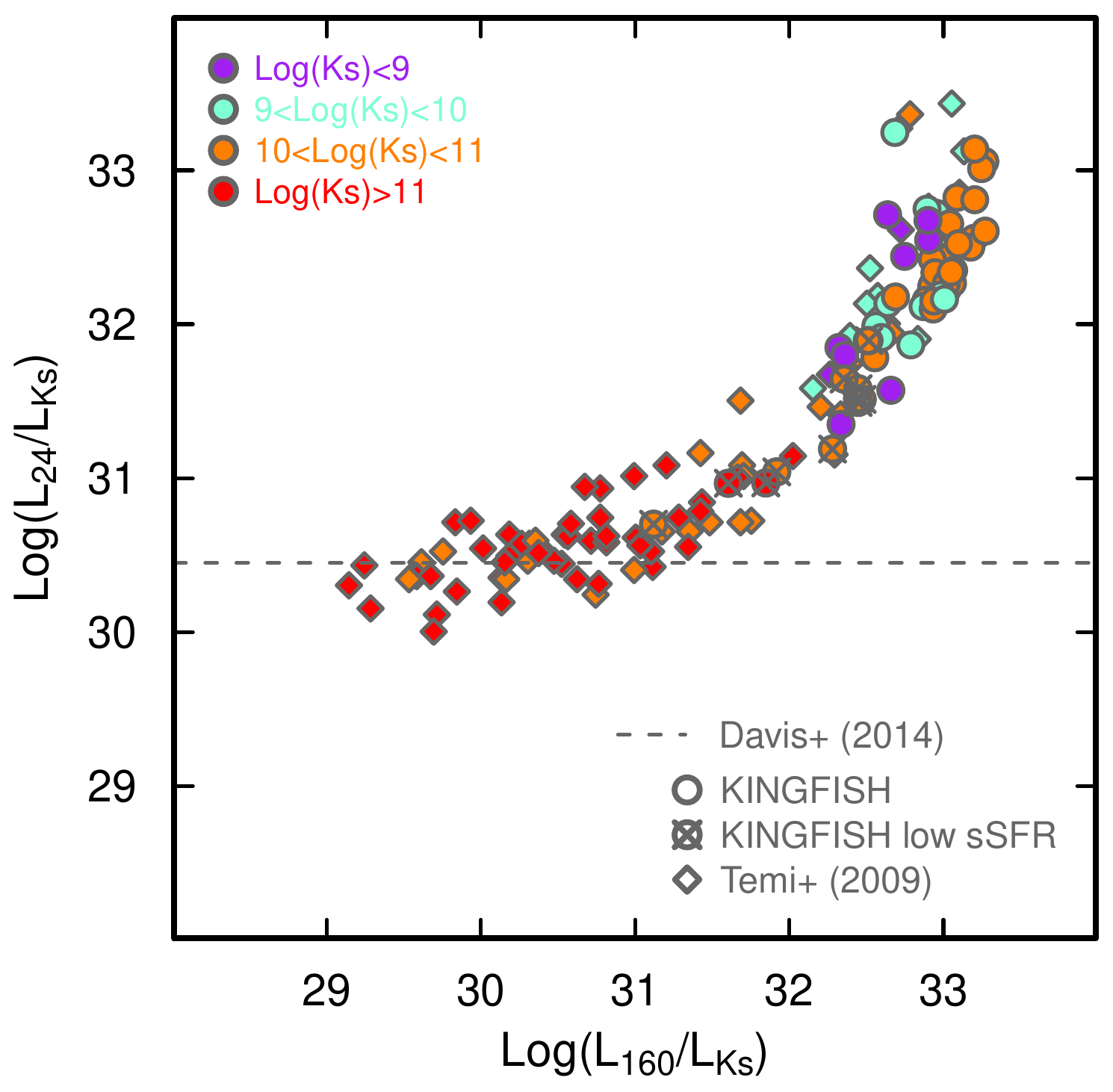} \\
\\
\includegraphics[width=0.48\textwidth]{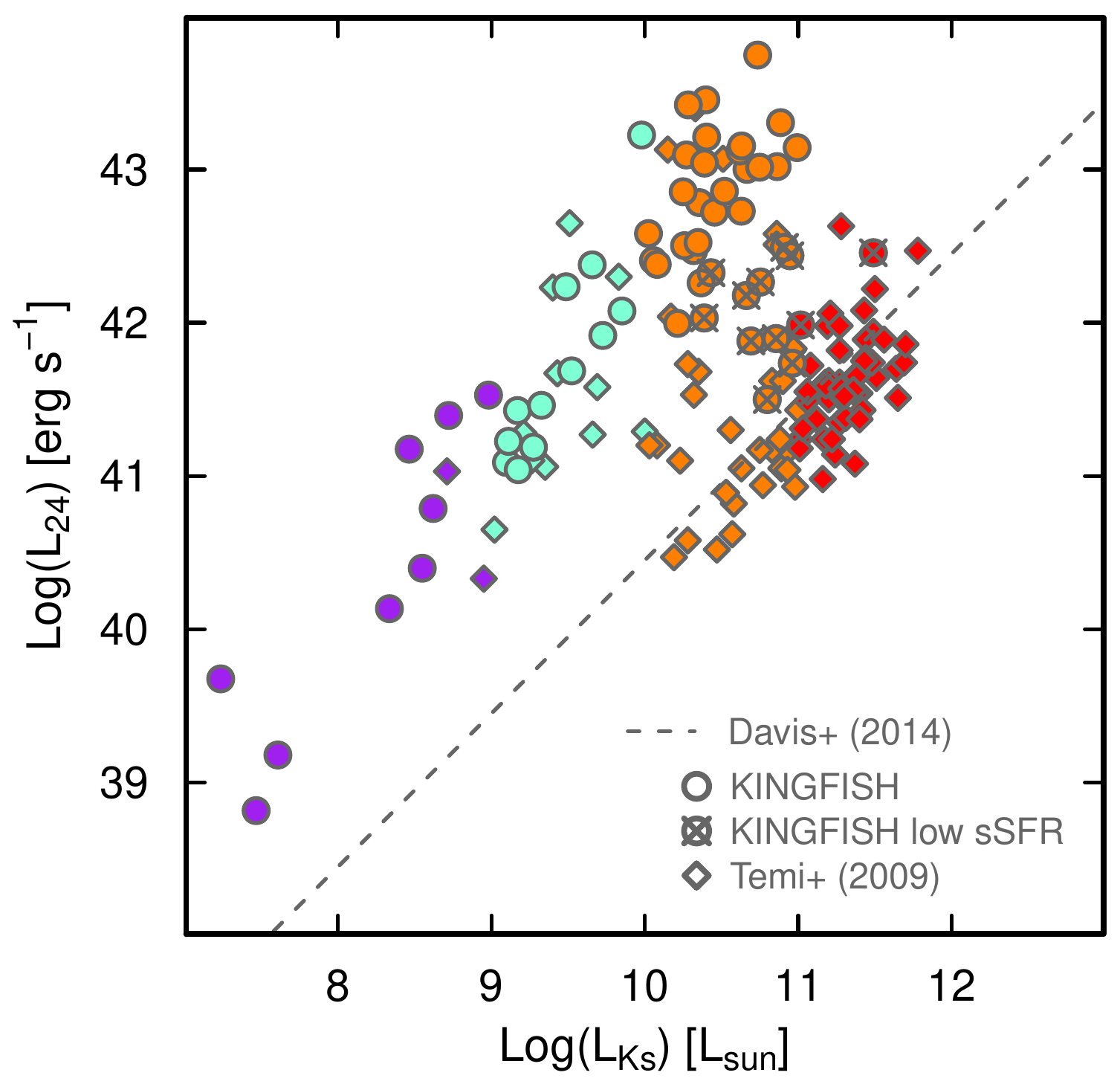}
}
\caption{Upper panel: Log(\lmipsone/\lks) vs. Log($L_{160}$/\lks\ luminosities
of the KINGFISH galaxies (shown as filled circles), 
together with the sample of ETGs from \citet{temi09b} shown as filled diamonds.
Following \citet{temi09a,temi09b} and \citet{davis14},
$L_{Ks}$ luminosities are in units of \lsun, and the IR luminosities in units of \ergs.
KINGFISH galaxies with low sSFR (as in previous figures) are shown with a $\times$
superimposed.
The color scale corresponds to bins of \lks\ as indicated in the upper left corner.
The horizontal dashed line corresponds to the quiescent stellar ratio of \lmipsone/\lks\
[Eqn. (\ref{eqn:l24lk})] as defined by \citet{davis14}.
It is evident that galaxies with low sSFR have \lmipsone/\lks\ ratios close to
the quiescent value.
Lower panel: Log(\lmipsone) vs. Log(\lks) with the \citet{davis14}
relation [Eqn. (\ref{eqn:l24lk})] shown as a dashed line.
Symbols are the same as in the upper panel.
}
\label{fig:l24lk}
\end{figure}

% DOES solve the hyperlink problem
% \vspace{-1.3\baselineskip}
%\vspace{2\baselineskip}

\subsection{SFR estimates revisited to account for older stars}
\label{sec:sfragain}

% DOES solve the hyperlink problem
%\vspace{-0.5\baselineskip}
As discussed above, the codes incorporate different approaches
to the parametrization of the SFH:
\cig\ and \magp\ have a bi-modal SFH, while \gra\ relies on a single SF episode. 
As shown in Fig. \ref{fig:sfr},
our choice of recipe SFR compares best with SED-derived values
by \gra\ while, as mentioned above,
\cig\ and to some extent also \magp\ underestimate SFR 
relative to the recipe %reference values
for galaxies with low sSFR ($\la 3 \times 10^{-11}$\,yr$^{-1}$). 
This could be consistent with the idea that the smoother SFH of \gra\ (because of
the one-component SFH) is closer to the constant SFR assumption of the recipe value.
Indeed, \citet{boquien14} found evidence that the usual assumption of a constant SFH over 100\,Myr can 
cause discrepancies of $\sim$25\% on average compared to the true SFR.

On the other hand, SFRs in galaxies with low sSFRs are notoriously difficult to measure
\citep[e.g.,][]{schiminovich07,temi09a,temi09b,davis14}.
Such galaxies are typically early types (ETGs), and the KINGFISH sample
is no exception, even though there is not an exact one-to-one correspondence between
Hubble type and sSFR.
%As noted in Sect. \ref{sec:comparison:sfr},
The UV upturn caused by extreme Horizontal Giant Branch stars can be an important component
of UV flux in ETGs \citep[e.g.,][]{kaviraj07}. 
Also FUV and \ha\ may be produced by photoionization
from old stars, in particular pAGBs \citep[e.g.,][]{binette94}.
As pointed out by \citet{sarzi10}, the ionizing continuum of pAGBs is not comparable to  
that of a single O-star, but their large numbers in ETGs
make them the probable source of ionizing photons in this population. 

The TIR component of the FUV$+$TIR SFR recipe is also potentially problematic
because of a contribution from the low-mass evolved stellar population. %evolved stars.
This effect was noticed more than three decades ago with IRAS data, in which 
there was strong evidence for an increasing ``cirrus'' contamination in earlier Hubble types
\citep{helou86,sauvage92}.
The problem with TIR estimates of SFR because of dust heating by older stars 
is now well established \citep[e.g.,][]{walterbos87,perez06,kennicutt09,bendo10,bendo12,leroy12,boquien14,hayward14,delooze14,herreracamus15,viaene17}.
24\,\micron\ luminosities, \lmipsone, can also be affected by older stellar populations, but
in this case the contamination is from AGB circumnuclear dust shells
\citep[e.g.,][]{bressan98,bressan02,verley09}.

Thus we are left with the difficulty for low sSFR galaxies of how to calculate SFRs 
that better reflect the truth in order to compare with SED results.
Despite possible problems with \ha,
\citet{temi09a,temi09b} and \citet{davis14} advocate for ETGs the use of SFRs from \ha$+$24\,\micron\ luminosities
after the stellar contribution to the 24\,\micron\ emission is subtracted;
here we adopt this method and re-compute the SFRs for the KINGFISH sample.
Following \citet{temi09b}, we first calculated the \lmipsone, $L_{160}$, and 
$K_s$-band luminosities\footnote{We assume that the absolute magnitude of the Sun at $K_s$ band is 3.28\,mag
\citep[see][]{binney98,davis14}.},
\lks,
from the data in \citet{dale17}.
The ratios are shown in the upper panel of Fig. \ref{fig:l24lk}, where
the quiescent stellar component of 24\,\micron\ emission \citep[normalized to $K$ band, see][]{davis14} is plotted 
as a horizontal dashed line.
It is clear that galaxies with low sSFR \citep[the KINGFISH $\times$ symbols, and virtually all the galaxies from][]{temi09b}
have \lmipsone/\lks\ ratios close to the quiescent stellar value.
Fig. \ref{fig:l24lk} (lower panel) also illustrates the trend between \lmipsone\ and \lks,
emphasizing the clustering of the ETGs in \citet{temi09b} around the regression line.

To correct the 24\,\micron\ luminosities, we first need to subtract the quiescent component.
This approach was first proposed by \citet{temi09b} who used the galaxies shown in Fig. \ref{fig:l24lk}
to calibrate the 24\,\micron\ emission from ``passive'' stars;
\citet{davis14} applied the method to a different sample observed with the 22\,\micron\ WISE band (W4), and we use their calibration:
\begin{equation}
\log \left( \frac{{\rm L}_{22\,\mu{\rm m,passive}}}{\rm erg\,s^{-1}} \right) \,=\, \log \left( \frac{{\rm L}_{Ks}}{{\rm L}_\odot} \right) + 30.45
\label{eqn:l24lk}
\end{equation}
\noindent
with L$_{22}$ ($\approx$\lmipsone) in units of \ergs\ and \lks\ in \lsun.
The smaller constant (30.1) found by \citet{temi09b} is consistent with Eqn. (\ref{eqn:l24lk}) given that
their values of \lks\ are in the mean 0.29 ($\pm\,0.08$)\,dex 
larger than ours \citep[and those in][]{davis14}; we have evaluated this offset 
using the 9 KINGFISH galaxies in common with \citet{temi09b}.
Since the \citet{davis14} analysis relied on W4, rather than on MIPS\_24, we
have also checked that this does not introduce an additional discrepancy;
we find a difference between the KINGFISH 24\,\micron\ and 22\,\micron\ Log(fluxes)
of $-0.03\,\pm\,0.06$\,dex, and thus assume equality.
Once we have subtracted this quiescent stellar emission from the observed \lmipsone\
(${\rm L}_{24\,\mu{\rm m,cor}}\,=\,{\rm L}_{24\,\mu{\rm m,obs}}\,-\, {\rm L}_{24\,\mu{\rm m,passive}}$)
we recalculate the SFRs using the same approach as in Appendix \ref{app:sfr} for 24\,\micron$+$\ha,
but now with the corrected L$_{24\,\mu{\rm m,cor}}$.

\begin{figure*}[t!]
\includegraphics[width=\textwidth]{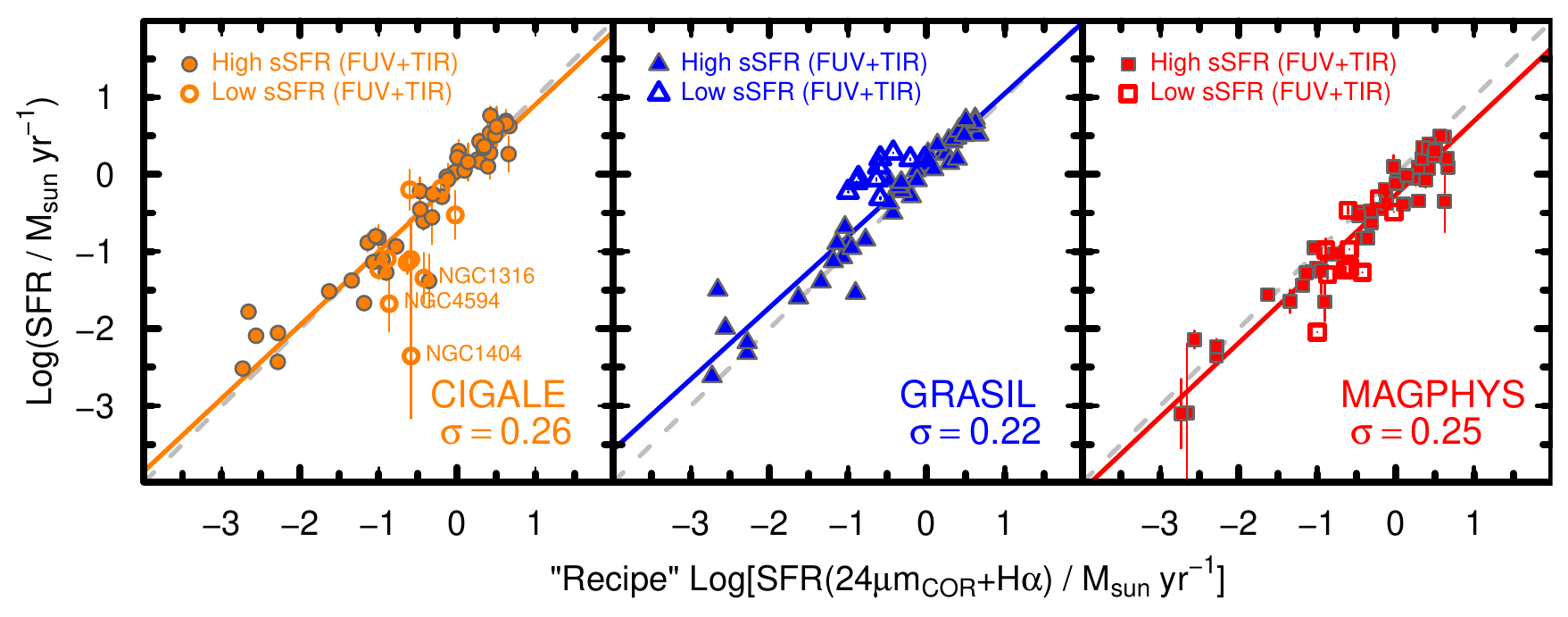}
\vspace{-\baselineskip}
\caption{SED-derived SFR plotted vs. SFRs determined from L$_{24\,\mu{\rm m,cor}}+{\rm L}_{{\rm H}\alpha}$. 
As in previous figures, filled symbols correspond to high specific SFR,  
and open ones to low specific SFR [as calculated with SFR(FUV$+$TIR)].
This figure is the same as Fig. \ref{fig:sfr}, but here the SFRs from \ha$+$24\,\micron\ luminosities
have been corrected as described in the text. The regression lines are as in Fig. \ref{fig:sfr};
the mean deviations for the fit of SED-derived quantities vs. the recipe [for SFR(L$_{24\,\mu{\rm m,cor}}+{\rm L}_{{\rm H}\alpha}$)] 
are shown by the $\sigma$ value in the lower right corner of each panel.
}
\label{fig:sfr_cor}
\end{figure*}

This comparison is shown in Fig. \ref{fig:sfr_cor} which is the same as Fig. \ref{fig:sfr} but with the 
24\,\micron\ luminosities now corrected for stellar emission according to Eqn. (\ref{eqn:l24lk}).
%The comparison is somewhat improved with fewer outliers, although for a few galaxies, \cig\ in particular seems
%xxx see George 
The comparison is not significantly changed, and in fact is slightly worse; 
the \rms\ deviation for the original SFR (uncorrected for stellar emission) inferred from \ha$+$24\,\micron\ 
is 0.25\,dex, 0.19\,dex, and 0.26\,dex, for \cig, \gra, and \magp, respectively (see Table \ref{tab:corr}).
For a few galaxies, \cig\ in particular seems
to find lower SFRs than what would be expected with the new 
estimates\footnote{The most extreme deviant using SFR(FUV$+$TIR), NGC\,584, as in Fig. \ref{fig:sfr}
has no \ha\ measurement, so we do not consider it further.}. 
For NGC\,1404, the new photometric SFR (and previous 24\,\micron$+$\ha\ estimate, see Table \ref{tab:app:parameters})
is almost certainly incorrect.
The \ha\ measurement \citep[see also][]{skibba11}
comes from the ``radial strip'' flux by \citet{moustakas10}, because there are no nuclear or circumnuclear fluxes, and
the resulting \ha\ luminosity is $>$5 times brighter than 0.02\,${\rm L}_{24\,\mu{\rm m,cor}}$ that is the other term in the
SFR calibration (see Appendix \ref{app:sfr}).
This seems unrealistic in such an ETG, so we do not consider this galaxy discrepant.
The three remaining problematic galaxies are NGC\,1316, NGC\,4569, and NGC\,4594, for which the new recipe SFR 
and the SED SFR by \cig\ differ by almost an order of magnitude.
Both NGC\,1316 (Fornax\,A) and NGC\,4569 host an AGN, but the 
nuclear \ha\ flux is $\sim$8\% and 24\%, respectively, 
of the circumnuclear emission \citep{moustakas10}, 
so the AGN is not dominating the \ha\ budget.
The ratio of ${\rm L}_{24\,\mu{\rm m,passive}}/{\rm L}_{24\,\mu{\rm m,obs}}$ for NGC\,1316 and NGC\,4594 is $\sim$30\%,
so not a huge correction; it is even smaller ($\sim$4\%) in NGC\,4569\footnote{NGC\,4569 is not really an ETG, 
but rather an \hi-deficient Virgo cluster galaxy suffering from gas
removal by ram-pressure stripping \citep{boselli16}.}.
In all these galaxies the contribution from \ha\ is 2--3 times lower than from 24\,\micron, 
so the reason for the discrepancy is not clear. 
However,
it is likely that these early-type galaxies are in a ``quenching'' phase of their SFH, as discussed
further in Sect. \ref{sec:scaling:sfms}. 

\subsection{Extinction, dust emission, and geometry}
\label{sec:assumptions:dust}

Both \cig\ and \magp\ require an energy balance (see Table \ref{tab:models}), namely that the fraction of stellar 
radiation absorbed by dust is re-emitted in the IR.
%In the case of \cig, the dust attenuation curve % governing dust attenuation
%is a power-law modified starburst curve, with the slope (power-law index)
%left as a free parameter; there is a differential reddening factor to account
%for stars of different ages \citep[e.g.,][]{calzetti00}.
%A variable bump in the attenuation curve at 0.2175\,\micron\ is also included to take into
%account small-grain (PAH) extinction\footnote{Because we are dealing with global fluxes,
%we assume that the entire scattering area is within the aperture of the observations;
%thus an ``extinction" curve is essentially an absorption curve, in principle related to
%the geometry of the dust and the grain opacity.}.
%\magp, instead adopts the two-phase attenuation model proposed by \citet{charlot00},
%which considers increased attenuation in birth clouds relative to the more diffuse ISM.
In both cases (see Table \ref{tab:models} for details),
the form of the interstellar attenuation % extinction (absorption) 
curve used in the \cig\ and \magp\ models is unrelated to the dust emissivity, but rather
relies on a two-component dust model \citep[e.g.,][]{calzetti00,charlot00}
to account for the differential reddening between stellar populations of different ages. 
%and the dust is implicitly assumed to be a ``screen" of material between the observer
%and the sources emitting the radiation to be subsequently absorbed by 
%dust\footnote{\cig\ lets vary some parameters of the shape of the attenuation
%curve, but does not account for radiation transfer.}.
\magp\ uses a time-dependent attenuation law, while \cig\ 
lets vary some parameters of the shape of the attenuation curve, but neither
account for radiation transfer.
% Ultimately, in both models, the correction for attenuation is applied implicitly
% %assuming that the attenuation is a ``screen'' of material, and 
% assuming that the dust emission is optically thin.
For \cig, dust emission is defined by the \citet{draineli07} models, while
for \magp, dust is divided into two components, birth clouds and the ambient ISM,
and emission within these components is modeled as a combination of PAH templates and MBBs at different
temperatures; 
the dust power-law emissivities are different for the various components,
but with the same normalization at long wavelengths.

\gra, on the other hand, considers three components of stars and dust:
stars embedded within GMCs, stars having already emerged from their birth clouds, and diffuse
gas ($+$ cirrus-like dust).
Previous incarnations of \gra\ included dust emission from circumstellar dust shells around AGBs
as in \citet{bressan98,bressan02}, but here we use the \citet{bruzual03} stellar populations that are
devoid of circumstellar dust.
The geometry of each of these components is specified in the model,
and radiative transfer is performed separately for each of the components assuming the
\citet{laor93} dust opacities/emissivities.
Thus, for \gra, the effect of dust extinction is related, by definition, to dust emission because
of the self-consistent definition of dust properties in the \citet{laor93} dust opacity curve.
\gra\ systematically gives higher \mdust\ relative to the other codes, and also to DL07 and MBB
dust estimates.
Because \gra\ also includes the cool dust that shines at longer wavelengths, 
necessary to produce the dust extinction, this component may add mass relative to the 
warmer luminosity-weighted dust emission that dominates the SED.

While one or another approach may be more valid for starbursts or high-$z$ galaxy populations,
the KINGFISH galaxies studied here are equally well fit by all three models.
Thus, the assumption of optically thin dust, which obviates the need for radiative transfer,
does not seem to be a problem for this sample in terms of estimating \mdust.
This is because in these galaxies the bulk of the dust emits at longer wavelengths where the dust
is optically thin, and because the long-wavelength dust
emissivities adopted here are similar (see Sect. \ref{sec:comparison:mdust}).
Moreover, the three rather different attenuation %extinction 
curves also do not seem to introduce
significant discrepancies in the SED shapes, possibly because any variations
are compensated for by differences in \afuv.
Even though the assumptions made for dust attenuation
and emission in each of the codes are quite different, in the end they lead to similar results, at least
for the KINGFISH sample. 

The difference of \afuv\ predicted by the \citet{murphy11} or \citet[][]{hao11} formulations 
and those of the SED models may also depend on the implicit assumptions.
\cig, and \magp\ 
rely on attenuation curves whose fitted parameters account for geometry and extinction,
% all assume that the dust is a homogeneous slab or ``screen''
while \gra\ takes into account the geometry of the dust and performs the radiative transfer.
However, for all three models the estimated \afuv\ tends to be smaller than that given
following \citet{hao11}.

This is not surprising for two reasons: the first is the geometry of the attenuating dust
relative to the emission sources, and the second is the homogeneity of the medium.
For a given dust column ($\propto \tau_{\rm dust}$),
a screen geometry would be expected to give larger attenuation relative
to a mixed or more complex distribution of dust \citep{witt00},
so this could be one part of the explanation.
Another part
lies in the probable clumpiness of the dust distribution
\citep[e.g.,][]{natta84,witt96,gordon00}.
If the dust is not uniformly distributed within the absorbing region, then the
optical depth inferred from SED modeling would be smaller than that derived
by assuming a homogeneous medium as done by \citet{hao11}.
The homogeneous constant-density medium corresponds to the highest efficiency for
dust attenuation given a specified dust mass \citep{witt96}.
This effect is clearly seen in the three-dimensional radiative transfer models of
M\,51 by \citet{delooze14}; the larger the fraction of dust mass in dense clumps, the lower the inferred \afuv.
Since \gra\ takes the dust distribution explicitly into account through geometry,
this would explain the discrepancy in \afuv\ relative to \citet{murphy11} or \citet{hao11}.
\cig\ and \magp\ also consider complex attenuation curves (see Sect. \ref{sec:codes}),
and so implicitly also account for different dust distributions rather than homogeneous
ones. 

\subsection{Metallicity }
\label{sec:assumptions:metallicity}

The \cig\ models used in this work adopt solar-metallicity SSPs,
while \magp\ considers a range in metallicity for the SSPs (see Table \ref{tab:models}). 
\gra\ instead models the metallicity evolution and gas content through CHE\_EVO, and relates the
dust mass necessary for the SED's best-fit shape to the hydrogen gas mass.
Consequently, metallicity is varied also (albeit indirectly) in the \gra\ models,
through its relation to the dust-to-gas mass ratio, assumed to vary linearly with metallicity\footnote{Linear 
scaling is potentially a problem at low metallicities \logoh$\la$8.0 \citep{remyruyer14}.}.
Despite these significant differences in treatment of metallicity, there seem to be
no salient differences in the quality of the SED shape relative to the observed SED.
It is also true that the KINGFISH sample does not probe metallicities below $\sim$20\%\,\zsun,
so it could be that lower metallicities are required to significantly reshape the SED.
We are pursuing possible reasons for this in a future paper.

\begin{figure}
%\vspace{\baselineskip}
%\includegraphics[width=0.48\textwidth]{BC03KINGFISH_SFRvsMstar_1panel_cigale_magphys_grasil_GREY-crop.pdf}
%\includegraphics[width=0.48\textwidth]{BC03KINGFISH_SFRvsMstar_1panel_cigale_magphys_grasil_WISE_June2018-crop.pdf}
\includegraphics[height=0.48\textwidth]{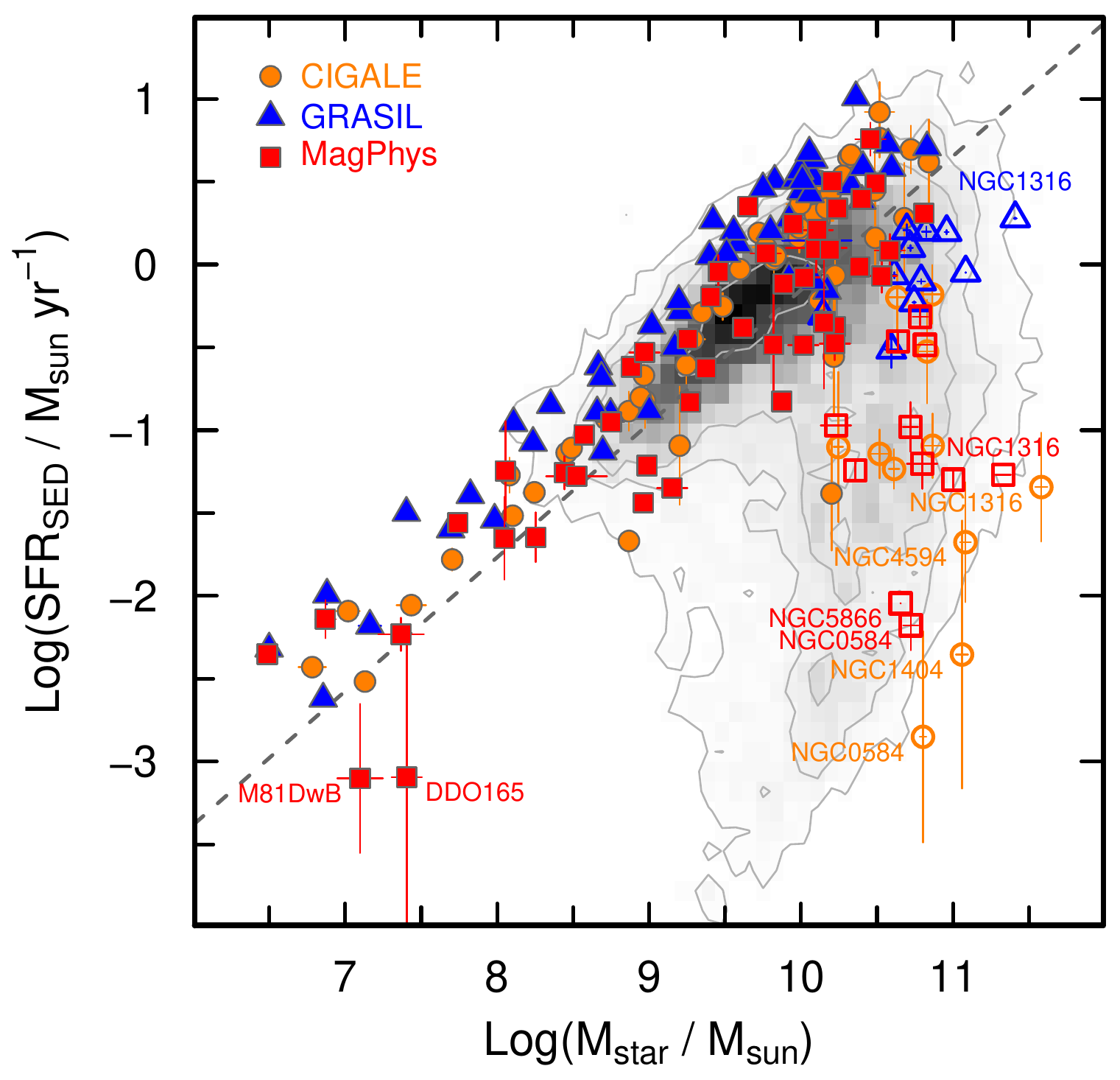}
\caption{SED-derived SFR vs. \mstar\ in logarithmic space superimposed on the
%deep 
GSWLC sample shown in gray-scale from \citet[][see text for details]{salim16}.
%The $\sigma$ values of the best-fit robust correlations are shown in the upper left corner,
%and the robust regressions for each SED-fitting algorithm are shown as solid lines.
As in previous figures, filled symbols correspond to high specific SFR (Log(sSFR/yr$^{-1}$) $> -10.6$), 
and open ones to low specific SFR (Log(sSFR/yr$^{-1}$) $\leq -10.6$).
The (gray) dashed line corresponds to the SFMS relation
found by \citet{hunt16} for nearby galaxies.
}
\label{fig:sfrvsmstar}
\end{figure}

\section{Scaling relations in the Local Universe}
\label{sec:scaling}

In what follows, we examine the derived quantities given by SED fitting in the context of several
well-established scaling relations.
Such scaling relations constrain
the observed parameter space of galaxy populations, and may give important insight
into the assumptions behind the SED models.

\subsection{The star-formation ``main sequence''}
\label{sec:scaling:sfms}

It is well known that \mstar\ and SFR are related both in local galaxies 
and at high redshift through the ``star formation main sequence'' \citep[e.g.,][]{brinchmann04,salim07,noeske07,karim11,elbaz11}.
We have investigated whether the KINGFISH SED results show a similar trend 
in Fig. \ref{fig:sfrvsmstar} where Log(SFR) is plotted against Log(\mstar).
%The solid lines show the best-fit regressions while 
The dashed (gray) line shows the SFMS relation derived by \citet{hunt16} for galaxies in the Local Universe
(including KINGFISH galaxies but with recipe-derived quantities); the
slope of this relation, $\sim$0.8 (SFR$\,\propto\,$\mstar$^{0.8}$), is consistent with 
the value found by \citet{elbaz07} of $\sim$0.77 for a local comparison sample 
\citep[see also compilation in][]{leitner12}.
As in previous figures, open symbols correspond to KINGFISH galaxies with 
low sSFR ($\la 3 \times 10^{-11}$\,yr$^{-1}$). 

\begin{figure}
%\vspace{\baselineskip}
\vbox{
\includegraphics[height=0.47\textwidth]{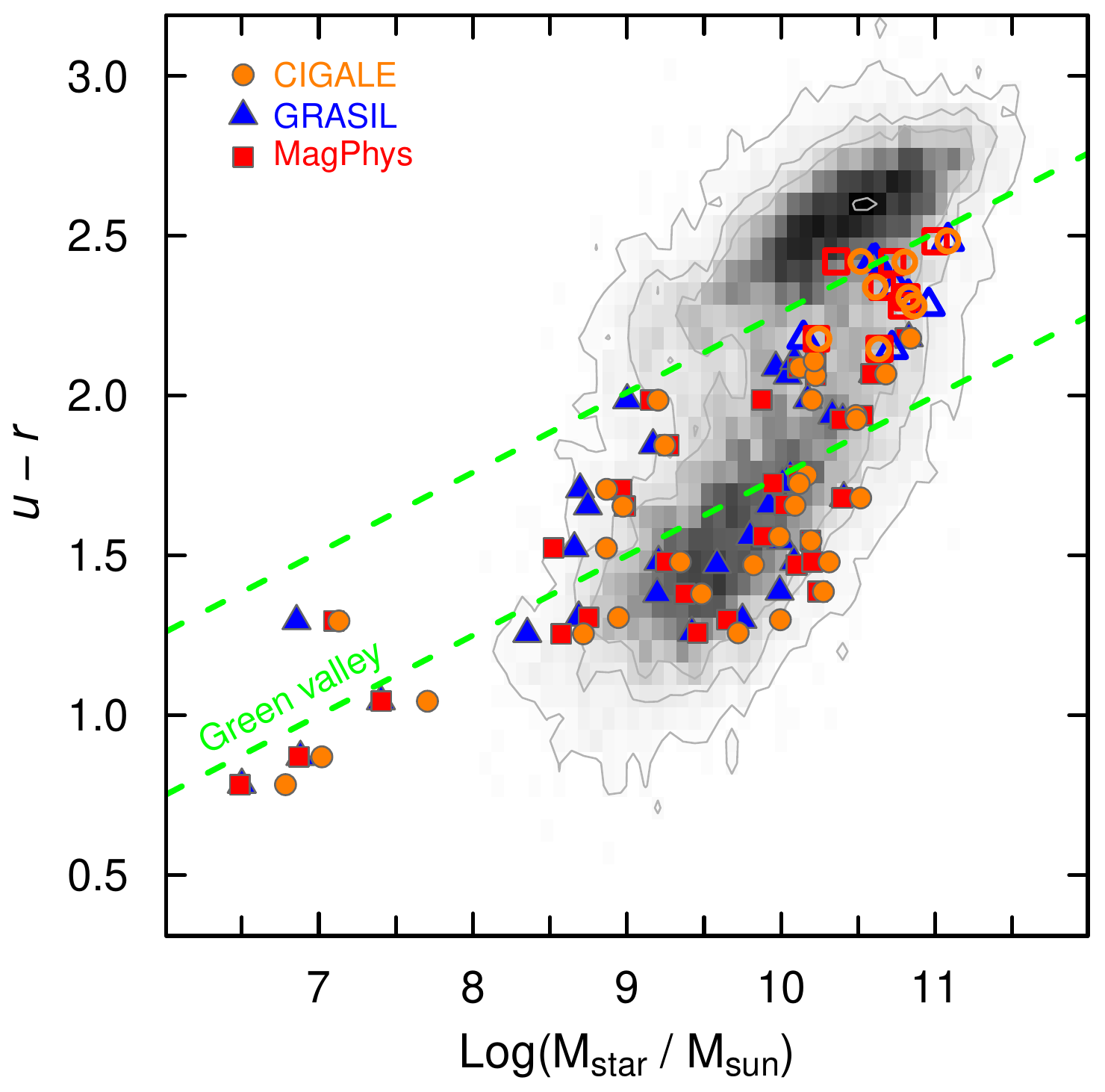} \\
\\
\includegraphics[height=0.47\textwidth]{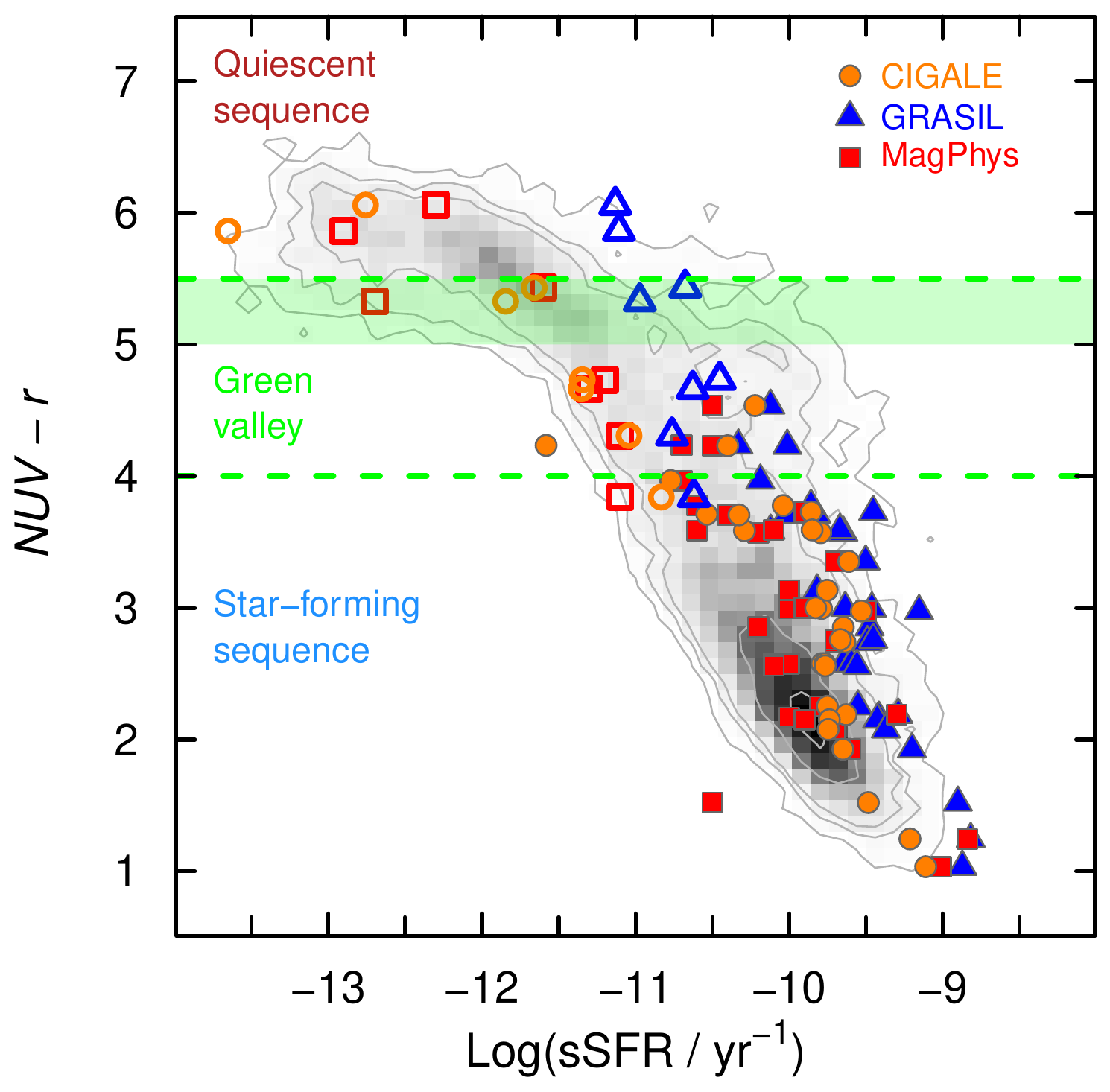}
}
\caption{ Colors of KINGFISH galaxies plotted against 
the logarithm of stellar mass given by the respective SED-fitting algorithms (top panel)
and the SED-derived logarithm of sSFR (bottom, with units of yr$^{-1}$);
the top panel shows SDSS $u-r$, and the bottom NUV$-r$. 
In both panels,
the KINGFISH galaxies are  superimposed on the GSWLC sample, %\citet{gallazzi05}, \citep{salim16}
taking only those galaxies with $0.015 \leq z \leq 0.06$.
%The $\sigma$ values of the best-fit robust correlations are shown in the upper left corner,
%and the robust regressions for each SED-fitting algorithm are shown as solid lines.
As in previous figures, filled symbols correspond to high specific SFR (Log(sSFR/yr$^{-1}$) $> -10.6$), 
and open ones to low specific SFR (Log(sSFR/yr$^{-1}$) $\leq -10.6$).
In the top panel, the (green) dashed lines correspond to the separation of the ``green valley'' from the upper
(red) and lower (blue) loci of SDSS galaxies as given by
\citet{schawinski14}.  %, reported to the $z\,=\,0.1$ offset for the K-correction.
In the bottom, we have included the NUV$-r$ color range for the ``green valley" transition proposed by \citet{salim14},
together with the limit for ETG SF activity of NUV$-r$\,=\,5.5 given by \citet{kaviraj07}.
The green shaded area marks the (uncertain) boundary between star-forming
and quiescent ETGs.
}
\label{fig:sdss_redsequence}
\end{figure}

Also shown in Fig. \ref{fig:sfrvsmstar} is the GALEX SDSS WISE Legacy Catalog (GSWLC) deep sample 
from \citet{salim16}; the plotted points have been limited to the ``Main Galaxy Sample'' (MGS),
and in redshift to $0.015 \leq z \leq 0.06$, and there is no K-correction applied to the data.
For the GSWLC sample,
the SFRs are derived by SED fitting using a different version of the \cig\
code than we use here, in particular, a SFH comprising two-component declining exponentials.
Moreover, we have adopted the infrared refinement of \citet{salim18}
that takes into account the WISE 22\,\micron\ photometry to constrain SFR;
there are no longer-wavelength constraints on the SED fitting.
Here, and in subsequent figures, the GSWLC gray scales correspond to galaxy number
densities within the sample, with outer contours delimiting 99.99\%. 

Most of the galaxies having disagreements between SED-derived SFRs and 
the recipe values are ETGs, and thus possibly in a quenching
(or already quiescent) phase of their SFH.
This is seen clearly in Fig. \ref{fig:sfrvsmstar} with the superposition of the KINGFISH galaxies
on the GSWLC locus below the main sequence %\citet{salim16} sample 
having low SFRs at high \mstar; galaxies (virtually all early-type) falling into this category are plotted
with open symbols (because of their low sSFR) and labeled in Fig.  \ref{fig:sfrvsmstar}.

We test further the idea that these galaxies are transitioning
into a more quiescent SFH phase in the upper panel of Fig. \ref{fig:sdss_redsequence} where we have plotted
SDSS $u-r$ colors against \mstar.
Again the KINGFISH galaxies are superimposed on the GSWLC \citep{salim16,salim18} %from the SDSS sample of \citet{gallazzi05}.
where, as before, the redshift range is limited to $0.015 \leq z \leq 0.06$, and only MGS galaxies are considered. %but unlike the sample from \citet{salim16},
%here the SDSS colors have been K-corrected to $z\,=\,0.1$ as noted in the figure caption.
The (green) dashed lines, taken from \citet{schawinski14}, delimit the transition
green valley regime from the ``red sequence'' to the ``blue cloud''.
% (taking into account the different K corrections of the two samples).
Virtually all the galaxies in which SED-derived SFRs differ from the recipe values
are upper ``green-valley'' or ``red-sequence'' galaxies, at the massive end of the transition from bluer,
star-forming ones.

However, it is well established that optical colors are less sensitive to low levels of SFR than the UV 
\citep[e.g.,][]{wyder07,schawinski07,kaviraj07,salim14}.
NUV$-r$ is particularly suited for examining weak SF because NUV traces young stars and
$r$ is a proxy for stellar mass. 
The lower panel of Fig. \ref{fig:sdss_redsequence} thus plots
observed NUV$-r$ against sSFRs as estimated by the SED-fitting codes. 
As in previous figures, the KINGFISH galaxies are superimposed on the GSWLC sample, again
limited to MGS galaxies and a redshift range of $0.015 \leq z \leq 0.06$.
The NUV$-r$ color correlates well with sSFR, proving to be an effective diagnostic of the transition from star-forming
galaxy populations to more passive ones \citep[e.g.,][]{salim14}.
ETGs with an NUV$-r$ color $\la$ 5.5 are
very likely to have experienced recent star formation, even when considering the contamination by UV upturn
\citep{kaviraj07}, while galaxies with colors redder than this have
very little molecular gas \citep{saintonge11} and are almost certainly non-star-forming quiescent
systems \citep{schawinski07}. 
The SFRs from \cig\ and \magp\ are consistent with the GSWLC, and the observed NUV$-r$ colors
seem to indicate that the galaxies with particularly low sSFR (as determined by SED fitting), are in a quiescent phase of their SFH.
On the other hand, \gra\ finds sSFRs that are higher for these galaxies but not inconsistently with what could
be expected given their NUV$-r$ colors.

As discussed above 
(Sect. \ref{sec:comparison:sfr}, Sect. \ref{sec:sfragain}), 
galaxies with very low sSFR are difficult to model
because of the potential similarity/degeneracies in SEDs in this parameter range.
Such difficulties are also seen in the comparisons with reference quantities shown in
Figs. \ref{fig:sfr}, \ref{fig:fuv}, and \ref{fig:sfr_cor} 
where parameter estimations show discrepancies with SFR and \lfuv\ relative to some of the models.
The essence of the problem is the SFH, and how we can ascertain observationally 
whether or not galaxies are already in the quenching phase. 

\subsection{Dust mass, star-formation rate, and stellar mass}
\label{sec:scaling:duststarssfr}

\begin{figure}
\vspace{\baselineskip}
\includegraphics[width=0.48\textwidth]{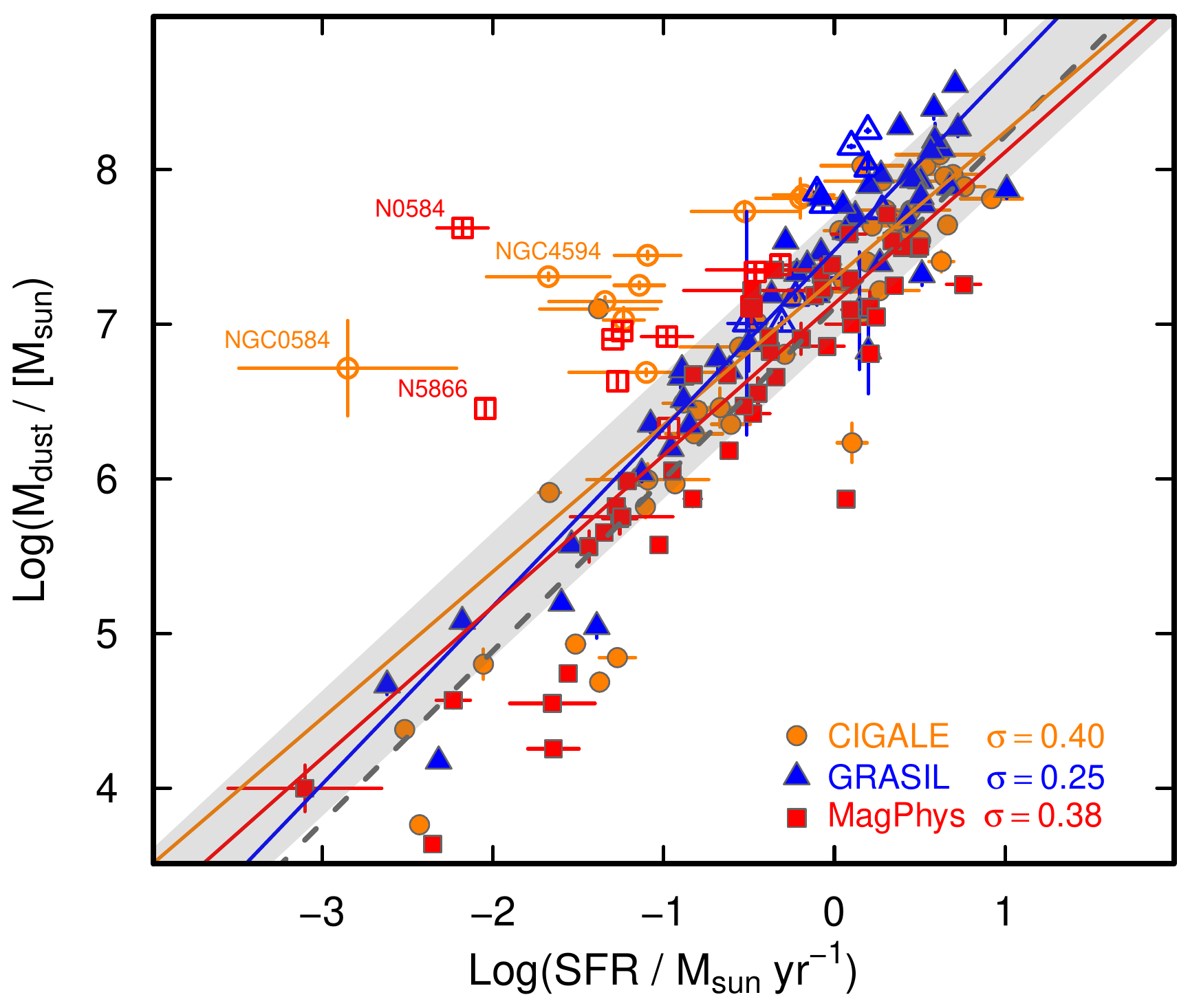}
\caption{SED-derived \mdust\ vs. SFR  in logarithmic space.
The $\sigma$ values of the best-fit robust correlations are shown in the lower right corner,
and the robust regressions for each SED-fitting algorithm are shown as solid lines.
The (gray) dashed one corresponds to the relation
given by \citet{dacunha10} for SDSS galaxies. 
The gray area illustrates the $\pm\,1\sigma$ range  around the mean slope:
here $\sigma$ corresponds to the %using the %minimum 
mean \rms\ of the three individual fits, and the mean slope
to the mean of the three individual slopes. 
As in previous figures, filled symbols correspond to high specific SFR (Log(sSFR/yr$^{-1}$) $> -10.6$), 
and open ones to low specific SFR (Log(sSFR/yr$^{-1}$) $\leq -10.6$).
}
\label{fig:mdustvssfr}
\end{figure}

Using \magp, \citet{dacunha10} found that \mdust\ and SFR are also tightly correlated in a large sample of SDSS galaxies with
IR photometry from IRAS.
We have explored this scaling relation 
in the KINGFISH galaxies using quantities derived from our SED fitting.
This correlation is seen not only with \magp, but also with \cig\ and \gra\ as shown in
Fig. \ref{fig:mdustvssfr} where the \mdust-SFR correlation is illustrated
(only the 58 galaxies with sufficient IR photometry are plotted);
the \citet{dacunha10} relation is given by a (gray) dashed line and the best-fit robust KINGFISH correlations
(for each algorithm separately) by solid ones.
\mdust\ and SFR are fairly well correlated in the KINGFISH galaxies with a scatter of
$\sim$0.4--0.5\,dex. 
%\textcolor{green}{
%[An important difference in the \cig\ modelling is that it includes SFHs with a strong diminution of star formation in the recent past. This allows the possibility of having very low SFR to model passive galaxies now forming few to no stars at all. However, the determination of very level of star formation is particularly difficult. An SFR of $10^{-7}$ M\sun/yr will give an SED very similar to that obtained with an SFR of $10^{-3}$ M\sun/yr as in either case, older stellar populations will contribute a large fraction of total dust heating.]
%}

\begin{figure}
\vspace{\baselineskip}
\includegraphics[width=0.48\textwidth]{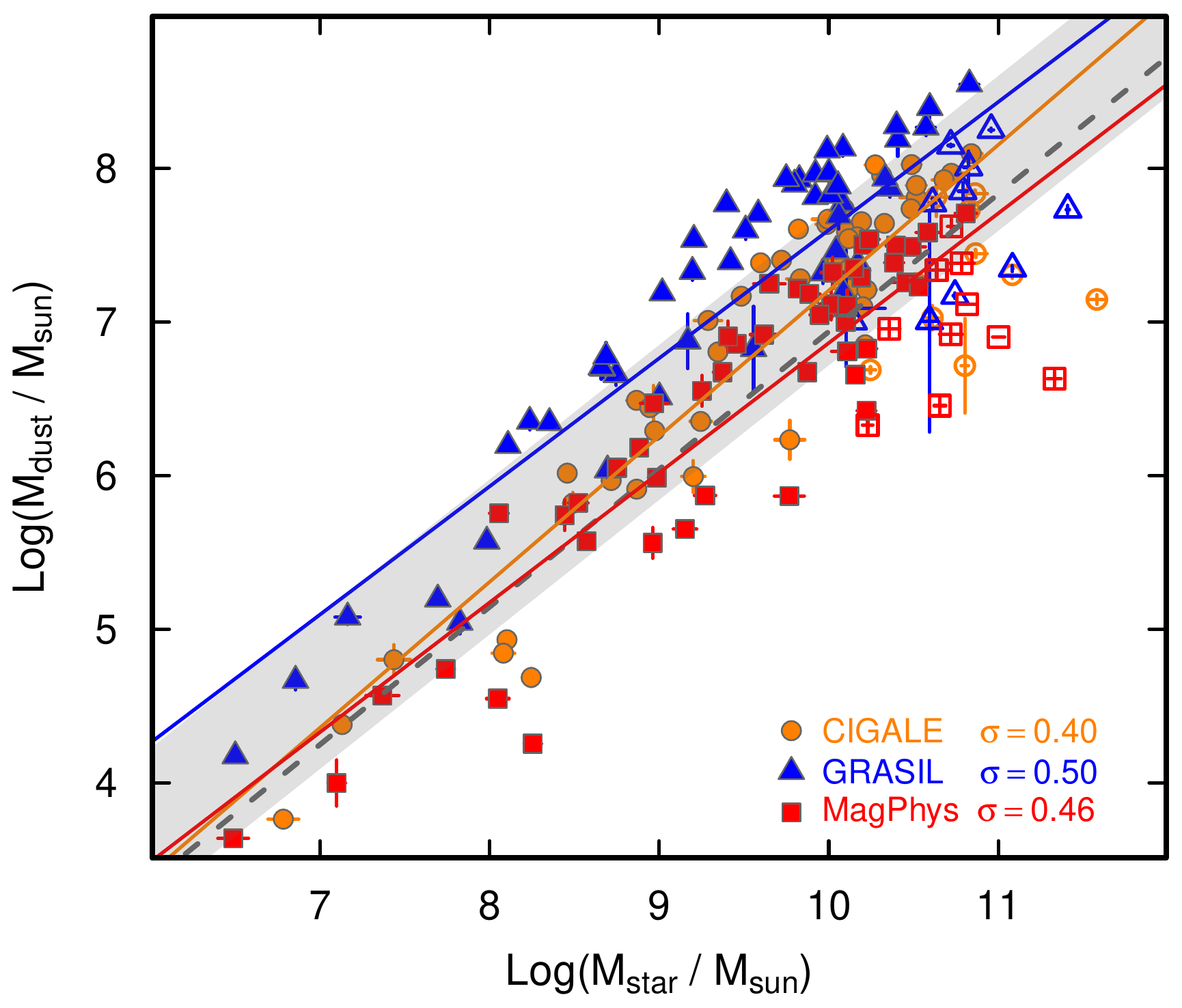}
\caption{SED-derived \mdust\ vs. \mstar\  in logarithmic space.
The $\sigma$ values of the best-fit robust correlation are shown in the lower right corner. 
The robust correlations are shown as solid lines, and the 
(gray) dashed one corresponds to the relation
given by \citet{dacunha10} for SDSS galaxies, reported to \mstar\ through the SFMS
by \citet{hunt16}.
The gray area %illustrates the $\pm\,1\sigma$ range (using the %minimum 
%mean \rms\ of the three fits) around the mean slope. 
is defined the same way as in Fig. \ref{fig:mdustvssfr}.
As in previous figures, filled symbols correspond to high specific SFR (Log(sSFR/yr$^{-1}$) $> -10.6$), 
and open ones to low specific SFR (Log(sSFR/yr$^{-1}$) $\leq -10.6$).
}
\label{fig:mdustvsmstar}
\end{figure}

Over four orders of magnitude in \mdust\ and SFR,
the scatter is smaller than that found for the KINGFISH SFMS, and is probably suggesting 
something fundamental about the relation of dust mass, gas mass, and SFR
as discussed by \citet{dacunha10}.
% 8/9/2018 as Ilse points out, this is a false statement.
%The offsets relative to the \citet{dacunha10} relation for \gra\ are due mainly to the differences in
%opacities with respect to \magp.
Again, galaxies with low sSFR are problematic, emerging as galaxies whose SFRs are too low 
for the inferred dust content;
low levels of SFR are difficult to constrain observationally
since evolved stars are expected to dominate the dust heating.
% ------------------------------------
% 27/7/2018 seems trivial at this point
% For both the \gra\ and the \magp\ models,
% the power-law indices are consistent with that of $\sim$1.1 found by \citet{dacunha10}: 
% $1.13\,\pm\,0.06$ for \gra\ and  $1.04\,\pm\,0.09$ for \magp;
% for \cig, the index is shallower: $0.90\,\pm\,0.08$.
% This latter difference with \cig\ is due to galaxies with low sSFR which
% artificially induce a shallower slope:
% the \cig\ fit omitting these (12) galaxies has an index of  $1.12\,\pm\,0.05$, and 
% the fit has an \rms\ scatter of 0.28\,dex.
% Indeed, another problem in Fig. \ref{fig:mdustvssfr}
% is that our regressions are performed with equal weighting for all
% data points; the low-sSFR galaxies tend to have derived quantities with larger uncertainties,
% so the flat slope for \cig\ is in truth also a consequence of our fitting technique.
% ------------------------------------
%{\bf slope for \cig\ is off because of the SFRs, need again to discuss this as above?} \textcolor{green}{[I think it would be useful to have 2 slopes, with including all galaxies as now, and one without galaxies with log sSFR<-10.6. Also I think the uncertainties should be included in the fit, otherwise galaxies at low SFR have a weight that strongly overestimated as they have large error bars.]}

\begin{figure*}[!h]
\vspace{\baselineskip}
\includegraphics[width=\textwidth]{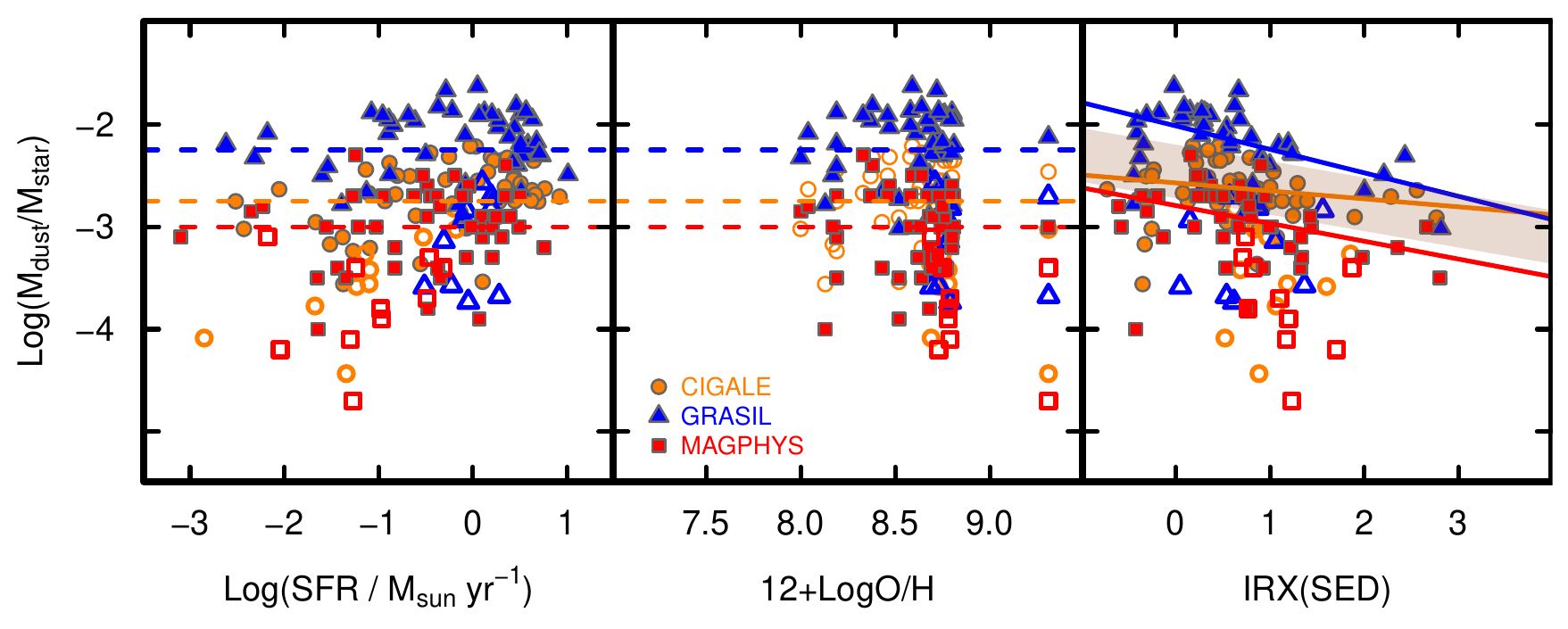}
\caption{SED-derived \mdust/\mstar\ ratios in logarithmic space plotted against
(SED-derived) SFR, \logoh, and (SED-derived) IRX.
%The $\sigma$ values of the best-fit robust correlation are shown in the upper left corner of each panel. 
The middle panel shows the PP04N2 calibration for \logoh\ as described in the text.
The dashed horizontal lines in the left and middle panels
show the means of Log(\mdust/\mstar) for each SED-fitting algorithm 
(dark orange: \cig; blue: \gra; red: \magp).
The fits of (Log) \mdust/\mstar\ vs. IRX for the individual SED algorithms are shown as colored
solid lines in the right panel; the gray region gives the $\pm 1\sigma$ interval around
the mean of the three individual regressions.
As in previous figures, open symbols correspond to  
galaxies with low sSFR
[Log(sSFR/yr$^{-1}$)\,$\leq\,-10.6$], and filled symbols to high sSFR
[Log(sSFR/yr$^{-1}$)\,$>\,-10.6$].
}
\label{fig:mdustmstarvsother}
\end{figure*}

In their metal census in star-forming galaxies at $z\sim 0$, 
\citet{peeples14} find a correlation between \mdust\ and \mstar\ using a dataset
similar to the KINGFISH sample studied here.
We reassess this correlation based on our SED-fitting results in 
Fig. \ref{fig:mdustvsmstar}, where \mdust\ is plotted against \mstar\ in logarithmic space.
The different SED algorithms give similar slopes ($\sim 0.8-0.9$), although \gra\ is
slightly shallower ($\sim 0.7$).
These regressions are consistent with that found by \citet[][]{peeples14}:
Log\,\mdust\,=\,0.86\,Log\,\mstar\ $- 1.31$.
With the expression for gas-mass fraction as a function of \mstar\ by \citet{peeples14}, 
this expression gives 
gas-to-dust ratios of between $\sim$80 and 200 for a galaxy with \mstar$\sim 10^{10.5}$\,\msun.

Because of the relatively strong correlations of both \mdust\ and \mstar\ with SFR
(see Figs. \ref{fig:sfrvsmstar}, \ref{fig:mdustvssfr}),
we might expect the relative dust content, as measured by dust-to-stellar mass ratios, to depend on SFR.
Dust content is also thought to depend on metallicity (as measured by its emission-line proxy O/H), and on IRX, 
the logarithm of the ratio between \ltir\ and (observed) \lfuv; 
thus \mdust/\mstar\ could also correlate with these quantities.
These trends are shown in Fig. \ref{fig:mdustmstarvsother} where we have plotted
the different SED-fitting algorithms with different symbols as before.
Here we have taken the metallicities 
%from \citet{hunt16} 
from \citet{aniano18}
where the original determinations by
\citet{moustakas10} \citep[see also][]{kennicutt11} have been converted to the nitrogen calibration
of \citet[][hereafter PP04N2]{pettini04} according to the prescriptions of \citet{kewley08}.
%These values are also used by \citet{aniano18}, and for more details see that paper.
For more details, see \citet{aniano18}.

Overall, there appears to be little dependence of \mdust/\mstar\ on either SFR
or O/H in these galaxies.
However, there is a weak trend of \mdust/\mstar\ with IRX, with \rms\ deviations
of $\sim$0.4\,dex.
%, at least when the galaxies
%with low sSFR are omitted from the fit [Log(sSFR)\,$>\,-10.6$\,yr$^{-1}$].
The individual slopes of \gra\ and \magp\ are consistent, but the \cig\ slope is
shallower ($\sim -0.27$ for \gra, \magp\ and $-0.12$ for \cig)
(see right panel of Fig. \ref{fig:mdustmstarvsother});
the mean relation (averaged over the three SED algorithms) is
Log(\mdust/\mstar)\,=\,$-2.5\,-\,0.25$\,IRX.
If only the high sSFR points are included in the fit, the slope is shallower
($-0.18$) and the scatter is smaller (0.18--0.26\,dex). %(\gra\ has 0.18; \magp\ 0.26; \cig 0.27)
Thus, the SED fitting of the KINGFISH galaxies implies that
the dust-to-stellar mass ratio decreases with IRX, but not very steeply and with large scatter;
for more than three orders of magnitude of change in IRX, the \mdust/\mstar\ ratio
decreases by only a factor of $\sim$10 (not considering the low sSFR objects).

\begin{figure*}[h!]
\vspace{\baselineskip}
\includegraphics[width=\textwidth]{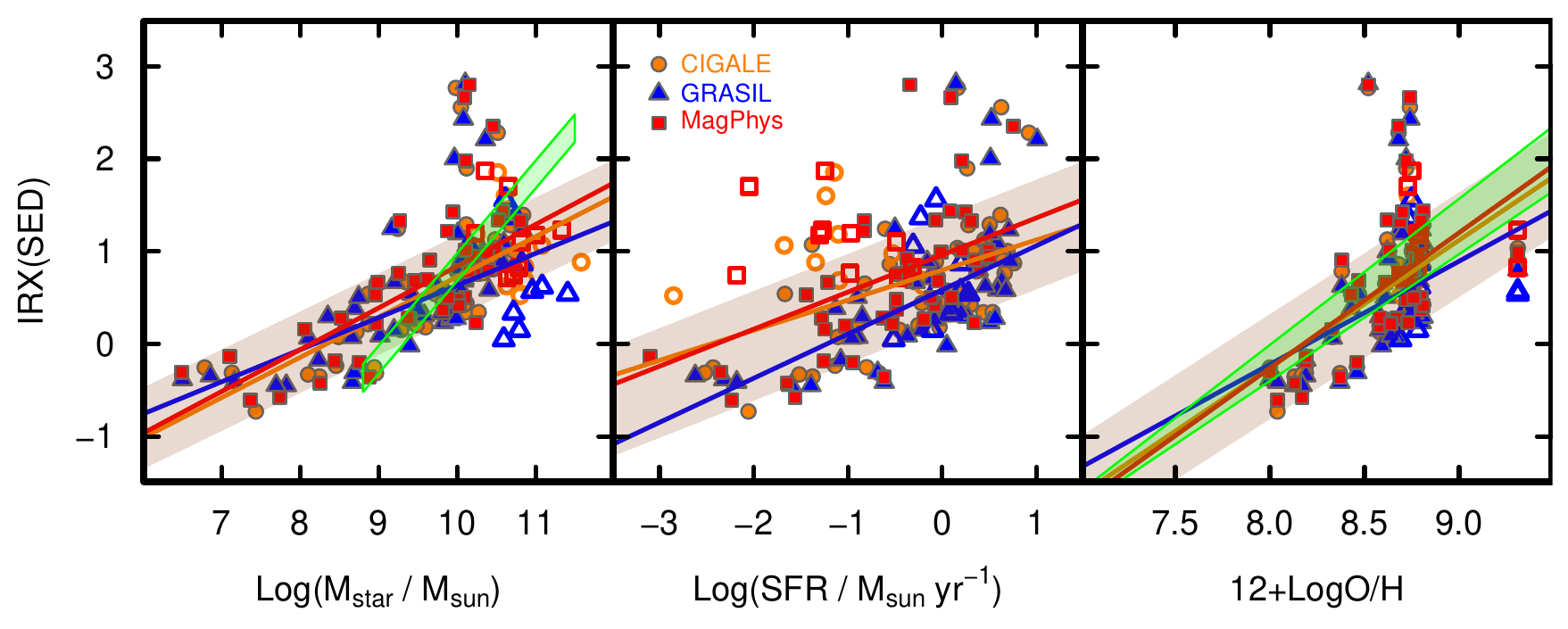}
\caption{IRX (from SED fitting) plotted against \mstar, SFR, and \logoh.
In each panel, the individual best-fit regressions are shown by colored solid lines,
and the gray regions denote the $\pm 1\sigma$ interval around the mean regression.
The left panel includes the ``consensus'' relation for galaxies at redshifts
$z\sim2-3$ found by \citet{bouwens16}, and shown as a green shaded region.
As in Fig. \ref{fig:mdustmstarvsother}, the right panel shows the PP04N2 calibration for
\logoh\ as described in the text.
Also shown in the right panel is the correlation between IRX and O/H found for normal
star-forming galaxies by \citet{cortese06} and for starbursts by \citet{heckman98};
the region enclosed between these two relations is green-shaded,
and is slightly steeper than the mean relation.
As in previous figures, open symbols correspond to  
galaxies with low sSFR
[Log(sSFR/yr$^{-1}$)\,$\leq\,-10.6$], and filled symbols to high sSFR
[Log(sSFR/yr$^{-1}$)\,$>\,-10.6$].
%The two KINGFISH galaxies with IRX $>$ 2 are NGC\,1266 (with IRX\,=\,3.3)
%and NGC\,1482 (IRX\,=\,2.8), both early type, S0's.
The three KINGFISH galaxies with IRX $>$ 2 are NGC\,1266 (IRX\,=\,3.3),
NGC\,1482 (IRX\,=\,2.8), and NGC\,2146 (IRX\,=\,2.6); the first two
are early type S0's and NGC\,2146 is a luminous IR galaxy. 
%with \ltir\,=\,$1.3\times10^{11}$\,\lsun. 
}
\label{fig:irxvsother}
\end{figure*}

\subsection{Infrared-to-ultraviolet luminosity ratio, IRX}
\label{sec:scaling:irx}

Attenuation of UV light is also expected to depend on relative dust
content, and IRX is one way to quantify this attenuation
\citep[e.g.,][]{kong04,cortese06,boquien09,hao11,boquien16,viaene16}.
However, IRX is somewhat dependent on the age of the dust-heating populations,
so may vary with other parameters besides dust content.
In Fig. \ref{fig:irxvsother}, 
%we compare IRX for KINGFISH galaxies (using
%the G13 formulation for \ltir\ as above) with \mstar, SFR, and O/H.
we compare IRX from the SED fitting of KINGFISH galaxies with 
the PP04N2 O/H calibration as in Fig. \ref{fig:mdustmstarvsother}, and SED-derived \mstar\ and SFR.
The left panel of Fig. \ref{fig:irxvsother} shows the correlation of IRX with \mstar\
(Pearson correlation coefficient $\rho$\,=\,0.6--0.7).
Although the formal dispersion
is high $\sim$0.6\,dex, it is mostly due to the three outliers with IRX$>$2:
NGC\,1266, an S0 galaxy with a molecular outflow \citep{pellegrini13};
NGC\,1482, an S0 galaxy with a dusty wind \citep{mccormick13};
and NGC\,2146, a luminous IR galaxy with \ltir\,=\,$1.3\times10^{11}$\,\lsun\
and a powerful outflow in atomic, ionized, and molecular gas \citep{kreckel14}.

That IRX, a measure of dust attenuation, is roughly correlated with \mstar\ is probably not
surprising given the relation between visual extinction \av\ and \mstar\ found
by \citet{garn10}.
A similar relation between UV attenuation and \mstar\ is evident over 
a wide range of redshifts
\citep[e.g.,][]{pannella09,whitaker14,pannella15,bouwens16}.
%which shows little evolution at low stellar masses \citep[$\la 3\times10^{10}$\,\msun,][]{whitaker14}.
The ``consensus relation" found by \citet{bouwens16} for galaxies at $z\sim2-3$
is also shown as a (green) shaded region in the left panel of Fig. \ref{fig:irxvsother}; 
with unit slope, it is steeper than the trends given by our SED-fitting
algorithms, and could indicate selection effects at high redshift given that the KINGFISH
sample probes more than two orders of magnitude lower in \mstar.
It could also point to different geometries for high-$z$ galaxies compared to local ones.
We find a mean regression of IRX\,=\,$-3.4\,+\,0.41$\,Log(\mstar).
Over the mass ranges probed by the \citet{bouwens16} study, there is no strong evidence for evolution,
at least to $z\sim3$, consistently with the conclusions of \citet{whitaker14} who noted
little evolution at low stellar masses $\la 3\times10^{10}$\,\msun.

The middle panel of Fig. \ref{fig:irxvsother} illustrates the trend of IRX with SFR;
the correlation is weaker than with \mstar\ ($\rho$\,=\,0.4--0.6), 
although excluding the low sSFR galaxies [with Log(sSFR/yr$^{-1}$)\,$\leq - 10.6$]
would improve the tightness of the trend.
The steepest power-law index is given by the \gra\ fits (0.49), and the shallowest by \cig\ (0.31);
the mean regression (averaged over the three fitting algorithms) is IRX\,=\,$0.78\,+\,0.39$\,Log(SFR) 
that is reflecting the increase of dust content with SFR
(e.g., Fig. \ref{fig:mdustvssfr}).

Because of the tendency of dust content to increase with metallicity, many previous studies
have examined the trend of IRX and metallicity in nearby galaxies
\citep[e.g.,][]{heckman98,cortese06,johnson07,boquien09}.
The correlation of IRX and metallicity shown in the right panel of Fig. \ref{fig:irxvsother}
is thus not a new result although here we confirm it with % a new metallicity calibration for
the KINGFISH sample, albeit with large spread at Solar metallicity.
The regressions found by \citet[][slope $\sim$1.4]{cortese06}
for normal star-forming galaxies and by \citet[][slope $\sim$1.2]{heckman98} for starbursts are shown as solid (green) lines,
enclosing the green-shaded region.
We find similar trends with power-law indices ranging from $\sim$1.2 (\gra) to 1.4 (\cig) and 1.5 (\magp).
Given the different metallicity calibrations and the previous lack of \hers\ data that would be
expected to lower the IR contribution, the agreement is fairly good between our determination 
and previous ones.
Here the scatter is high, $\sim$0.5--0.6\,dex ($\rho\,=\,0.4-0.5$), but again mostly due to the three
outliers at high IRX.
The mean regression averaged over the three SED algorithms is:
IRX\,=\,0.28\,+\,1.4\,(\logoh - 8.0)
In conclusion, for the KINGFISH galaxies IRX is at least approximately related
to \mstar, SFR, and O/H, as might be expected given that dust attenuation should
grow with the increase of each of these quantities. 

\subsection{Inferring stellar masses from IRAC and WISE W1 luminosities}
\label{sec:stellarmassrecipe}

The SED-derived \mstar\ values can be used to derive mass-to-light ratios
and thus a new recipe for stellar masses and M/L ratios in the mid-infrared,
from \upsiirac, based on IRAC 3.6\,\micron\ luminosities, or equivalently \upsiwise\ based on WISE W1. 
The super-linear power-law index for the trend of SED-\mstar\ vs. \mstar\ derived with
a constant $\Upsilon_*$ ratio indicates that the $\Upsilon_*$ ratio increases with increasing \liracone,
similar to the trend found by \citet{wen13} with \lwiseone.
As we argued in Appendix \ref{app:mstar}, IRAC 3.6\,\micron\ 
and WISE W1 photometry is virtually indistinguishable, and here we analyze only \lwiseone\
in order to compare with the GSWLC \citep{salim16,salim18}.
To better assess non-linearity in the luminosity dependence of \upsiwise\ (or \upsiirac),
we have fit the M/L ratio \upsiwise\ as a function of luminosity; % of  the SED-derived \mstar\ values and W1 luminosities \lwiseone;
thus in the case of constant M/L ratio, we would expect a slope of zero.
Instead, we find the following 
best-fit regressions (where \lwiseone\ %\liracone
is given\footnote{%Like in Appendix \ref{app:mstar}, 
%we have taken \lsun(3.6) to
%be 1.4$\times 10^{32}$\,\ergs\ as in \citet{cook14}.} in \lsun\ and \mstar\ in \msun):
We have taken \lsun(W1) to
be 1.68$\times 10^{32}$\,\ergs\ \citep[see also][]{cook14}, %as in \citet{cook14};
assuming that the W1 Solar (Vega) magnitude is
3.24 \citep{norris14,jarrett13},
and that the W1 zero-point calibration is 309.5\,Jy \citep{jarrett13}.} in $L_{{\rm W1,}\odot}$ and \mstar\ in \msun):
{\small
\begin{eqnarray}
%\log[M_{\rm star}(\texttt{CIGALE})]\,\,&=\,(1.05\,\pm\,0.01)\,\log(L_{3.6}) - (1.09\,\pm\,0.15) ; \nonumber \\
%\log[M_{\rm star}(\texttt{GRASIL})]\,\,&=\,(1.10\,\pm\,0.03)\,\log(L_{3.6}) - (1.84\,\pm\,0.26) ; \nonumber \\
%\log[M_{\rm star}(\texttt{MAGPHYS})]&=\,(1.07\,\pm\,0.02)\,\log(L_{3.6}) - (1.44\,\pm\,0.25) . \nonumber 
\log[M_{\rm star}(\texttt{CIGALE})/L_{\mathrm W1})]\,\,= & \nonumber \\
	\,(0.050\,\pm\,0.013)\,\log(L_{\mathrm W1}) - (1.05\,\pm\,0.14) ; \nonumber \\
\log[M_{\rm star}(\texttt{GRASIL})/L_{\mathrm W1})]\,\,= & \nonumber \\
	\,(0.102\,\pm\,0.026)\,\log(L_{\mathrm W1}) - (1.76\,\pm\,0.26) ; \nonumber \\
\log[M_{\rm star}(\texttt{MAGPHYS})]/L_{\mathrm W1})]\,= & \nonumber \\
	\,(0.079\,\pm\,0.024)\,\log(L_{\mathrm W1}) - (1.46\,\pm\,0.25) \quad . 
\label{eqn:mstar}
\end{eqnarray}
}

\begin{figure}
\vspace{\baselineskip}
\includegraphics[width=0.48\textwidth]{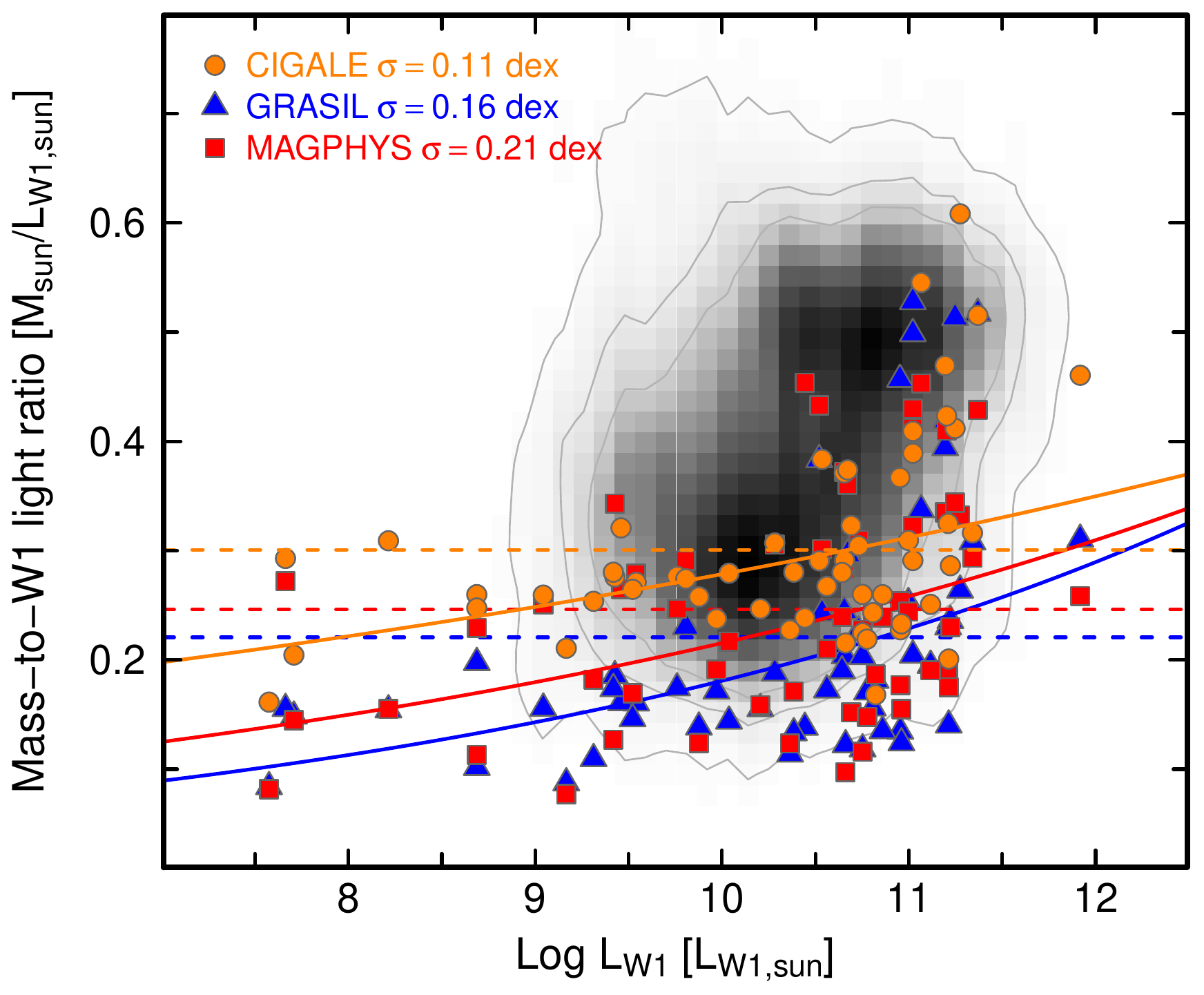}
\caption{SED-derived stellar masses with observed WISE W1 luminosities for mass-to-W1 light ratios
plotted against observed W1 luminosity;
the underlying gray scale gives the GSWLC sample from \citet[][see text for details]{salim16}.
Squares (orange) show \cig\ \mstar\ values, triangles (blue) \gra, and circles (red) \magp.
The robust regressions for each SED-fitting algorithm are shown as solid curves, and
the $\sigma$ values are given in the upper left corner.
The horizontal dashed lines show the mean of \upsiwise\ for \mstar\ values from
the three fitting codes:
0.30 (\cig), 0.22 (\gra), and 0.25 (\magp).
}
\label{fig:masstolight}
\end{figure}

\vspace{-0.5\baselineskip}
\noindent
Figure \ref{fig:masstolight} shows the mass-to-light ratio inferred from \mstar\ from the SED algorithms
divided by the W1 luminosity \lwiseone, 
%against \liracone, 
together with the best-fit regressions given in Eqn. (\ref{eqn:mstar}).
As seen in Fig. \ref{fig:masstolight}, these expressions reproduce the SED-derived \mstar\ values with \rms\ deviations of
0.11\,dex, 0.16\,dex, and 0.21\,dex for \cig, \gra, and \magp, respectively;
moreover, the slopes ($+$1) are identical to those given in Table \ref{tab:corr} for \liracone,
reinforcing the notion that IRAC 3.6\,\micron\ and WISE 3.4\,\micron\ photometry is indistinguishable.

The power-law slopes are significantly larger than zero %unity, %(except possibly for \cig), 
implying that to within the scatter, the M/L ratio at 3.4\,\micron\ depends on luminosity as also found
by \citet{wen13}.
To calculate \mstar\ with the \citet{wen13} formulation (see Appendix \ref{app:mstar}), 
we adopted their variation with Hubble type which assumes slopes between 1.03 and 1.04;
however,
values from the SED-fitting codes are better fit with larger slopes [see Table \ref{tab:corr}
and Eqn. (\ref{eqn:mstar})].
A comparably large slope connecting \mstar\ and \lwiseone\ was obtained by \citet{wen13} for the sample as a whole (1.12),
with active galaxy nuclei (AGN) having a steeper trend (1.13) than either
composite (star-forming and AGN hybrids with 1.08) or late Hubble types (1.03).

Several previous studies have found that a constant value of $\Upsilon^{[3.6]}\sim 0.5-0.7$ fits
SSP-derived stellar masses quite well 
\citep[e.g.,][]{oh08,eskew12,meidt12,mcgaugh14,meidt14,norris14,mcgaugh15,querejeta15}.
For %L(3.6\,\micron)\,=\,$10^{11}$\,\lsun(3.6), a typical luminosity
\lwiseone\,=\,$10^{11}$\,\lsun(W1) (see Fig. \ref{fig:masstolight}),
we would infer (with \cig) \upsiwise\,=\,0.3, roughly 2 times smaller.
%For higher luminosities, the M/L values are closer to previous work, $\sim$0.4.
%We have redone the fits using only galaxies with  L(3.6\,\micron)\,$\geq\,10^{11}$\,\lsun(3.6) 
%to assess whether the fits here are affected by the broad range in \mstar\ in the KINGFISH sample, extending
%the calibration to lower L(3.6\,\micron) luminosities.
%However, even for this luminous subset, %L(3.6)$\geq\,10^{11}$\lsun(3.6),
%the inferred $\Upsilon^{[3.6]}$ values are unaltered: $\Upsilon^{[3.6]}\,\sim\,0.2-0.3$. 
From dynamical considerations of the vertical force perpendicular to the disk
in 30 galaxies, \citet{martinsson13} find a mean $K$-band M/L ratio
$\Upsilon^{[K]}\,=\,0.31\,\pm\,0.07$.
Assuming $\Upsilon^{[K]}\,=\,1.29\,\Upsilon^{[3.6]}$ \citep{mcgaugh14}, 
this would give %$\Upsilon^{[3.6]}$
\upsiwise\,$\approx$\,\upsiirac\,=\,0.24,
consistent with what we have derived from SED fitting.
\citet{just15} analyzed a new sample of stars in the Milky Way and obtained a local
volumetric mass-to-light ratio $\Upsilon^{[K]}\,=\,0.31\,\pm\,0.02$, the same as found by \citet{martinsson13}.
\citet{ponomareva18} compared $\Upsilon^{[3.6]}$ from various methods, and found that
SED fitting and dynamical arguments tend to give lower $\Upsilon^{[3.6]}$ than values derived
from correlations with NIR color \citep[e.g.,][]{eskew12,meidt14,querejeta15}.
% xxx modify when/if we get GSWLC with colors xxx
Moreover, Fig. \ref{fig:masstolight} shows a steepening luminosity dependence of \upsiwise\
beginning around \lwiseone$\sim 3\times10^{10}$\,\lsun; thus the higher M/L ratios could also be
a function of more massive samples under consideration.
In any case,
because the reason for these discrepancies is not yet understood, 
the stellar mass scale is evidently pervaded by a systematic uncertainty of roughly a factor of two \citep[e.g.,][]{mcgaugh14}. 

Unlike global assessments of galaxy mass,
to ensure that region-by-region cumulative stellar masses agree with globally measured values,
resolved studies of stellar mass surface density require an approach that is linear with luminosity.
Here we attempt to furnish color-dependent recipes that can be used {\it within} galaxies,
rather than only between galaxies.
Our approach is similar to that of \citet{eskew12,meidt14,querejeta15}, but here we incorporate
the vast range of photometric bands available for the KINGFISH sample.
The idea is to compensate the non-linear slope of the \mstar\,--\,luminosity trend by exploiting 
the color-magnitude effect;
colors typically change with luminosity (reflecting trends with age and metallicity), 
thus implying a change in M/L.
We have investigated several single colors (ranging from FUV/NUV to W1-W3), 
%combinations of hybrid UV$+$NIR, MIR (GALEX, 2MASS, WISE), purely optical (SDSS), 
%hybrid optical-NIR (SDSS$+$2MASS), NIR-MIR (2MASS$+$WISE), or purely MIR (WISE); 
and have also assessed the improvement offered by introducing two colors rather than only one.
Judging from \citet{zibetti09}, one of the best colors for reducing scatter in M/L ratios
should be SDSS $g-i$; for the KINGFISH sample, the $g-i$ color does a good job of reducing the scatter in
the \upsiwise\ ratio (0.066\,dex w.r.t. 0.11\,dex for \cig\ \mstar), but this includes a residual non-linear slope with 
\lwiseone\ luminosity.
Imposing linearity for \upsiwise\ gives an increased \rms\ scatter for \upsiwise\ vs. $g-i$ of 0.075\,dex (\cig).
Other single colors we tested (in the AB system) under the necessity of imposing linearity with \lwiseone\ luminosity
include FUV-NUV, NUV-(W1, W3), NUV$-r$, NUV$-J$, $r-J$, $i-H$, $r-$(W1,W3), $J-H$, $J-$(W1, W2, W3), 
and W1$-$W3. 

\begin{figure*}[!h]
\resizebox{\textwidth}{!}{
\includegraphics[width=1.5\textwidth]{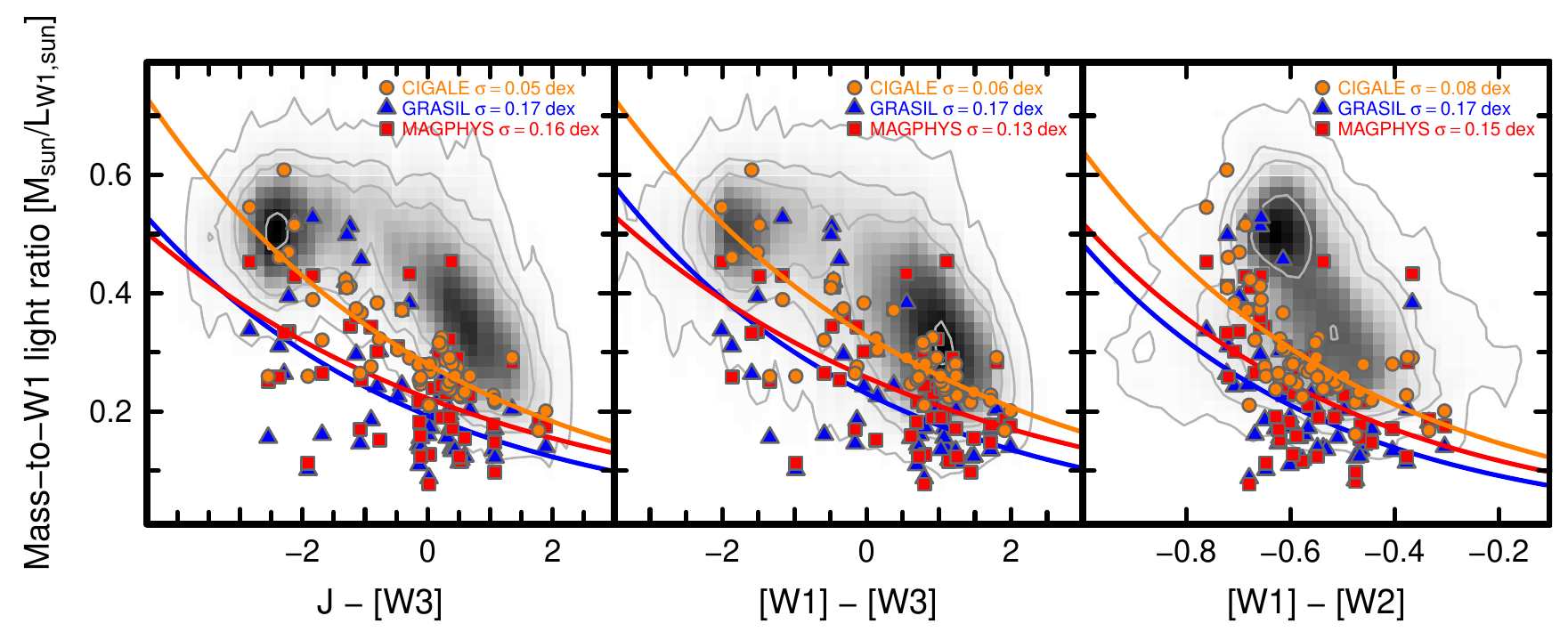}
}
\vspace{-0.6\baselineskip}
\caption{Mass-to-light ratios in the WISE W1 band of KINGFISH galaxies
plotted vs.  the J$-$[W3] color (left panel); [W1]$-$[W3] (middle); and [W1]$-$[W2] (right).
As in Fig. \ref{fig:masstolight}, the underlying gray scale corresponds to the GSWLC data
\citep{salim16}.
Legends in the upper right corners give the \rms\ deviation of the
robust best fits, shown as curves in each panel, of the M/L(W1) ratios vs. and colors.
The best fit \rms\ of 0.05\,dex (for \cig) is obtained for M/L as a function of J-[W3] (left panel), but [W1]$-$[W3]
is only 0.01\,dex worse (for \cig, see middle panel).
The fit of M/L with [W1]-[W2] (right panel) is the worst of all three colors shown here, but 
only by 0.03\,dex (for \cig, comparable for the other two algorithms).
All magnitudes are on the AB system.
\label{fig:masstolight_colors}
}
\end{figure*}

Fig. \ref{fig:masstolight_colors} shows \upsiwise\ plotted against the single colors that most reduced the \rms\ scatter
for the KINGFISH sample;
the best color is $J-$W3 (with \rms\ $\sigma$ of 0.05\,dex for \cig, left panel), followed closely by W1$-$W3 (\rms\ $\sigma$\,=\,0.06\,dex for \cig, middle).
\citet{meidt14} and \citet{querejeta15} have used W1$-$W2 to refine M/L ratios in the mid-infrared (\upsiirac, \upsiwise),
and we have compared this color with our best-fit results in the right panel of Fig. \ref{fig:masstolight_colors}.
Compared to the W3 colors ($J-$W3, W1$-$W3), W1$-$W2 gives a slightly worse fit to SED-derived \upsiwise\
(\rms\ $\sigma \sim$0.08\,dex for \cig).
Part of the reason for this could be simply the smaller dynamic range of the W1$-$W2 color: $\sim$0.4\,AB mag 
relative to $\sim$5\,AB mag for $J-$W3 and $\sim$4\,AB mag for W1$-$W3.
In other samples, the ranges in these colors tend to be even smaller, given that KINGFISH encompasses low mass,
blue, dwarf galaxies, often excluded by sensitivity considerations.
After experimenting with some additional colors (e.g., NUV$-$W1, NUV$-J$), with two colors
the improvement in the scatter of \upsiwise\ was marginal; 
we were unable to reduce the scatter below $\sim$0.05\,dex in any case.
M/L ratios derived from \cig\ \mstar\ are generally less noisy with color (and luminosity) than those
from either \gra\ or \magp; the reasons for this are not completely clear.
Summarizing, our best recipes for resolved studies of stellar masses within galaxies are given by
(see Fig. \ref{fig:masstolight_colors}):
{\small
\begin{eqnarray}
\log[M_{\rm star}(\texttt{CIGALE})/L_{\mathrm W1})]\,\,= & \nonumber \\
	\,(-0.093\,\pm\,0.007)\,J-{\mathrm W3} - (0.552\,\pm\,0.009) ; \nonumber \\
\log[M_{\rm star}(\texttt{GRASIL})/L_{\mathrm W1})]\,\,= & \nonumber \\
	\,(-0.096\,\pm\,0.022)\,J-{\mathrm W3} - (0.712\,\pm\,0.026) ; \nonumber \\
\log[M_{\rm star}(\texttt{MAGPHYS})]/L_{\mathrm W1})]\,= & \nonumber \\
	\,(-0.079\,\pm\,0.021)\,J-{\mathrm W3} - (0.653\,\pm\,0.024) \quad . 
\label{eqn:mstar_jw3}
\end{eqnarray}
}

\vspace{-0.5\baselineskip}
\noindent
(with \rms\ deviations of 0.05\,dex, 0.17\,dex, and 0.16\,dex for \cig, \gra, and \magp, respectively),
and
{\small
\begin{eqnarray}
\log[M_{\rm star}(\texttt{CIGALE})/L_{\mathrm W1})]\,\,= & \nonumber \\
	\,(-0.099\,\pm\,0.008)\,{\mathrm W1-W3} - (0.485\,\pm\,0.009) ; \nonumber \\
\log[M_{\rm star}(\texttt{GRASIL})/L_{\mathrm W1})]\,\,= & \nonumber \\
	\,(-0.113\,\pm\,0.022)\,{\mathrm W1-W3} - (0.637\,\pm\,0.025) ; \nonumber \\
\log[M_{\rm star}(\texttt{MAGPHYS})]/L_{\mathrm W1})]\,= & \nonumber \\
	\,(-0.090\,\pm\,0.023)\,{\mathrm W1-W3} - (0.589\,\pm\,0.026) \quad . 
\label{eqn:mstar_w1w3}
\end{eqnarray}
}

\vspace{-0.5\baselineskip}
\noindent
(with \rms\ deviations of 0.06\,dex, 0.17\,dex, and 0.14\,dex for \cig, \gra, and \magp, respectively).

Although the KINGFISH sample is much smaller in number than the SDSS collection adopted by
\citet{wen13}, it spans a large range of Hubble types and more than four orders of magnitude
in \mstar.
The detailed SED fitting done here may be a better representation of stellar mass, implying
a steeper variation of M/L ratio with \lwiseone\ (or \liracone) than previously determined.
This is borne out by the comparison with the large GSWLC sample \citep{salim16}, 
suggesting that sample selection is important because of the color dependence of the M/L ratio.
Nevertheless, the comparison with the GSWLC also suggests that simple power-law recipes relating \upsiwise\ (or \upsiirac)
to AB colors are insufficient to completely capture the behavior shown by the GSWLC:
M/L is apparently constant until a threshold where M/L decreases with increasing color.
Especially for extreme starbursts, it is important to subtract non-stellar emission (e.g., ionized
gas continuum, hot dust) from the flux as we have described in Appendix \ref{app:mstar}
\citep[see also, e.g.,][]{querejeta15}.
It is also essential to avoid application of the non-linear relations in Eqn. (\ref{eqn:mstar}) to resolved measurements
of stellar mass surface density.

\begin{figure}[!h]
\vspace{\baselineskip}
\includegraphics[height=0.5\textwidth]{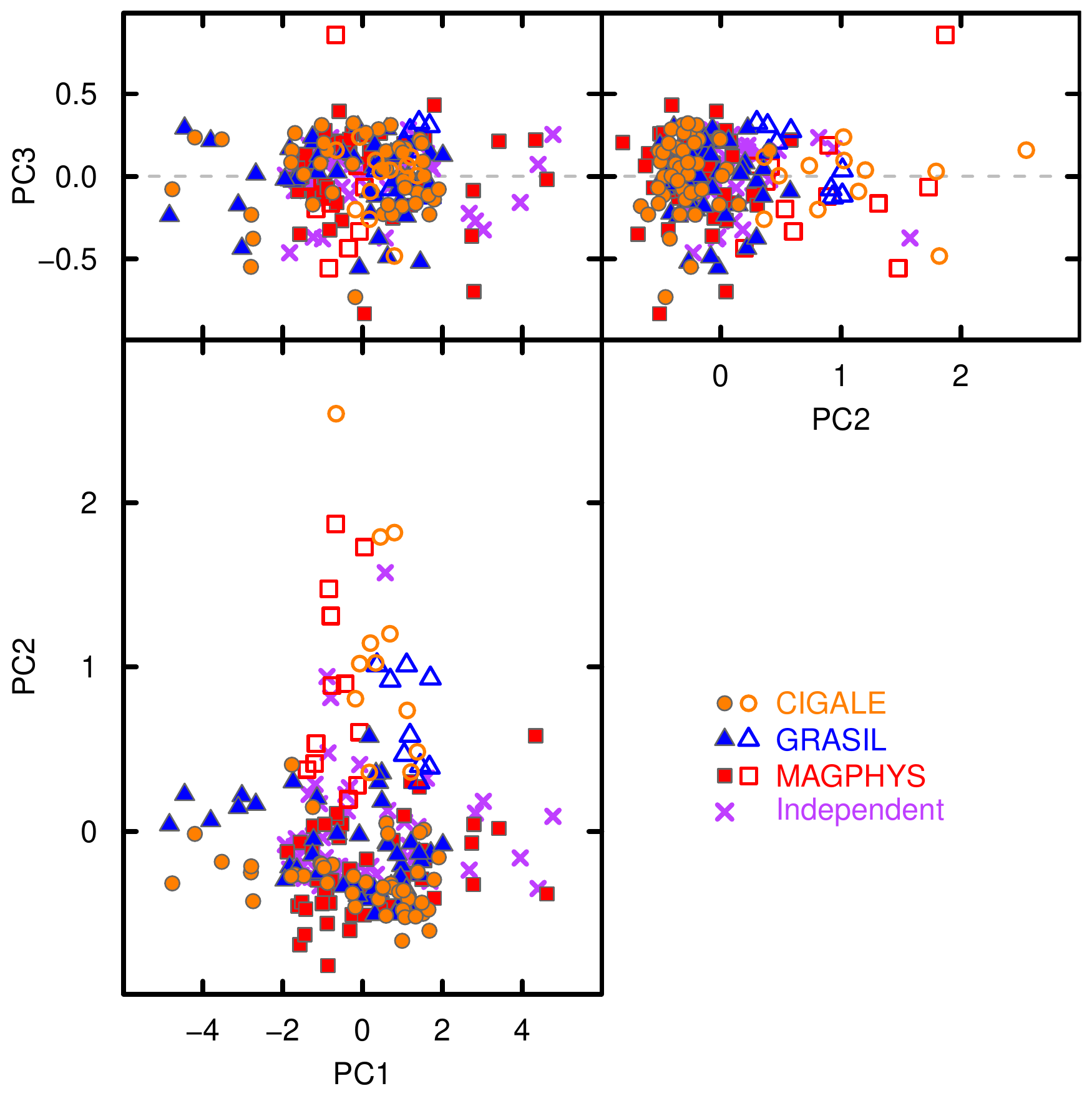} 
\caption{Different projections of the plane defined by Log(\mstar), Log(SFR), and Log(\mdust) for
KINGFISH galaxies: the edge-on projection is given in the top panels and the face-on 
in the bottom.
As in previous figures, open symbols correspond to  
galaxies with low sSFR
[Log(sSFR/yr$^{-1}$)\,$\leq\,-10.6$], and filled symbols to high sSFR
[Log(sSFR/yr$^{-1}$)\,$>\,-10.6$].
With 
$x_1\,=\,\log({\rm M_{\rm dust}}) - \langle\log ({\rm M_{\rm dust}}/M_\odot)\rangle$;
$x_2\,=\,\log({\rm SFR}) - \langle\log ({\rm SFR}/M_\odot\,{\rm yr}^{-1})\rangle$;
$x_3\,=\,\log({\rm M_{\rm star}}) - \langle\log ({\rm M_{\rm star}}/M_\odot)\rangle$;
and for \cig\ mean values 
$\langle\log ({\rm M_{\rm dust}}/M_\odot)\rangle\,=\,6.93$;
$\langle\log ({\rm SFR}/M_\odot\,{\rm yr}^{-1})\rangle\,=\,-0.44$;
$\langle\log ({\rm M_{\rm star}}/M_\odot)\rangle\,=\,9.76$;  
we find
PC1\,=\,$0.65\ x_1 + 0.48\ x_2 + 0.59\ x_3$; 
PC2\,=\,$0.01\ x_1 - 0.80\ x_2 + 0.60\ x_3$; 
PC3\,=\,$0.76\ x_1 - 0.37\ x_2 - 0.54\ x_3$. 
The PCAs for the different SED-fitting algorithms are similar.
}
\label{fig:pcadust_components}
\end{figure}

\begin{figure}
\vspace{\baselineskip}
\includegraphics[height=0.48\textwidth]{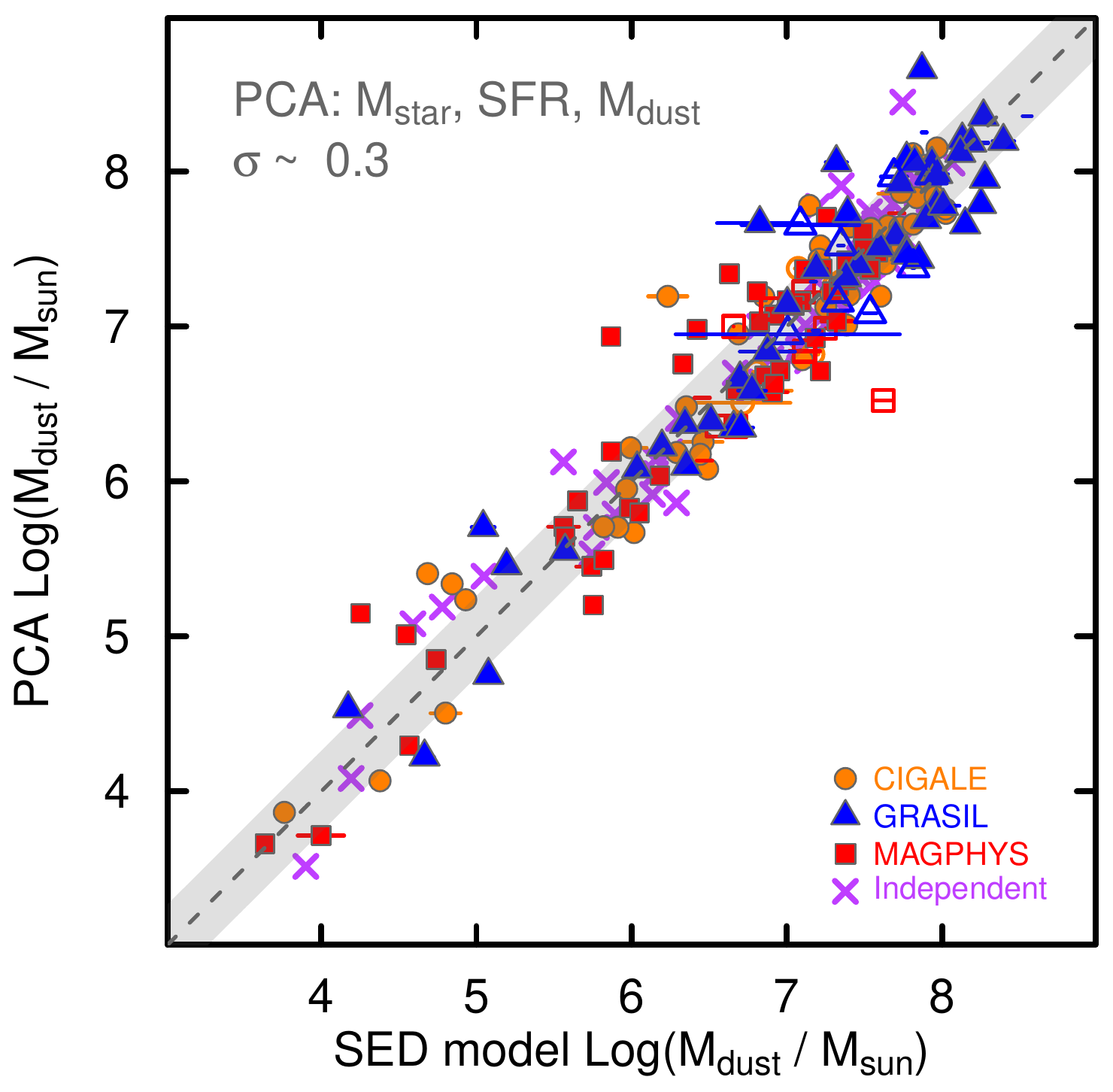} 
\caption{PCA-derived Log(\mdust) vs. model SED-derived and ``independently''-derived (here MBB) Log(\mdust) for KINGFISH galaxies.
The identity relation as a (gray) dashed line as described in the text.
and the $\sigma$ values of the four PCAs range from
%(0.27\,dex for \cig\ to 0.34\,dex for \gra), with mean $\sigma\,=\,0.3$;
(0.28\,dex for \cig\ to 0.4\,dex for \gra), with mean $\sigma\sim\,0.3$;
the gray region shows identity $\pm 1\sigma$.
As in previous figures, open symbols correspond to  
galaxies with low sSFR
[Log(sSFR/yr$^{-1}$)\,$\leq\,-10.6$], and filled symbols to high sSFR
[Log(sSFR/yr$^{-1}$)\,$>\,-10.6$].
Horizontal error bars show the uncertainties in the SED-fitted parameters (usually
smaller than the symbol size).
%$ \log({\mathrm M}_{\rm dust})\,=\,0.872\,\log({\rm SFR}) + 0.266\,\log({\mathrm M}_*) + 4.555$
%(see Eqn. \ref{eqn:mdustpca}).
}
\label{fig:pcadust_ind}
\end{figure}
 
\subsection{Principal component analysis}
\label{sec:pca}

Because of the mutual correlations of \mstar, \mdust, and SFR, it is likely that one or
more of them is just a secondary consequence of a fundamental, intrinsic, relation.
In this case,
these three variables could define a planar relation, based on just two parameters,
and it is important to know which of these three is the most fundamental in defining
the correlations.
To accomplish this, a PCA is an ideal tool.
A PCA essentially diagonalizes the three-dimensional covariance matrix, thus defining the 
``optimum projection'' of the parameter space which minimizes the covariance.
The orientation is defined by the eigenvectors, which by definition are mutually
orthogonal.
For a truly planar representation, we would expect most of the variation to be contained
in the first two eigenvectors; for the third, perpendicular, eigenvector, the variance
should be minimal.

We have performed a PCA on Log(\mstar), Log(\mdust), and Log(SFR) for each 
SED -fitting algorithm,
and one for the independently-determined recipe quantities. 
For the independently-determined quantities, we have adopted the \mstar\ values derived with the
\citet{wen13} formulation, \mdust\ DL07 values from \citet{aniano18}, and SFRs calculated with FUV+\ltir.
Results of the PCA of these variables show that they truly define a plane:
92\% of the variance is contained in the first eigenvector (E1), $\sim$5\% in the second (E2),
and only $\sim$3\% in the third (E3)\footnote{These numbers come from the independently-determined
quantities, but are similar for the other PCAs.}.
Figure \ref{fig:pcadust_components} illustrates the eigenvectors, and the different projections
of the plane:
E1 (PC1) has roughly equal contributions from all three parameters,
and E2 (PC2) has virtually no dependence on \mdust.
Interestingly, the galaxies with low sSFRs are the most discrepant from the main trends,
independently of the fitting algorithm.

The eigenvector containing the least variance, E3 (PC3), 
is dominated by \mdust.
Thus, by inverting the expression for E3, it is possible to calculate \mdust\ from SFR and \mstar, to an accuracy that
corresponds to the scatter of the PCA estimation.
The PCA inference of \mdust\ from \mstar\ and SFR is shown in Fig. \ref{fig:pcadust_ind}, where
we have compared \mdust\ that would be
%inferred from the PCA against the MBB \mdust\ values discussed in Sect. \ref{sec:comparison:mdust}.
derived from the PCA (as a function of \mstar\ and SFR) 
against the true (observed) values of \mdust\ 
using the values independently determined and from SED fitting. 
Results show that with knowledge of only \mstar\ and SFR for galaxies like those in the KINGFISH
sample, mainly main sequence galaxies, we can estimate \mdust\ to within a factor of 2
($\sigma \sim$0.26\,dex) through the equation:

\begin{equation}
%\log({\mathrm M}_{\rm dust})\,=\,0.872\,\log({\rm SFR}) + 0.266\,\log({\mathrm M}_*) + 4.555
% 28/11/2017
%\log({\mathrm M}_{\rm dust})\,=\,0.993\,\log({\rm SFR}) + 0.187\,\log({\mathrm M}_*) + 5.393
% 28/11/2017 with additional good Mdust data points (now missing only 3)
% NOT EXTREMELY STABLE
% \log({\mathrm M}_{\rm dust})\,=\,0.80\,\log({\rm SFR}) + 0.36\,\log({\mathrm M}_*) + 3.58
% 28/6/2018 now with Aniano et al. 2018 dust values looks better
\log({\mathrm M}_{\rm dust})\,=\,1.13\,\log({\rm SFR}) + 0.14\,\log({\mathrm M}_*) + 5.77 \quad ,
\label{eqn:mdustpca}
\end{equation}

\noindent
where \mdust\ and \mstar\ are in units of \msun, and SFR in \msunyr.
Eqn. (\ref{eqn:mdustpca}) is the equation resulting from the independent-parameter PCA with
\mdust\ from DL07 models \citep{aniano18}, 
while the PCA from MBB \mdust\ values is somewhat different ($\sigma \sim$0.31\,dex):
\begin{equation}
% MBB values
% 28/11/2017 with additional good Mdust data points (now missing only 3)
\log({\mathrm M}_{\rm dust})\,=\,0.80\,\log({\rm SFR}) + 0.36\,\log({\mathrm M}_*) + 3.58 \quad .
\label{eqn:mdustpca_mbb}
\end{equation}
\noindent
Those for the SED-fitting algorithms are close to these:  \\
% redone 2/2/2018
\\
\cig\ ($\sigma\,=\,0.29$\,dex): 
\vspace{-0.5\baselineskip}
\begin{equation}
\log({\mathrm M}_{\rm dust})\,=\,0.48\,\log({\rm SFR}) + 0.71\,\log({\mathrm M}_*) + 0.23
\label{eqn:mdustpca_cig}
\end{equation}
\gra\ ($\sigma\,=\,0.32$\,dex): 
\vspace{-0.5\baselineskip}
\begin{equation}
\log({\mathrm M}_{\rm dust})\,=\,1.16\,\log({\rm SFR}) + 0.072\,\log({\mathrm M}_*) + 6.76
\label{eqn:mdustpca_gra}
\end{equation}
\magp\ ($\sigma\,=\,0.36$\,dex): 
\vspace{-0.5\baselineskip}
\begin{equation}
\log({\mathrm M}_{\rm dust})\,=\,0.46\,\log({\rm SFR}) + 0.66\,\log({\mathrm M}_*) + 0.49 \quad .
\label{eqn:mdustpca_mag}
\end{equation}
\noindent
The seemingly innocuous differences in the behavior of the three SED-fitting codes emerge
strikingly in the PCA. 
In particular, \gra\ coefficients %are the most extreme, possibly because \gra\ tends to give lower \mstar\ and higher \mdust\ and SFRs than the other two codes.
are most similar to the PCA derived from the DL07 \mdust\ values [see Eqn. (\ref{eqn:mdustpca})], 
and \cig\ and \magp\ are consistent with the PCA with MBB values [Eqn. (\ref{eqn:mdustpca_mbb}].
%with differences possibly related to the lower \mstar\ and higher \mdust\ estimated by \gra\ relative to the independent values.
The PCAs of \cig\ and \magp\ are different from \gra, 
but mutually consistent, 
possibly because of the similarity in their underlying assumptions (see Table \ref{tab:models}).

Independently of the SED code,
we might expect that \mdust\ depends on metallicity (or its common proxy, oxygen abundance O/H),
so we have also performed a PCA on a set of four quantities, \mstar, \mdust, and SFR as before, but now
including \logoh.  
As in the previous section,
we have taken the values of \logoh\ from \citet{aniano18} %\citet{hunt16} 
converted to the nitrogen calibration of \citet[][]{pettini04} according to \citet{kewley08}.
In this case, the least variation is contained in the eigenvector dominated
by O/H, similar to other PCA analyses including \mstar, SFR, and O/H of galaxies
\citep[e.g.,][]{hunt12,bothwell16,hunt16}.
Thus, \logoh\ can be expressed as a linear combination of terms depending on \mstar, \mdust, and SFR
to the accuracy (mean dispersion) of the PCA, in this case 0.14\,dex. %0.144\,dex.

However, this mean dispersion is similar to that obtained by a 3-component PCA without \mdust,
namely with \mstar, SFR, and \logoh.
For a sample of $\sim$1000 galaxies up to $z\sim 3.7$ using the PP04N2 O/H calibration,
\citet{hunt16} find a mean dispersion of $\sim$0.16\,dex of such a PCA.
Performing a similar 3-component PCA analysis on the $\sim$60 KINGFISH galaxies
alone gives a mean dispersion of 0.15\,dex, %0.148\,dex, 
not significantly larger than with the 4-component PCA
including \mdust.
This is telling us that the addition of the \mdust\ parameter does not help to reduce the scatter of the PCA.
The correlations of \mdust\ with \mstar\ and SFR make \mdust\ superfluous in describing the scaling relations
with metallicity.
In fact, we have inverted the 4-component PCA to derive \mdust, even though the \mdust-dominated eigenvector
does not contain the least variation;
the result is an expression for \mdust\ which has the same dispersion as that without \logoh\ ($\sim$0.3\,dex).
Thus, for the KINGFISH sample, it seems that \mstar\ and SFR are sufficient to determine \mdust\ to within a factor of two.
Moreover, \mdust\ is not needed to determine \logoh\ to an accuracy of $\sim$0.14-0.15\,dex;
\mstar\ and SFR alone are also sufficient to describe metallicity. 
% 20/8/2018 from Chris:
% The underlying relationship could reflect a relation between \mstar\
% and gas mass, via a fairly constant gas-to-dust mass ratio; it would appear as an additional 
% correlation with SFR because the gas depletion time is relatively constant in the KINGFISH sample.
% Applying this relation to other samples will help clarify this.
% However, G2D varies in KINGFISH by almost a factor of 100 (see Aniano et al. 2018), so not constant.
Ultimately, at least for the KINGFISH galaxies, the relative importance of current star formation (SFR)
and past star formation (\mstar) essentially drive the observed dust content
and metallicity. %in local galaxies.

\section{Summary and conclusions}
\label{sec:conclusions}

We have fit the observed SEDs \citep{dale17} of the 61 galaxies from KINGFISH with three well-known models:
\cig\ \citep{noll09},
\gra\ \citep{silva98},
and \magp\ \citep{dacunha08}.
Although these codes differ in their approach to defining SFHs and dust attenuation, they all provide
excellent approximations to the shape of the observed SEDs with \rms\ deviations ranging from
(0.05--0.08\,dex); these values are comparable to the typical uncertainties in the fluxes
\citep{dale17}.
Nevertheless, the three algorithms show significantly different behavior in the 
% referee
mid-infrared:
in the 25--70\,\micron\ range where there are no observational constraints,
but also between 5\,\micron\ and 10\,\micron\ where the SED is constrained by observations
and dominated by PAH emission.
We summarize below the comparison of
the associated SED derived quantities with recipe-derived values of \mstar,
SFR, \mdust, and monochromatic luminosities. 
\begin{itemize}
\item
Stellar masses estimated with simple methods are fairly consistent with the SED-fitting results
to within $\la 0.2$\,dex (see Fig. \ref{fig:mstar}).
Nevertheless, the assumption of the ``standard'' \citep[e.g.,][]{mcgaugh14}
constant 3.6\,\micron\ M/L ratio results in super-linear power-law slopes relative to SED-inferred values,
and overestimates \mstar\ by $\sim$0.3--0.5\,dex. 
\item
Although there is generally good agreement between SED-derived SFRs and those estimated either from
FUV$+$TIR or from \ha$+$24\,\micron\ luminosities, 
in galaxies with low sSFRs ($\la 3\times10^{-11}$\,yr), recipe SFRs are larger than those from 
\cig\ and to some extent \magp. 
SFRs in galaxies without IR constraints can create some difficulties for \gra\
(see Fig. \ref{fig:sfr}) 
\item
The most salient difference among the three fitting codes is in the determination of \mdust;
\gra\ tends to give dust masses that are larger than either \cig\ or \magp\ 
(or the recipe values) by a factor of $\sim$0.3\,dex
(see Fig. \ref{fig:mdust}).
Because it is the only code that performs radiative transfer in realistic geometries, this
may be telling us that the usual methods of deriving \mdust\ are underestimating dust mass even in
``normal'' galaxies like the KINGFISH sample.
\item
Infrared luminosity \ltir\ is the most robust recipe estimate, consistent with all the SED-inferred
values to within 0.02--0.09\,dex (see Fig. \ref{fig:tir}). 
FUV luminosity \lfuv\ derived from photometry and corrected using IRX \citep[e.g.,][]{hao11,murphy11}
is within $0.08-0.13$\,dex of the \lfuv\ from the SED (Fig. \ref{fig:fuv}), although
the recipe estimate of FUV extinction \afuv\ is too high compared with all three SED codes 
(see Fig. \ref{fig:afuv}).
This is almost certainly due to a clumpy dust distribution 
%that would tend to make dust attenuation less efficient, 
that, for a given IRX value, would reduce the effective attenuation,
relative to the uniform dust screens implicitly assumed by the IRX recipes.
\end{itemize}

We have explored scaling relations based on the derived quantities from SED fitting, and confirm
previously established relations including the SFMS, the correlation between \mdust\ and SFR
\citep[e.g.,][]{dacunha10},
between \mdust\ and \mstar\ \citep[e.g.,][]{peeples14},
and various scalings of IRX including \mstar, SFR, and O/H
(see Fig. \ref{fig:irxvsother}).
%IRX can be calculated quite accurately using the ratio of monochromatic 160\,\micron\ luminosity
%$L_{160}$ to \lfuv\ [see Eqn. (\ref{eqn:irx}],
%a relation that may be useful at high redshift. 
Galaxies with low sSFRs tend to be either on the red sequence as quenched systems or in a pre-quenching phase
of their SFHs, as reflected by their UV-optical colors 
and discrepancies between recipe and model SFRs.
As seen in Fig. \ref{fig:sdss_redsequence},
these disagreements occur primarily in galaxies with red
NUV$-r \ga 5$, where the correlation between NUV$-r$ and sSFR
begins to degrade and flatten. 

We have established a new expression for \mstar\ depending on %\liracone\ 
\lwiseone\ and colors that is accurate
to 0.06--0.17\,dex [see Eqns. (\ref{eqn:mstar}, \ref{eqn:mstar_jw3}, \ref{eqn:mstar_w1w3})]. 
In addition, to further investigate possible dependencies among the fundamental quantities, 
we have computed a PCA of the KINGFISH sample using \mstar, SFR, O/H, and \mdust.
The result is that both O/H and \mdust\ can be expressed to within good accuracy using
only \mstar\ and SFR.
The PCA of \mstar, SFR, and \mdust\ is to our knowledge a new result, and enables estimating
dust mass to within a factor of 2 using only \mstar\ and SFR
% $\log({\mathrm M}_{\rm dust})\,=\,0.80\,\log({\rm SFR}) + 0.36\,\log({\mathrm M}_*) + 3.58$
[see Eqns. (\ref{eqn:mdustpca}, \ref{eqn:mdustpca_mbb})].

Overall, our results suggest that there are two main challenges to global SED fitting of galaxies.
The first is the problem of assessing dust mass and the dust properties that shape attenuation curves.
Dust {\it luminosity} drives the infrared shape of the SED, but absorption and attenuation are governed by dust {\it mass} and also strongly affected
by geometry and dust inhomogeneities. 
The absorbing (and scattering) dust is not necessarily the same dust as the dust that dominates the emission of the long-wavelength SED.
A galaxy's inclination is also crucial because the lines of sight in the outer regions include cooler dust
that may not be detectable at low inclination.
Inferring dust properties from SED fitting requires a large spectral range in photometry, but
even then, accurate dust masses are difficult to obtain;
this is mainly because of the temperature mixing along the line of
sight \citep[e.g.,][]{hunt15a}, but also because of the lack of consensus about dust opacities
(see Sect. \ref{sec:comparison:mdust}).

A second challenge is the inherent degeneracy of using SED fitting to derive fundamental
properties of galaxies such as SFRs and SFHs.
We have shown that galaxies with low sSFRs are problematic, and the lack of diagnostic power of the SED
gets translated into problems with \lfuv\ and attenuation as measured by \afuv\
(see e.g., Figs. \ref{fig:sfr}, \ref{fig:fuv}, \ref{fig:afuv}).
Evidence shows that most of the problematic galaxies with low sSFRs are in a quenching or
pre-quenching phase
(see e.g., Figs. \ref{fig:sfrvsmstar}, \ref{fig:sdss_redsequence}).
Thus an important, possibly the most crucial, aspect of SED fitting is 
the approach to SFHs,
and consequent degeneracies in connecting a specific SFH with a specific form of the SED. 
There is an ambiguity of heating sources for dust (young vs. old stars),
and in the MIR spectral regime, there are mixed contributions of 
ionized gas, stellar photospheres, and hot dust, both stochastically- and bulk-heated. 
These aspects of the emerging SED are dependent on the 
evolutionary phase of the galaxy as determined by its SFH.
The different approach of \gra\ may be an advantage particularly in the case of low sSFRs, because the shape of the SED is
not directly connected with the fitted parameters (see Table \ref{tab:models}).

Although \cig, \gra, and \magp\ are rather different in their approaches to fitting SEDs,
they are all extremely successful in reproducing the observed SED shapes.
Throughout the paper, we have emphasized that
the three codes give generally similar estimates of the fundamental quantities 
\mstar, SFR, \mdust, dust optical depth, and monochromatic luminosities.
The implication is that in some sense the problem is overdetermined, that is the number of
%constraints used 
parameters necessary
to construct a SED model exceeds the number of unknown quantities defining its shape.
Thus, either the SED fitting is not altogether sensitive to the specific underlying physics or 
there are ``hidden'' dependencies among the fundamental quantities.
Indeed, these emerge as scaling relations that are observed broadly among all galaxy types.

Given the amount of already available new FUV, IR, and mm data, together with observations of atomic and molecular gas 
\citep[e.g.,][]{salim16,devis17,orellana17},
it is paramount to establish the systematics of different SED models.
The models tested here are expected to remain at the state-of-the-art for many years to come, 
given their current success in fitting panchromatic galaxy SEDs.
Their further application to larger datasets containing galaxies with more extreme properties has been,
and will continue to be, an important
tool for understanding galaxy evolution both in the nearby and distant universe.

\section*{Acknowledgments}
% referee
We thank the anonymous referee for a very timely report and constructive comments.
We thank Paolo Serra for insights into star-formation rates for early-type galaxies,
and Anna Gallazzi for kindly passing us her SDSS sample in digital form for comparison.
We are also grateful to Michael Brown for helpful input,
and Elisabete da Cunha for her careful comments on the manuscript in advance of publication.
SB, GLG, LKH, AR, and LS acknowledge 
funding by an Italian research grant, PRIN-INAF/2012, and 
SB, GLG, LKH, LS, and SZ by the INAF PRIN-SKA 2017 program 1.05.01.88.04.
MB was supported by the FONDECYT regular project 1170618 and the
MINEDUC-UA projects codes ANT 1655 and ANT 1656.
IDL gratefully acknowledges the support of the Flemish Fund for Scientific
Research (FWO-Vlaanderen).
RN acknowledges partial support by FONDECYT grant No. 3140436, and
MR support by Spanish MEC Grant AYA-2014-53506-P.
This research has made use of the NASA/IPAC Extragalactic Database (NED) which
is operated by the Jet Propulsion Laboratory, California Institute of
Technology, under contract with the National Aeronautics and Space
Administration. 

\bibliographystyle{aa}
\bibliography{kingfish_seds}

\providecommand{\noopsort}[1]{}
\begin{thebibliography}{168}
\expandafter\ifx\csname natexlab\endcsname\relax\def\natexlab#1{#1}\fi

\bibitem[{{Aniano} {et~al.}(2012){Aniano}, {Draine}, {Calzetti}, {Dale},
  {Engelbracht}, {Gordon}, {Hunt}, {Kennicutt}, {Krause}, {Leroy}, {Rix},
  {Roussel}, {Sandstrom}, {Sauvage}, {Walter}, {Armus}, {Bolatto}, {Crocker},
  {Donovan Meyer}, {Galametz}, {Helou}, {Hinz}, {Johnson}, {Koda}, {Montiel},
  {Murphy}, {Skibba}, {Smith}, \& {Wolfire}}]{aniano12}
{Aniano}, G., {Draine}, B.~T., {Calzetti}, D., {et~al.} 2012, \apj, 756, 138

\bibitem[{{Aniano} {et~al.}(2018){Aniano}, {Draine}, {Hunt}, {Sandstrom}, \&
  et~al.}]{aniano18}
{Aniano}, G., {Draine}, B.~T., {Hunt}, L.~K., {Sandstrom}, K.~M., \& et~al.
  2018, in prep.

\bibitem[{{Bendo} {et~al.}(2012){Bendo}, {Boselli}, {Dariush}, {Pohlen},
  {Roussel}, {Sauvage}, {Smith}, {Wilson}, {Baes}, {Cooray}, {Clements},
  {Cortese}, {Foyle}, {Galametz}, {Gomez}, {Lebouteiller}, {Lu}, {Madden},
  {Mentuch}, {O'Halloran}, {Page}, {Remy}, {Schulz}, \& {Spinoglio}}]{bendo12}
{Bendo}, G.~J., {Boselli}, A., {Dariush}, A., {et~al.} 2012, \mnras, 419, 1833

\bibitem[{{Bendo} {et~al.}(2010){Bendo}, {Wilson}, {Pohlen}, {Sauvage}, {Auld},
  {Baes}, {Barlow}, {Bock}, {Boselli}, {Bradford}, {Buat}, {Castro-Rodriguez},
  {Chanial}, {Charlot}, {Ciesla}, {Clements}, {Cooray}, {Cormier}, {Cortese},
  {Davies}, {Dwek}, {Eales}, {Elbaz}, {Galametz}, {Galliano}, {Gear}, {Glenn},
  {Gomez}, {Griffin}, {Hony}, {Isaak}, {Levenson}, {Lu}, {Madden},
  {O'Halloran}, {Okumura}, {Oliver}, {Page}, {Panuzzo}, {Papageorgiou},
  {Parkin}, {Perez-Fournon}, {Rangwala}, {Rigby}, {Roussel}, {Rykala},
  {Sacchi}, {Schulz}, {Schirm}, {Smith}, {Spinoglio}, {Stevens}, {Sundar},
  {Symeonidis}, {Trichas}, {Vaccari}, {Vigroux}, {Wozniak}, {Wright}, \&
  {Zeilinger}}]{bendo10}
{Bendo}, G.~J., {Wilson}, C.~D., {Pohlen}, M., {et~al.} 2010, \aap, 518, L65

\bibitem[{{Berta} {et~al.}(2013){Berta}, {Lutz}, {Santini}, {Wuyts}, {Rosario},
  {Brisbin}, {Cooray}, {Franceschini}, {Gruppioni}, {Hatziminaoglou}, {Hwang},
  {Le Floc'h}, {Magnelli}, {Nordon}, {Oliver}, {Page}, {Popesso}, {Pozzetti},
  {Pozzi}, {Riguccini}, {Rodighiero}, {Roseboom}, {Scott}, {Symeonidis},
  {Valtchanov}, {Viero}, \& {Wang}}]{berta13}
{Berta}, S., {Lutz}, D., {Santini}, P., {et~al.} 2013, \aap, 551, A100

\bibitem[{{Bianchi}(2007)}]{bianchi07}
{Bianchi}, S. 2007, \aap, 471, 765

\bibitem[{{Bianchi}(2013)}]{bianchi13}
{Bianchi}, S. 2013, \aap, 552, A89

\bibitem[{{Binette} {et~al.}(1994){Binette}, {Magris}, {Stasi{\'n}ska}, \&
  {Bruzual}}]{binette94}
{Binette}, L., {Magris}, C.~G., {Stasi{\'n}ska}, G., \& {Bruzual}, A.~G. 1994,
  \aap, 292, 13

\bibitem[{{Binney} \& {Merrifield}(1998)}]{binney98}
{Binney}, J. \& {Merrifield}, M. 1998, {Galactic Astronomy}

\bibitem[{{Boquien} {et~al.}(2012){Boquien}, {Buat}, {Boselli}, {Baes},
  {Bendo}, {Ciesla}, {Cooray}, {Cortese}, {Eales}, {Gavazzi}, {Gomez},
  {Lebouteiller}, {Pappalardo}, {Pohlen}, {Smith}, \& {Spinoglio}}]{boquien12}
{Boquien}, M., {Buat}, V., {Boselli}, A., {et~al.} 2012, \aap, 539, A145

\bibitem[{{Boquien} {et~al.}(2014){Boquien}, {Buat}, \& {Perret}}]{boquien14}
{Boquien}, M., {Buat}, V., \& {Perret}, V. 2014, \aap, 571, A72

\bibitem[{{Boquien} {et~al.}(2018){Boquien}, {Burgarella}, {Roehlly}, {Buat},
  {Ciesla}, {Corre}, {Inoue}, \& {Salas}}]{boquien18}
{Boquien}, M., {Burgarella}, D., {Roehlly}, Y., {et~al.} 2018, \aap, in press
  [\eprint[arXiv]{1811.03094}]

\bibitem[{{Boquien} {et~al.}(2009){Boquien}, {Calzetti}, {Kennicutt}, {Dale},
  {Engelbracht}, {Gordon}, {Hong}, {Lee}, \& {Portouw}}]{boquien09}
{Boquien}, M., {Calzetti}, D., {Kennicutt}, R., {et~al.} 2009, \apj, 706, 553

\bibitem[{{Boquien} {et~al.}(2016){Boquien}, {Kennicutt}, {Calzetti}, {Dale},
  {Galametz}, {Sauvage}, {Croxall}, {Draine}, {Kirkpatrick}, {Kumari}, {Hunt},
  {De Looze}, {Pellegrini}, {Rela{\~n}o}, {Smith}, \& {Tabatabaei}}]{boquien16}
{Boquien}, M., {Kennicutt}, R., {Calzetti}, D., {et~al.} 2016, \aap, 591, A6

\bibitem[{{Boselli} {et~al.}(2016){Boselli}, {Cuillandre}, {Fossati},
  {Boissier}, {Bomans}, {Consolandi}, {Anselmi}, {Cortese}, {C{\^o}t{\'e}},
  {Durrell}, {Ferrarese}, {Fumagalli}, {Gavazzi}, {Gwyn}, {Hensler}, {Sun}, \&
  {Toloba}}]{boselli16}
{Boselli}, A., {Cuillandre}, J.~C., {Fossati}, M., {et~al.} 2016, \aap, 587,
  A68

\bibitem[{{Bothwell} {et~al.}(2016){Bothwell}, {Maiolino}, {Peng}, {Cicone},
  {Griffith}, \& {Wagg}}]{bothwell16}
{Bothwell}, M.~S., {Maiolino}, R., {Peng}, Y., {et~al.} 2016, \mnras, 455, 1156

\bibitem[{{Bouwens} {et~al.}(2016){Bouwens}, {Aravena}, {Decarli}, {Walter},
  {da Cunha}, {Labb{\'e}}, {Bauer}, {Bertoldi}, {Carilli}, {Chapman}, {Daddi},
  {Hodge}, {Ivison}, {Karim}, {Le Fevre}, {Magnelli}, {Ota}, {Riechers},
  {Smail}, {van der Werf}, {Weiss}, {Cox}, {Elbaz}, {Gonzalez-Lopez},
  {Infante}, {Oesch}, {Wagg}, \& {Wilkins}}]{bouwens16}
{Bouwens}, R.~J., {Aravena}, M., {Decarli}, R., {et~al.} 2016, \apj, 833, 72

\bibitem[{{Bressan} {et~al.}(1998){Bressan}, {Granato}, \& {Silva}}]{bressan98}
{Bressan}, A., {Granato}, G.~L., \& {Silva}, L. 1998, \aap, 332, 135

\bibitem[{{Bressan} {et~al.}(2002){Bressan}, {Silva}, \& {Granato}}]{bressan02}
{Bressan}, A., {Silva}, L., \& {Granato}, G.~L. 2002, \aap, 392, 377

\bibitem[{{Brinchmann} {et~al.}(2004){Brinchmann}, {Charlot}, {White},
  {Tremonti}, {Kauffmann}, {Heckman}, \& {Brinkmann}}]{brinchmann04}
{Brinchmann}, J., {Charlot}, S., {White}, S.~D.~M., {et~al.} 2004, \mnras, 351,
  1151

\bibitem[{{Brown} {et~al.}(2014b){Brown}, {Jarrett}, \& {Cluver}}]{brown14b}
{Brown}, M.~J.~I., {Jarrett}, T.~H., \& {Cluver}, M.~E. 2014b, \pasa, 31, HASH

\bibitem[{{Brown} {et~al.}(2014a){Brown}, {Moustakas}, {Smith}, {da Cunha},
  {Jarrett}, {Imanishi}, {Armus}, {Brandl}, \& {Peek}}]{brown14a}
{Brown}, M.~J.~I., {Moustakas}, J., {Smith}, J.-D.~T., {et~al.} 2014a, \apjs,
  212, 18

\bibitem[{{Bruzual} \& {Charlot}(2003)}]{bruzual03}
{Bruzual}, G. \& {Charlot}, S. 2003, \mnras, 344, 1000

\bibitem[{{Buat} {et~al.}(2005){Buat}, {Iglesias-P{\'a}ramo}, {Seibert},
  {Burgarella}, {Charlot}, {Martin}, {Xu}, {Heckman}, {Boissier}, {Boselli},
  {Barlow}, {Bianchi}, {Byun}, {Donas}, {Forster}, {Friedman}, {Jelinski},
  {Lee}, {Madore}, {Malina}, {Milliard}, {Morissey}, {Neff}, {Rich},
  {Schiminovitch}, {Siegmund}, {Small}, {Szalay}, {Welsh}, \& {Wyder}}]{buat05}
{Buat}, V., {Iglesias-P{\'a}ramo}, J., {Seibert}, M., {et~al.} 2005, \apjl,
  619, L51

\bibitem[{{Buat} {et~al.}(2012){Buat}, {Noll}, {Burgarella}, {Giovannoli},
  {Charmandaris}, {Pannella}, {Hwang}, {Elbaz}, {Dickinson}, {Magdis}, {Reddy},
  \& {Murphy}}]{buat12}
{Buat}, V., {Noll}, S., {Burgarella}, D., {et~al.} 2012, \aap, 545, A141

\bibitem[{{Burgarella} {et~al.}(2011){Burgarella}, {Heinis}, {Magdis}, {Auld},
  {Blain}, {Bock}, {Brisbin}, {Buat}, {Chanial}, {Clements}, {Cooray}, {Eales},
  {Franceschini}, {Giovannoli}, {Glenn}, {Gonz{\'a}lez Solares}, {Griffin},
  {Hwang}, {Ilbert}, {Marchetti}, {Mortier}, {Oliver}, {Page}, {Papageorgiou},
  {Pearson}, {P{\'e}rez-Fournon}, {Pohlen}, {Rawlings}, {Raymond},
  {Rigopoulou}, {Rodighiero}, {Roseboom}, {Rowan-Robinson}, {Scott}, {Seymour},
  {Smith}, {Symeonidis}, {Tugwell}, {Vaccari}, {Vieira}, {Viero}, {Vigroux},
  {Wang}, \& {Wright}}]{burgarella11}
{Burgarella}, D., {Heinis}, S., {Magdis}, G., {et~al.} 2011, \apjl, 734, L12

\bibitem[{{Calura} {et~al.}(2009){Calura}, {Pipino}, {Chiappini}, {Matteucci},
  \& {Maiolino}}]{calura09}
{Calura}, F., {Pipino}, A., {Chiappini}, C., {Matteucci}, F., \& {Maiolino}, R.
  2009, \aap, 504, 373

\bibitem[{{Calzetti} {et~al.}(2000){Calzetti}, {Armus}, {Bohlin}, {Kinney},
  {Koornneef}, \& {Storchi-Bergmann}}]{calzetti00}
{Calzetti}, D., {Armus}, L., {Bohlin}, R.~C., {et~al.} 2000, \apj, 533, 682

\bibitem[{{Calzetti} {et~al.}(2010){Calzetti}, {Wu}, {Hong}, {Kennicutt},
  {Lee}, {Dale}, {Engelbracht}, {van Zee}, {Draine}, {Hao}, {Gordon},
  {Moustakas}, {Murphy}, {Regan}, {Begum}, {Block}, {Dalcanton}, {Funes}, {Gil
  de Paz}, {Johnson}, {Sakai}, {Skillman}, {Walter}, {Weisz}, {Williams}, \&
  {Wu}}]{calzetti10}
{Calzetti}, D., {Wu}, S.-Y., {Hong}, S., {et~al.} 2010, \apj, 714, 1256

\bibitem[{{Chabrier}(2003)}]{chabrier03}
{Chabrier}, G. 2003, \pasp, 115, 763

\bibitem[{{Charlot} \& {Fall}(2000)}]{charlot00}
{Charlot}, S. \& {Fall}, S.~M. 2000, \apj, 539, 718

\bibitem[{{Ciesla} {et~al.}(2014){Ciesla}, {Boquien}, {Boselli}, {Buat},
  {Cortese}, {Bendo}, {Heinis}, {Galametz}, {Eales}, {Smith}, {Baes},
  {Bianchi}, {De Looze}, {di Serego Alighieri}, {Galliano}, {Hughes}, {Madden},
  {Pierini}, {R{\'e}my-Ruyer}, {Spinoglio}, {Vaccari}, {Viaene}, \&
  {Vlahakis}}]{ciesla14}
{Ciesla}, L., {Boquien}, M., {Boselli}, A., {et~al.} 2014, \aap, 565, A128

\bibitem[{{Ciesla} {et~al.}(2016){Ciesla}, {Boselli}, {Elbaz}, {Boissier},
  {Buat}, {Charmandaris}, {Schreiber}, {B{\'e}thermin}, {Baes}, {Boquien}, {De
  Looze}, {Fern{\'a}ndez-Ontiveros}, {Pappalardo}, {Spinoglio}, \&
  {Viaene}}]{ciesla16}
{Ciesla}, L., {Boselli}, A., {Elbaz}, D., {et~al.} 2016, \aap, 585, A43

\bibitem[{{Conroy} {et~al.}(2010){Conroy}, {White}, \& {Gunn}}]{conroy10}
{Conroy}, C., {White}, M., \& {Gunn}, J.~E. 2010, \apj, 708, 58

\bibitem[{{Cook} {et~al.}(2014){Cook}, {Dale}, {Johnson}, {Van Zee}, {Lee},
  {Kennicutt}, {Calzetti}, {Staudaher}, \& {Engelbracht}}]{cook14}
{Cook}, D.~O., {Dale}, D.~A., {Johnson}, B.~D., {et~al.} 2014, \mnras, 445, 899

\bibitem[{{Cortese} {et~al.}(2006){Cortese}, {Boselli}, {Buat}, {Gavazzi},
  {Boissier}, {Gil de Paz}, {Seibert}, {Madore}, \& {Martin}}]{cortese06}
{Cortese}, L., {Boselli}, A., {Buat}, V., {et~al.} 2006, \apj, 637, 242

\bibitem[{{\noopsort{Cunha}da Cunha} {et~al.}(2008){\noopsort{Cunha}da Cunha},
  {Charlot}, \& {Elbaz}}]{dacunha08}
{\noopsort{Cunha}da Cunha}, E., {Charlot}, S., \& {Elbaz}, D. 2008, \mnras,
  388, 1595

\bibitem[{{\noopsort{Cunha}da Cunha} {et~al.}(2010){\noopsort{Cunha}da Cunha},
  {Eminian}, {Charlot}, \& {Blaizot}}]{dacunha10}
{\noopsort{Cunha}da Cunha}, E., {Eminian}, C., {Charlot}, S., \& {Blaizot}, J.
  2010, \mnras, 403, 1894

\bibitem[{{Dale} {et~al.}(2012){Dale}, {Aniano}, {Engelbracht}, {Hinz},
  {Krause}, {Montiel}, {Roussel}, {Appleton}, {Armus}, {Beir{\~a}o}, {Bolatto},
  {Brandl}, {Calzetti}, {Crocker}, {Croxall}, {Draine}, {Galametz}, {Gordon},
  {Groves}, {Hao}, {Helou}, {Hunt}, {Johnson}, {Kennicutt}, {Koda}, {Leroy},
  {Li}, {Meidt}, {Miller}, {Murphy}, {Rahman}, {Rix}, {Sandstrom}, {Sauvage},
  {Schinnerer}, {Skibba}, {Smith}, {Tabatabaei}, {Walter}, {Wilson}, {Wolfire},
  \& {Zibetti}}]{dale12}
{Dale}, D.~A., {Aniano}, G., {Engelbracht}, C.~W., {et~al.} 2012, \apj, 745, 95

\bibitem[{{Dale} {et~al.}(2017){Dale}, {Cook}, {Roussel}, {Turner}, {Armus},
  {Bolatto}, {Boquien}, {Brown}, {Calzetti}, {De Looze}, {Galametz}, {Gordon},
  {Groves}, {Jarrett}, {Helou}, {Herrera-Camus}, {Hinz}, {Hunt}, {Kennicutt},
  {Murphy}, {Rest}, {Sandstrom}, {Smith}, {Tabatabaei}, \& {Wilson}}]{dale17}
{Dale}, D.~A., {Cook}, D.~O., {Roussel}, H., {et~al.} 2017, \apj, 837, 90

\bibitem[{{Dale} \& {Helou}(2002)}]{dale02}
{Dale}, D.~A. \& {Helou}, G. 2002, \apj, 576, 159

\bibitem[{{Davis} {et~al.}(2014){Davis}, {Young}, {Crocker}, {Bureau}, {Blitz},
  {Alatalo}, {Emsellem}, {Naab}, {Bayet}, {Bois}, {Bournaud}, {Cappellari},
  {Davies}, {de Zeeuw}, {Duc}, {Khochfar}, {Krajnovi{\'c}}, {Kuntschner},
  {McDermid}, {Morganti}, {Oosterloo}, {Sarzi}, {Scott}, {Serra}, \&
  {Weijmans}}]{davis14}
{Davis}, T.~A., {Young}, L.~M., {Crocker}, A.~F., {et~al.} 2014, \mnras, 444,
  3427

\bibitem[{{De Looze} {et~al.}(2014){De Looze}, {Fritz}, {Baes}, {Bendo},
  {Cortese}, {Boquien}, {Boselli}, {Camps}, {Cooray}, {Cormier}, {Davies}, {De
  Geyter}, {Hughes}, {Jones}, {Karczewski}, {Lebouteiller}, {Lu}, {Madden},
  {R{\'e}my-Ruyer}, {Spinoglio}, {Smith}, {Viaene}, \& {Wilson}}]{delooze14}
{De Looze}, I., {Fritz}, J., {Baes}, M., {et~al.} 2014, \aap, 571, A69

\bibitem[{{De Vis} {et~al.}(2017){De Vis}, {Dunne}, {Maddox}, {Gomez}, {Clark},
  {Bauer}, {Viaene}, {Schofield}, {Baes}, {Baker}, {Bourne}, {Driver}, {Dye},
  {Eales}, {Furlanetto}, {Ivison}, {Robotham}, {Rowlands}, {Smith}, {Smith},
  {Valiante}, \& {Wright}}]{devis17}
{De Vis}, P., {Dunne}, L., {Maddox}, S., {et~al.} 2017, \mnras, 464, 4680

\bibitem[{{Dole} {et~al.}(2006){Dole}, {Lagache}, {Puget}, {Caputi},
  {Fern{\'a}ndez-Conde}, {Le Floc'h}, {Papovich}, {P{\'e}rez-Gonz{\'a}lez},
  {Rieke}, \& {Blaylock}}]{dole06}
{Dole}, H., {Lagache}, G., {Puget}, J.-L., {et~al.} 2006, \aap, 451, 417

\bibitem[{{Draine}(2003)}]{draine03}
{Draine}, B.~T. 2003, \araa, 41, 241

\bibitem[{{Draine} {et~al.}(2014){Draine}, {Aniano}, {Krause}, {Groves},
  {Sandstrom}, {Braun}, {Leroy}, {Klaas}, {Linz}, {Rix}, {Schinnerer},
  {Schmiedeke}, \& {Walter}}]{draine14}
{Draine}, B.~T., {Aniano}, G., {Krause}, O., {et~al.} 2014, \apj, 780, 172

\bibitem[{{Draine} {et~al.}(2007){Draine}, {Dale}, {Bendo}, {Gordon}, {Smith},
  {Armus}, {Engelbracht}, {Helou}, {Kennicutt}, {Li}, {Roussel}, {Walter},
  {Calzetti}, {Moustakas}, {Murphy}, {Rieke}, {Bot}, {Hollenbach}, {Sheth}, \&
  {Teplitz}}]{draine07}
{Draine}, B.~T., {Dale}, D.~A., {Bendo}, G., {et~al.} 2007, \apj, 663, 866

\bibitem[{{Draine} \& {Li}(2007)}]{draineli07}
{Draine}, B.~T. \& {Li}, A. 2007, \apj, 657, 810

\bibitem[{{Dunne} {et~al.}(2000){Dunne}, {Eales}, {Edmunds}, {Ivison},
  {Alexander}, \& {Clements}}]{dunne00}
{Dunne}, L., {Eales}, S., {Edmunds}, M., {et~al.} 2000, \mnras, 315, 115

\bibitem[{{Elbaz} {et~al.}(2007){Elbaz}, {Daddi}, {Le Borgne}, {Dickinson},
  {Alexander}, {Chary}, {Starck}, {Brandt}, {Kitzbichler}, {MacDonald},
  {Nonino}, {Popesso}, {Stern}, \& {Vanzella}}]{elbaz07}
{Elbaz}, D., {Daddi}, E., {Le Borgne}, D., {et~al.} 2007, \aap, 468, 33

\bibitem[{{Elbaz} {et~al.}(2011){Elbaz}, {Dickinson}, {Hwang},
  {D{\'{\i}}az-Santos}, {Magdis}, {Magnelli}, {Le Borgne}, {Galliano},
  {Pannella}, {Chanial}, {Armus}, {Charmandaris}, {Daddi}, {Aussel}, {Popesso},
  {Kartaltepe}, {Altieri}, {Valtchanov}, {Coia}, {Dannerbauer}, {Dasyra},
  {Leiton}, {Mazzarella}, {Alexander}, {Buat}, {Burgarella}, {Chary}, {Gilli},
  {Ivison}, {Juneau}, {Le Floc'h}, {Lutz}, {Morrison}, {Mullaney}, {Murphy},
  {Pope}, {Scott}, {Brodwin}, {Calzetti}, {Cesarsky}, {Charlot}, {Dole},
  {Eisenhardt}, {Ferguson}, {F{\"o}rster Schreiber}, {Frayer}, {Giavalisco},
  {Huynh}, {Koekemoer}, {Papovich}, {Reddy}, {Surace}, {Teplitz}, {Yun}, \&
  {Wilson}}]{elbaz11}
{Elbaz}, D., {Dickinson}, M., {Hwang}, H.~S., {et~al.} 2011, \aap, 533, A119

\bibitem[{{Eskew} {et~al.}(2012){Eskew}, {Zaritsky}, \& {Meidt}}]{eskew12}
{Eskew}, M., {Zaritsky}, D., \& {Meidt}, S. 2012, \aj, 143, 139

\bibitem[{Fox(2008)}]{fox08}
Fox, J. 2008, Applied Regression Analysis and Generalized Linear Models (SAGE
  Publications)

\bibitem[{{Franceschini} {et~al.}(2008){Franceschini}, {Rodighiero}, \&
  {Vaccari}}]{franceschini08}
{Franceschini}, A., {Rodighiero}, G., \& {Vaccari}, M. 2008, \aap, 487, 837

\bibitem[{{Galametz} {et~al.}(2013){Galametz}, {Kennicutt}, {Calzetti},
  {Aniano}, {Draine}, {Boquien}, {Brandl}, {Croxall}, {Dale}, {Engelbracht},
  {Gordon}, {Groves}, {Hao}, {Helou}, {Hinz}, {Hunt}, {Johnson}, {Li},
  {Murphy}, {Roussel}, {Sandstrom}, {Skibba}, \& {Tabatabaei}}]{galametz13}
{Galametz}, M., {Kennicutt}, R.~C., {Calzetti}, D., {et~al.} 2013, \mnras, 431,
  1956

\bibitem[{{Galliano} {et~al.}(2011){Galliano}, {Hony}, {Bernard}, {Bot},
  {Madden}, {Roman-Duval}, {Galametz}, {Li}, {Meixner}, {Engelbracht},
  {Lebouteiller}, {Misselt}, {Montiel}, {Panuzzo}, {Reach}, \&
  {Skibba}}]{galliano11}
{Galliano}, F., {Hony}, S., {Bernard}, J.-P., {et~al.} 2011, \aap, 536, A88

\bibitem[{{Garn} \& {Best}(2010)}]{garn10}
{Garn}, T. \& {Best}, P.~N. 2010, \mnras, 409, 421

\bibitem[{{Giovannoli} {et~al.}(2011){Giovannoli}, {Buat}, {Noll},
  {Burgarella}, \& {Magnelli}}]{giovannoli11}
{Giovannoli}, E., {Buat}, V., {Noll}, S., {Burgarella}, D., \& {Magnelli}, B.
  2011, \aap, 525, A150

\bibitem[{{Gordon} {et~al.}(2000){Gordon}, {Clayton}, {Witt}, \&
  {Misselt}}]{gordon00}
{Gordon}, K.~D., {Clayton}, G.~C., {Witt}, A.~N., \& {Misselt}, K.~A. 2000,
  \apj, 533, 236

\bibitem[{{Grossi} {et~al.}(2015){Grossi}, {Hunt}, {Madden}, {Hughes}, {Auld},
  {Baes}, {Bendo}, {Bianchi}, {Bizzocchi}, {Boquien}, {Boselli}, {Clemens},
  {Corbelli}, {Cortese}, {Davies}, {De Looze}, {di Serego Alighieri}, {Fritz},
  {Pappalardo}, {Pierini}, {R{\'e}my-Ruyer}, {Smith}, {Verstappen}, {Viaene},
  \& {Vlahakis}}]{grossi15}
{Grossi}, M., {Hunt}, L.~K., {Madden}, S.~C., {et~al.} 2015, \aap, 574, A126

\bibitem[{{Hao} {et~al.}(2011){Hao}, {Kennicutt}, {Johnson}, {Calzetti},
  {Dale}, \& {Moustakas}}]{hao11}
{Hao}, C.-N., {Kennicutt}, R.~C., {Johnson}, B.~D., {et~al.} 2011, \apj, 741,
  124

\bibitem[{Hastings(1970)}]{Hastings_1970}
Hastings, W.~K. 1970, Biometrika, 57, 97

\bibitem[{{Hauser} \& {Dwek}(2001)}]{hauser01}
{Hauser}, M.~G. \& {Dwek}, E. 2001, \araa, 39, 249

\bibitem[{{Hayward} {et~al.}(2014){Hayward}, {Lanz}, {Ashby}, {Fazio},
  {Hernquist}, {Mart{\'{\i}}nez-Galarza}, {Noeske}, {Smith}, {Wuyts}, \&
  {Zezas}}]{hayward14}
{Hayward}, C.~C., {Lanz}, L., {Ashby}, M.~L.~N., {et~al.} 2014, \mnras, 445,
  1598

\bibitem[{{Hayward} \& {Smith}(2015)}]{hayward15}
{Hayward}, C.~C. \& {Smith}, D.~J.~B. 2015, \mnras, 446, 1512

\bibitem[{{Heckman} {et~al.}(1998){Heckman}, {Robert}, {Leitherer}, {Garnett},
  \& {van der Rydt}}]{heckman98}
{Heckman}, T.~M., {Robert}, C., {Leitherer}, C., {Garnett}, D.~R., \& {van der
  Rydt}, F. 1998, \apj, 503, 646

\bibitem[{{Helou}(1986)}]{helou86}
{Helou}, G. 1986, \apjl, 311, L33

\bibitem[{{Herrera-Camus} {et~al.}(2015){Herrera-Camus}, {Bolatto}, {Wolfire},
  {Smith}, {Croxall}, {Kennicutt}, {Calzetti}, {Helou}, {Walter}, {Leroy},
  {Draine}, {Brandl}, {Armus}, {Sandstrom}, {Dale}, {Aniano}, {Meidt},
  {Boquien}, {Hunt}, {Galametz}, {Tabatabaei}, {Murphy}, {Appleton}, {Roussel},
  {Engelbracht}, \& {Beirao}}]{herreracamus15}
{Herrera-Camus}, R., {Bolatto}, A.~D., {Wolfire}, M.~G., {et~al.} 2015, \apj,
  800, 1

\bibitem[{{Hunt} {et~al.}(2016){Hunt}, {Dayal}, {Magrini}, \&
  {Ferrara}}]{hunt16}
{Hunt}, L., {Dayal}, P., {Magrini}, L., \& {Ferrara}, A. 2016, \mnras, 463,
  2002

\bibitem[{{Hunt} {et~al.}(2012){Hunt}, {Magrini}, {Galli}, {Schneider},
  {Bianchi}, {Maiolino}, {Romano}, {Tosi}, \& {Valiante}}]{hunt12}
{Hunt}, L., {Magrini}, L., {Galli}, D., {et~al.} 2012, \mnras, 427, 906

\bibitem[{{Hunt} {et~al.}(2015{\natexlab{a}}){Hunt}, {Draine}, {Bianchi},
  {Gordon}, {Aniano}, {Calzetti}, {Dale}, {Helou}, {Hinz}, {Kennicutt},
  {Roussel}, {Wilson}, {Bolatto}, {Boquien}, {Croxall}, {Galametz}, {Gil de
  Paz}, {Koda}, {Mu{\~n}oz-Mateos}, {Sandstrom}, {Sauvage}, {Vigroux}, \&
  {Zibetti}}]{hunt15a}
{Hunt}, L.~K., {Draine}, B.~T., {Bianchi}, S., {et~al.} 2015{\natexlab{a}},
  \aap, 576, A33

\bibitem[{{Hunt} {et~al.}(2015{\natexlab{b}}){Hunt}, {Garc{\'{\i}}a-Burillo},
  {Casasola}, {Caselli}, {Combes}, {Henkel}, {Lundgren}, {Maiolino}, {Menten},
  {Testi}, \& {Weiss}}]{hunt15b}
{Hunt}, L.~K., {Garc{\'{\i}}a-Burillo}, S., {Casasola}, V., {et~al.}
  2015{\natexlab{b}}, \aap, 583, A114

\bibitem[{{Iglesias-P{\'a}ramo} {et~al.}(2007){Iglesias-P{\'a}ramo}, {Buat},
  {Hern{\'a}ndez-Fern{\'a}ndez}, {Xu}, {Burgarella}, {Takeuchi}, {Boselli},
  {Shupe}, {Rowan-Robinson}, {Babbedge}, {Conrow}, {Fang}, {Farrah},
  {Gonz{\'a}lez-Solares}, {Lonsdale}, {Smith}, {Surace}, {Barlow}, {Forster},
  {Friedman}, {Martin}, {Morrissey}, {Neff}, {Schiminovich}, {Seibert},
  {Small}, {Wyder}, {Bianchi}, {Donas}, {Heckman}, {Lee}, {Madore}, {Milliard},
  {Rich}, {Szalay}, {Welsh}, \& {Yi}}]{iglesias07}
{Iglesias-P{\'a}ramo}, J., {Buat}, V., {Hern{\'a}ndez-Fern{\'a}ndez}, J.,
  {et~al.} 2007, \apj, 670, 279

\bibitem[{{Jarrett} {et~al.}(2013){Jarrett}, {Masci}, {Tsai}, {Petty},
  {Cluver}, {Assef}, {Benford}, {Blain}, {Bridge}, {Donoso}, {Eisenhardt},
  {Koribalski}, {Lake}, {Neill}, {Seibert}, {Sheth}, {Stanford}, \&
  {Wright}}]{jarrett13}
{Jarrett}, T.~H., {Masci}, F., {Tsai}, C.~W., {et~al.} 2013, \aj, 145, 6

\bibitem[{{Johnson} {et~al.}(2007){Johnson}, {Schiminovich}, {Seibert},
  {Treyer}, {Martin}, {Barlow}, {Forster}, {Friedman}, {Morrissey}, {Neff},
  {Small}, {Wyder}, {Bianchi}, {Donas}, {Heckman}, {Lee}, {Madore}, {Milliard},
  {Rich}, {Szalay}, {Welsh}, \& {Yi}}]{johnson07}
{Johnson}, B.~D., {Schiminovich}, D., {Seibert}, M., {et~al.} 2007, \apjs, 173,
  392

\bibitem[{{Just} {et~al.}(2015){Just}, {Fuchs}, {Jahrei{\ss}}, {Flynn},
  {Dettbarn}, \& {Rybizki}}]{just15}
{Just}, A., {Fuchs}, B., {Jahrei{\ss}}, H., {et~al.} 2015, \mnras, 451, 149

\bibitem[{{Karim} {et~al.}(2011){Karim}, {Schinnerer},
  {Mart{\'{\i}}nez-Sansigre}, {Sargent}, {van der Wel}, {Rix}, {Ilbert},
  {Smol{\v c}i{\'c}}, {Carilli}, {Pannella}, {Koekemoer}, {Bell}, \&
  {Salvato}}]{karim11}
{Karim}, A., {Schinnerer}, E., {Mart{\'{\i}}nez-Sansigre}, A., {et~al.} 2011,
  \apj, 730, 61

\bibitem[{{Kaviraj} {et~al.}(2007){Kaviraj}, {Schawinski}, {Devriendt},
  {Ferreras}, {Khochfar}, {Yoon}, {Yi}, {Deharveng}, {Boselli}, {Barlow},
  {Conrow}, {Forster}, {Friedman}, {Martin}, {Morrissey}, {Neff},
  {Schiminovich}, {Seibert}, {Small}, {Wyder}, {Bianchi}, {Donas}, {Heckman},
  {Lee}, {Madore}, {Milliard}, {Rich}, \& {Szalay}}]{kaviraj07}
{Kaviraj}, S., {Schawinski}, K., {Devriendt}, J.~E.~G., {et~al.} 2007, \apjs,
  173, 619

\bibitem[{{Kennicutt} {et~al.}(2011){Kennicutt}, {Calzetti}, {Aniano},
  {Appleton}, {Armus}, {Beir{\~a}o}, {Bolatto}, {Brandl}, {Crocker}, {Croxall},
  {Dale}, {Donovan Meyer}, {Draine}, {Engelbracht}, {Galametz}, {Gordon},
  {Groves}, {Hao}, {Helou}, {Hinz}, {Hunt}, {Johnson}, {Koda}, {Krause},
  {Leroy}, {Li}, {Meidt}, {Montiel}, {Murphy}, {Rahman}, {Rix}, {Roussel},
  {Sandstrom}, {Sauvage}, {Schinnerer}, {Skibba}, {Smith}, {Srinivasan},
  {Vigroux}, {Walter}, {Wilson}, {Wolfire}, \& {Zibetti}}]{kennicutt11}
{Kennicutt}, R.~C., {Calzetti}, D., {Aniano}, G., {et~al.} 2011, \pasp, 123,
  1347

\bibitem[{{Kennicutt}(1998)}]{kennicutt98}
{Kennicutt}, Jr., R.~C. 1998, \apj, 498, 541

\bibitem[{{Kennicutt} {et~al.}(2003){Kennicutt}, {Armus}, {Bendo}, {Calzetti},
  {Dale}, {Draine}, {Engelbracht}, {Gordon}, {Grauer}, {Helou}, {Hollenbach},
  {Jarrett}, {Kewley}, {Leitherer}, {Li}, {Malhotra}, {Regan}, {Rieke},
  {Rieke}, {Roussel}, {Smith}, {Thornley}, \& {Walter}}]{kennicutt03}
{Kennicutt}, Jr., R.~C., {Armus}, L., {Bendo}, G., {et~al.} 2003, \pasp, 115,
  928

\bibitem[{{Kennicutt} {et~al.}(2009){Kennicutt}, {Hao}, {Calzetti},
  {Moustakas}, {Dale}, {Bendo}, {Engelbracht}, {Johnson}, \&
  {Lee}}]{kennicutt09}
{Kennicutt}, Jr., R.~C., {Hao}, C.-N., {Calzetti}, D., {et~al.} 2009, \apj,
  703, 1672

\bibitem[{{Kewley} \& {Ellison}(2008)}]{kewley08}
{Kewley}, L.~J. \& {Ellison}, S.~L. 2008, \apj, 681, 1183

\bibitem[{{King}(1962)}]{king62}
{King}, I. 1962, \aj, 67, 471

\bibitem[{{Kong} {et~al.}(2004){Kong}, {Charlot}, {Brinchmann}, \&
  {Fall}}]{kong04}
{Kong}, X., {Charlot}, S., {Brinchmann}, J., \& {Fall}, S.~M. 2004, \mnras,
  349, 769

\bibitem[{{Kreckel} {et~al.}(2014){Kreckel}, {Armus}, {Groves}, {Lyubenova},
  {D{\'{\i}}az-Santos}, {Schinnerer}, {Appleton}, {Croxall}, {Dale}, {Hunt},
  {Beir{\~a}o}, {Bolatto}, {Calzetti}, {Donovan Meyer}, {Draine}, {Hinz},
  {Kennicutt}, {Meidt}, {Murphy}, {Smith}, {Tabatabaei}, \&
  {Walter}}]{kreckel14}
{Kreckel}, K., {Armus}, L., {Groves}, B., {et~al.} 2014, \apj, 790, 26

\bibitem[{{Kroupa}(2001)}]{kroupa01}
{Kroupa}, P. 2001, \mnras, 322, 231

\bibitem[{{Laor} \& {Draine}(1993)}]{laor93}
{Laor}, A. \& {Draine}, B.~T. 1993, \apj, 402, 441

\bibitem[{{Leitner}(2012)}]{leitner12}
{Leitner}, S.~N. 2012, \apj, 745, 149

\bibitem[{{Leroy} {et~al.}(2012){Leroy}, {Bigiel}, {de Blok}, {Boissier},
  {Bolatto}, {Brinks}, {Madore}, {Munoz-Mateos}, {Murphy}, {Sandstrom},
  {Schruba}, \& {Walter}}]{leroy12}
{Leroy}, A.~K., {Bigiel}, F., {de Blok}, W.~J.~G., {et~al.} 2012, \aj, 144, 3

\bibitem[{Li(2006)}]{li06}
Li, G. 2006, Robust Regression (John Wiley \& Sons, Inc.), 281--343

\bibitem[{{Lo Faro} {et~al.}(2013){Lo Faro}, {Franceschini}, {Vaccari},
  {Silva}, {Rodighiero}, {Berta}, {Bock}, {Burgarella}, {Buat}, {Cava},
  {Clements}, {Cooray}, {Farrah}, {Feltre}, {Gonz{\'a}lez Solares}, {Hurley},
  {Lutz}, {Magdis}, {Magnelli}, {Marchetti}, {Oliver}, {Page}, {Popesso},
  {Pozzi}, {Rigopoulou}, {Rowan-Robinson}, {Roseboom}, {Scott}, {Smith},
  {Symeonidis}, {Wang}, \& {Wuyts}}]{lofaro13}
{Lo Faro}, B., {Franceschini}, A., {Vaccari}, M., {et~al.} 2013, \apj, 762, 108

\bibitem[{{Lotz} {et~al.}(2000){Lotz}, {Ferguson}, \& {Bohlin}}]{lotz00}
{Lotz}, J.~M., {Ferguson}, H.~C., \& {Bohlin}, R.~C. 2000, \apj, 532, 830

\bibitem[{{Magdis} {et~al.}(2012){Magdis}, {Daddi}, {B{\'e}thermin}, {Sargent},
  {Elbaz}, {Pannella}, {Dickinson}, {Dannerbauer}, {da Cunha}, {Walter},
  {Rigopoulou}, {Charmandaris}, {Hwang}, \& {Kartaltepe}}]{magdis12}
{Magdis}, G.~E., {Daddi}, E., {B{\'e}thermin}, M., {et~al.} 2012, \apj, 760, 6

\bibitem[{{Maraston}(2005)}]{maraston05}
{Maraston}, C. 2005, \mnras, 362, 799

\bibitem[{{Martinsson} {et~al.}(2013){Martinsson}, {Verheijen}, {Westfall},
  {Bershady}, {Andersen}, \& {Swaters}}]{martinsson13}
{Martinsson}, T.~P.~K., {Verheijen}, M.~A.~W., {Westfall}, K.~B., {et~al.}
  2013, \aap, 557, A131

\bibitem[{{McCormick} {et~al.}(2013){McCormick}, {Veilleux}, \&
  {Rupke}}]{mccormick13}
{McCormick}, A., {Veilleux}, S., \& {Rupke}, D.~S.~N. 2013, \apj, 774, 126

\bibitem[{{McGaugh} \& {Schombert}(2014)}]{mcgaugh14}
{McGaugh}, S.~S. \& {Schombert}, J.~M. 2014, \aj, 148, 77

\bibitem[{{McGaugh} \& {Schombert}(2015)}]{mcgaugh15}
{McGaugh}, S.~S. \& {Schombert}, J.~M. 2015, \apj, 802, 18

\bibitem[{{Meidt} {et~al.}(2012){Meidt}, {Schinnerer}, {Knapen}, {Bosma},
  {Athanassoula}, {Sheth}, {Buta}, {Zaritsky}, {Laurikainen}, {Elmegreen},
  {Elmegreen}, {Gadotti}, {Salo}, {Regan}, {Ho}, {Madore}, {Hinz}, {Skibba},
  {Gil de Paz}, {Mu{\~n}oz-Mateos}, {Men{\'e}ndez-Delmestre}, {Seibert}, {Kim},
  {Mizusawa}, {Laine}, \& {Comer{\'o}n}}]{meidt12}
{Meidt}, S.~E., {Schinnerer}, E., {Knapen}, J.~H., {et~al.} 2012, \apj, 744, 17

\bibitem[{{Meidt} {et~al.}(2014){Meidt}, {Schinnerer}, {van de Ven},
  {Zaritsky}, {Peletier}, {Knapen}, {Sheth}, {Regan}, {Querejeta},
  {Mu{\~n}oz-Mateos}, {Kim}, {Hinz}, {Gil de Paz}, {Athanassoula}, {Bosma},
  {Buta}, {Cisternas}, {Ho}, {Holwerda}, {Skibba}, {Laurikainen}, {Salo},
  {Gadotti}, {Laine}, {Erroz-Ferrer}, {Comer{\'o}n}, {Men{\'e}ndez-Delmestre},
  {Seibert}, \& {Mizusawa}}]{meidt14}
{Meidt}, S.~E., {Schinnerer}, E., {van de Ven}, G., {et~al.} 2014, \apj, 788,
  144

\bibitem[{{Metropolis} {et~al.}(1953){Metropolis}, {Rosenbluth}, {Rosenbluth},
  {Teller}, \& {Teller}}]{Metropolis+1953}
{Metropolis}, N., {Rosenbluth}, A.~W., {Rosenbluth}, M.~N., {Teller}, A.~H., \&
  {Teller}, E. 1953, \jcp, 21, 1087

\bibitem[{{Micha{\l}owski} {et~al.}(2014){Micha{\l}owski}, {Hayward}, {Dunlop},
  {Bruce}, {Cirasuolo}, {Cullen}, \& {Hernquist}}]{michalowski14}
{Micha{\l}owski}, M.~J., {Hayward}, C.~C., {Dunlop}, J.~S., {et~al.} 2014,
  \aap, 571, A75

\bibitem[{{Micha{\l}owski} {et~al.}(2008){Micha{\l}owski}, {Hjorth}, {Castro
  Cer{\'o}n}, \& {Watson}}]{michalowski08}
{Micha{\l}owski}, M.~J., {Hjorth}, J., {Castro Cer{\'o}n}, J.~M., \& {Watson},
  D. 2008, \apj, 672, 817

\bibitem[{{Moustakas} {et~al.}(2010){Moustakas}, {Kennicutt}, {Tremonti},
  {Dale}, {Smith}, \& {Calzetti}}]{moustakas10}
{Moustakas}, J., {Kennicutt}, Jr., R.~C., {Tremonti}, C.~A., {et~al.} 2010,
  \apjs, 190, 233

\bibitem[{{Mu{\~n}oz-Mateos} {et~al.}(2011){Mu{\~n}oz-Mateos}, {Boissier}, {Gil
  de Paz}, {Zamorano}, {Kennicutt}, {Moustakas}, {Prantzos}, \&
  {Gallego}}]{munoz09}
{Mu{\~n}oz-Mateos}, J.~C., {Boissier}, S., {Gil de Paz}, A., {et~al.} 2011,
  \apj, 731, 10

\bibitem[{{Murphy} {et~al.}(2011){Murphy}, {Condon}, {Schinnerer}, {Kennicutt},
  {Calzetti}, {Armus}, {Helou}, {Turner}, {Aniano}, {Beir{\~a}o}, {Bolatto},
  {Brandl}, {Croxall}, {Dale}, {Donovan Meyer}, {Draine}, {Engelbracht},
  {Hunt}, {Hao}, {Koda}, {Roussel}, {Skibba}, \& {Smith}}]{murphy11}
{Murphy}, E.~J., {Condon}, J.~J., {Schinnerer}, E., {et~al.} 2011, \apj, 737,
  67

\bibitem[{{Natta} \& {Panagia}(1984)}]{natta84}
{Natta}, A. \& {Panagia}, N. 1984, \apj, 287, 228

\bibitem[{{Nikutta}(2012)}]{Nikutta2012phd}
{Nikutta}, R. 2012, PhD thesis, University of Kentucky

\bibitem[{{Noeske} {et~al.}(2007){Noeske}, {Weiner}, {Faber}, {Papovich},
  {Koo}, {Somerville}, {Bundy}, {Conselice}, {Newman}, {Schiminovich}, {Le
  Floc'h}, {Coil}, {Rieke}, {Lotz}, {Primack}, {Barmby}, {Cooper}, {Davis},
  {Ellis}, {Fazio}, {Guhathakurta}, {Huang}, {Kassin}, {Martin}, {Phillips},
  {Rich}, {Small}, {Willmer}, \& {Wilson}}]{noeske07}
{Noeske}, K.~G., {Weiner}, B.~J., {Faber}, S.~M., {et~al.} 2007, \apjl, 660,
  L43

\bibitem[{{Noll} {et~al.}(2009){Noll}, {Burgarella}, {Giovannoli}, {Buat},
  {Marcillac}, \& {Mu{\~n}oz-Mateos}}]{noll09}
{Noll}, S., {Burgarella}, D., {Giovannoli}, E., {et~al.} 2009, \aap, 507, 1793

\bibitem[{{Norris} {et~al.}(2014){Norris}, {Meidt}, {Van de Ven}, {Schinnerer},
  {Groves}, \& {Querejeta}}]{norris14}
{Norris}, M.~A., {Meidt}, S., {Van de Ven}, G., {et~al.} 2014, \apj, 797, 55

\bibitem[{{Oh} {et~al.}(2008){Oh}, {de Blok}, {Walter}, {Brinks}, \&
  {Kennicutt}}]{oh08}
{Oh}, S.-H., {de Blok}, W.~J.~G., {Walter}, F., {Brinks}, E., \& {Kennicutt},
  Jr., R.~C. 2008, \aj, 136, 2761

\bibitem[{{Orellana} {et~al.}(2017){Orellana}, {Nagar}, {Elbaz},
  {Calder{\'o}n-Castillo}, {Leiton}, {Ibar}, {Magnelli}, {Daddi}, {Messias},
  {Cerulo}, \& {Slater}}]{orellana17}
{Orellana}, G., {Nagar}, N.~M., {Elbaz}, D., {et~al.} 2017, \aap, 602, A68

\bibitem[{{Osterbrock} \& {Ferland}(2006)}]{osterbrock06}
{Osterbrock}, D.~E. \& {Ferland}, G.~J. 2006, {Astrophysics of gaseous nebulae
  and active galactic nuclei}

\bibitem[{{Pannella} {et~al.}(2009){Pannella}, {Carilli}, {Daddi}, {McCracken},
  {Owen}, {Renzini}, {Strazzullo}, {Civano}, {Koekemoer}, {Schinnerer},
  {Scoville}, {Smol{\v c}i{\'c}}, {Taniguchi}, {Aussel}, {Kneib}, {Ilbert},
  {Mellier}, {Salvato}, {Thompson}, \& {Willott}}]{pannella09}
{Pannella}, M., {Carilli}, C.~L., {Daddi}, E., {et~al.} 2009, \apjl, 698, L116

\bibitem[{{Pannella} {et~al.}(2015){Pannella}, {Elbaz}, {Daddi}, {Dickinson},
  {Hwang}, {Schreiber}, {Strazzullo}, {Aussel}, {Bethermin}, {Buat},
  {Charmandaris}, {Cibinel}, {Juneau}, {Ivison}, {Le Borgne}, {Le Floc'h},
  {Leiton}, {Lin}, {Magdis}, {Morrison}, {Mullaney}, {Onodera}, {Renzini},
  {Salim}, {Sargent}, {Scott}, {Shu}, \& {Wang}}]{pannella15}
{Pannella}, M., {Elbaz}, D., {Daddi}, E., {et~al.} 2015, \apj, 807, 141

\bibitem[{{Pappalardo} {et~al.}(2016){Pappalardo}, {Bizzocchi}, {Fritz},
  {Boselli}, {Boquien}, {Boissier}, {Baes}, {Ciesla}, {Bianchi}, {Clemens},
  {Viaene}, {Bendo}, {De Looze}, {Smith}, \& {Davies}}]{pappalardo16}
{Pappalardo}, C., {Bizzocchi}, L., {Fritz}, J., {et~al.} 2016, \aap, 589, A11

\bibitem[{{Peeples} {et~al.}(2014){Peeples}, {Werk}, {Tumlinson},
  {Oppenheimer}, {Prochaska}, {Katz}, \& {Weinberg}}]{peeples14}
{Peeples}, M.~S., {Werk}, J.~K., {Tumlinson}, J., {et~al.} 2014, \apj, 786, 54

\bibitem[{{Pellegrini} {et~al.}(2013){Pellegrini}, {Smith}, {Wolfire},
  {Draine}, {Crocker}, {Croxall}, {van der Werf}, {Dale}, {Rigopoulou},
  {Wilson}, {Schinnerer}, {Groves}, {Kreckel}, {Sandstrom}, {Armus},
  {Calzetti}, {Murphy}, {Walter}, {Koda}, {Bayet}, {Beirao}, {Bolatto},
  {Bradford}, {Brinks}, {Hunt}, {Kennicutt}, {Knapen}, {Leroy}, {Rosolowsky},
  {Vigroux}, \& {Hopwood}}]{pellegrini13}
{Pellegrini}, E.~W., {Smith}, J.~D., {Wolfire}, M.~G., {et~al.} 2013, \apjl,
  779, L19

\bibitem[{{Pereira-Santaella} {et~al.}(2015){Pereira-Santaella},
  {Alonso-Herrero}, {Colina}, {Miralles-Caballero}, {P{\'e}rez-Gonz{\'a}lez},
  {Arribas}, {Bellocchi}, {Cazzoli}, {D{\'{\i}}az-Santos}, \& {Piqueras
  L{\'o}pez}}]{pereira15}
{Pereira-Santaella}, M., {Alonso-Herrero}, A., {Colina}, L., {et~al.} 2015,
  \aap, 577, A78

\bibitem[{{P{\'e}rez-Gonz{\'a}lez} {et~al.}(2006){P{\'e}rez-Gonz{\'a}lez},
  {Kennicutt}, {Gordon}, {Misselt}, {Gil de Paz}, {Engelbracht}, {Rieke},
  {Bendo}, {Bianchi}, {Boissier}, {Calzetti}, {Dale}, {Draine}, {Jarrett},
  {Hollenbach}, \& {Prescott}}]{perez06}
{P{\'e}rez-Gonz{\'a}lez}, P.~G., {Kennicutt}, Jr., R.~C., {Gordon}, K.~D.,
  {et~al.} 2006, \apj, 648, 987

\bibitem[{{Pettini} \& {Pagel}(2004)}]{pettini04}
{Pettini}, M. \& {Pagel}, B.~E.~J. 2004, \mnras, 348, L59

\bibitem[{{Planck Collaboration} {et~al.}(2016){Planck Collaboration}, {Ade},
  {Aghanim}, {Alves}, {Aniano}, {Arnaud}, {Ashdown}, {Aumont}, {Baccigalupi},
  {Banday}, {Barreiro}, {Bartolo}, {Battaner}, {Benabed}, {Benoit-L{\'e}vy},
  {Bernard}, {Bersanelli}, {Bielewicz}, {Bonaldi}, {Bonavera}, {Bond},
  {Borrill}, {Bouchet}, {Boulanger}, {Burigana}, {Butler}, {Calabrese},
  {Cardoso}, {Catalano}, {Chamballu}, {Chiang}, {Christensen}, {Clements},
  {Colombi}, {Colombo}, {Couchot}, {Crill}, {Curto}, {Cuttaia}, {Danese},
  {Davies}, {Davis}, {de Bernardis}, {de Rosa}, {de Zotti}, {Delabrouille},
  {Dickinson}, {Diego}, {Dole}, {Donzelli}, {Dor{\'e}}, {Douspis}, {Draine},
  {Ducout}, {Dupac}, {Efstathiou}, {Elsner}, {En{\ss}lin}, {Eriksen},
  {Falgarone}, {Finelli}, {Forni}, {Frailis}, {Fraisse}, {Franceschi},
  {Frejsel}, {Galeotta}, {Galli}, {Ganga}, {Ghosh}, {Giard}, {Gjerl{\o}w},
  {Gonz{\'a}lez-Nuevo}, {G{\'o}rski}, {Gregorio}, {Gruppuso}, {Guillet},
  {Hansen}, {Hanson}, {Harrison}, {Henrot-Versill{\'e}},
  {Hern{\'a}ndez-Monteagudo}, {Herranz}, {Hildebrandt}, {Hivon}, {Holmes},
  {Hovest}, {Huffenberger}, {Hurier}, {Jaffe}, {Jaffe}, {Jones},
  {Keih{\"a}nen}, {Keskitalo}, {Kisner}, {Kneissl}, {Knoche}, {Kunz},
  {Kurki-Suonio}, {Lagache}, {Lamarre}, {Lasenby}, {Lattanzi}, {Lawrence},
  {Leonardi}, {Levrier}, {Liguori}, {Lilje}, {Linden-V{\o}rnle},
  {L{\'o}pez-Caniego}, {Lubin}, {Mac{\'{\i}}as-P{\'e}rez}, {Maffei}, {Maino},
  {Mandolesi}, {Maris}, {Marshall}, {Martin}, {Mart{\'{\i}}nez-Gonz{\'a}lez},
  {Masi}, {Matarrese}, {Mazzotta}, {Melchiorri}, {Mendes}, {Mennella},
  {Migliaccio}, {Miville-Desch{\^e}nes}, {Moneti}, {Montier}, {Morgante},
  {Mortlock}, {Munshi}, {Murphy}, {Naselsky}, {Natoli}, {N{\o}rgaard-Nielsen},
  {Novikov}, {Novikov}, {Oxborrow}, {Pagano}, {Pajot}, {Paladini}, {Paoletti},
  {Pasian}, {Perdereau}, {Perotto}, {Perrotta}, {Pettorino}, {Piacentini},
  {Piat}, {Plaszczynski}, {Pointecouteau}, {Polenta}, {Ponthieu}, {Popa},
  {Pratt}, {Prunet}, {Puget}, {Rachen}, {Reach}, {Rebolo}, {Reinecke},
  {Remazeilles}, {Renault}, {Ristorcelli}, {Rocha}, {Roudier},
  {Rubi{\~n}o-Mart{\'{\i}}n}, {Rusholme}, {Sandri}, {Santos}, {Scott},
  {Spencer}, {Stolyarov}, {Sudiwala}, {Sunyaev}, {Sutton}, {Suur-Uski},
  {Sygnet}, {Tauber}, {Terenzi}, {Toffolatti}, {Tomasi}, {Tristram}, {Tucci},
  {Umana}, {Valenziano}, {Valiviita}, {Van Tent}, {Vielva}, {Villa}, {Wade},
  {Wandelt}, {Wehus}, {Ysard}, {Yvon}, {Zacchei}, \& {Zonca}}]{planck16}
{Planck Collaboration}, {Ade}, P.~A.~R., {Aghanim}, N., {et~al.} 2016, \aap,
  586, A132

\bibitem[{{Ponomareva} {et~al.}(2018){Ponomareva}, {Verheijen}, {Papastergis},
  {Bosma}, \& {Peletier}}]{ponomareva18}
{Ponomareva}, A.~A., {Verheijen}, M.~A.~W., {Papastergis}, E., {Bosma}, A., \&
  {Peletier}, R.~F. 2018, \mnras, 474, 4366

\bibitem[{{Pozzetti} \& {Mannucci}(2000)}]{pozzetti00}
{Pozzetti}, L. \& {Mannucci}, F. 2000, \mnras, 317, L17

\bibitem[{{Querejeta} {et~al.}(2015){Querejeta}, {Meidt}, {Schinnerer},
  {Cisternas}, {Mu{\~n}oz-Mateos}, {Sheth}, {Knapen}, {van de Ven}, {Norris},
  {Peletier}, {Laurikainen}, {Salo}, {Holwerda}, {Athanassoula}, {Bosma},
  {Groves}, {Ho}, {Gadotti}, {Zaritsky}, {Regan}, {Hinz}, {Gil de Paz},
  {Menendez-Delmestre}, {Seibert}, {Mizusawa}, {Kim}, {Erroz-Ferrer}, {Laine},
  \& {Comer{\'o}n}}]{querejeta15}
{Querejeta}, M., {Meidt}, S.~E., {Schinnerer}, E., {et~al.} 2015, \apjs, 219, 5

\bibitem[{{R{\'e}my-Ruyer} {et~al.}(2014){R{\'e}my-Ruyer}, {Madden},
  {Galliano}, {Galametz}, {Takeuchi}, {Asano}, {Zhukovska}, {Lebouteiller},
  {Cormier}, {Jones}, {Bocchio}, {Baes}, {Bendo}, {Boquien}, {Boselli},
  {DeLooze}, {Doublier-Pritchard}, {Hughes}, {Karczewski}, \&
  {Spinoglio}}]{remyruyer14}
{R{\'e}my-Ruyer}, A., {Madden}, S.~C., {Galliano}, F., {et~al.} 2014, \aap,
  563, A31

\bibitem[{{R{\'e}my-Ruyer} {et~al.}(2015){R{\'e}my-Ruyer}, {Madden},
  {Galliano}, {Lebouteiller}, {Baes}, {Bendo}, {Boselli}, {Ciesla}, {Cormier},
  {Cooray}, {Cortese}, {De Looze}, {Doublier-Pritchard}, {Galametz}, {Jones},
  {Karczewski}, {Lu}, \& {Spinoglio}}]{remyruyer15}
{R{\'e}my-Ruyer}, A., {Madden}, S.~C., {Galliano}, F., {et~al.} 2015, \aap,
  582, A121

\bibitem[{{Rich} {et~al.}(2005){Rich}, {Salim}, {Brinchmann}, {Charlot},
  {Seibert}, {Kauffmann}, {Lee}, {Yi}, {Barlow}, {Bianchi}, {Byun}, {Donas},
  {Forster}, {Friedman}, {Heckman}, {Jelinsky}, {Madore}, {Malina}, {Martin},
  {Milliard}, {Morrissey}, {Neff}, {Schiminovich}, {Siegmund}, {Small},
  {Szalay}, {Welsh}, \& {Wyder}}]{rich05}
{Rich}, R.~M., {Salim}, S., {Brinchmann}, J., {et~al.} 2005, \apjl, 619, L107

\bibitem[{{Roussel} {et~al.}(2006){Roussel}, {Helou}, {Smith}, {Draine},
  {Hollenbach}, {Moustakas}, {Spoon}, {Kennicutt}, {Rieke}, {Walter}, {Armus},
  {Dale}, {Sheth}, {Bendo}, {Engelbracht}, {Gordon}, {Meyer}, {Regan}, \&
  {Murphy}}]{roussel06}
{Roussel}, H., {Helou}, G., {Smith}, J.~D., {et~al.} 2006, \apj, 646, 841

\bibitem[{{Saintonge} {et~al.}(2011){Saintonge}, {Kauffmann}, {Kramer},
  {Tacconi}, {Buchbender}, {Catinella}, {Fabello}, {Graci{\'a}-Carpio}, {Wang},
  {Cortese}, {Fu}, {Genzel}, {Giovanelli}, {Guo}, {Haynes}, {Heckman},
  {Krumholz}, {Lemonias}, {Li}, {Moran}, {Rodriguez-Fernandez}, {Schiminovich},
  {Schuster}, \& {Sievers}}]{saintonge11}
{Saintonge}, A., {Kauffmann}, G., {Kramer}, C., {et~al.} 2011, \mnras, 415, 32

\bibitem[{{Salim}(2014)}]{salim14}
{Salim}, S. 2014, Serbian Astronomical Journal, 189, 1

\bibitem[{{Salim} {et~al.}(2018){Salim}, {Boquien}, \& {Lee}}]{salim18}
{Salim}, S., {Boquien}, M., \& {Lee}, J.~C. 2018, \apj, 859, 11

\bibitem[{{Salim} {et~al.}(2016){Salim}, {Lee}, {Janowiecki}, {da Cunha},
  {Dickinson}, {Boquien}, {Burgarella}, {Salzer}, \& {Charlot}}]{salim16}
{Salim}, S., {Lee}, J.~C., {Janowiecki}, S., {et~al.} 2016, \apjs, 227, 2

\bibitem[{{Salim} {et~al.}(2007){Salim}, {Rich}, {Charlot}, {Brinchmann},
  {Johnson}, {Schiminovich}, {Seibert}, {Mallery}, {Heckman}, {Forster},
  {Friedman}, {Martin}, {Morrissey}, {Neff}, {Small}, {Wyder}, {Bianchi},
  {Donas}, {Lee}, {Madore}, {Milliard}, {Szalay}, {Welsh}, \& {Yi}}]{salim07}
{Salim}, S., {Rich}, R.~M., {Charlot}, S., {et~al.} 2007, \apjs, 173, 267

\bibitem[{{Santini} {et~al.}(2014){Santini}, {Maiolino}, {Magnelli}, {Lutz},
  {Lamastra}, {Li Causi}, {Eales}, {Andreani}, {Berta}, {Buat}, {Cooray},
  {Cresci}, {Daddi}, {Farrah}, {Fontana}, {Franceschini}, {Genzel}, {Granato},
  {Grazian}, {Le Floc'h}, {Magdis}, {Magliocchetti}, {Mannucci}, {Menci},
  {Nordon}, {Oliver}, {Popesso}, {Pozzi}, {Riguccini}, {Rodighiero}, {Rosario},
  {Salvato}, {Scott}, {Silva}, {Tacconi}, {Viero}, {Wang}, {Wuyts}, \&
  {Xu}}]{santini14}
{Santini}, P., {Maiolino}, R., {Magnelli}, B., {et~al.} 2014, \aap, 562, A30

\bibitem[{{Sarzi} {et~al.}(2010){Sarzi}, {Shields}, {Schawinski}, {Jeong},
  {Shapiro}, {Bacon}, {Bureau}, {Cappellari}, {Davies}, {de Zeeuw}, {Emsellem},
  {Falc{\'o}n-Barroso}, {Krajnovi{\'c}}, {Kuntschner}, {McDermid}, {Peletier},
  {van den Bosch}, {van de Ven}, \& {Yi}}]{sarzi10}
{Sarzi}, M., {Shields}, J.~C., {Schawinski}, K., {et~al.} 2010, \mnras, 402,
  2187

\bibitem[{{Sauvage} \& {Thuan}(1992)}]{sauvage92}
{Sauvage}, M. \& {Thuan}, T.~X. 1992, \apjl, 396, L69

\bibitem[{{Schawinski} {et~al.}(2007){Schawinski}, {Kaviraj}, {Khochfar},
  {Yoon}, {Yi}, {Deharveng}, {Boselli}, {Barlow}, {Conrow}, {Forster},
  {Friedman}, {Martin}, {Morrissey}, {Neff}, {Schiminovich}, {Seibert},
  {Small}, {Wyder}, {Bianchi}, {Donas}, {Heckman}, {Lee}, {Madore}, {Milliard},
  {Rich}, \& {Szalay}}]{schawinski07}
{Schawinski}, K., {Kaviraj}, S., {Khochfar}, S., {et~al.} 2007, \apjs, 173, 512

\bibitem[{{Schawinski} {et~al.}(2014){Schawinski}, {Urry}, {Simmons},
  {Fortson}, {Kaviraj}, {Keel}, {Lintott}, {Masters}, {Nichol}, {Sarzi},
  {Skibba}, {Treister}, {Willett}, {Wong}, \& {Yi}}]{schawinski14}
{Schawinski}, K., {Urry}, C.~M., {Simmons}, B.~D., {et~al.} 2014, \mnras, 440,
  889

\bibitem[{{Schiminovich} {et~al.}(2007){Schiminovich}, {Wyder}, {Martin},
  {Johnson}, {Salim}, {Seibert}, {Treyer}, {Budav{\'a}ri}, {Hoopes},
  {Zamojski}, {Barlow}, {Forster}, {Friedman}, {Morrissey}, {Neff}, {Small},
  {Bianchi}, {Donas}, {Heckman}, {Lee}, {Madore}, {Milliard}, {Rich}, {Szalay},
  {Welsh}, \& {Yi}}]{schiminovich07}
{Schiminovich}, D., {Wyder}, T.~K., {Martin}, D.~C., {et~al.} 2007, \apjs, 173,
  315

\bibitem[{{Schlafly} \& {Finkbeiner}(2011)}]{schlafly11}
{Schlafly}, E.~F. \& {Finkbeiner}, D.~P. 2011, \apj, 737, 103

\bibitem[{{Schmidt}(1959)}]{schmidt59}
{Schmidt}, M. 1959, \apj, 129, 243

\bibitem[{{Silva}(1999)}]{silva99}
{Silva}, L. 1999, PhD thesis, Sissa

\bibitem[{{Silva} {et~al.}(1998){Silva}, {Granato}, {Bressan}, \&
  {Danese}}]{silva98}
{Silva}, L., {Granato}, G.~L., {Bressan}, A., \& {Danese}, L. 1998, \apj, 509,
  103

\bibitem[{{Skibba} {et~al.}(2011){Skibba}, {Engelbracht}, {Dale}, {Hinz},
  {Zibetti}, {Crocker}, {Groves}, {Hunt}, {Johnson}, {Meidt}, {Murphy},
  {Appleton}, {Armus}, {Bolatto}, {Brandl}, {Calzetti}, {Croxall}, {Galametz},
  {Gordon}, {Kennicutt}, {Koda}, {Krause}, {Montiel}, {Rix}, {Roussel},
  {Sandstrom}, {Sauvage}, {Schinnerer}, {Smith}, {Walter}, {Wilson}, \&
  {Wolfire}}]{skibba11}
{Skibba}, R.~A., {Engelbracht}, C.~W., {Dale}, D., {et~al.} 2011, \apj, 738, 89

\bibitem[{{Smith} \& {Hancock}(2009)}]{smith09}
{Smith}, B.~J. \& {Hancock}, M. 2009, \aj, 138, 130

\bibitem[{{Smith} {et~al.}(2012){Smith}, {Dunne}, {da Cunha}, {Rowlands},
  {Maddox}, {Gomez}, {Bonfield}, {Charlot}, {Driver}, {Popescu}, {Tuffs},
  {Dunlop}, {Jarvis}, {Seymour}, {Symeonidis}, {Baes}, {Bourne}, {Clements},
  {Cooray}, {De Zotti}, {Dye}, {Eales}, {Scott}, {Verma}, {van der Werf},
  {Andrae}, {Auld}, {Buttiglione}, {Cava}, {Dariush}, {Fritz}, {Hopwood},
  {Ibar}, {Ivison}, {Kelvin}, {Madore}, {Pohlen}, {Rigby}, {Robotham},
  {Seibert}, \& {Temi}}]{smith12}
{Smith}, D.~J.~B., {Dunne}, L., {da Cunha}, E., {et~al.} 2012, \mnras, 427, 703

\bibitem[{{Smith} {et~al.}(2007){Smith}, {Draine}, {Dale}, {Moustakas},
  {Kennicutt}, {Helou}, {Armus}, {Roussel}, {Sheth}, {Bendo}, {Buckalew},
  {Calzetti}, {Engelbracht}, {Gordon}, {Hollenbach}, {Li}, {Malhotra},
  {Murphy}, \& {Walter}}]{smith07}
{Smith}, J.~D.~T., {Draine}, B.~T., {Dale}, D.~A., {et~al.} 2007, \apj, 656,
  770

\bibitem[{{Speagle} {et~al.}(2014){Speagle}, {Steinhardt}, {Capak}, \&
  {Silverman}}]{speagle14}
{Speagle}, J.~S., {Steinhardt}, C.~L., {Capak}, P.~L., \& {Silverman}, J.~D.
  2014, \apjs, 214, 15

\bibitem[{{Takagi} {et~al.}(2003){Takagi}, {Vansevicius}, \&
  {Arimoto}}]{takagi03}
{Takagi}, T., {Vansevicius}, V., \& {Arimoto}, N. 2003, \pasj, 55, 385

\bibitem[{{Temi} {et~al.}({2009}{\natexlab{a}}){Temi}, {Brighenti}, \&
  {Mathews}}]{temi09a}
{Temi}, P., {Brighenti}, F., \& {Mathews}, W.~G. {2009}{\natexlab{a}}, \apj,
  695, 1

\bibitem[{{Temi} {et~al.}({2009}{\natexlab{b}}){Temi}, {Brighenti}, \&
  {Mathews}}]{temi09b}
{Temi}, P., {Brighenti}, F., \& {Mathews}, W.~G. {2009}{\natexlab{b}}, \apj,
  707, 890

\bibitem[{{Trotta}(2008)}]{Trotta2008}
{Trotta}, R. 2008, Contemporary Physics, 49, 71

\bibitem[{{Verley} {et~al.}(2009){Verley}, {Corbelli}, {Giovanardi}, \&
  {Hunt}}]{verley09}
{Verley}, S., {Corbelli}, E., {Giovanardi}, C., \& {Hunt}, L.~K. 2009, \aap,
  493, 453

\bibitem[{{Viaene} {et~al.}(2016){Viaene}, {Baes}, {Bendo}, {Boquien},
  {Boselli}, {Ciesla}, {Cortese}, {De Looze}, {Eales}, {Fritz}, {Karczewski},
  {Madden}, {Smith}, \& {Spinoglio}}]{viaene16}
{Viaene}, S., {Baes}, M., {Bendo}, G., {et~al.} 2016, \aap, 586, A13

\bibitem[{{Viaene} {et~al.}(2017){Viaene}, {Baes}, {Tamm}, {Tempel}, {Bendo},
  {Blommaert}, {Boquien}, {Boselli}, {Camps}, {Cooray}, {De Looze}, {De Vis},
  {Fern{\'a}ndez-Ontiveros}, {Fritz}, {Galametz}, {Gentile}, {Madden}, {Smith},
  {Spinoglio}, \& {Verstocken}}]{viaene17}
{Viaene}, S., {Baes}, M., {Tamm}, A., {et~al.} 2017, \aap, 599, A64

\bibitem[{{Viaene} {et~al.}(2014){Viaene}, {Fritz}, {Baes}, {Bendo},
  {Blommaert}, {Boquien}, {Boselli}, {Ciesla}, {Cortese}, {De Looze}, {Gear},
  {Gentile}, {Hughes}, {Jarrett}, {Karczewski}, {Smith}, {Spinoglio}, {Tamm},
  {Tempel}, {Thilker}, \& {Verstappen}}]{viaene14}
{Viaene}, S., {Fritz}, J., {Baes}, M., {et~al.} 2014, \aap, 567, A71

\bibitem[{{Walterbos} \& {Schwering}(1987)}]{walterbos87}
{Walterbos}, R.~A.~M. \& {Schwering}, P.~B.~W. 1987, \aap, 180, 27

\bibitem[{{Wen} {et~al.}(2013){Wen}, {Wu}, {Zhu}, {Lam}, {Wu}, {Wicker}, \&
  {Zhao}}]{wen13}
{Wen}, X.-Q., {Wu}, H., {Zhu}, Y.-N., {et~al.} 2013, \mnras, 433, 2946

\bibitem[{{Whitaker} {et~al.}(2014){Whitaker}, {Franx}, {Leja}, {van Dokkum},
  {Henry}, {Skelton}, {Fumagalli}, {Momcheva}, {Brammer}, {Labb{\'e}},
  {Nelson}, \& {Rigby}}]{whitaker14}
{Whitaker}, K.~E., {Franx}, M., {Leja}, J., {et~al.} 2014, \apj, 795, 104

\bibitem[{{Witt} \& {Gordon}(1996)}]{witt96}
{Witt}, A.~N. \& {Gordon}, K.~D. 1996, \apj, 463, 681

\bibitem[{{Witt} \& {Gordon}(2000)}]{witt00}
{Witt}, A.~N. \& {Gordon}, K.~D. 2000, \apj, 528, 799

\bibitem[{{Wu} {et~al.}(2010){Wu}, {Helou}, {Armus}, {Cormier}, {Shi}, {Dale},
  {Dasyra}, {Smith}, {Papovich}, {Draine}, {Rahman}, {Stierwalt}, {Fadda},
  {Lagache}, \& {Wright}}]{wu10}
{Wu}, Y., {Helou}, G., {Armus}, L., {et~al.} 2010, \apj, 723, 895

\bibitem[{{Wyder} {et~al.}(2007){Wyder}, {Martin}, {Schiminovich}, {Seibert},
  {Budav{\'a}ri}, {Treyer}, {Barlow}, {Forster}, {Friedman}, {Morrissey},
  {Neff}, {Small}, {Bianchi}, {Donas}, {Heckman}, {Lee}, {Madore}, {Milliard},
  {Rich}, {Szalay}, {Welsh}, \& {Yi}}]{wyder07}
{Wyder}, T.~K., {Martin}, D.~C., {Schiminovich}, D., {et~al.} 2007, \apjs, 173,
  293

\bibitem[{{Zibetti} {et~al.}(2009){Zibetti}, {Charlot}, \& {Rix}}]{zibetti09}
{Zibetti}, S., {Charlot}, S., \& {Rix}, H.-W. 2009, \mnras, 400, 1181

\end{thebibliography}
% the above works on mac, but not on linux CentOS
% this has percent preceded by \
% 12/7/2016 edited mnras.bst to eliminate adsurl
% (see mnras_orig.bst, and line 906
%\bibliography{kingfish_seds_percent}

\clearpage

\appendix
\section{Best-fit SED results}
\label{app:seds}

The physical quantities from the best-fit SED models are reported in 
Tables \ref{tab:app:cigale}, \ref{tab:app:grasil}, and \ref{tab:app:magphys}
for \cig, \gra, and \magp, respectively.
The best-fit SEDs where each
model is plotted together with the multiwavelength photometry
are shown in Fig. \ref{fig:n5457_sed} for NGC\,5457 (M\,101) in the
main text, and here in Fig. \ref{fig:all_seds} for the remaining galaxies.

\begin{table*} 
\caption{\cig\ quantities for KINGFISH sample }
\resizebox{0.89\textwidth}{!}{
\begin{tabular}{lrrrccrcc}
\hline 
\multicolumn{1}{c}{Galaxy} &
\multicolumn{1}{c}{\rms} &
\multicolumn{1}{c}{Log(\mstar)} &
\multicolumn{1}{c}{Log(SFR)} &
\multicolumn{1}{c}{Log(\mdust)} &
\multicolumn{1}{c}{Log(\ltir)} &
\multicolumn{1}{c}{Log(\lfuv)} &
\multicolumn{1}{c}{$A_V$} &
\multicolumn{1}{c}{\afuv} \\
& \multicolumn{1}{c}{(dex)} &
\multicolumn{1}{c}{(\msun)} &
\multicolumn{1}{c}{(\msunyr)} &
\multicolumn{1}{c}{(\msun)} &
\multicolumn{1}{c}{(\lsun)} &
\multicolumn{1}{c}{(\lsun)} &
\multicolumn{1}{c}{(mag)} &
\multicolumn{1}{c}{(mag)} \\
\hline 
%\input ../../statistics/kingfish_SEDs/cigale_params_wise.tex
% 8/11/2018
DDO\,053     &  0.224 & 6.784 & -2.430 & 3.763  & 7.012 & 7.413 & 0.072 & 0.362 \\
DDO\,154     &  0.086 & 7.019 & -2.092 & 5.056  & 6.980 & 7.753 & 0.030 & 0.139 \\
DDO\,165     &  0.065 & 7.705 & -1.780 & 5.594  & 7.395 & 8.058 & 0.031 & 0.171 \\
Ho\,I        &  0.072 & 7.435 & -2.056 & 4.801  & 6.937 & 7.730 & 0.037 & 0.157 \\
Ho\,II       &  0.161 & 8.245 & -1.375 & 4.686  & 7.862 & 8.332 & 0.054 & 0.293 \\
IC\,0342     &  0.161 & 10.104 & 0.437 & 7.589  & 10.197 & 10.260 & 0.189 & 1.009 \\
IC\,2574     &  0.259 & 8.457 & -1.136 & 6.016  & 8.230 & 8.613 & 0.072 & 0.368 \\
M81\,Dw\,B   &  0.052 & 7.130 & -2.517 & 4.379  & 6.562 & 6.996 & 0.076 & 0.306 \\
NGC\,0337    &  0.035 & 9.721 & 0.191 & 7.403  & 10.026 & 10.034 & 0.200 & 1.369 \\
NGC\,0584    &  0.107 & 10.802 & -2.851 & 6.716  & 8.528 & 8.033 & 0.017 & 0.064 \\
NGC\,0628    &  0.022 & 9.821 & 0.031 & 7.605  & 9.843 & 9.861 & 0.183 & 1.175 \\
NGC\,0855    &  0.056 & 8.867 & -1.669 & 5.912  & 8.523 & 8.427 & 0.203 & 1.109 \\
NGC\,0925    &  0.095 & 9.598 & -0.028 & 7.386  & 9.599 & 9.735 & 0.133 & 0.790 \\
NGC\,1097    &  0.008 & 10.722 & 0.694 & 7.968  & 10.620 & 10.427 & 0.502 & 2.026 \\
NGC\,1266    &  0.066 & 9.984 & 0.163 & 7.074  & 10.330 & 10.039 & 1.199 & 6.189 \\
NGC\,1291    &  0.045 & 10.866 & -1.094 & 7.446  & 9.370 & 8.876 & 0.080 & 0.476 \\
NGC\,1316    &  0.080 & 11.581 & -1.343 & 7.147  & 9.796 & 8.991 & 0.047 & 0.195 \\
NGC\,1377    &  0.172 & 9.770 & 0.104 & 6.233  & 10.122 & 9.981 & 0.593 & 2.669 \\
NGC\,1404    &  0.034 & 11.059 & -2.354 & 6.959  & 8.878 & 8.435 & 0.025 & 0.101 \\
NGC\,1482    &  0.041 & 10.048 & 0.627 & 7.405  & 10.683 & 10.487 & 1.159 & 5.903 \\
NGC\,1512    &  0.049 & 10.119 & -0.217 & 7.346  & 9.541 & 9.482 & 0.210 & 0.950 \\
NGC\,2146    &  0.068 & 10.516 & 0.920 & 7.812  & 11.071 & 10.879 & 1.151 & 5.227 \\
NGC\,2798    &  0.128 & 10.121 & 0.265 & 7.216  & 10.536 & 10.349 & 0.894 & 4.278 \\
NGC\,2841    &  0.030 & 10.862 & -0.180 & 7.836  & 10.018 & 9.677 & 0.218 & 1.237 \\
NGC\,2915    &  0.154 & 8.102 & -1.516 & 4.931  & 7.558 & 8.004 & 0.078 & 0.291 \\
NGC\,2976    &  0.023 & 8.974 & -0.821 & 6.291  & 8.904 & 8.848 & 0.258 & 1.487 \\
NGC\,3049    &  0.054 & 9.346 & -0.290 & 6.805  & 9.528 & 9.490 & 0.263 & 1.573 \\
NGC\,3077    &  0.042 & 9.201 & -1.093 & 5.994  & 8.880 & 8.728 & 0.293 & 1.389 \\
NGC\,3184    &  0.048 & 10.090 & 0.303 & 7.739  & 9.951 & 9.963 & 0.221 & 1.090 \\
NGC\,3190    &  0.047 & 10.517 & -1.143 & 7.251  & 9.830 & 8.758 & 0.409 & 1.951 \\
NGC\,3198    &  0.051 & 9.989 & 0.221 & 7.635  & 9.892 & 9.881 & 0.248 & 1.153 \\
NGC\,3265    &  0.052 & 9.245 & -0.606 & 6.353  & 9.370 & 9.197 & 0.493 & 2.677 \\
NGC\,3351    &  0.035 & 10.224 & -0.067 & 7.209  & 9.847 & 9.643 & 0.347 & 1.703 \\
NGC\,3521    &  0.013 & 10.679 & 0.279 & 7.927  & 10.523 & 10.191 & 0.549 & 2.380 \\
NGC\,3621    &  0.044 & 9.832 & 0.052 & 7.281  & 9.876 & 9.850 & 0.254 & 1.372 \\
NGC\,3627    &  0.014 & 10.485 & 0.447 & 7.737  & 10.388 & 10.175 & 0.454 & 2.288 \\
NGC\,3773    &  0.068 & 8.717 & -0.933 & 5.968  & 8.723 & 8.886 & 0.115 & 0.727 \\
NGC\,3938    &  0.025 & 10.194 & 0.429 & 7.653  & 10.235 & 10.229 & 0.217 & 1.264 \\
NGC\,4236    &  0.161 & 8.966 & -0.670 & 6.461  & 8.594 & 9.036 & 0.055 & 0.314 \\
NGC\,4254    &  0.011 & 10.312 & 0.645 & 7.956  & 10.565 & 10.475 & 0.354 & 2.031 \\
NGC\,4321    &  0.011 & 10.516 & 0.763 & 7.890  & 10.493 & 10.364 & 0.386 & 1.858 \\
NGC\,4536    &  0.033 & 10.167 & 0.339 & 7.555  & 10.291 & 10.149 & 0.471 & 2.147 \\
NGC\,4559    &  0.069 & 9.484 & -0.253 & 7.168  & 9.396 & 9.515 & 0.142 & 0.804 \\
NGC\,4569    &  0.016 & 10.201 & -1.382 & 7.100  & 9.711 & 9.297 & 0.338 & 1.646 \\
NGC\,4579    &  0.061 & 10.829 & -0.525 & 7.728  & 9.969 & 9.484 & 0.242 & 1.228 \\
NGC\,4594    &  0.002 & 11.082 & -1.676 & 7.308  & 9.494 & 8.580 & 0.083 & 0.378 \\
NGC\,4625    &  0.046 & 8.866 & -0.885 & 6.490  & 8.703 & 8.814 & 0.151 & 0.814 \\
NGC\,4631    &  0.081 & 9.995 & 0.365 & 7.670  & 10.316 & 10.235 & 0.444 & 1.758 \\
NGC\,4725    &  0.059 & 10.634 & -0.199 & 7.814  & 9.829 & 9.604 & 0.194 & 0.999 \\
NGC\,4736    &  0.000 & 10.215 & -0.556 & 6.852  & 9.746 & 9.452 & 0.310 & 1.415 \\
NGC\,4826    &  0.023 & 10.245 & -1.101 & 6.688  & 9.588 & 9.077 & 0.320 & 1.682 \\
NGC\,5055    &  0.019 & 10.488 & 0.161 & 8.025  & 10.238 & 9.995 & 0.414 & 1.983 \\
NGC\,5398    &  0.167 & 8.490 & -1.107 & 5.820  & 8.542 & 8.734 & 0.111 & 0.669 \\
NGC\,5408    &  0.216 & 8.081 & -1.271 & 4.844  & 8.164 & 8.570 & 0.067 & 0.368 \\
NGC\,5457    &  0.072 & 10.274 & 0.536 & 8.023  & 10.280 & 10.336 & 0.183 & 1.000 \\
NGC\,5474    &  0.026 & 8.944 & -0.802 & 6.443  & 8.630 & 9.025 & 0.050 & 0.357 \\
NGC\,5713    &  0.050 & 10.117 & 0.505 & 7.542  & 10.491 & 10.348 & 0.527 & 2.872 \\
NGC\,5866    &  0.038 & 10.612 & -1.234 & 7.028  & 9.646 & 8.682 & 0.219 & 1.593 \\
NGC\,6946    &  0.083 & 10.328 & 0.663 & 7.642  & 10.540 & 10.480 & 0.297 & 1.762 \\
NGC\,7331    &  0.011 & 10.841 & 0.619 & 8.096  & 10.678 & 10.366 & 0.606 & 2.705 \\
NGC\,7793    &  0.080 & 9.288 & -0.452 & 7.010  & 9.222 & 9.367 & 0.138 & 0.740 \\

\hline 
\label{tab:app:cigale} 
\end{tabular} 
}
\end{table*}

\begin{table*} 
\caption{\gra\ quantities for KINGFISH sample } %("DISK" models are the "NSD" templates) }
\resizebox{0.89\textwidth}{!}{
\begin{tabular}{lcrrrccrcc}
\hline 
\multicolumn{1}{c}{Galaxy} &
\multicolumn{1}{c}{Geometry} &
\multicolumn{1}{c}{\rms} &
\multicolumn{1}{c}{Log(\mstar)} &
\multicolumn{1}{c}{Log(SFR)} &
\multicolumn{1}{c}{Log(\mdust)} &
\multicolumn{1}{c}{Log(\ltir)} &
\multicolumn{1}{c}{Log(\lfuv)} &
\multicolumn{1}{c}{$A_V$} &
\multicolumn{1}{c}{\afuv} \\
&& \multicolumn{1}{c}{(dex)} &
\multicolumn{1}{c}{(\msun)} &
\multicolumn{1}{c}{(\msunyr)} &
\multicolumn{1}{c}{(\msun)} &
\multicolumn{1}{c}{(\lsun)} &
\multicolumn{1}{c}{(\lsun)} &
\multicolumn{1}{c}{(mag)} &
\multicolumn{1}{c}{(mag)} \\
\hline 
DDO\,053     &  NSD  & 0.111 & 6.500 & -2.318 & 4.174  & 6.951 & 7.439 & 0.037 & 0.260 \\
DDO\,154     &  NSS  & 0.054 & 6.880 & -1.992 & 5.240  & 7.431 & 7.854 & 0.070 & 0.405 \\
DDO\,165     &  NSS  & 0.068 & 7.403 & -1.499 & 5.712  & 8.331 & 8.321 & 0.188 & 1.134 \\
Ho\,I        &  NSS  & 0.145 & 7.163 & -2.180 & 5.078  & 7.043 & 7.556 & 0.022 & 0.253 \\
Ho\,II       &  NSS  & 0.090 & 7.824 & -1.392 & 5.044  & 7.847 & 8.397 & 0.019 & 0.253 \\
IC\,0342     &  NSD  & 0.108 & 9.826 & 0.506 & 7.926  & 10.186 & 10.236 & 0.161 & 0.704 \\
IC\,2574     &  NSS  & 0.105 & 8.236 & -1.077 & 6.353  & 8.308 & 8.684 & 0.056 & 0.503 \\
M81\,Dw\,B   &  NSS  & 0.097 & 6.855 & -2.620 & 4.666  & 6.601 & 7.070 & 0.023 & 0.316 \\
NGC\,0337    &  NSD  & 0.058 & 9.421 & 0.266 & 7.392  & 10.048 & 10.005 & 0.179 & 1.245 \\
NGC\,0584    &  NSS  & 0.375 & 10.594 & -0.514 & 7.006  & 9.049 & 9.274 & 0.011 & 0.681 \\
NGC\,0628    &  NSD  & 0.064 & 9.584 & 0.122 & 7.702  & 9.819 & 9.835 & 0.113 & 0.801 \\
NGC\,0855    &  NSD  & 0.083 & 8.695 & -1.129 & 6.036  & 8.551 & 8.582 & 0.056 & 1.039 \\
NGC\,0925    &  NSS  & 0.079 & 9.397 & 0.050 & 7.768  & 9.576 & 9.839 & 0.106 & 0.610 \\
NGC\,1097    &  NSD  & 0.048 & 10.573 & 0.724 & 8.268  & 10.591 & 10.433 & 0.260 & 1.846 \\
NGC\,1266    &  NSS  & 0.118 & 10.104 & 0.146 & 7.087  & 10.495 & 9.882 & 1.451 & 5.503 \\
NGC\,1291    &  NSS  & 0.155 & 10.790 & -0.102 & 7.852  & 9.427 & 9.600 & 0.019 & 0.804 \\
NGC\,1316    &  NSS  & 0.224 & 11.410 & 0.279 & 7.732  & 10.080 & 10.068 & 0.021 & 1.316 \\
NGC\,1377    &  NSS  & 0.081 & 9.557 & 0.199 & 6.826  & 10.102 & 9.950 & 0.455 & 3.922 \\
NGC\,1404    &  NSD  & 0.052 & 10.696 & 0.206 & 7.259  & 10.149 & 9.974 & 0.160 & 3.690 \\
NGC\,1482    &  NSS  & 0.086 & 10.081 & 0.522 & 7.772  & 10.650 & 10.219 & 1.226 & 5.002 \\
NGC\,1512    &  NSS  & 0.062 & 9.921 & -0.082 & 7.815  & 9.505 & 9.599 & 0.125 & 0.895 \\
NGC\,2146    &  NSD  & 0.092 & 10.361 & 1.010 & 7.871  & 11.081 & 10.743 & 1.121 & 4.677 \\
NGC\,2798    &  NSD  & 0.069 & 9.966 & 0.513 & 7.320  & 10.573 & 10.272 & 0.948 & 4.251 \\
NGC\,2841    &  NSS  & 0.103 & 10.958 & 0.196 & 8.251  & 9.956 & 9.929 & 0.057 & 1.356 \\
NGC\,2915    &  NSS  & 0.067 & 7.694 & -1.599 & 5.195  & 7.590 & 8.141 & 0.018 & 0.256 \\
NGC\,2976    &  NSD  & 0.073 & 8.746 & -0.901 & 6.657  & 8.845 & 8.812 & 0.108 & 1.181 \\
NGC\,3049    &  NSS  & 0.056 & 9.205 & -0.286 & 7.537  & 9.534 & 9.487 & 0.172 & 1.549 \\
NGC\,3077    &  NSS  & 0.070 & 9.001 & -0.884 & 6.511  & 8.849 & 8.799 & 0.093 & 1.725 \\
NGC\,3184    &  NSD  & 0.066 & 9.921 & 0.273 & 7.963  & 9.920 & 9.936 & 0.117 & 0.767 \\
NGC\,3190    &  NSD  & 0.104 & 10.612 & -0.065 & 7.773  & 9.821 & 9.600 & 0.194 & 3.347 \\
NGC\,3198    &  NSD  & 0.052 & 9.798 & 0.202 & 7.897  & 9.860 & 9.874 & 0.118 & 0.868 \\
NGC\,3265    &  NSS  & 0.063 & 9.169 & -0.501 & 6.875  & 9.394 & 9.259 & 0.282 & 2.781 \\
NGC\,3351    &  NSS  & 0.055 & 10.043 & -0.078 & 7.463  & 9.810 & 9.673 & 0.208 & 1.693 \\
NGC\,3521    &  NSS  & 0.104 & 10.596 & 0.583 & 8.396  & 10.476 & 10.269 & 0.315 & 2.211 \\
NGC\,3621    &  NSD  & 0.064 & 9.509 & 0.065 & 7.598  & 9.816 & 9.801 & 0.193 & 0.880 \\
NGC\,3627    &  NSS  & 0.080 & 10.332 & 0.474 & 7.935  & 10.366 & 10.178 & 0.343 & 2.022 \\
NGC\,3773    &  NSS  & 0.059 & 8.353 & -0.848 & 6.344  & 8.748 & 8.849 & 0.118 & 0.993 \\
NGC\,3938    &  NSD  & 0.065 & 10.001 & 0.443 & 7.970  & 10.178 & 10.167 & 0.127 & 0.848 \\
NGC\,4236    &  NSS  & 0.088 & 8.666 & -0.615 & 6.699  & 8.620 & 9.139 & 0.023 & 0.262 \\
NGC\,4254    &  NSD  & 0.089 & 10.084 & 0.631 & 8.131  & 10.516 & 10.373 & 0.315 & 1.345 \\
NGC\,4321    &  NSS  & 0.098 & 10.407 & 0.589 & 8.189  & 10.415 & 10.318 & 0.324 & 1.215 \\
NGC\,4536    &  NSD  & 0.067 & 10.059 & 0.424 & 7.691  & 10.288 & 10.132 & 0.311 & 1.907 \\
NGC\,4559    &  NSD  & 0.049 & 9.196 & -0.220 & 7.328  & 9.397 & 9.509 & 0.081 & 0.655 \\
NGC\,4569    &  NSS  & 0.102 & 10.172 & -0.159 & 7.382  & 9.681 & 9.606 & 0.068 & 2.034 \\
NGC\,4579    &  NSS  & 0.094 & 10.825 & 0.197 & 8.007  & 9.988 & 9.903 & 0.055 & 1.958 \\
NGC\,4594    &  NSS  & 0.240 & 11.083 & -0.048 & 7.349  & 9.646 & 9.726 & 0.029 & 1.742 \\
NGC\,4625    &  NSS  & 0.065 & 8.658 & -0.893 & 6.705  & 8.663 & 8.836 & 0.146 & 0.611 \\
NGC\,4631    &  NSS  & 0.059 & 9.749 & 0.460 & 7.933  & 10.296 & 10.200 & 0.506 & 1.321 \\
NGC\,4725    &  NSS  & 0.099 & 10.719 & 0.099 & 8.149  & 9.774 & 9.829 & 0.071 & 0.974 \\
NGC\,4736    &  NSS  & 0.062 & 10.086 & -0.102 & 7.190  & 9.726 & 9.627 & 0.150 & 1.519 \\
NGC\,4826    &  NSS  & 0.074 & 10.144 & -0.308 & 7.004  & 9.589 & 9.450 & 0.108 & 2.301 \\
NGC\,5055    &  NSS  & 0.093 & 10.400 & 0.384 & 8.276  & 10.203 & 10.055 & 0.290 & 1.634 \\
NGC\,5398    &  NSS  & 0.111 & 8.108 & -0.957 & 6.196  & 8.536 & 8.773 & 0.084 & 0.751 \\
NGC\,5408    &  NSS  & 0.212 & 7.983 & -1.540 & 5.572  & 8.313 & 8.208 & 0.234 & 2.135 \\
NGC\,5457    &  NSD  & 0.048 & 9.991 & 0.563 & 8.118  & 10.265 & 10.288 & 0.115 & 0.748 \\
NGC\,5474    &  NSS  & 0.075 & 8.686 & -0.684 & 6.776  & 8.625 & 9.069 & 0.035 & 0.335 \\
NGC\,5713    &  NSS  & 0.072 & 10.009 & 0.507 & 7.824  & 10.464 & 10.196 & 0.665 & 2.312 \\
NGC\,5866    &  NSS  & 0.112 & 10.744 & -0.228 & 7.172  & 9.724 & 9.538 & 0.057 & 2.949 \\
NGC\,6946    &  NSD  & 0.101 & 10.054 & 0.676 & 7.886  & 10.497 & 10.433 & 0.247 & 1.287 \\
NGC\,7331    &  NSD  & 0.090 & 10.829 & 0.707 & 8.549  & 10.638 & 10.374 & 0.354 & 2.423 \\
NGC\,7793    &  NSD  & 0.058 & 9.018 & -0.368 & 7.189  & 9.201 & 9.345 & 0.109 & 0.578 \\

\hline 
\label{tab:app:grasil} 
\end{tabular} 
}
\end{table*}

\begin{table*} 
\caption{\magp\ quantities for KINGFISH sample }
\resizebox{0.89\textwidth}{!}{
\begin{tabular}{lrrrccrcc}
\hline 
\multicolumn{1}{c}{Galaxy} &
\multicolumn{1}{c}{\rms} &
\multicolumn{1}{c}{Log(\mstar)} &
\multicolumn{1}{c}{Log(SFR)} &
\multicolumn{1}{c}{Log(\mdust)} &
\multicolumn{1}{c}{Log(\ltir)} &
\multicolumn{1}{c}{Log(\lfuv)} &
\multicolumn{1}{c}{$A_V$} &
\multicolumn{1}{c}{\afuv} \\
& \multicolumn{1}{c}{(dex)} &
\multicolumn{1}{c}{(\msun)} &
\multicolumn{1}{c}{(\msunyr)} &
\multicolumn{1}{c}{(\msun)} &
\multicolumn{1}{c}{(\lsun)} &
\multicolumn{1}{c}{(\lsun)} &
\multicolumn{1}{c}{(mag)} &
\multicolumn{1}{c}{(mag)} \\
\hline 
DDO\,053     &  0.096 & 6.488 & -2.352 & 3.638  & 7.039 & 7.467 & 0.060 & 0.304 \\
DDO\,154     &  0.059 & 6.870 & -2.140 & 4.105  & 7.369 & 7.738 & 0.018 & 0.334 \\
DDO\,165     &  0.063 & 7.405 & -3.095 & 4.305  & 8.289 & 8.264 & 0.190 & 1.095 \\
Ho\,I        &  0.086 & 7.368 & -2.232 & 4.568  & 7.045 & 7.710 & 0.039 & 0.143 \\
Ho\,II       &  0.072 & 8.254 & -1.646 & 4.254  & 7.870 & 8.370 & 0.062 & 0.189 \\
IC\,0342     &  0.115 & 9.817 & -0.483 & 7.217  & 10.147 & 10.136 & 0.350 & 0.901 \\
IC\,2574     &  0.073 & 8.443 & -1.257 & 5.743  & 8.358 & 8.680 & 0.073 & 0.341 \\
M81\,Dw\,B   &  0.071 & 7.098 & -3.102 & 3.998  & 6.604 & 6.883 & 0.054 & 0.359 \\
NGC\,0337    &  0.061 & 9.457 & -0.043 & 6.857  & 10.076 & 9.975 & 0.342 & 1.449 \\
NGC\,0584    &  0.099 & 10.722 & -2.178 & 7.622  & 8.900 & 8.307 & 0.024 & 0.365 \\
NGC\,0628    &  0.041 & 10.100 & 0.100 & 7.000  & 9.855 & 9.774 & 0.382 & 1.058 \\
NGC\,0855    &  0.057 & 8.962 & -1.438 & 5.562  & 8.633 & 8.525 & 0.148 & 1.060 \\
NGC\,0925    &  0.049 & 9.406 & -0.194 & 6.906  & 9.591 & 9.628 & 0.204 & 0.758 \\
NGC\,1097    &  0.041 & 10.490 & 0.490 & 7.490  & 10.619 & 10.359 & 0.479 & 1.877 \\
NGC\,1266    &  0.103 & 10.157 & -0.343 & 6.657  & 10.385 & 9.571 & 1.660 & 4.969 \\
NGC\,1291    &  0.066 & 10.720 & -0.980 & 6.920  & 9.395 & 8.843 & 0.062 & 0.534 \\
NGC\,1316    &  0.098 & 11.330 & -1.270 & 6.630  & 9.890 & 8.878 & 0.041 & 0.542 \\
NGC\,1377    &  0.112 & 9.768 & 0.068 & 5.868  & 10.139 & 9.936 & 0.631 & 2.373 \\
NGC\,1404    &  0.081 & 10.796 & -1.204 & 6.796  & 9.610 & 8.538 & 0.099 & 0.484 \\
NGC\,1482    &  0.056 & 10.094 & 0.094 & 7.094  & 10.648 & 9.980 & 1.713 & 4.992 \\
NGC\,1512    &  0.041 & 10.015 & -0.485 & 7.115  & 9.531 & 9.420 & 0.179 & 0.921 \\
NGC\,2146    &  0.055 & 10.457 & 0.757 & 7.257  & 11.060 & 10.607 & 1.687 & 4.749 \\
NGC\,2798    &  0.059 & 10.109 & 0.209 & 6.809  & 10.548 & 10.151 & 1.319 & 3.953 \\
NGC\,2841    &  0.057 & 10.784 & -0.316 & 7.384  & 9.979 & 9.607 & 0.171 & 1.130 \\
NGC\,2915    &  0.069 & 7.741 & -1.559 & 4.741  & 7.615 & 8.258 & 0.033 & 0.162 \\
NGC\,2976    &  0.038 & 8.986 & -1.214 & 5.986  & 8.906 & 8.760 & 0.268 & 1.301 \\
NGC\,3049    &  0.053 & 9.252 & -0.448 & 6.552  & 9.618 & 9.471 & 0.418 & 1.543 \\
NGC\,3077    &  0.047 & 9.152 & -1.348 & 5.652  & 8.868 & 8.601 & 0.231 & 1.318 \\
NGC\,3184    &  0.056 & 10.022 & -0.078 & 7.322  & 9.893 & 9.891 & 0.127 & 1.034 \\
NGC\,3190    &  0.047 & 10.357 & -1.243 & 6.957  & 9.793 & 8.633 & 0.312 & 1.777 \\
NGC\,3198    &  0.044 & 9.884 & -0.116 & 7.184  & 9.878 & 9.802 & 0.246 & 1.060 \\
NGC\,3265    &  0.036 & 9.270 & -0.830 & 5.870  & 9.417 & 9.050 & 0.676 & 2.400 \\
NGC\,3351    &  0.038 & 10.225 & -0.375 & 6.825  & 9.844 & 9.548 & 0.301 & 1.575 \\
NGC\,3521    &  0.049 & 10.584 & 0.084 & 7.584  & 10.477 & 9.933 & 0.518 & 2.234 \\
NGC\,3621    &  0.053 & 9.618 & -0.382 & 6.918  & 9.829 & 9.693 & 0.359 & 1.400 \\
NGC\,3627    &  0.047 & 10.530 & -0.070 & 7.230  & 10.376 & 9.878 & 0.553 & 2.097 \\
NGC\,3773    &  0.045 & 8.572 & -1.028 & 5.572  & 8.768 & 8.873 & 0.116 & 0.752 \\
NGC\,3938    &  0.049 & 10.191 & 0.091 & 7.291  & 10.164 & 10.072 & 0.218 & 1.149 \\
NGC\,4236    &  0.070 & 8.882 & -0.618 & 6.182  & 8.628 & 9.087 & 0.054 & 0.246 \\
NGC\,4254    &  0.043 & 10.202 & 0.502 & 7.502  & 10.514 & 10.288 & 0.652 & 1.804 \\
NGC\,4321    &  0.059 & 10.396 & 0.396 & 7.496  & 10.449 & 10.288 & 0.282 & 1.592 \\
NGC\,4536    &  0.050 & 10.108 & 0.208 & 7.108  & 10.333 & 10.135 & 0.492 & 2.037 \\
NGC\,4559    &  0.045 & 9.373 & -0.627 & 6.673  & 9.403 & 9.429 & 0.224 & 0.771 \\
NGC\,4569    &  0.046 & 9.873 & -0.827 & 6.673  & 9.661 & 9.039 & 0.261 & 1.493 \\
NGC\,4579    &  0.053 & 10.815 & -0.485 & 7.115  & 9.980 & 9.393 & 0.210 & 1.289 \\
NGC\,4594    &  0.070 & 11.002 & -1.298 & 6.902  & 9.527 & 8.665 & 0.065 & 0.783 \\
NGC\,4625    &  0.050 & 8.523 & -1.277 & 5.823  & 8.654 & 8.670 & 0.186 & 0.737 \\
NGC\,4631    &  0.064 & 9.650 & 0.350 & 7.250  & 10.344 & 10.229 & 0.421 & 1.978 \\
NGC\,4725    &  0.056 & 10.637 & -0.463 & 7.337  & 9.813 & 9.494 & 0.171 & 0.983 \\
NGC\,4736    &  0.029 & 10.222 & -0.478 & 6.422  & 9.702 & 9.478 & 0.177 & 1.317 \\
NGC\,4826    &  0.050 & 10.229 & -0.971 & 6.329  & 9.563 & 8.944 & 0.244 & 1.445 \\
NGC\,5055    &  0.035 & 10.385 & -0.015 & 7.385  & 10.205 & 9.916 & 0.354 & 1.756 \\
NGC\,5398    &  0.088 & 8.054 & -1.246 & 5.754  & 8.604 & 8.735 & 0.059 & 0.725 \\
NGC\,5408    &  0.077 & 8.048 & -1.652 & 4.548  & 8.279 & 8.391 & 0.063 & 0.615 \\
NGC\,5457    &  0.051 & 10.238 & 0.338 & 7.538  & 10.221 & 10.298 & 0.164 & 0.762 \\
NGC\,5474    &  0.040 & 8.751 & -0.949 & 6.051  & 8.623 & 8.950 & 0.075 & 0.319 \\
NGC\,5713    &  0.061 & 9.946 & 0.246 & 7.046  & 10.477 & 10.143 & 0.765 & 2.728 \\
NGC\,5866    &  0.051 & 10.655 & -2.045 & 6.455  & 9.677 & 8.273 & 0.198 & 0.745 \\
NGC\,6946    &  0.067 & 10.151 & -0.349 & 7.351  & 10.493 & 10.293 & 0.560 & 1.451 \\
NGC\,7331    &  0.048 & 10.808 & 0.308 & 7.708  & 10.628 & 10.186 & 0.531 & 2.215 \\
NGC\,7793    &  0.058 & 8.970 & -0.530 & 6.470  & 9.213 & 9.291 & 0.178 & 0.742 \\

\hline 
\label{tab:app:magphys} 
\end{tabular} 
}
\end{table*}

\renewcommand\thefigure{\thesection.\arabic{figure}}    
\setcounter{figure}{0}   
\begin{figure*}
%\vspace{\baselineskip}
%\includegraphics[width=0.91\textwidth]{SED_KINGFISH_plot1.pdf}
%\includegraphics[width=0.91\textwidth]{SED_KINGFISH_plot1_new.pdf}
%\includegraphics[width=0.91\textwidth]{SED_KINGFISH_plot1_jan2018.pdf}
% \includegraphics[width=0.91\textwidth]{SED_KINGFISH_plot1_july2018.pdf}
% corrects mistake for uncertainty 70um NGC584
% \includegraphics[width=0.91\textwidth]{SED_KINGFISH_plot1_july2018_new.pdf}
\includegraphics[width=0.91\textwidth]{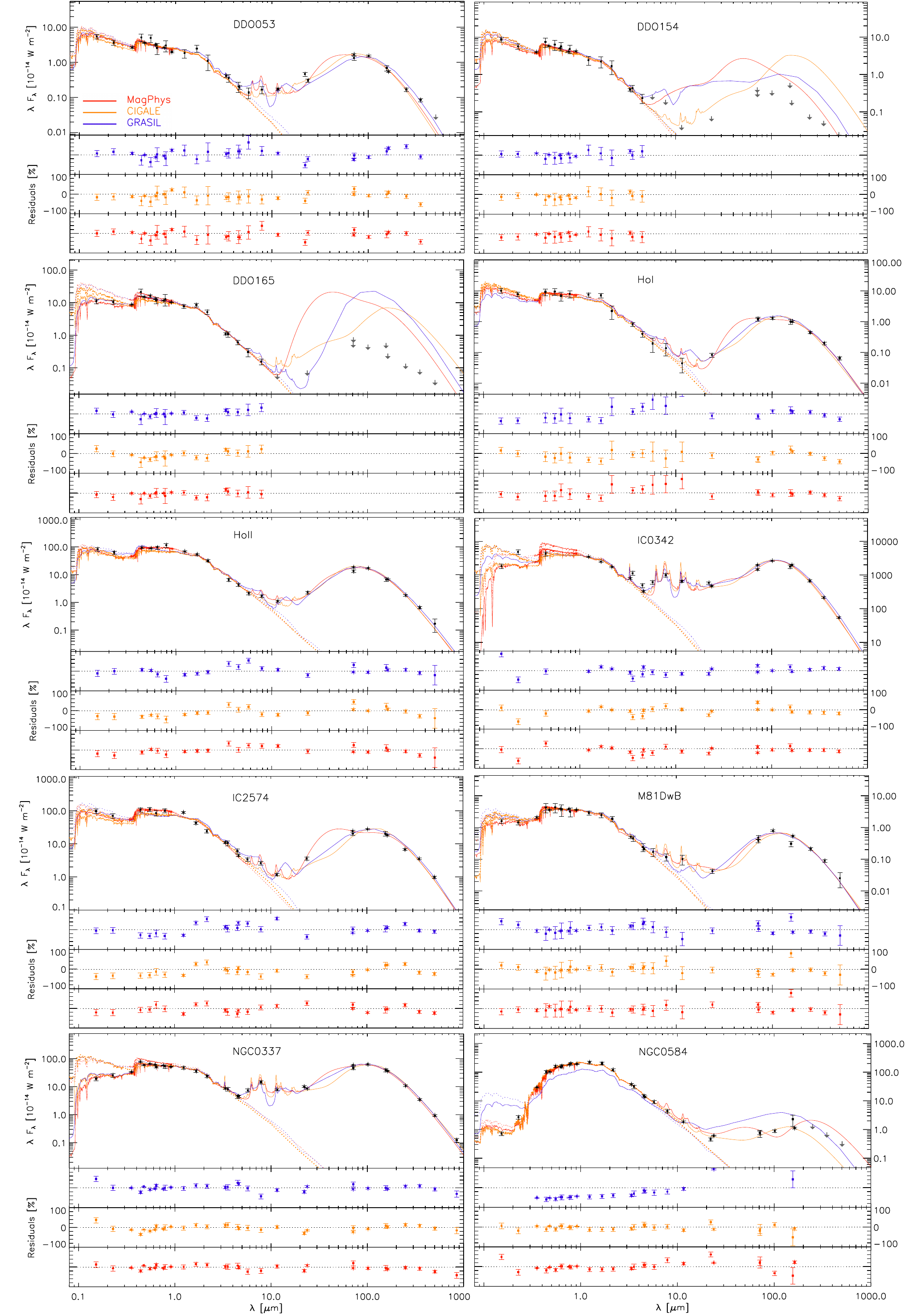}
\vspace{-0.6\baselineskip}
\caption{Panchromatic SEDs for the KINGFISH galaxies based on the photometry measurements from \citet{dale17}
overlaid with the best-fitting SED models inferred from the SED fitting tools 
%MagPhys 
\magp\ (red curve), %CIGALE 
% referee
\cig\ (dark-orange curve) and %GRASIL 
% \cig\ (green curve) and %GRASIL 
\gra\ (blue curve). The dashed curves represent the (unattenuated) intrinsic model emission for each SED fitting method (using the same color coding). 
The bottom part of each panel shows the residuals for each of these models compared to the observed fluxes in each waveband.
Gray arrows points show the upper limits when available. 
}
\label{fig:all_seds}
\end{figure*}

\setcounter{figure}{0}   
\begin{figure*}
\vspace{\baselineskip}
\includegraphics[width=0.91\textwidth]{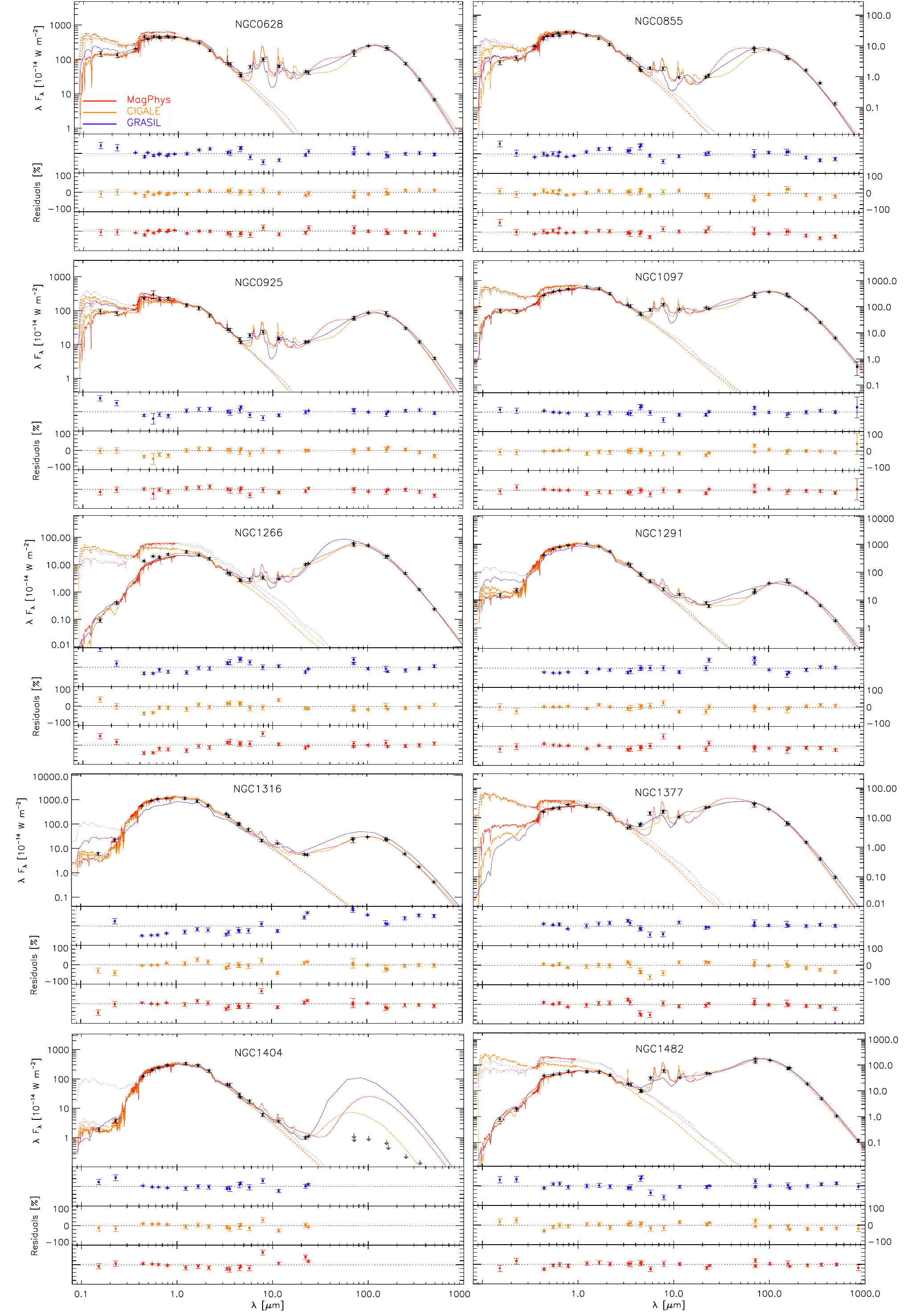}
\caption{Continued.}
\end{figure*}

\setcounter{figure}{0}   
\begin{figure*}
\vspace{\baselineskip}
\includegraphics[width=0.91\textwidth]{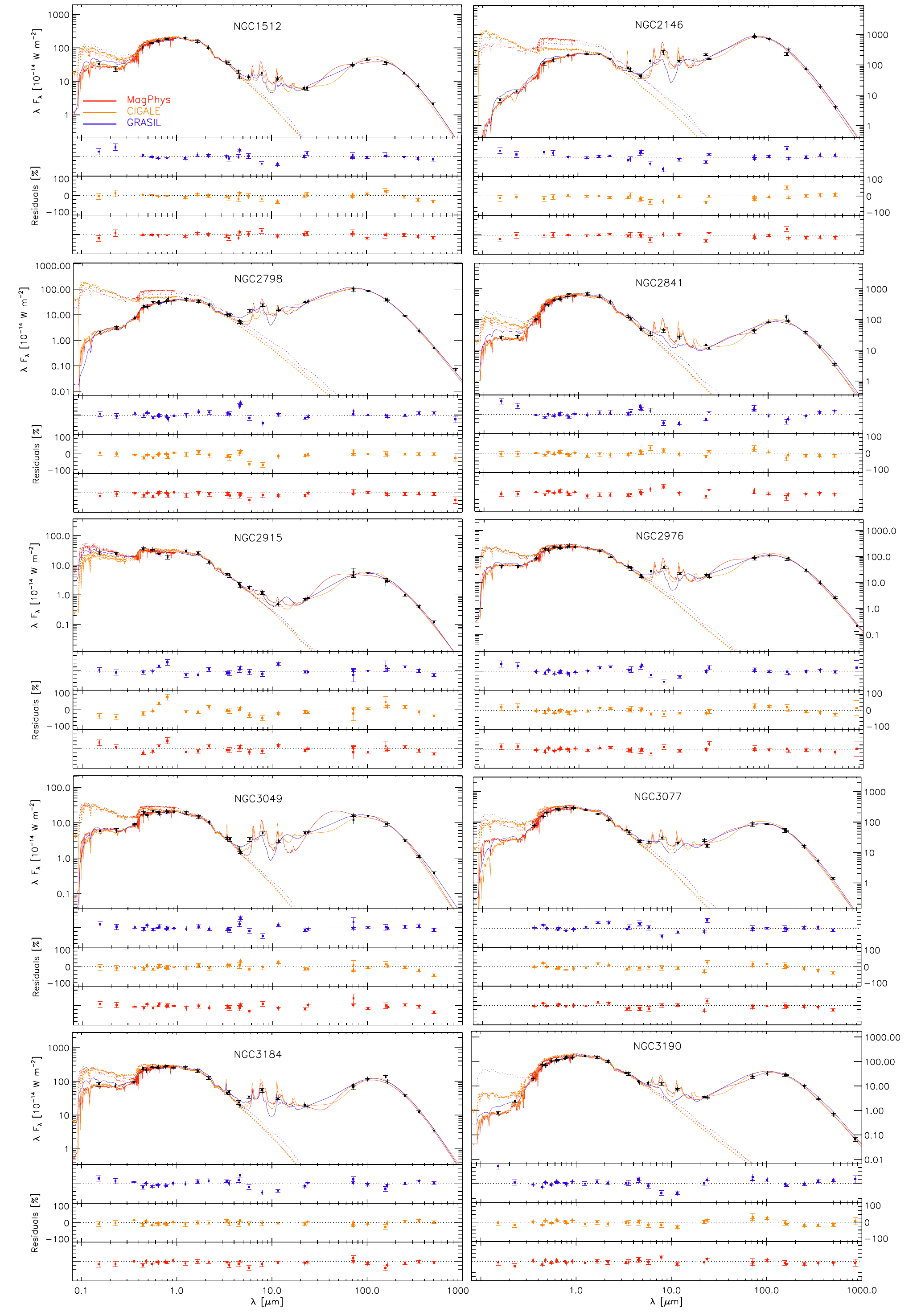}
\caption{Continued.}
\end{figure*}

\setcounter{figure}{0}   
\begin{figure*}
\vspace{\baselineskip}
\includegraphics[width=0.91\textwidth]{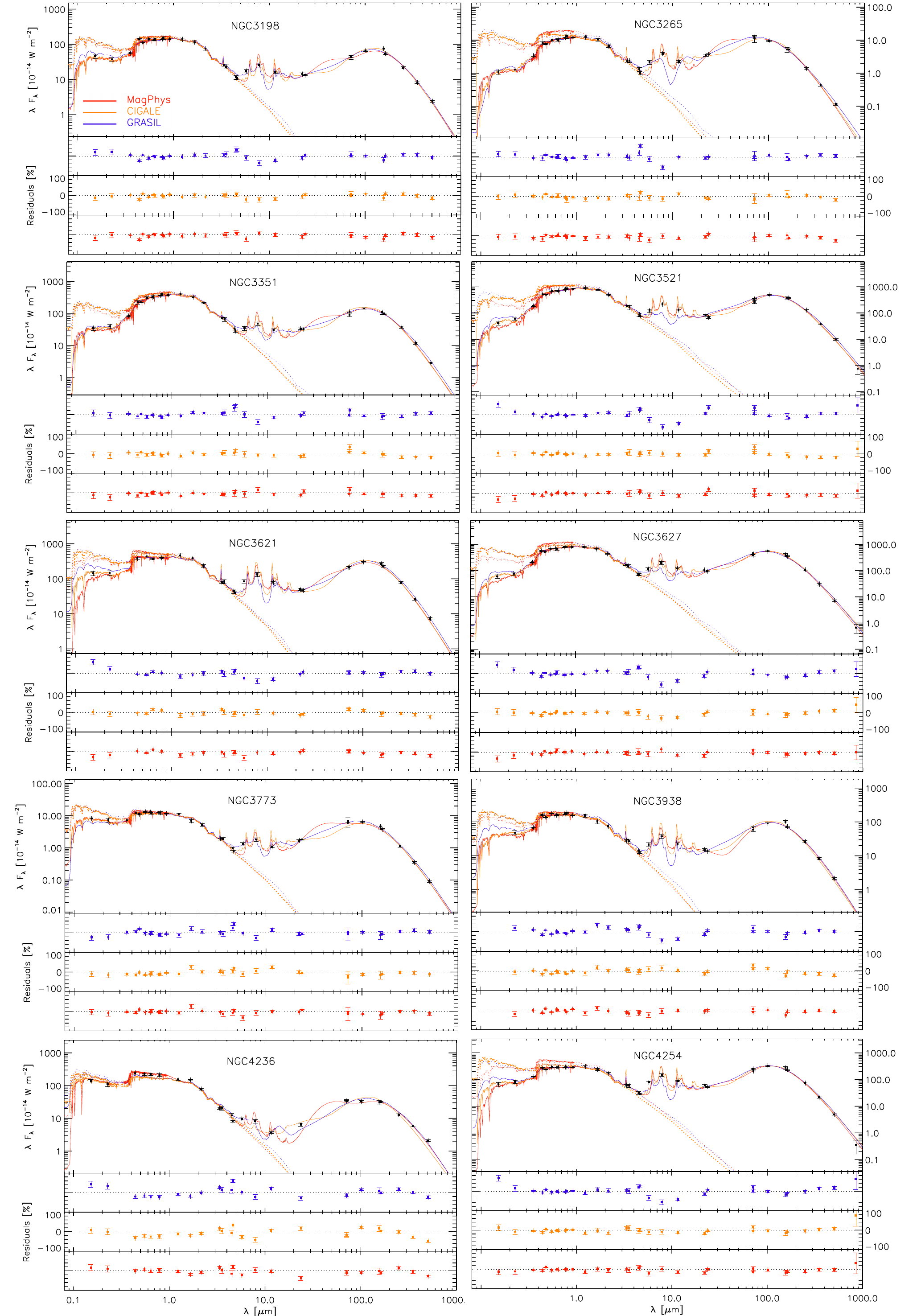}
\caption{Continued.}
\end{figure*}

\setcounter{figure}{0}   
\begin{figure*}
\vspace{\baselineskip}
\includegraphics[width=0.91\textwidth]{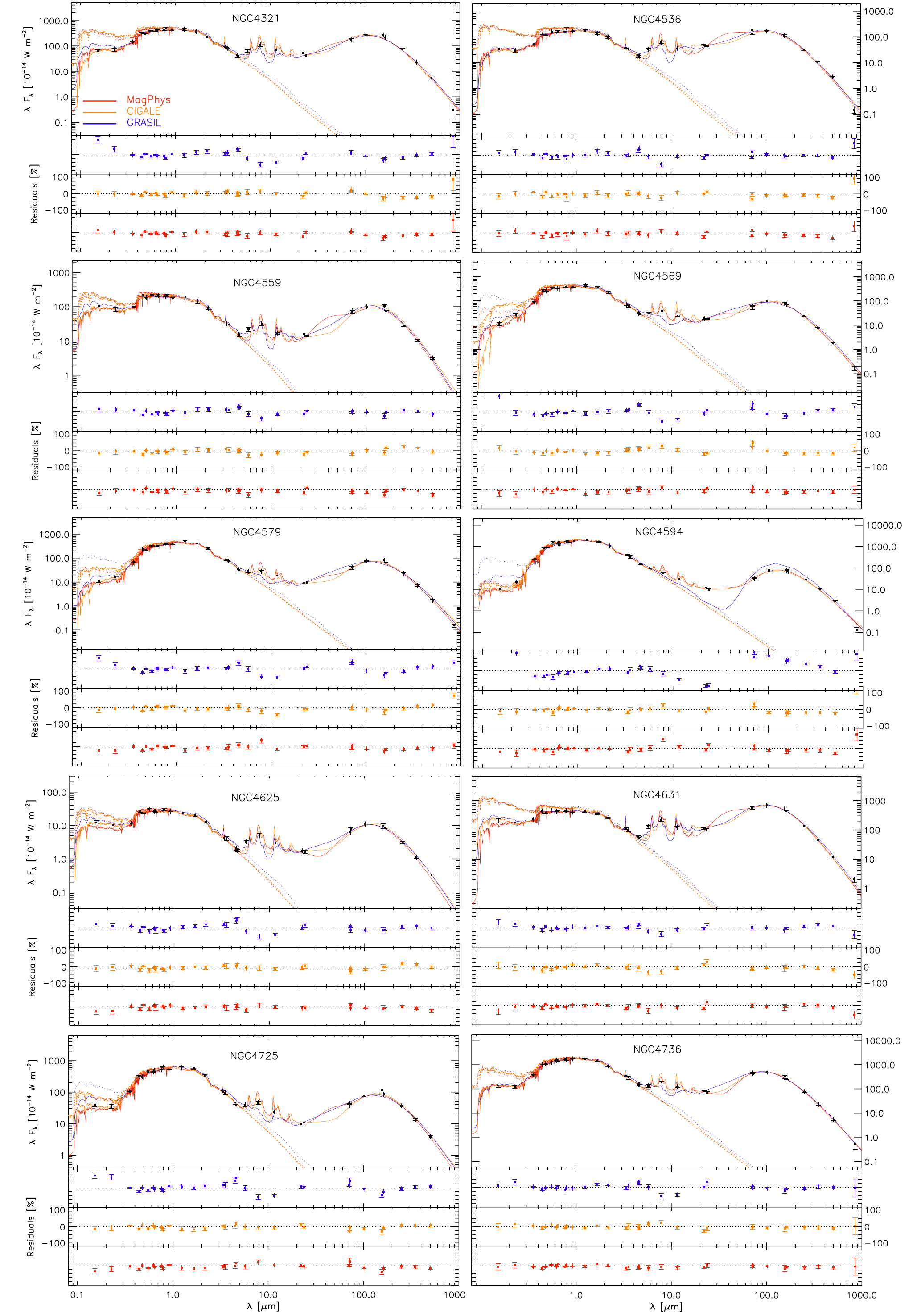}
\caption{Continued.}
\end{figure*}

\setcounter{figure}{0}   
\begin{figure*}
\vspace{\baselineskip}
\includegraphics[width=0.91\textwidth]{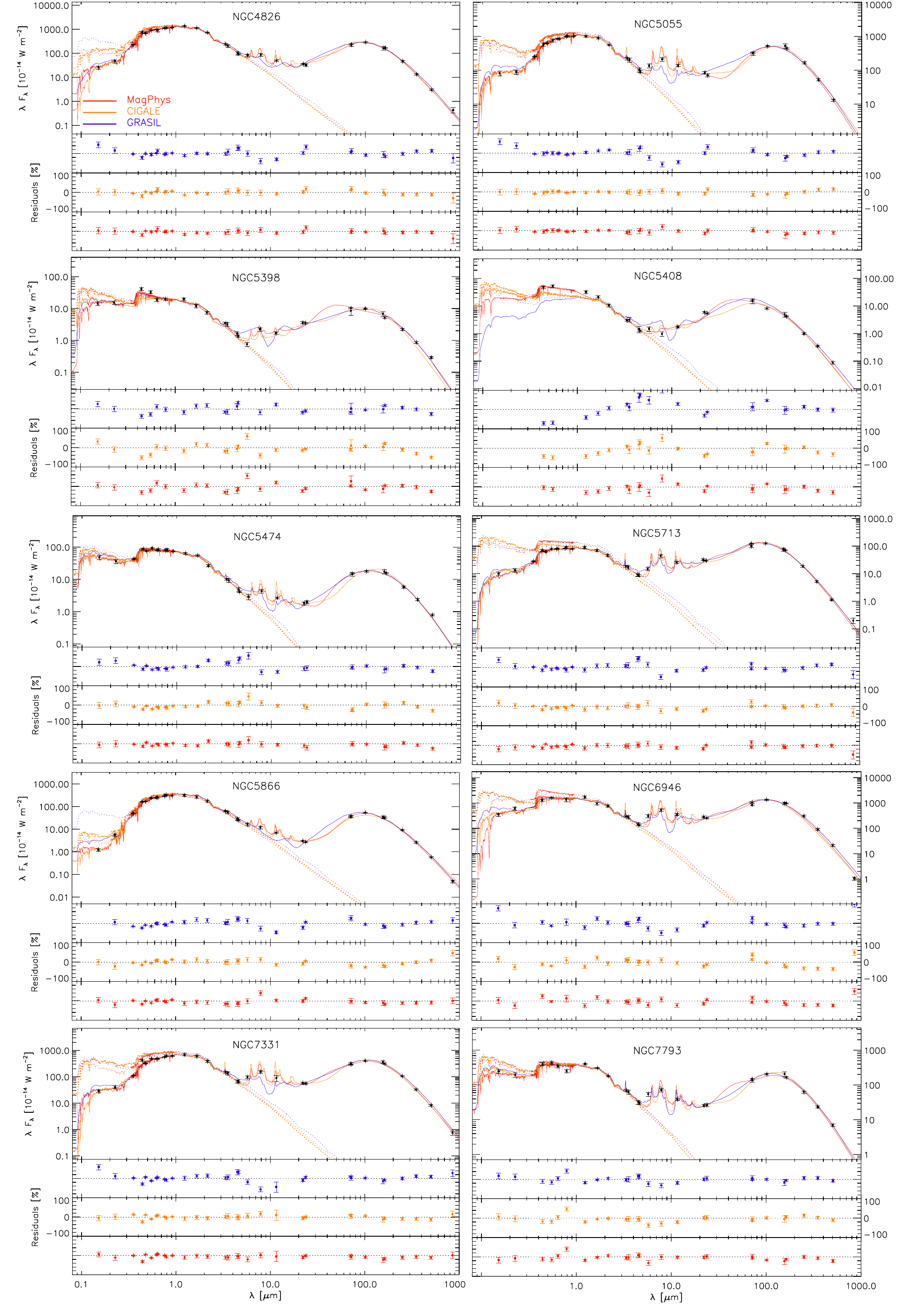}
\caption{Continued.}
\end{figure*}

%\setcounter{figure}{0}   
%\begin{figure*}
%\vspace{\baselineskip}
%%\includegraphics[height=0.27\textwidth]{SED_KINGFISH_plot7.pdf}
%%\includegraphics[height=0.27\textwidth]{SED_KINGFISH_plot7_new.pdf}
%%\includegraphics[height=0.27\textwidth]{SED_KINGFISH_plot7_jan2018.pdf}
%\caption{Continued.}
%\end{figure*}

\section{Description of reference quantities }
\label{app:ref}

Here we describe in detail our choices for the % photometric
inference of the six reference or recipe quantities
introduced in Sect. \ref{sec:comparison:reference}.
All photometry from \citet{dale17} has been corrected for foreground
Galactic extinction according to \av\ measurements by
\citet{schlafly11} and the extinction curve of \citet{draine03}.
A summary of the methods is given in Table \ref{tab:app:methods}, and
the values of the computed quantities are reported in Table \ref{tab:app:parameters}.

\subsection{Reference stellar mass}
\label{app:mstar}

The availability of data at near- to mid-infrared (NIR, MIR) wavelengths, both from
\spit/IRAC and WISE, has prompted the widespread use of 3.4 or 3.6\,\micron\
luminosities to measure stellar mass. At these wavelengths, the mass-to-light (M/L\,=\,$\Upsilon_*$) ratios of stellar populations are
relatively constant, independently of metallicity and age \citep{eskew12,meidt14,norris14,mcgaugh14}.  
We have relied on two formulations for estimating \mstar\ from 3.4-3.6 luminosities:
the first by \citet{wen13} is based on WISE W1 (3.4\,\micron) photometry and calibrated
to the stellar masses from the Sloan Digital Sky Survey (SDSS) value-added catalogs.
We have used IRAC 3.6\,\micron\ luminosities interchangeably with WISE W1 3.4\,\micron\ values.
For the 59 KINGFISH galaxies with both W1 and IRAC photometry the mean ratio
is 1.07\,$\pm$\,0.09. 
\citet{grossi15} find for 23 spiral galaxies a mean flux ratio $F_{3.4}/F_{3.6}\,=\,1.02\,\pm\,0.035$, and
% including \citet{brown14} data also for dwarf irregulars, 
using data from \citet{brown14a}, \citet{hunt15b} obtain
a mean flux ratio  $F_{3.4}/F_{3.6}\,=\,0.98\,\pm\,0.061$.
Thus, within the uncertainties of the photometry, WISE W1 3.4\,\micron\ and IRAC 3.6\,\micron\
photometry is virtually indistinguishable.
% Although this method adopts W1 3.4\,\micron\ data rather than IRAC 3.6\,\micron,
% the two bands are very similar.
% Using data from \citet{brown14}, \citet{grossi15} find for spiral galaxies
% a mean flux ratio $F_{3.4}/F_{3.6}\,=\,1.02\,\pm\,0.035$;
% including \citet{brown14} data also for dwarf irregulars, \citet{hunt15} obtain
% a mean flux ratio  $F_{3.4}/F_{3.6}\,=\,0.98\,\pm\,0.061$.
% Thus, as in \citet{hunt15}, we assume a W1/IRAC(3.6) ratio of unity, with an implied 
% uncertainty of 4-6\%.

For the estimates of the recipe stellar masses, we combine IRAC 3.6\,\micron\ luminosities
and the \citet{wen13} approach based on Hubble type, divided into early- and late-type galaxies;
we also apply their suggested correction for low metallicity (\logoh $\leq$ 8.2) amounting to a multiplicative factor of
0.8. The \citet{kroupa01} IMF used by \citet{wen13} was converted to \citet{chabrier03} according to the
formulation of \citet{speagle14}. 
The second method for calculating stellar mass assumes a constant $\Upsilon_*$ value %M/L ratio 
at 3.6\,\micron, as found by \citet{eskew12,meidt14,norris14,mcgaugh14}.
Here we adopt the \citet{mcgaugh14} $\Upsilon_*$ (in solar units at 3.6\,\micron) $\Upsilon^{[3.6]}\,=\,0.47$, 
assuming that \lsun(3.6\,\micron)\,=\,$1.4\times10^{32}$\,\ergs\ as given by \citet{cook14}.

However, before applying either method, we first estimate the non-stellar
continuum at these wavelengths and subtract it.
Such contamination can be very important in dwarf galaxies, especially in those with
high SFRs \citep[e.g.,][]{smith09}.
The contribution from the ionized gas continuum to the 3.4-3.6\,\micron\ flux was estimated
from the SFR (see Sect. \ref{app:sfr}) using the emission coefficients from \citet{osterbrock06}. 
We did not attempt to subtract emission from hot dust, since globally its contribution in disk galaxies
is typically small \citep[$\la$10\%,][]{meidt12}.
For the KINGFISH galaxies, our estimate of the fraction of nebular continuum in the 3.6\,\micron\
IRAC band ranges from 0 to 2\%, so is a very small correction.

\subsection{Reference star-formation rate}
\label{app:sfr}

To estimate SFRs, we used Eqn. (18) by \citet{murphy11} based on \lfuv\ and \ltir;
these quantities were available for 50 galaxies.
Otherwise, we preferred the SFR estimate from \lfuv\ \citep[Eqn. (3) in][]{murphy11} which was possible
only for DDO\,154 and DDO\,165 (these galaxies are missing also MIPS 24\,\micron\ and longer-wavelength detections,
and DDO\,154 has no detections at all beyond IRAC 4.5\,\micron).
As the last choice, we took SFR from \ltir\ \citep[Eqn. (4) in][]{murphy11}
which assumes that only the FUV radiation up to the Balmer decrement is reprocessed by dust;
SFR(TIR) is used for 9 galaxies (IC\,342, NGC\,1377, NGC\,2146, NGC\,3049, NGC\,3077, NGC\,393, NGC\,4254,
NGC\,4321, and NGC\,5408).
In all cases, the \citet{kroupa01} IMF adopted by \citet{murphy11} was converted to Chabrier
according to \citet{speagle14}.

In order to obviate possible problems with FUV$+$TIR derived SFRs, we also
calculated SFRs inferred from \ha\ and 24\micron\ luminosities 
using \ha\ fluxes corrected for Galactic extinction and \nii\ contamination from
\citet{kennicutt09} or \citet{moustakas10}.
%by \citet{kennicutt11} using the calibration of \citet{calzetti10}.
To convert these quantities to SFRs, 
we adopted the constants from \citet{calzetti10} \citep[which are within 1\% of those used by][]{murphy11},
after adjusting them to an electron temperature of $\sim$7000\,K 
\citep[to calibrate \ha][uses T\,=\,10\,000\,K]{murphy11} in order to minimize the offset with the
SFRs inferred from \lfuv$+$\ltir.
When \ha\ is unavailable in \citet{kennicutt09} or \citet{moustakas10}
(i.e., for NGC\,855, NGC\,1266, NGC\,1316, NGC\,1404), we have taken SFR estimates from \citet{kennicutt11}
or \citet{skibba11} %using the same calibration,
after correcting to the same distance scale  as \citet{kennicutt11}.
The assumed IMF \citep{kroupa01} was converted to \citet{chabrier03}, as before according to \citet{speagle14}.
These \ha$+$24\micron\ SFRs are available for 60 galaxies.

\subsection{Reference dust masses}
\label{app:mdust}

Although numerous studies have inferred dust masses, \mdust, of the KINGFISH galaxies
\citep[e.g.,][]{draine07,dacunha08,noll09,munoz09,dale12,remyruyer15}, they are all based on
models, either the ones scrutinized here or others \citep[e.g.,][]{draineli07,galliano11}.
To compare the values of \mdust\ found here through SED fitting, we
prefer to minimize discrepancies induced by differences in the assumptions made by models. 
Thus, we adopted the dust masses calculated according to \citet{bianchi13} who performed single-temperature
modified blackbody (MBB) fits to the KINGFISH galaxies, and assessed differences caused by
different dust opacities assumed by various groups.
New dust masses were calculated with the same methods as in \citet{bianchi13}, but using the updated
\hers\ fluxes \citep[see][]{dale17} %(see Dale et al. 2016) 
and the revised \hers\ filter transmission curves;
as in \citet{bianchi13}, the dust opacities are taken from the DL07 models.
These new \mdust\ values are, on average, 0.83 times those found by \citet{bianchi13},
with most of the change due to the updated flux values.
%Figure \ref{fig:mdust} compares the SED-derived \mdust\ values to 
%and Table \ref{tab:corr} gives the best-fit (robust) regressions. 
%Six KINGFISH galaxies are missing the requisite IR {\bf detections} to calculate \mdust:
%DDO\,53, DDO\,154, DDO\,165, M\,81\,DwB, NGC\,584, and NGC\,1404.
Three KINGFISH galaxies are missing the requisite IR detections to infer \mdust:
DDO\,154, DDO\,165, and NGC\,1404.

For completeness, we also include in the comparison the updated \mdust\ values taken from \citet[][their Table 10]{aniano18}.
These values are derived using the DL07 models presented by \citet{aniano12},
but have been renormalized taking into account the post-Planck results \citep{planck16}.

\subsection{Reference luminosities and attenuation}
\label{app:lums}

We have calculated \ltir\ as suggested by DL07 based on \spit\ photometry
and by \citet[][G13]{galametz13} by combining \spit\ and \hers.
DL07 gives an analytical expression for \ltir\ based on luminosities at IRAC 8\,\micron, and the MIPS bands at 24, 70, and 160\,\micron. 
The expression is calibrated on their models and describes the modeled \ltir\ to within $\sim\,$10\%.
From G13, we took the formulation (from their Table 3) for \ltir\ based on the linear combination with the lowest RMS error,
normalized to the mean values of global flux density.
To optimize the choice of indicator,
we also considered the one based on the largest number of detections for the KINGFISH galaxies (somewhat fewer galaxies were detected
with SPIRE).
With these constraints, the best G13 recipe, also calibrated on the DL07 models, is based on MIPS 24\,\micron, and two PACS bands, 70 and 100\,\micron\ 
(see Table \ref{tab:app:methods}).

The FUV luminosity, \lfuv\,=\,$\nu_{\rm FUV}\,\ell_{\rm FUV}$ ($\lambda$\,=\,0.15\,\micron), is calculated from the observed FUV fluxes corrected for extinction,
%according to the formulation of \afuv\ from \citet{hao11}
according to \citet{murphy11},
based on IRX, the logarithm (base 10) of the ratio of \ltir\ to observed FUV luminosity, 
\lfuv \footnote{Here we have used \ltir\ from the formulation of G13.}.
The constant 0.43 relating \ltir\ to \lfuv\ of \citet{murphy11} is close to the
value of 0.46 found by \citet{hao11}, and the two estimates give similar results.
%As noted by \citet{hao11}, 
IRX is a relatively robust indicator of dust attenuation because it is
based on energy balance arguments, and is almost independent of dust properties and dust geometry relative
to heating sources \citep[e.g.,][]{buat05,hao11}.
The FUV attenuation, \afuv\ ($\lambda$\,=\,0.15\,\micron) is 
taken accordingly from \citet{murphy11}, again using \ltir\ from the G13 formulation
%accordingly based on the 
%formulation of \citet{hao11} with IRX\footnote{Here again we use \ltir\ derived using the G13 formulation.} 
%%based on a combination of MIP24 and \hers\ PACS 70 and 100\,\micron\ luminosities 
(see Table \ref{tab:app:methods}). 

%%%%%%%%
\begin{table*} 
%\begin{center} 
\caption{Summary of methods for independently-derived quantities}
\begin{tabular}{ccc}
\hline 
\multicolumn{1}{c}{Parameter} &
\multicolumn{1}{c}{Method} &
\multicolumn{1}{c}{Reference} \\
\hline 
\\
Stellar mass, \mstar       &  3.6\,\micron\ luminosity$^{\mathrm a}$   & \citet{wen13} \\
Stellar mass, \mstar       &  3.6\,\micron\ luminosity, constant $\Upsilon^{[3.6]}$                 & \citet{mcgaugh14} \\
Star-formation rate, SFR   &  FUV$+$TIR$^{\mathrm b}$                  & \citet{murphy11} \\
Star-formation rate, SFR   &  \ha$+$24\,\micron                        & \citet{kennicutt09},\\
& & \citet{moustakas10}, \\
& & \citet{skibba11}, see text \\
%Dust mass, \mdust          &  Modified blackbody fits                  & \citet{bianchi13} \\
Dust mass, \mdust          &  Modified blackbody fits                  & As in \citet[][]{bianchi13}, see text \\
Dust mass, \mdust          &  DL07 model fits                          & \citet[][]{aniano18} \\
Total IR luminosity, \ltir &  \spit\ data$^{\mathrm c}$         & \citet{draineli07} \\
Total IR luminosity, \ltir &  \spit$+$\hers\ data$^{\mathrm d}$        & \citet{galametz13} \\
%FUV attenuation, \afuv     &  IRX, ratio of IR and observed FUV fluxes & \citet{hao11} \\
FUV attenuation, \afuv     &  IRX, logarithm of ratio of IR and observed FUV fluxes & \citet{murphy11} \\
\\
\hline 
\label{tab:app:methods} 
\end{tabular} 
%\end{center}
\vspace{-\baselineskip}
\begin{flushleft}
$^{\mathrm a}$ ~Assuming IRAC flux = WISE W1 flux ($\pm$5\%); \\
$^{\mathrm b}$ ~If not available, then SFR(FUV), or as last choice SFR(TIR); \\ 
$^{\mathrm c}$ ~Based on a linear combination IRAC 8\,\micron, MIPS 24\,\micron, MIPS 70\,\micron, and 
MIPS 160\,\micron\ luminosities;  \\ 
$^{\mathrm d}$ ~Based on a linear combination MIPS 24\,\micron, PACS 70\,\micron, and PACS 100\,\micron\
luminosities. \\ 
\end{flushleft}
\end{table*}

\begin{table*} 
\caption{Independently-derived quantities for KINGFISH sample$^{\mathrm a}$ }
\resizebox{0.97\textwidth}{!}{
\begin{tabular}{lrrrrrrccrcc}
\hline 
\multicolumn{1}{c}{Galaxy} &
\multicolumn{1}{c}{MW \av} &
\multicolumn{1}{c}{Hubble} &
\multicolumn{1}{c}{Distance} &
\multicolumn{1}{c}{12$+$} &
\multicolumn{1}{c}{Log(\mstar)$^{\mathrm b}$} &
\multicolumn{1}{c}{Log(SFR)} &
\multicolumn{1}{c}{Log(SFR)} &
\multicolumn{1}{c}{Log(\mdust)} &
\multicolumn{1}{c}{Log(\ltir)} &
\multicolumn{1}{c}{Log(\lfuv)} &
\multicolumn{1}{c}{\afuv} \\
& \multicolumn{1}{c}{(mag)} &
\multicolumn{1}{c}{type} &
\multicolumn{1}{c}{(Mpc)} & 
\multicolumn{1}{c}{log(O/H)} &
\multicolumn{1}{c}{(\msun)} &
\multicolumn{1}{c}{(FUV$+$TIR)} &
\multicolumn{1}{c}{(\ha$+$24\micron)} &
\multicolumn{1}{c}{(\msun)} &
\multicolumn{1}{c}{(\lsun)} &
\multicolumn{1}{c}{(\lsun)} &
\multicolumn{1}{c}{(mag)} \\
& & & & 
\multicolumn{1}{c}{(PP04N2)} &
& 
\multicolumn{1}{c}{(\msunyr)} &
\multicolumn{1}{c}{(\msunyr)} \\
\hline 
DDO\,053     &  0.10 &  10 &  3.61 &  8.00 & 6.919 & -2.372 & -2.285  & 3.954 & 7.088 & 7.347 & 0.231 \\
DDO\,154     &  0.03 &  10 &  4.30 &  8.02 & 7.156 & -2.097 & -2.562  & \multicolumn{1}{c}{$-$} & \multicolumn{1}{c}{$-$} & 7.701 & \multicolumn{1}{c}{$-$} \\
DDO\,165     &  0.07 &  10 &  4.57 &  8.04 & 7.625 & -1.934 & -2.658  & \multicolumn{1}{c}{$-$} & \multicolumn{1}{c}{$-$} & 7.863 & \multicolumn{1}{c}{$-$} \\
Ho\,I        &  0.14 &  10 &  3.90 &  8.04 & 7.356 & -2.068 & -2.282  & 4.901 & 7.004 & 7.686 & 0.093 \\
Ho\,II       &  0.09 &  10 &  3.05 &  8.13 & 8.070 & -1.363 & -1.342  & 4.762 & 7.951 & 8.383 & 0.160 \\
IC\,0342     &  1.53 &  6 &  3.28 &  8.80 & 10.550 & 0.320 & 0.272  & 7.644 & 10.216 & 9.798 & 0.819 \\
IC\,2574     &  0.10 &  9 &  3.79 &  8.19 & 8.577 & -1.071 & -1.068  & 5.889 & 8.341 & 8.645 & 0.210 \\
M81\,Dw\,B   &  0.22 &  10 &  3.60 &  8.19 & 7.028 & -2.888 & -2.729  & 4.331 & 6.693 & 6.810 & 0.308 \\
NGC\,0337    &  0.31 &  7 &  19.30 &  8.47 & 9.927 & 0.090 & 0.256  & 7.284 & 10.115 & 9.351 & 1.359 \\
NGC\,0584    &  0.12 &  -4 &  20.80 &  8.69 & 10.782 & -1.053 & \multicolumn{1}{c}{$-$}  & 7.216 & 8.554 & 7.989 & 1.029 \\
NGC\,0628    &  0.19 &  5 &  7.20 &  8.80 & 10.015 & -0.028 & -0.027  & 7.443 & 9.862 & 9.380 & 0.906 \\
NGC\,0855    &  0.19 &  -5 &  9.73 &  8.43 & 9.121 & -1.368 & -1.182  & 5.913 & 8.614 & 7.935 & 1.212 \\
NGC\,0925    &  0.21 &  7 &  9.12 &  8.59 & 9.776 & -0.143 & -0.131  & 7.335 & 9.590 & 9.398 & 0.556 \\
NGC\,1097    &  0.07 &  3 &  14.20 &  8.75 & 10.796 & 0.601 & 0.625  & 7.935 & 10.662 & 9.647 & 1.841 \\
NGC\,1266    &  0.27 &  -2 &  30.60 &  8.52 & 10.263 & 0.253 & 0.296  & 7.039 & 10.478 & 7.429 & 6.708 \\
NGC\,1291    &  0.04 &  1 &  10.40 &  8.78 & 10.940 & -0.495 & -0.799  & 7.270 & 9.372 & 8.706 & 1.190 \\
NGC\,1316    &  0.06 &  -2 &  21.00 &  9.31 & 11.657 & -0.082 & -0.296  & 7.113 & 9.882 & 8.925 & 1.724 \\
NGC\,1377    &  0.08 &  -1 &  24.60 &  8.52 & 10.083 & 0.299 & 0.394  & 6.306 & 10.191 & \multicolumn{1}{c}{$-$} & \multicolumn{1}{c}{$-$} \\
NGC\,1404    &  0.03 &  -4 &  20.20 &  8.78 & 11.066 & -1.413 & -0.528  & \multicolumn{1}{c}{$-$} & \multicolumn{1}{c}{$-$} & 8.384 & \multicolumn{1}{c}{$-$} \\
NGC\,1482    &  0.11 &  -2 &  22.60 &  8.74 & 10.592 & 0.548 & 0.667  & 7.463 & 10.733 & 8.097 & 5.679 \\
NGC\,1512    &  0.03 &  2 &  11.60 &  8.72 & 10.150 & -0.314 & -0.464  & 7.346 & 9.517 & 9.151 & 0.752 \\
NGC\,2146    &  0.26 &  2 &  17.20 &  8.68 & 10.793 & 1.001 & 0.900  & 7.771 & 11.120 & 8.815 & 4.859 \\
NGC\,2798    &  0.06 &  1 &  25.80 &  8.72 & 10.444 & 0.450 & 0.657  & 7.208 & 10.629 & 8.657 & 4.040 \\
NGC\,2841    &  0.04 &  3 &  14.10 &  9.31 & 10.802 & 0.072 & -0.181  & 7.836 & 9.957 & 9.199 & 1.349 \\
NGC\,2915    &  0.75 &  10 &  3.78 &  8.17 & 8.200 & -1.651 & -1.629  & 4.912 & 7.634 & 8.077 & 0.157 \\
NGC\,2976    &  0.20 &  5 &  3.55 &  8.61 & 9.054 & -1.068 & -0.999  & 6.362 & 8.897 & 8.197 & 1.248 \\
NGC\,3049    &  0.10 &  2 &  19.20 &  8.72 & 9.530 & -0.437 & -0.191  & 6.817 & 9.612 & 8.828 & 1.395 \\
NGC\,3077    &  0.18 &  10 &  3.83 &  8.64 & 9.245 & -0.925 & -1.027  & 6.064 & 8.884 & \multicolumn{1}{c}{$-$} & \multicolumn{1}{c}{$-$} \\
NGC\,3184    &  0.05 &  6 &  11.70 &  8.81 & 10.263 & 0.137 & 0.024  & 7.575 & 9.947 & 9.548 & 0.794 \\
NGC\,3190    &  0.07 &  1 &  19.30 &  8.75 & 10.720 & -0.251 & -0.602  & 7.358 & 9.807 & 7.960 & 3.737 \\
NGC\,3198    &  0.03 &  5 &  14.10 &  8.76 & 10.111 & 0.055 & 0.007  & 7.546 & 9.892 & 9.460 & 0.837 \\
NGC\,3265    &  0.07 &  -5 &  19.60 &  8.69 & 9.566 & -0.651 & -0.427  & 6.355 & 9.442 & 8.121 & 2.501 \\
NGC\,3351    &  0.08 &  3 &  9.33 &  8.77 & 10.208 & -0.166 & -0.112  & 7.269 & 9.883 & 8.979 & 1.620 \\
NGC\,3521    &  0.16 &  4 &  11.20 &  8.81 & 10.791 & 0.441 & 0.425  & 7.975 & 10.524 & 9.216 & 2.471 \\
NGC\,3621    &  0.22 &  7 &  6.55 &  8.75 & 10.002 & -0.038 & 0.096  & 7.330 & 9.858 & 9.287 & 1.038 \\
NGC\,3627    &  0.09 &  3 &  9.38 &  8.62 & 10.587 & 0.369 & 0.389  & 7.633 & 10.450 & 9.223 & 2.291 \\
NGC\,3773    &  0.07 &  -2 &  12.40 &  8.58 & 9.038 & -0.949 & -0.774  & 5.919 & 8.811 & 8.595 & 0.581 \\
NGC\,3938    &  0.06 &  5 &  17.90 &  8.68 & 10.396 & 0.472 & 0.288  & 7.735 & 10.213 & \multicolumn{1}{c}{$-$} & \multicolumn{1}{c}{$-$} \\
NGC\,4236    &  0.04 &  8 &  4.45 &  8.37 & 9.027 & -0.773 & -0.886  & 6.446 & 8.602 & 8.926 & 0.201 \\
NGC\,4254    &  0.11 &  5 &  14.40 &  8.79 & 10.551 & 0.574 & 0.568  & 7.881 & 10.586 & 9.633 & 1.716 \\
NGC\,4321    &  0.07 &  4 &  14.30 &  8.76 & 10.684 & 0.509 & 0.428  & 7.955 & 10.488 & 9.596 & 1.597 \\
NGC\,4536    &  0.05 &  4 &  14.50 &  8.63 & 10.301 & 0.291 & 0.337  & 7.537 & 10.362 & 9.338 & 1.860 \\
NGC\,4559    &  0.05 &  6 &  6.98 &  8.58 & 9.595 & -0.312 & -0.307  & 7.007 & 9.430 & 9.205 & 0.590 \\
NGC\,4569    &  0.13 &  2 &  9.86 &  8.80 & 10.252 & -0.347 & -0.346  & 7.149 & 9.735 & 8.544 & 2.213 \\
NGC\,4579    &  0.11 &  3 &  16.40 &  8.79 & 10.766 & 0.023 & -0.008  & 7.607 & 10.032 & 8.977 & 1.924 \\
NGC\,4594    &  0.14 &  1 &  9.08 &  8.79 & 11.104 & -0.458 & -0.747  & 7.291 & 9.521 & 8.431 & 1.997 \\
NGC\,4625    &  0.05 &  9 &  9.30 &  8.67 & 8.963 & -1.009 & -1.136  & 6.283 & 8.720 & 8.521 & 0.563 \\
NGC\,4631    &  0.05 &  7 &  7.62 &  8.38 & 10.219 & 0.380 & 0.346  & 7.624 & 10.355 & 9.589 & 1.363 \\
NGC\,4725    &  0.03 &  2 &  11.90 &  8.71 & 10.599 & -0.027 & -0.566  & 7.724 & 9.771 & 9.282 & 0.916 \\
NGC\,4736    &  0.05 &  2 &  4.66 &  8.68 & 10.256 & -0.212 & -0.311  & 6.827 & 9.777 & 8.976 & 1.426 \\
NGC\,4826    &  0.11 &  2 &  5.27 &  8.78 & 10.205 & -0.472 & -0.573  & 6.724 & 9.634 & 8.354 & 2.409 \\
NGC\,5055    &  0.05 &  4 &  7.94 &  9.31 & 10.566 & 0.250 & 0.146  & 7.843 & 10.248 & 9.209 & 1.890 \\
NGC\,5398    &  0.18 &  8 &  7.66 &  8.33 & 8.675 & -1.160 & -0.949  & 5.903 & 8.619 & 8.421 & 0.562 \\
NGC\,5408    &  0.19 &  10 &  4.80 &  8.19 & 8.256 & -1.562 & -0.903  & 4.980 & 8.265 & \multicolumn{1}{c}{$-$} & \multicolumn{1}{c}{$-$} \\
NGC\,5457    &  0.02 &  6 &  6.70 &  8.73 & 10.489 & 0.513 & 0.419  & 7.877 & 10.301 & 10.008 & 0.665 \\
NGC\,5474    &  0.03 &  6 &  6.80 &  8.46 & 9.043 & -0.820 & -1.034  & 6.383 & 8.647 & 8.851 & 0.259 \\
NGC\,5713    &  0.11 &  4 &  21.40 &  8.70 & 10.348 & 0.431 & 0.485  & 7.501 & 10.558 & 9.153 & 2.691 \\
NGC\,5866    &  0.04 &  -2 &  15.30 &  8.73 & 10.772 & -0.354 & -0.918  & 6.943 & 9.784 & 7.967 & 3.664 \\
NGC\,6946    &  0.94 &  6 &  6.80 &  8.75 & 10.572 & 0.535 & 0.627  & 7.863 & 10.566 & 9.705 & 1.538 \\
NGC\,7331    &  0.25 &  3 &  14.50 &  8.80 & 10.931 & 0.635 & 0.508  & 8.125 & 10.665 & 9.281 & 2.643 \\
NGC\,7793    &  0.05 &  7 &  3.91 &  8.64 & 9.410 & -0.472 & -0.469  & 6.876 & 9.232 & 9.079 & 0.518 \\

\hline 
\label{tab:app:parameters} 
\end{tabular} 
}
\begin{flushleft}
$^{\mathrm a}$~$-$ corresponds to unavailable data, so missing quantity.\\
$^{\mathrm b}$~Values according to the \citet{wen13} formulation.\\
\end{flushleft}
\end{table*}

\null
\label{lastpage}

\end{document}